\documentclass[a4paper,11pt,twoside,openright]{book} 
\pdfoutput=1
\usepackage[textwidth=390pt,voffset=30pt]{geometry}
\usepackage[utf8]{inputenc}						% characters and fonts encoding
\usepackage[T1]{fontenc} 
\usepackage[italian,english]{babel}				% language
\usepackage{indentfirst} 						% indent the first paragraph
\usepackage{emptypage}							% real empty page
\usepackage{amsmath,amssymb,amsfonts,amsthm,bm,cancel,latexsym}	% math package
\usepackage[pdftex,colorlinks,linkcolor=blue,citecolor=blue,urlcolor=blue,hyperfootnotes=false]{hyperref} % PDF version
\usepackage[square,comma,sort&compress,numbers]{natbib}	% package for multiple reference citations
\usepackage{quoting}							% quoting package

\usepackage{graphicx} 							% figures package with subfigure command
\usepackage{subfigure}							
\newcommand{\goodgap}{%							% optimal gap for aligned figures in subfigure mode
	\hspace{\subfigtopskip}%
	\hspace{\subfigbottomskip}}
\graphicspath{{Images/}}						% images path

\usepackage{booktabs,siunitx}					% packages for tables	
\usepackage{multirow}							% package for tables multirow command: use \multirow{number of rows}{*}{text}

\usepackage{titlesec}							% title style package
\titleformat{\chapter}[display]{\huge\bfseries}{\huge Chapter \huge\thechapter}{0.1em}{\titlerule\vspace{1em}}[\vspace{0.1em}]	% redefinition of chapter title 

\usepackage{fancyhdr}							% page style package
\DeclareMathOperator{\e}{e}						% define exponential operator
\DeclareMathOperator{\ord}{o}					% define order operator
\DeclareMathOperator{\re}{Re}					% define real part operator
\DeclareMathOperator{\sgn}{sgn}					% define sign operator

\begin{document}

	\frontmatter
	% Frontispiece

\thispagestyle{empty}
\setlength{\parskip}{0.6ex}

\begin{titlepage}
	\begin{center}
		{\bf{\large UNIVERSIT\`A DEGLI STUDI DI TRENTO}}\\
		\vspace{0.3 cm}
		Facolt\`a di Scienze Matematiche, Fisiche e Naturali\\
		\vspace{0.3 cm}
		Dipartimento di Fisica
	\end{center}
	\vspace{0.1 cm}
	\begin{figure} [h]
		\begin{center}
			\includegraphics[width=0.2\textwidth]{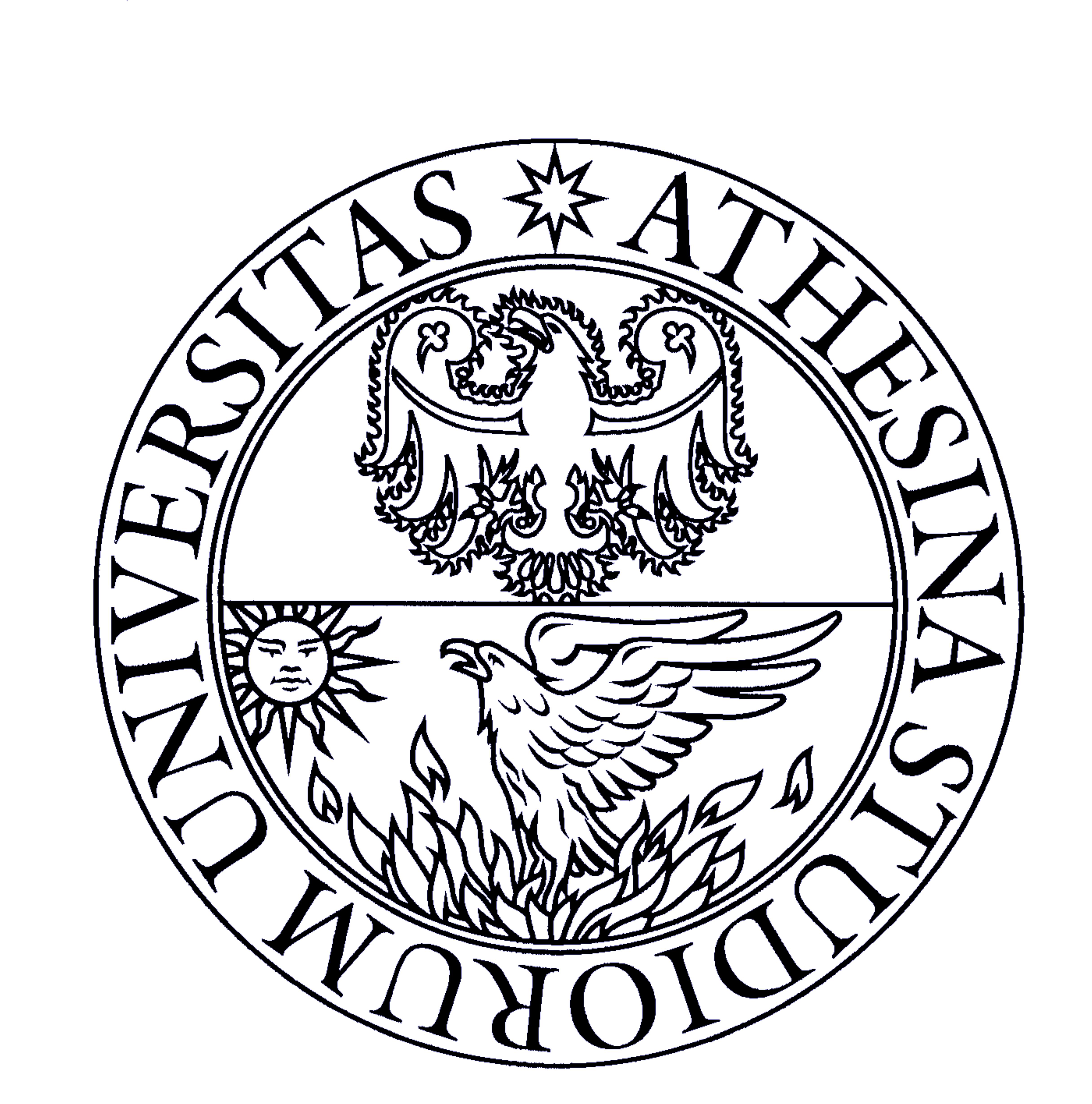}
		\end{center}
	\end{figure}
	\ \hrulefill \
	\
	\
	\vspace{0.2 cm}
	\begin{center}
		Tesi di Dottorato di Ricerca in Fisica\\
		Ph.D. Thesis in Physics
	\end{center}
	\
	\vspace{1 cm}
	\begin{center}
		{\bf{\Large\textsc{From Hypernuclei to Hypermatter: \\ a Quantum Monte Carlo Study of Strangeness \\ in Nuclear Structure and Nuclear Astrophysics}\par}}
	\end{center}
	\
	\vspace{2 cm}
	\flushleft
	Supervisor:\hspace{\stretch{2}}
	Candidate:\\
	Prof. Francesco Pederiva \hspace{\stretch{1}}
	Diego Lonardoni\\
	\
	\vspace{2 cm}
	\begin{center}
		\textsc{Dottorato di Ricerca in Fisica, XXVI ciclo}\\
		Trento, November 8th, 2013
	\end{center}
\end{titlepage}

\cleardoublepage

	% Quotation

\thispagestyle{empty}

\vspace*{2cm}

\begin{quotation}
	\emph{Tutta la materia di cui siamo fatti noi l’hanno costruita le stelle, tutti gli elementi dall’idrogeno all’uranio sono stati fatti nelle reazioni nucleari che avvengono nelle supernove, cioè queste stelle molto più grosse del Sole che alla fine della loro vita esplodono e sparpagliano nello spazio il risultato di tutte le reazioni nucleari avvenute al loro interno. Per cui noi siamo veramente figli delle stelle.}
	\flushright{Margherita Hack}
	\flushright{Intervista su \href{http://www.cortocircuito.re.it/intervista-a-margherita-hack/}{\emph{Cortocircuito}}}
\end{quotation}

\vspace*{2cm}

\begin{quotation}
	\emph{Il computer non è una macchina intelligente che aiuta le persone stupide, anzi, è una macchina stupida che funziona solo nelle mani delle persone intelligenti.}
	\flushright{Umberto Eco}
	\flushright{dalla prefazione a Claudio Pozzoli, \emph{Come scrivere una tesi di laurea con il personal computer}, Rizzoli}
\end{quotation}

	% Table of contents

\hypersetup{linkcolor=black}

\cleardoublepage
\fancyhead[LO]{\emph{Table of Contents}}				% redefinition of left and right header fields for this chapter
\fancyhead[RE]{\emph{Table of Contents}}
\pdfbookmark[0]{Table of Contents}{Table of Contents}
\tableofcontents

% List of figures

\cleardoublepage
\fancyhead[LO]{\emph{List of Figures}}					% redefinition of left and right header fields for this chapter
\fancyhead[RE]{\emph{List of Figures}}
\pdfbookmark[0]{List of Figures}{List of Figures}
\listoffigures

% List of tables

\cleardoublepage
\fancyhead[LO]{\emph{List of Tables}}					% redefinition of left and right header fields for this chapter
\fancyhead[RE]{\emph{List of Tables}}
\pdfbookmark[0]{List of Tables}{List of Tables}
\listoftables

\newpage
\phantom{Empty page}

	% Introduction

\chapter{Introduction}
\label{chap:introduction}

\fancyhead[LO]{\emph{Introduction}}					% redefinition of left and right header fields for this chapter
\fancyhead[RE]{\emph{Introduction}} 
\hypersetup{linkcolor=blue}
\renewcommand{\thefigure}{\emph{i}.\arabic{figure}}

\begin{figure}[htb]
	\begin{center}
		\includegraphics[width=0.65\linewidth]{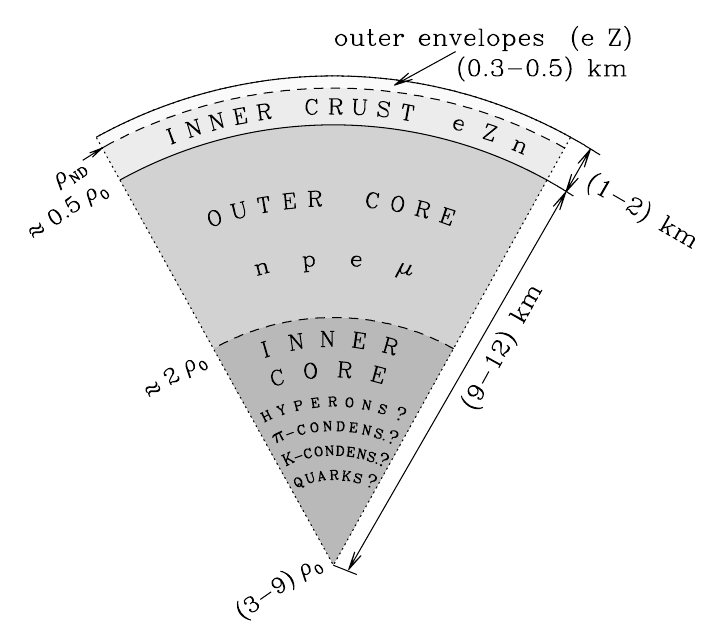}
		\caption[Neutron star structure]{Schematic structure of a neutron star. Stellar parameters strongly depend on the equation of state of the core. 
			Figure taken from Ref.~\cite{Haensel:2006}}
		\label{fig:NS_structure}
	\end{center}
\end{figure}

Neutron stars (NS) are among the densest objects in the Universe, with central densities several times larger than the nuclear saturation density $\rho_0=0.16~\text{fm}^{-3}$. As soon as the density significantly exceeds this value, the structure and composition of the NS core become uncertain.
Moving from the surface towards the interior of the star, the stellar matter undergoes a number of transitions, Fig.~\ref{fig:NS_structure}. From electrons and neutron rich ions in the outer envelopes, the composition is supposed to change to the $npe\mu$ matter in the outer core, a degenerate gas of neutrons, protons, electrons and muons. At densities larger than $\sim2\rho_0$ the $npe\mu$ assumption can be invalid due to the appearance of new hadronic degrees of freedom or exotic phases.

In the pioneering work of 1960~\cite{Ambartsumyan:1960}, Ambartsumyan and Saakyan reported the first theoretical evidence of hyperons in the core of a NS.
Contrary to terrestrial conditions, where hyperons are unstable and decay into nucleons through the weak interaction, the equilibrium conditions in a NS can make the inverse process happen. At densities of the order $2\div3\rho_0$, the nucleon chemical potential is large enough to make the conversion of nucleons into hyperons energetically favorable. This conversion reduces the Fermi pressure exerted by the baryons, and makes the equation of state (EoS) softer. As a consequence, the maximum mass of the star is typically reduced.~\nocite{Shapiro:1983}

Nowadays many different approaches of hyperonic matter are available, but there is no general agreement among the predicted results for the EoS and the maximum mass of a NS including hyperons. Some classes of methods extended to the hyperonic sector predict that the appearance of hyperons at around $2\div3\rho_0$ leads to a strong softening of EoS and consequently to a large reduction of the maximum mass. Other approaches, instead, indicate much weaker effects as a consequence of the presence of strange baryons in the core of the star.

The situation has recently become even more controversial as a result of the latest astrophysical observations. Until 2010, the value of $1.4\div1.5M_\odot$ for the maximum mass of a NS, inferred from precise neutron star mass determinations~\cite{Thorsett:1999}, was considered the canonical limit. First neutron star matter calculations with the inclusion of hyperons seemed to better agree with this value compared to the case of pure nucleonic EoS, that predicts relatively large maximum masses ($1.8\div2.4M_\odot$)~\cite{Akmal:1998}. The recent measurements of unusually high masses of the millisecond pulsars PSR J1614-2230 ($1.97(4)M_\odot$)~\cite{Demorest:2010} and PSR J1903+0327 ($2.01(4)M_\odot$)~\cite{Antoniadis:2013}, rule out almost all these results, making uncertain the appearance of strange baryons in high-density matter. However, in the last three years new models compatible with the recent observations have been proposed, but many inconsistency still remain. The solution of this problem, known as \emph{hyperon puzzle}, is far to be understood. 

The difficulty of correctly describe the effect of strange baryons in the nuclear medium, is that one needs a precise solution of a many-body problem for a very dense system with strong and complicated interactions which are often poorly known. 

The determination of a realistic interaction among hyperons and nucleons capable to reconcile the terrestrial measurements on hypernuclei and the NS observations is still an unsolved question. The amount of data available for nucleon-nucleon scattering and binding energies is enough to build satisfactory models of nuclear forces, either purely phenomenological or built on the basis of an effective field theory. Same approaches have been used to derive potentials for the hyperon-nucleon and hyperon-hyperon interaction, but the accuracy of these models is far from that of the non strange counterparts. The main reason of this is the lack of experimental information due the impossibility to collect hyperon-neutron and hyperon-hyperon scattering data. This implies that interaction models must be fitted mostly on binding energies (and possibly excitations) of hypernuclei. In the last years several measurements of the energy of hypernuclei became available. These can be used to validate or to constrain the hyperon-nucleon interactions within the framework of many-body systems. The ultimate goal is then to constrain these forces by reproducing at best the experimental energies of hypernuclei from light systems made of few particles up to heavier systems.

The method used to accurately solve the many-body Schr\"odinger equation represents the second part of the problem. Accurate calculations are indeed limited to very few nucleons. The exact Faddeev-Yakubovsky equation approach has been applied up to four particle systems~\cite{Glockle:1993}. Few nucleon systems can be accurately described by means of techniques based on shell models calculations like the No-Core Shell Model~\cite{Navratil:2009}, on the Hyperspherical Harmonics approach~\cite{Barnea:2001,Bacca:2002,Bacca:2004,Barnea:2004,Deflorian:2013} or on QuantumMonte Carlo methods, like the Variational Monte Carlo~\cite{Wiringa:1991,Wiringa:1992} or Green Function Monte Carlo~\cite{Pieper:2005,Lusk:2010,Lovato:2013,Gandolfi:2011}. These methods have been proven to solve the nuclear Schr\"odinger equation in good agreement with the Faddeev-Yakubovsky method~\cite{Kamada:2001}. For heavier nuclei, Correlated Basis Function theory~\cite{AriasdeSaavedra:2007}, Cluster Variational Monte Carlo~\cite{Pieper:1990,Pieper:1992} and Coupled Cluster Expansion~\cite{Heisenberg:1999,Hagen:2010} are typically adopted. In addition, the class of method which includes the Brueckner-Goldstone~\cite{Day:1967} and the Hartree-Fock~\cite{Vautherin:1972} algorithms is widely used, also for nuclear matter calculations. The drawback of these many-body methods is that they modify the original Hamiltonian to a more manageable form, often introducing uncontrolled approximations in the algorithm. In absence of an exact method for solving the many-body Schr\"odinger equation for a large number of nucleons, the derivation of model interactions and their applicability in different regimes is subject to an unpleasant degree of arbitrariness.

In this work we address the problem of the hyperon-nucleon interaction from a Quantum Monte Carlo point of view. We discuss the application of the Auxiliary Field Diffusion Monte Carlo (AFDMC) algorithm to study a non relativistic Hamiltonian based on a phenomenological hyperon-nucleon interaction with explicit two- and three-body components. The method was originally developed for nuclear systems~\cite{Schmidt:1999} and it has been successfully applied to the study of nuclei~\cite{Gandolfi:2006,Gandolfi:2007,Gandolfi:2008}, neutron drops~\cite{Pederiva:2004,Gandolfi:2011,Maris:2013}, nuclear matter~\cite{Gandolfi:2007_SNM,Gandolfi:2010} and neutron matter~\cite{Sarsa:2003,Gandolfi:2009,Gandolfi:2009_gap,Gandolfi:2012}. We have extended this ab-initio algorithm in order to include the lightest of the strange baryons, the $\Lambda$~particle. By studying the ground state properties of single and double $\Lambda$~hypernuclei, information about the employed microscopic hyperon-nucleon interaction are deduced. 

The main outcome of the study on finite strange systems is that only the inclusion of explicit $\Lambda NN$ terms provides the necessary repulsion to realistically describe the separation energy of a $\Lambda$~hyperon in hypernuclei of intermediate masses~\cite{Lonardoni:2013_PRC(R),Lonardoni:2013_HYP2012,Lonardoni:2013_PRC}. The analysis of single particle densities confirms the importance of the inclusion of the $\Lambda NN$ contribution. On the ground of this observation, the three-body hyperon-nucleon interaction has been studied in detail. By refitting the coefficients in the potential, it has been possible to reproduce at the same time the available experimental data accessible with AFDMC calculations in a medium-heavy mass range~\cite{Lonardoni:2013_PRC}. Other details of the hypernuclear force, like the charge symmetry breaking contribution and the effect of a $\Lambda\Lambda$ interaction, have been successfully analyzed. The AFDMC study of $\Lambda$~hypernuclei results thus in a realistic phenomenological hyperon-nucleon interaction accurate in describing the ground state physics of medium-heavy mass hypernuclei.

The large repulsive contribution induced by the three-body $\Lambda NN$ term, makes very clear the fact that the lack of an accurate Hamiltonian might be responsible for the unrealistic predictions of the EoS, that would tend to rule out the appearance of strange baryons in high-density matter. We speculate that the application of the developed hyperon-nucleon interaction to the study of the homogeneous medium would lead to a stiffer EoS for the $\Lambda$~neutron matter. This fact might eventually reconcile the physically expected onset of hyperons in the inner core of a NS with the observed masses of order $2M_\odot$.

First steps in this direction have been taken. The study of $\Lambda$~neutron matter at fixed $\Lambda$ fraction shows that the repulsive nature of the three-body hyperon-nucleon interaction is still active and relevant at densities larger than the saturation density. The density threshold for the appearance of $\Lambda$~hyperons has then been derived and the EoS has been computed. Very preliminary results suggest a rather stiff EoS even in the presence of hyperons, implying a maximum mass above the observational limit. The study of hypermatter is still work in progress.

\vspace{2cm}
\hypersetup{linkcolor=black}
\noindent The present work is organized as follows:

\begin{description}
	\item[Chapter~\ref{chap:strangeness}:] a general overview about strangeness in nuclear systems, from hypernuclei to neutron stars, is reported
		with reference to the terrestrial experiments and astronomical observations.
	\item[Chapter~\ref{chap:hamiltonians}:] a description of nuclear and hypernuclear non-relativistic Hamiltonians is presented, with particular attention to the
		hyperon-nucleon sector in the two- and three-body channels.
	\item[Chapter~\ref{chap:method}:] the Auxiliary Field Diffusion Monte Carlo method is discussed in its original form for nuclear systems and in the newly 
		developed version with the inclusion of strange degrees of freedom, both for finite and infinite systems.
	\item[Chapter~\ref{chap:results_finite}:] the analysis and set up of a realistic hyperon-nucleon interaction are reported in 
		connection with the AFDMC results for the hyperon separation energy. Qualitative information are also deduced from single particle densities and root mean square radii 
		for single and double $\Lambda$~hypernuclei.
	\item[Chapter~\ref{chap:results_infinite}:] using the interaction developed for finite strange systems, first Quantum Monte Carlo calculations on 
		$\Lambda$~neutron matter are presented and the implications of the obtained results for the properties of neutron stars are explored.
	\item[Chapter~\ref{chap:conclusion}:] the achievements of this work are finally summarized and future perspective are discussed.
\end{description}

\newpage
\phantom{Empty page}

	\mainmatter
	% Chapter 1: Strangeness in nuclear systems

\chapter{Strangeness in nuclear systems}
\label{chap:strangeness}

\fancyhead[LO]{\emph{\nouppercase{\rightmark}}}		% back to main page style preferences for left and right header fields
\fancyhead[RE]{\emph{\nouppercase{\leftmark}}}
\hypersetup{linkcolor=blue}
\renewcommand{\thefigure}{\arabic{chapter}.\arabic{figure}}

Hyperons are baryons containing one or more strange quarks. They have masses larger than nucleons and lifetimes characteristic of the weak decay. The $\Lambda$ and $\Omega$ hyperons belong to an isospin singlet, the $\Sigma$s to an isospin triplet and the $\Xi$ particles to an isospin doublet. In Tab.~\ref{tab:hyperons} we report the list of hyperons (excluding resonances and unnatural parity states~\cite{Beringer:2012}), with their main properties. The isospin doublet of nucleons is also shown for comparison. 
\renewcommand{\arraystretch}{1.4}
\begin{table}[hb]
	\begin{center}
		\begin{tabular*}{\linewidth}{@{\hspace{0.5em}\extracolsep{\fill}}lccc S[table-format=4.7] S[table-format=1.8] l@{\extracolsep{\fill}\hspace{0.5em}}}
			\toprule
			\toprule
		   	              Baryon     & qqq & $S$ & $I$ & {$m$~[\rm{MeV}]} & {$\tau$~[$10^{-10}$~s]} & Decay mode       \\
			\midrule                          
			\hspace{1.3em}$p$	     & uud & \multirow{2}{*}{$0$} & \multirow{2}{*}{$\displaystyle\frac{1}{2}$} & 938.27205(2) & {$\sim10^{32}$~y} & many   \\
			\hspace{1.3em}$n$	     & udd &      &   &    939.56538(2)     &       {808(1)~s} & $p\,e\,\bar\nu_e$    \\[0.8em]
			\hspace{1.3em}$\Lambda$	 & uds & $-1$ & 0 &    1115.683(6) &       2.63(2)    & $p\,\pi^-, n\,\pi^0$ \\[0.8em]
			\hspace{1.3em}$\Sigma^+$ & uus & \multirow{3}{*}{$-1$} & \multirow{3}{*}{1} & 1189.37(7) & 0.802(3) & $p\,\pi^0, n\,\pi^+$ \\
			\hspace{1.3em}$\Sigma^0$ & uds &      &   &    1192.64(2)  &       7.4(7)$\hspace{-1cm}\times10^{-10}$ & $\Lambda\,\gamma$ \\
			\hspace{1.3em}$\Sigma^-$ & dds &      &   &    1197.45(3)  &       1.48(1)    & $n\,\pi^-$           \\[0.8em]
			\hspace{1.3em}$\Xi^0$	 & uss & \multirow{2}{*}{$-2$} & \multirow{2}{*}{$\displaystyle\frac{1}{2}$} & 1314.9(2) & 2.90(9) & $\Lambda\,\pi^0$   \\
			\hspace{1.3em}$\Xi^-$	 & dss &      &   &    1321.71(7)  &       1.64(2)    & $\Lambda\,\pi^-$     \\[0.8em]
			\hspace{1.3em}$\Omega^-$ & sss & $-3$ & 0 &    1672.5(3)   &       0.82(1)    & $\Lambda\,K^-, \Xi^0\,\pi^-, \Xi^-\,\pi^0$ \\
			\bottomrule
			\bottomrule
		\end{tabular*}
		\caption[Nucleon and hyperon properties]
			{Nucleon and hyperon properties: quark components, strangeness, isospin, mass, mean life and principal decay 	
			modes~\cite{Beringer:2012}.}
		\label{tab:hyperons}
	\end{center}
\end{table}
\renewcommand{\arraystretch}{1.0}

In the non strange nuclear sector many information are available for nucleon-nucleon scattering. The Nijmegen $NN$ scattering database~\cite{Bergervoet:1990,Stoks:1993} includes 1787~$pp$ and 2514~$np$ data in the range $0\div350$~MeV. Due to the instability of hyperons in the vacuum and the impossibility to collect hyperon-neutron and hyperon-hyperon scattering data, the available information in the strange nuclear sector are instead very limited. Although many events have been reported both in the low and high energy regimes~\cite{Gibson:1995}, the standard set employed in the modern hyperon-nucleon interactions (see for example Ref.~\cite{Schulze:2013}) comprises 35 selected $\Lambda p$ low energy scattering data~\cite{deSwart:1971} and some $\Lambda N$ and $\Sigma N$ data at higher energies~\cite{Kadyk:1971}. In addition there are the recently measured $\Sigma^+ p$ cross sections of the KEK-PS E289 experiment~\cite{Ahn:2005}, for a total of 52 $YN$ scattering data.

The very limited experimental possibilities of exploring hyperon-nucleon and hyperon-hyperon interactions in elementary scattering experiments, makes the detailed study of hypernuclei essential to understand the physics in the strange sector. In the next, we will present a summary of the available hypernuclei experimental data. These information are the key ingredient to develop realistic hyperon-nucleon and hyperon-hyperon interactions, as described in the next chapters. The theoretical evidence of the appearance of hyperons in the core of a NS and the problem of the hyperon puzzle will then be discussed, following the results of many-body calculations for the available models of hypermatter.

\section{Hyperons in finite nuclei}
\label{sec:hyp}

In high-energy nuclear reactions strange hadrons are produced abundantly, and they are strongly involved in the reaction process. When hyperons are captured by nuclei, hypernuclei are formed, which can live long enough in comparison with nuclear reaction times. Extensive efforts have been devoted to the study of hypernuclei. Among many strange nuclear systems, the single $\Lambda$~hypernucleus is the most investigated one~\cite{Hashimoto:2006}. 

\begin{figure}[htb]
	\centering
	\includegraphics[width=0.73\linewidth]{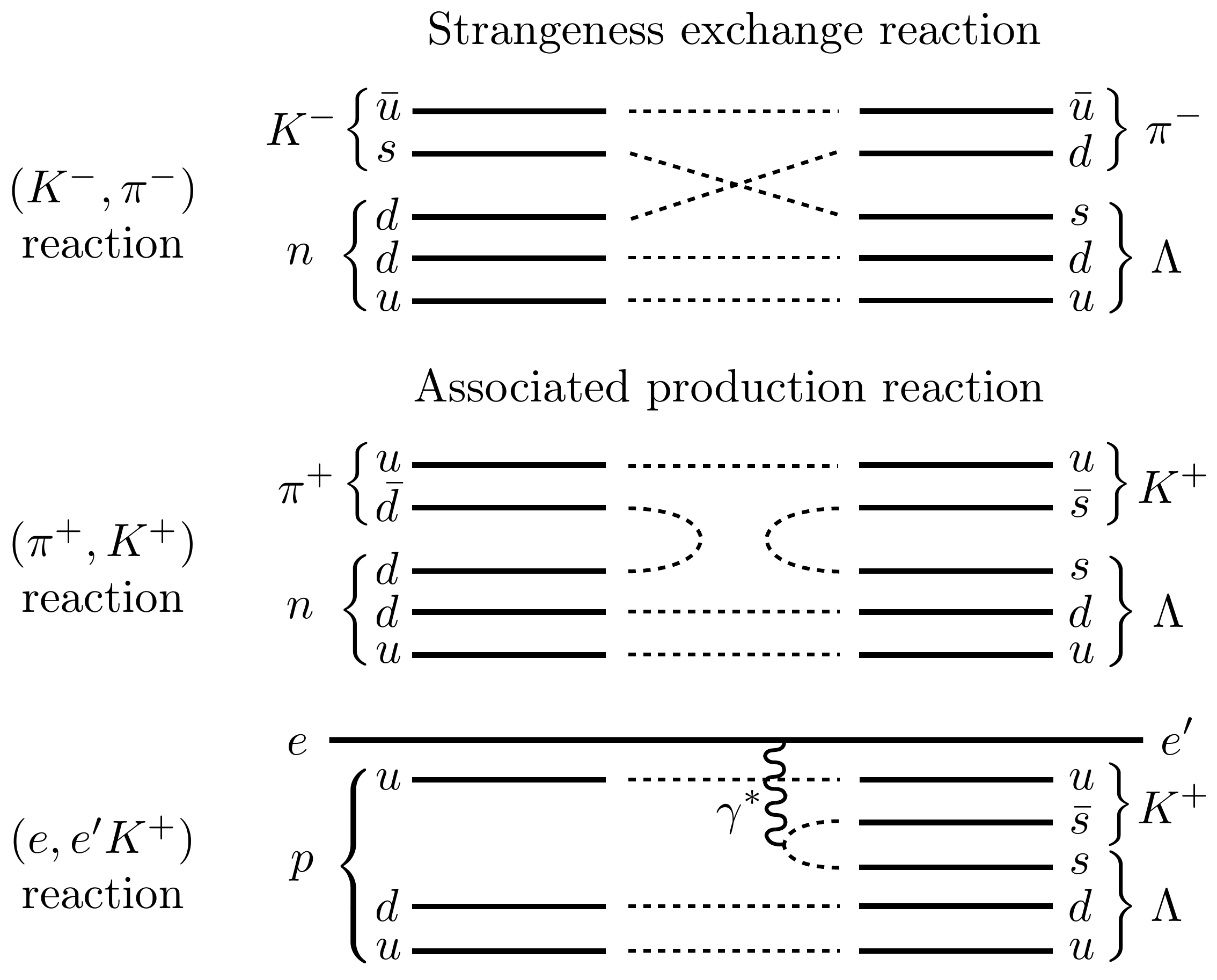}
	\caption[Strangeness producing reactions]{Schematic presentation of three strangeness producing reactions used in the study of $\Lambda$~hypernuclei.}
	\label{fig:reactions}
\end{figure}

The history of hypernuclear experimental research (see Refs.~\cite{Davis:2005,Dalitz:2005,Hashimoto:2006} for a complete review) celebrates this year the sixtieth anniversary, since the publication of the discovery of hypernuclei by Danysz and Pniewski in 1953~\cite{Danysz:1953}. Their first event was an example of $^3_\Lambda$H decaying via
\begin{align}
	^3_\Lambda\text{H}\longrightarrow\,^3\text{He}+\pi^-\;,
\end{align}
confirming that the bound particle was a $\Lambda$~hyperon. The event was observed in an emulsion stack as a consequence of nuclear multifragmentation induced by cosmic rays. This first evidence opened the study of light $\Lambda$~hypernuclei ($A<16$) by emulsion experiments, by means of cosmic ray observations at the beginning and then through proton and pion beams, although the production rates were low and there was much background. In the early 70's, the advent of kaon beam at CERN and later at Brookhaven National Laboratory (BNL), opened the possibility of spectroscopic studies of hypernuclei, including excited states, by means of the $(K^-,\pi^-)$ reaction (see Fig.~\ref{fig:reactions}). A third stage, which featured the use of the $(\pi^+,K^+)$ reaction, began in the mid 1980's at the Alternating Gradient Synchrotron (AGS) of BNL first, and then at the proton synchrotron (PS) of the High Energy Accelerator Organization (KEK) in Japan. Here, the superconducting kaon spectrometer (SKS) played a key role in exploring $\Lambda$~hypernuclear spectroscopy by the $(\pi^+,K^+)$ reaction. $\gamma$-ray spectroscopy developed reaching unprecedented resolution through the use of a germanium detector array, the Hyperball, and the high quality and high intensity electron beams available at the Thomas Jefferson National Accelerator Facility (JLab). This permitted the first successful $(e,e' K^+)$ hypernuclear spectroscopy measurement (an historical review of hypernuclear spectroscopy with electron beams can be found in Ref.~\cite{Nakamura:2013_HYP2012}. The detailed analysis of $\Lambda$~hypernuclei spectroscopy is reported in Ref.~\cite{Hashimoto:2006}). 

\begin{figure}[htb]
	\centering
	\includegraphics[width=\linewidth]{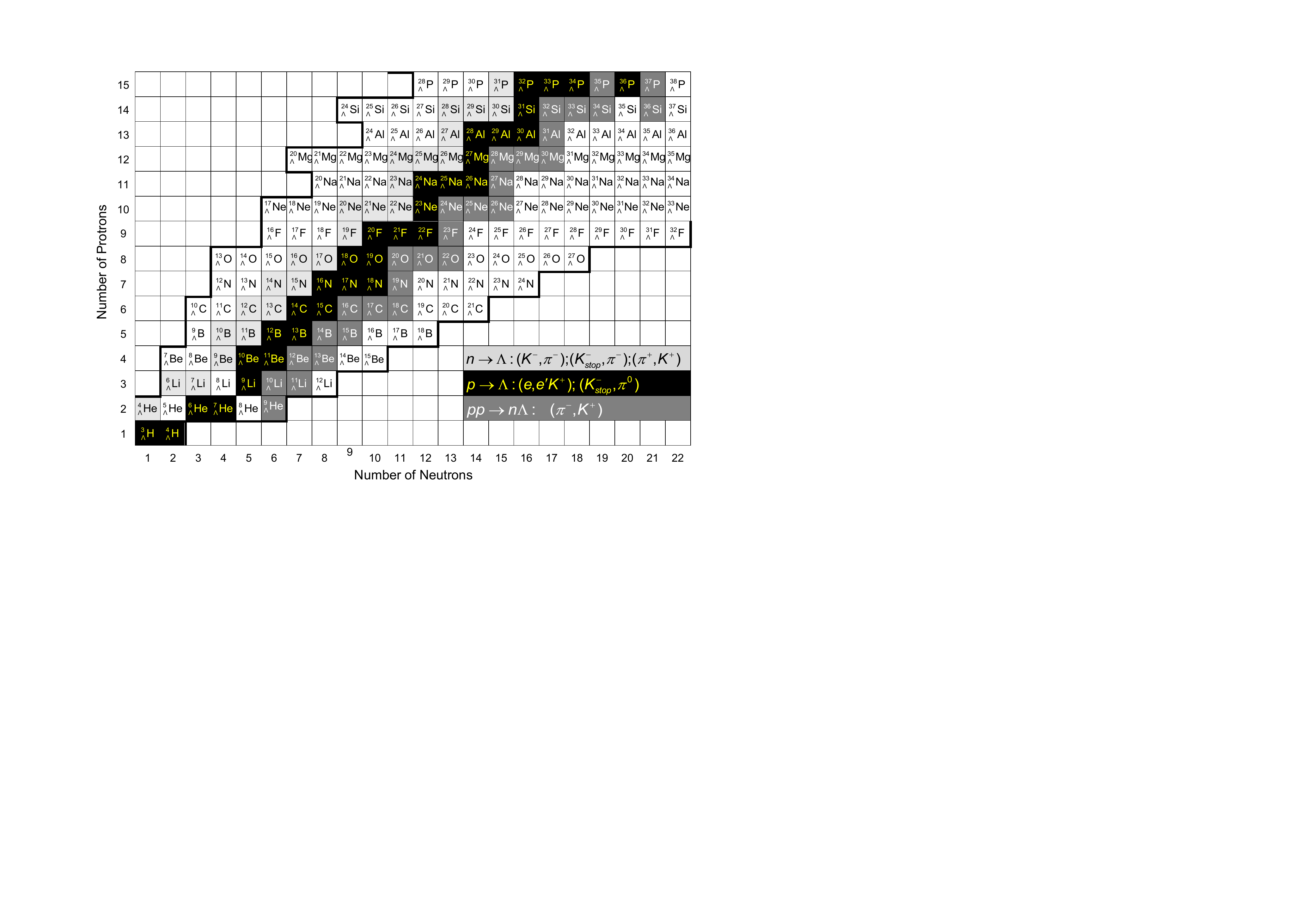}
	\caption[$\Lambda$~hypernuclei accessible via different experimental reactions]
		{$\Lambda$~hypernuclei accessible by experiments for different production channels. 
		The boundaries at the neutron and proton rich side mark the predicted drip lines by a nuclear mass formula extended to strange nuclei.
		Figure taken from Ref.~\cite{Pochodzalla:2011}.}
	\label{fig:exp_reactions}
\end{figure}

With the development of new facilities, like the japanese J-PARC (Proton Accelerator Research Complex), other reaction channels for the production of neutron rich $\Lambda$~hypernuclei became available. The candidates are the single charge exchange (SCX) reactions $(K^-,\pi^0)$ and $(\pi^-,K^0)$, and double charge exchange (DCX) reactions $(\pi^-,K^+)$ and $(K^-,\pi^+)$. Fig.~\ref{fig:exp_reactions} nicely illustrates the complementarity of the various production mechanisms and thus the need to study hypernuclei with different reactions. Moreover, during the last 20 years of research, great progress has been made in the investigation of multifragmentation reactions associated with heavy ion collisions (see for instance~\cite{Ogul:2011} and reference therein). This gives the opportunity to apply the same reactions for the production of hypernuclei too~\cite{Botvina:2007,Topor:2010}. On the other hand, it was noticed that the absorption of hyperons in spectator regions of peripheral relativistic ion collisions is a promising way to produce hypernuclei~\cite{Wakai:1988,Gaitanos:2009}. Also, central collisions of relativistic heavy ions can lead to the production of light hypernuclei~\cite{Steinheimer:2012}. Recent experiments have confirmed observations of hypernuclei in such reactions, in both peripheral~\cite{Saito:2012,Botvina:2012} and central collisions~\cite{STAR:2010}.

At the time of writing, many laboratories are making extensive efforts in the study of $\Lambda$~hypernuclei. The status of the art together with future prospects can be found in Refs.~\cite{Tamura:2012,Tamura:2013_HYP2012,TakahashiT:2013_HYP2012} for the J-PARC facility and in Ref.~\cite{Lea:2013_HYP2012} for the ALICE (A Large Ion Collider Experiment) experiment at the LHC. Ref.~\cite{Garibaldi:2013_HYP2012} reports the status of the JLab's Hall A program. In Ref.~\cite{Esser:2013_HYP2012} future prospects for the the PANDA (antiProton ANihilation at DArmstadt) project at FAIR (Facility for Antiproton ad Ion Research) and the hypernuclear experiments using the KAOS spectrometer at MAMI (Mainz Microtron) can be found. Last results from the FINUDA (FIsica NUcleare a DA$\Phi$NE) collaboration at DA$\Phi$NE, Italy, are reported in Ref.~\cite{Feliciello:2013}. Recent interest has been also focused on the $S=-2$ sector with the study of double $\Lambda$~hypernuclei~\cite{Harada:2013_HYP2012} and the $S=-3$ sector with the search for $\Omega$~hypernuclei~\cite{TakahashiH:2013_HYP2012}.

\begin{figure}[!b]
	\centering
	\includegraphics[width=\linewidth]{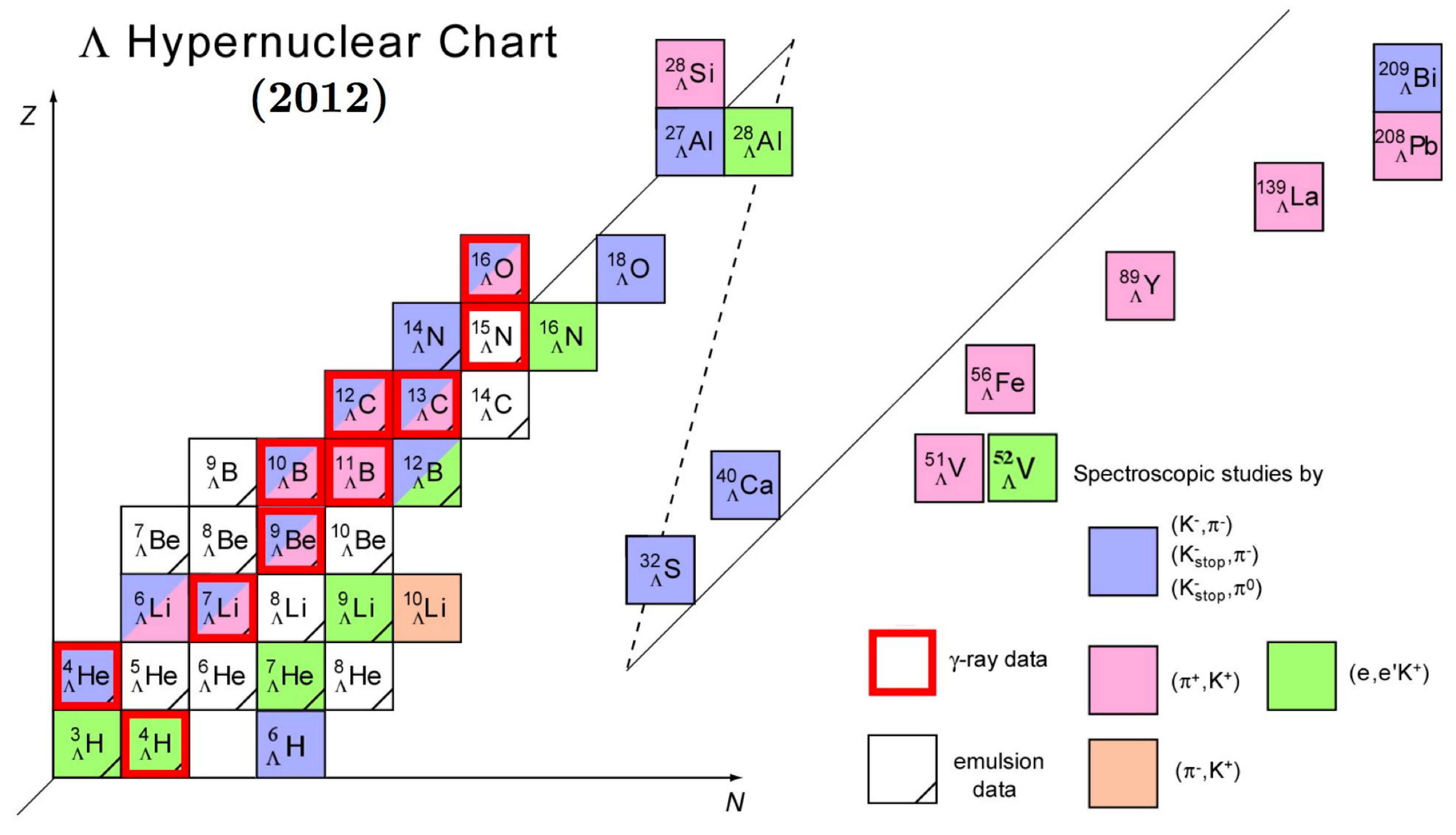}
	\caption[$\Lambda$~hypernuclear chart]{$\Lambda$~hypernuclear chart presented at the 
		\href{http://icc.ub.edu/congress/HYP2012/}{XI International Conference on Hypernuclear and Strange Particle Physics (HYP2012)}, October 2012, Spain. 
		The figure has been updated from Ref.~\cite{Hashimoto:2006}.}
	\label{fig:hyperchart}
\end{figure}

So far, there is no evidence for $\Lambda p$ and $^3_\Lambda$He bound states. Only very recently the possible evidence of the three-body system $\Lambda nn$ has been reported~\cite{Rappold:2013_PRC(R)}. The first well established weakly bound systems is $^3_\Lambda$H, with hyperon separation energy $B_\Lambda$ (the energy difference between the $A-1$ nucleus and the $A$ hypernucleus, being $A$ the total number of baryons) of $0.13(5)$~MeV~\cite{Juric:1973}. Besides the very old experimental results~\cite{Juric:1973,Cantwell:1974,Prowse:1966}, several measurements of single $\Lambda$~hypernuclei became available in the last years trough the many techniques described above~\cite{Pile:1991,Hasegawa:1996,Yuan:2006,Cusanno:2009,Agnello:2010,Agnello:2012_H6L,Nakamura:2013,Feliciello:2013}. The update determination of the lifetime of $_\Lambda^3$H and $_\Lambda^4$H has been recently reported~\cite{Rappold:2013} and new proposals for the search of exotic $\Lambda$~hypernuclei are constantly discussed (see for example the search for $_\Lambda^9$He~\cite{Agnello:2012}). One of the results of this investigation is the compilation of the $\Lambda$~hypernuclear chart reported in Fig.~\ref{fig:hyperchart}. Although the extensive experimental studies in the $S=-1$ strangeness sector, the availability of information for hypernuclei is still far from the abundance of data for the non strange sector.

It is interesting to observe that with the increase of $A$, there is an orderly increase of $B_\Lambda$ with the number of particles, of the order of 1~MeV/nucleon (see Tab.~\ref{tab:BL} or the mentioned experimental references).
Many stable hypernuclei with unstable cores appears, as for example $^6_\Lambda$He, $^8_\Lambda$He, $^7_\Lambda$Be and $^9_\Lambda$Be. These evidences testify that the presence of a $\Lambda$~particle inside a nucleus has a glue like effect, increasing the binding energy and stability of the system. This should be reflected by the attractive behavior of the $\Lambda$-nucleon interaction, at least in the low density regime of hypernuclei. 

For $\Sigma$~hypernuclei, the situation is quite different. Up to now, only one bound $\Sigma$~hypernucleus, $^4_\Sigma$He, was detected~\cite{Nagae:1998}, despite extensive searches. The analysis of experimental data suggests a dominant $\Sigma$-nucleus repulsion inside the nuclear surface and a weak attraction outside the nucleus. In the case of $\Xi$~hypernuclei, although there is no definitive data for any $\Xi$~hypernucleus at present, several experimental results suggest that $\Xi$-nucleus interactions are weakly attractive~\cite{Khaustov:2000}. No experimental indication exists for $\Omega$~hypernuclei. It is a challenge to naturally explain the net attraction in $\Lambda$- and $\Xi$-nucleus potentials and at the same time the dominant repulsion in $\Sigma$-nucleus potentials. 

In addition to single hyperon nuclei, the binding energies of few double $\Lambda$~hypernuclei ($^{\;\;\,6}_{\Lambda\Lambda}$He~\cite{Takahashi:2001,Nakazawa:2010,Ahn:2013}, $^{\;10}_{\Lambda\Lambda}$Be, $^{\;12}_{\Lambda\Lambda}$Be and $^{\;12}_{\Lambda\Lambda}$Be~\cite{Danysz:1963,Nakazawa:2010}, $^{\;13}_{\Lambda\Lambda}$B~\cite{Nakazawa:2010}) have been measured. The indication is that of a weakly attractive $\Lambda\Lambda$ interaction, which reinforces the glue like role of $\Lambda$~hyperons inside nuclei.

From the presented picture it is clear that experimental hypernuclear physics has become a very active field of research. However there is still lack of information, even in the most investigated sector of $\Lambda$~hypernuclei. Due to the technical difficulties in performing scattering experiments involving hyperons and nucleons, the present main goal is the extension of the $\Lambda$~hypernuclear chart to the proton and neutron drip lines and for heavier systems. Parallel studies on $\Sigma$, $\Xi$ and double~$\Lambda$~hypernuclei have been and will we be funded in order to try to complete the scheme. This will hopefully provide the necessary information for the development of realistic hyperon-nucleon and hyperon-hyperon interactions.

\section{Hyperons in neutron stars}
\label{sec:ns}

The matter in the outer core of a NS is supposed to be composed by a degenerate gas of neutrons, protons, electrons and muons, the $npe\mu$ matter, under $\beta$~equilibrium. Given the energy density
\begin{align}
	\mathcal E(\rho_n,\rho_p,\rho_e,\rho_\mu)=\mathcal E_N(\rho_n,\rho_p)+\mathcal E_e(\rho_e)+\mathcal E_\mu(\rho_\mu) \;,
	\label{eq:E_npemu}
\end{align}
where $\mathcal E_N$ is the nucleon contribution, the equilibrium condition at a given baryon density $\rho_b$ corresponds to the minimum of $\mathcal E$ under the constraints
\begin{subequations}
	\begin{align}
		\mbox{fixed baryon density:\qquad}  & \rho_n+\rho_p-\rho_b=0 \;,	\\[0.5em]
		\mbox{electrical neutrality:\qquad} & \rho_e+\rho_\mu-\rho_p=0 \;.	
	\end{align}
	\label{eq:eq_constraints}
\end{subequations}
The result is the set of conditions
\begin{subequations}
	\label{eq:chem_pot}
	\begin{align}
		\mu_n&=\mu_p+\mu_e \;,\\[0.5em]
		\mu_\mu&=\mu_e \;,
	\end{align}
\end{subequations}
where $\mu_j=\partial\mathcal E/\partial\rho_j$ with $j=n,p,e,\mu$ are the chemical potentials. These relations express the equilibrium with respect to the weak interaction processes
\begin{align}
	\begin{array}{rclrcl}
		n & \longrightarrow & p+e+\bar\nu_e \;,    \qquad\quad & p+e   & \longrightarrow & n+\nu_e   \;,\\[0.5em]
		n & \longrightarrow & p+\mu+\bar\nu_\mu \;,\qquad\quad & p+\mu & \longrightarrow & n+\nu_\mu \;.	
	\end{array}
	\label{eq:npemu_eq}
\end{align}
(Neutrino do not affect the matter thermodynamics so their chemical potential is set to zero). Eqs.~(\ref{eq:chem_pot}) supplemented by the constraints (\ref{eq:eq_constraints}) form a closed system which determines the equilibrium composition of the $npe\mu$ matter. Once the equilibrium is set, the energy and pressure as a function of the baryon density can be derived and thus the EoS is obtained.

Given the EoS, the structure of a non rotating NS can be fully determined by solving the Tolman-Oppenheimer-Volkoff (TOV) equations~\cite{Oppenheimer:1939,Lattimer:2004}
\begin{subequations}
	\begin{align}
		\frac{dP(r)}{dr}&=-G\frac{\Bigl[\mathcal E(r)+P(r)\Bigr]\Bigl[m(r)+4\pi r^3 P(r)\Bigr]}{r^2\Bigl[1-\frac{2Gm(r)}{r}\Bigr]}\;,\label{eq:TOV_1} \\[0.5em]
		\frac{dm(r)}{dr}&=4\pi r^2 \mathcal E(r) \;,\label{eq:TOV_2}
	\end{align}
	\label{eq:TOV}
\end{subequations}
which describe the hydrostatic equilibrium of a static spherically symmetric star. $\mathcal E(r)$ and $P(r)$ are the energy density and the pressure of the matter, $m(r)$ is the gravitational mass enclosed within a radius $r$, and $G$ is the Gravitational constant. In the stellar interior $P>0$ and $dP/dr<0$. The condition $P(R)=0$ fixes the stellar radius $R$. Outside the star for $r>R$, we have $P=0$ and $\mathcal E=0$. Eq.~(\ref{eq:TOV_2}) gives thus $m(r>R)=M=const$, which is  total gravitational mass. Starting with a central energy density $\mathcal E_c=\mathcal E(r=0)$ and using the above conditions, the TOV equations can be numerically solved and the mass-radius relation $M=M(R)$ is obtained. It can be shown~\cite{Haensel:2006}, that the relativistic corrections to the Newtonian law $dP(r)/dr=-Gm\mathcal E(r)/r^2$ included in Eq.~(\ref{eq:TOV_1}) give an upperbound to the $M(R)$ relation, i.e. there exists a maximum mass for a NS in hydrostatic equilibrium. It is important to note that, given the EoS, the mass-radius relation is univocally determined. Any modification made on the EoS will lead to a change in the $M(R)$ curve and thus in the allowed maximum mass.

For $\rho_b\gtrsim2\rho_0$, the inner core is thought to have the same $npe\mu$ composition of the outer core. However, since at high densities the nucleon gas will be highly degenerate, hyperons with energies lower than a threshold value will become stable, because the nucleon arising from their decay cannot find a place in phase space in accordance to the Pauli principle~\cite{Ambartsumyan:1960}. Thus, beyond a density threshold we have to take into account the contribution of hyperons to the $\beta$~equilibrium. Eq.~(\ref{eq:E_npemu}) becomes a function of $\rho_b$ (baryons: nucleons and hyperons) and $\rho_l$ (leptons: electrons and muons). Given the baryon density and imposing electrical neutrality conditions, the equilibrium equations now read:
\begin{subequations}
	\label{eq:chem_pot_Y}
	\begin{align}
		Q_b=-1\,:           && \mu_{b^-}&=\mu_n+\mu_e && \Rightarrow && \mu_{\Omega^-}&&\hspace{-0.7cm}=\mu_{\Xi^-}   =\mu_{\Sigma^-}=\mu_n+\mu_e \;,			\\[0.5em]
		Q_b=\phantom{+}0\,: && \mu_{b^0}&=\mu_n       && \Rightarrow && \mu_{\Xi^0}   &&\hspace{-0.7cm}=\mu_{\Sigma^0}=\mu_\Lambda   =\mu_n \;,	\\[0.5em]
		Q_b=+1\,:           && \mu_{b^+}&=\mu_n-\mu_e && \Rightarrow && \mu_{\Sigma^+}&&\hspace{-0.7cm}=\mu_p         =\mu_n-\mu_e \;,
	\end{align}
\end{subequations}  
where $Q_b$ is the electric charge of a baryon. As soon as the neutron chemical potential becomes sufficiently large, energetic neutrons can decay via weak strangeness nonconserving reactions into $\Lambda$~hyperons, leading to a $\Lambda$ Fermi sea.

\begin{figure}[htb]
	\centering
	\includegraphics[width=0.9\linewidth]{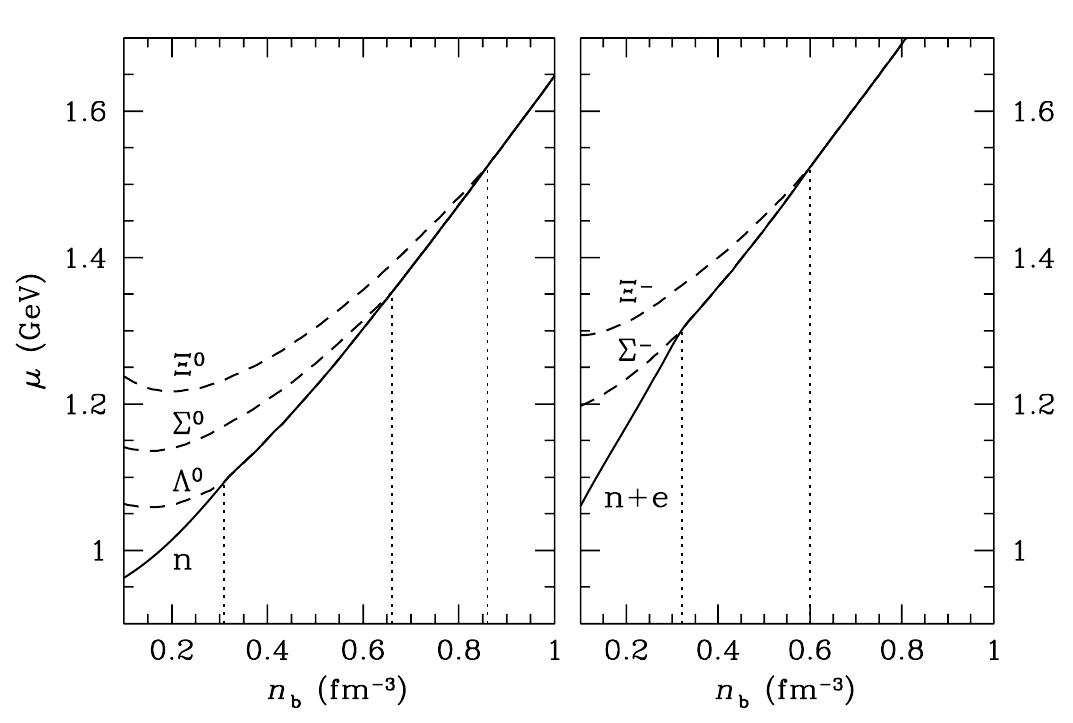}
	\caption[Hyperon and nucleon chemical potentials]
		{Threshold chemical potentials of neutral hyperons and neutron (left panel), and of negatively charged hyperons and the sum 
		$\mu_n+\mu_e$ (right panel) versus baryon density. Vertical dotted lines mark the thresholds for the creation of new hyperons. Dashed lines show  
		the minimum chemical potential $\mu_Y^0$ of unstable hyperons before the thresholds. Figure taken from Ref.~\cite{Haensel:2006}.}
	\label{fig:chemicalpot}
\end{figure}

We can derive the hyperons threshold densities $\rho_Y$ by calculating the minimum increase of the energy of the matter produced by adding a single strange particle at a fixed pressure. This can be done by considering the energy of the matter with an admixture of given hyperons and by calculating numerically the limit of the derivative
\begin{align}
	\lim_{\rho_Y\rightarrow0}\,\frac{\partial\mathcal E}{\partial\rho_Y}\bigg|_{eq}\!=\mu_Y^0\;.
\end{align}
Consider for example the lightest $\Lambda$~hyperon. As long as $\mu_\Lambda^0>\mu_n$, the strange baryon cannot survive because the system will lower its energy via an exothermic reaction $\Lambda+N\longrightarrow n+N$. However, $\mu_n$ increases with growing $\rho_b$ and the functions $\mu_\Lambda^0(\rho_b)$ and $\mu_n^0(\rho_b)$ intersect at some $\rho_b=\rho_\Lambda^{th}$ (the left panel in Fig.~\ref{fig:chemicalpot}). For $\rho_b>\rho_\Lambda^{th}$ the $\Lambda$~hyperons become stable in dense matter because their decay is blocked by the Pauli principle. 

Although the $\Lambda$~particle is the lightest among hyperons, one expects the $\Sigma^-$ to appear via
\begin{align}
	n+e^-\longrightarrow\Sigma^-+\nu_e
\end{align}
at densities lower than the $\Lambda$~threshold, even thought the $\Sigma^-$ is more massive. This is because the negatively charged hyperons appear in the ground state of matter when their masses equal $\mu_n+\mu_e$, while the neutral hyperon $\Lambda$ appears when its mass equals $\mu_n$. Since the electron chemical potential in matter is typically larger (ultrarelativistic degenerate electrons $\mu_e\sim E_{F_e}\sim \hbar c (3\pi^2\rho_e)^{1/3}>120~\text{MeV~for~}\rho_e\sim5\%\rho_0$) than the mass difference $m_{\Sigma^-}-m_\Lambda=81.76~\mbox{MeV}$, the $\Sigma^-$ will appear at lower densities. However, in typical neutron matter calculations with the inclusion of strange degrees of freedom, only $\Lambda$, $\Sigma^0$ and $\Xi^0$ hyperons are taken into account due to charge conservation.

The formation of hyperons softens the EoS because high energy neutrons are replaced by more massive low energy hyperons which can be accommodated in lower momentum states. There is thus a decrease in the kinetic energy that produces lower pressure. The softening of the EoS of the inner core of a NS induced by the presence of hyperons is generic effect. However, its magnitude is strongly model dependent.

\begin{figure}[htb]
	\centering
	\includegraphics[width=0.7\linewidth]{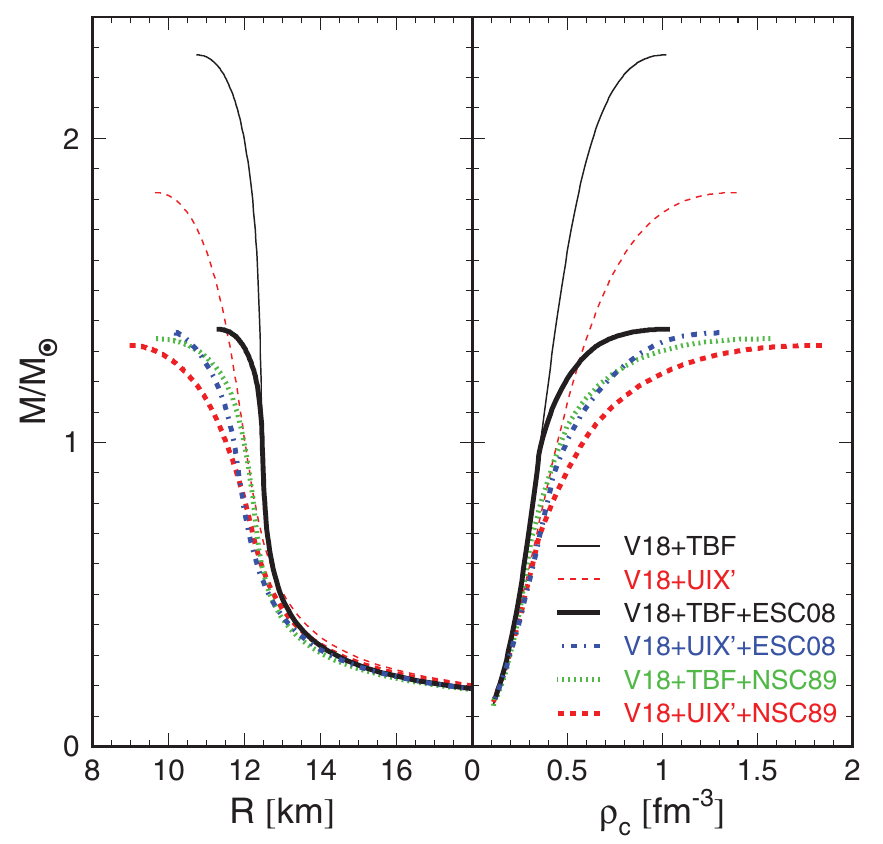}
	\caption[Neutron star mass-radius relation: Schulze 2011]
		{Mass-radius and mass-central density relations for different NS EoS obtained in Brueckner-Hartree-Fock calculations of hypernuclear matter. 
		V18+TBF and V18+UIX' refer to purely nuclear matter EoS built starting from two- and three-body nucleon-nucleon potentials (see \S~\ref{sec:nuc_int}).
		The other curves are obtained adding two different hyperon-nucleon forces among the Nijmegen models to the previous nucleonic EoS.
		For more details see the original paper~\cite{Schulze:2011}.}
	\label{fig:Schulze2011}
\end{figure}

\begin{figure}[htb]
	\centering
	\includegraphics[width=0.8\linewidth]{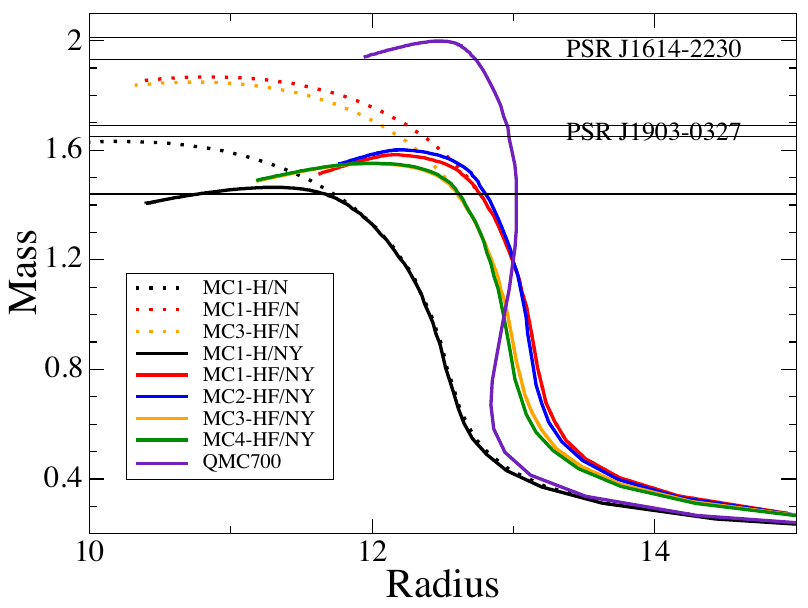}
	\caption[Neutron star mass-radius relation: Massot 2012]
		{Neutron star mass as a function of the circumferential radius. QMC700 and MC\emph{i}-H(F)/N refer to EoS based on quark-meson coupling model and chiral model 
		in the Hartee(Fock) approximation without hyperons. In the MC\emph{i}-H(F)/NY models also hyperons are taken into account.
		The canonical maximum mass limit of $\sim1.45M_\odot$ and the mass of the two heavy millisecond pulsars 
		PSR J1903+0327 ($1.67(2)M_\odot$) and PSR J1614-2230 ($1.97(4)M_\odot$) are shown.
		Details on the potentials and method adopted can be found in Ref.~\cite{Massot:2012}.} 
	\label{fig:Massot2012}
\end{figure}

Calculations based on the extension to the hyperonic sector of the Hartree-Fock (HF)~\cite{Dapo:2010,Massot:2012} and Brueckner-Hartree-Fock (BHF)~\cite{Schulze:2011,Vidana:2011} methods, do all agree that the appearance of hyperons around $2\div3\rho_0$ leads to a strong softening of the EoS. Consequently, the expected maximum mass is largely reduced, as shown for instance in Fig.~\ref{fig:Schulze2011} and Fig.~\ref{fig:Massot2012}. The addition of the hyperon-nucleon force to the pure nucleonic Hamiltonian, lowers the maximum mass of a value between $0.4M_\odot$ and more than $1M_\odot$. From the pure nucleonic case of $M_{\max}>1.8M_\odot$, the limit for hypernuclear matter is thus reduced to the range $1.4M_\odot<M_{\max}<1.6M_\odot$. These results, although compatible with the canonical limit of $1.4\div1.5M_\odot$, cannot be consistent with the recent observations of $2M_\odot$ millisecond pulsars~\cite{Demorest:2010,Antoniadis:2013}. 

It is interesting to note that the hyperonic $M_{\max}$ weakly depends on the details of the employed nucleon-nucleon interaction and even less on the hypernuclear forces. In Ref.~\cite{Dapo:2010} the interaction used for the nuclear sector is an analytic parametrization fitted to energy of symmetric matter obtained from variational calculations with the Argonne V18 nucleon-nucleon interaction (see \S~\ref{sec:nuc_int}) including three-body forces and relativistic boost corrections. Refs.~\cite{Schulze:2011} and \cite{Vidana:2011} adopted the bare $NN$ Argonne V18 supplemented with explicit three-nucleon forces or phenomenological density-dependent contact terms that account for the effect of nucleonic and hyperonic three-body interactions. The hypernuclear forces employed in these work belong to the class of Nijmegen potentials (see \S~\ref{chap:hamiltonians}). Finally, in Ref.~\cite{Massot:2012} chiral Lagrangian and quark-meson coupling models of hyperon matter have been employed. Despite the differences in the potentials used in the strange and non strange sectors, the outcomes of these works give the same qualitative and almost quantitative picture about the reduction of $M_{\max}$ due to the inclusions of strange baryons. Therefore, the (B)HF results seem to be rather robust and thus, many doubts arise about the real appearance of hyperons in the inner core of NSs.

\begin{figure}[htb]
	\centering
	\includegraphics[width=\linewidth]{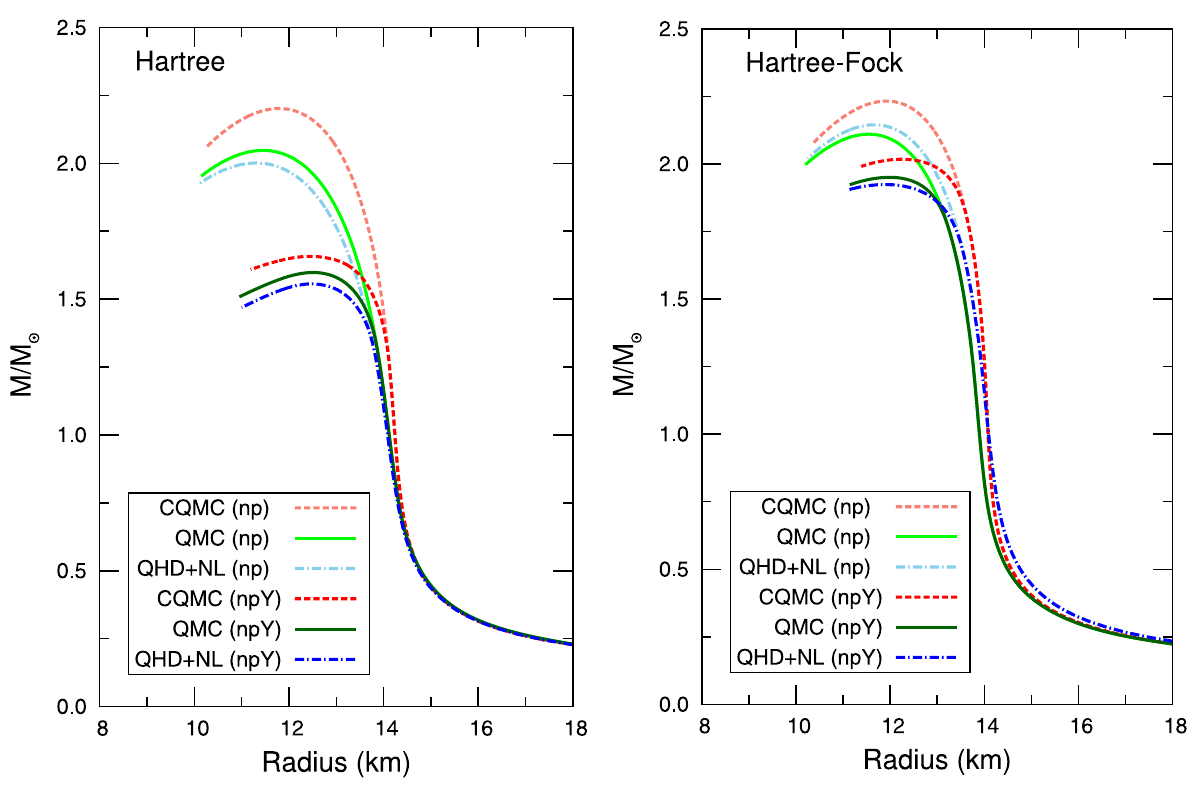}
	\caption[Neutron star mass-radius relation: Miyatsu 2012]
		{Neutron star mass-radius relations in Hartree (left panel) and Hartree-Fock (right panel) calculations.
		CQMC, QMC and QHD+NL denote the chiral quark-meson coupling, quark-meson coupling and non linear quantum hadrodynamics employed potentials, with (npY) and without hyperons (np).
		For details see Ref.~\cite{Miyatsu:2012}.} 
	\label{fig:Miyatsu2012}
\end{figure}

\begin{figure}[p]
	\centering
	\includegraphics[width=0.9\linewidth]{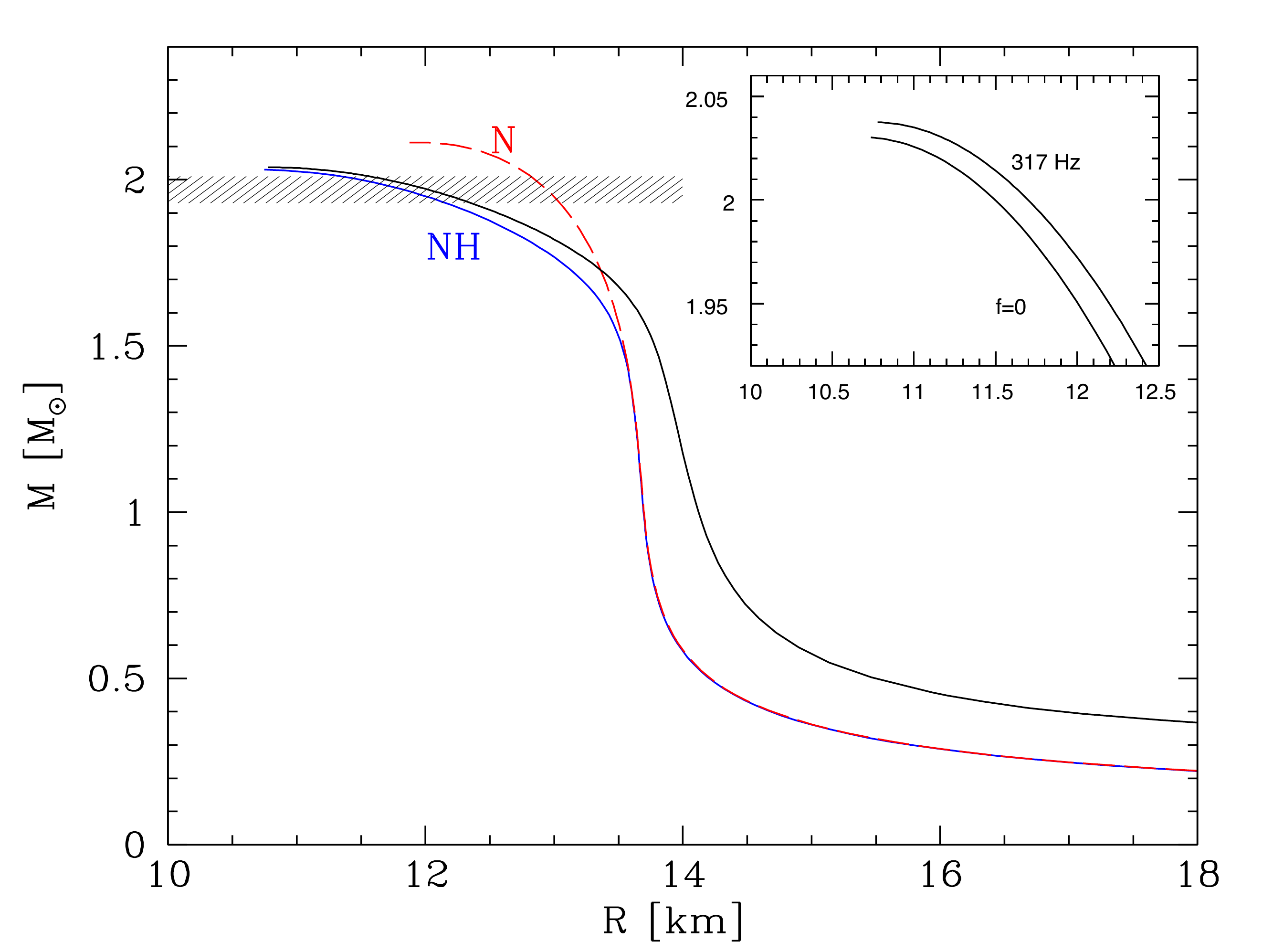}
	\caption[Neutron star mass-radius relation: Bednarek 2012]
		{Stellar mass versus circumferential radius in non linear relativistic mean field model. The purely nucleon case is denoted with N, the nucleon+hyperon case with NH.
		In the inset, the effect of rotation at $f=317$~Hz on the mass-radius relation near $M_{\max}$. 
		The dashed region refers to the mass of the pulsar PSR J1614-2230. All the details are reported in Ref.~\cite{Bednarek:2012}.} 
	\label{fig:Bednarek2012}
\end{figure}

Other approaches, such as relativistic Hartree-Fock~\cite{Miyatsu:2012,Miyatsu:2013,Gupta:2013}, standard, density-dependent and nonlinear Relativistic Mean Field models
~\cite{Bednarek:2012,Weissenborn:2012,Tsubakihara:2013_HYP2012,Jiang:2012,Mallick:2013} and Relativistic Density Functional Theory with density-dependent couplings~\cite{Colucci:2013}, indicate much weaker effects as a consequence of the presence of strange baryons in the core of NSs, as shown for example in Fig.~\ref{fig:Miyatsu2012} and Fig.~\ref{fig:Bednarek2012}. In all these works, it was possible to find a description of hypernuclear matter, within the models analyzed, that produces stiff EoS, supporting a $2M_\odot$ neutron star. Same conclusion has been reported in Ref.~\cite{Bonanno:2012} where the EoS of matter including hyperons and deconfined quark matter has been constructed on the basis of relativistic mean-field nuclear functional at low densities and effective Nambu-Jona-Lasinio model of quark matter. The results of this class of calculations seem to reconcile the onset of hyperons in the inner core of a NS with the observed masses of order $2M_\odot$.

This inconsistency among different calculations and between the theoretical results and the observational constraints, at present is still an open question. For example, given the theoretical evidence about the appearance of hyperons in the inner core of a NS, the results of all available (B)HF calculations seem to be in contradiction with the picture drawn by the relativistic mean field models. On one hand there should be uncontrolled approximations on the method used to solve the many-body Hamiltonian. On the other hand the employed hypernuclear interactions might not be accurate enough in describing the physics of the infinite nuclear medium with strange degrees of freedom. For instance, as reported in Refs.~\cite{Vidana:2013_HYP2012,Tsubakihara:2013_HYP2012}, one of the possible solutions to improve the hyperon-nucleon interactions might be the inclusion of explicit three-body forces in the models. These should involve one or more hyperons (i.e., hyperon-nucleon-nucleon, hyperon-hyperon-nucleon or hyperon-hyperon-hyperon interactions) and they could eventually provide the additional repulsion needed to make the EoS stiffer and, therefore the maximum mass compatible with the current observational limits. On the grounds of this observation, we decided to revisit the problem focusing on a systematic construction of a realistic, though phenomenological hyperon-nucleon interaction with explicit two- and three-body components (\S~\ref{chap:hamiltonians}) by means of Quantum Monte Carlo calculations (\S~\ref{chap:method}). 

\newpage
\phantom{Empty page}

					% chapter 1: Strangeness in nuclear systems
	% Chapter 2: Hamiltonians

\chapter{Hamiltonians}
\label{chap:hamiltonians}

The properties of nuclear systems arise from the interactions between the individual constituents. In order to understand these properties, the starting point is the determination of the Hamiltonian to be used in the description of such systems. In principle the nuclear Hamiltonian should be directly derived from Quantum Chromodynamics (QCD). Many efforts have been done in the last years~\cite{Savage:2012,Beane:2012,Beane:2013}, but this goal is still far to be achieved. 

The problem with such derivation is that QCD is non perturbative in the low-temperature regime characteristic of nuclear physics, which makes direct solutions very difficult. Moving from the real theory to effective models, the structure of a nuclear Hamiltonian can be determined phenomenologically and then fitted to exactly reproduce the properties of few-nucleon systems. In this picture, the degrees of freedom are the baryons, which are considered as non relativistic point-like particles interacting by means of phenomenological potentials. These potentials describe both short and the long range interactions, typically via one-boson and two-meson exchanges, and they have been fitted to exactly reproduce the properties of few-nucleon systems~\cite{Carlson:1998}. In more details, different two-body phenomenological forms have been proposed and fitted on the nucleon-nucleon ($NN$) scattering data of the Nijmegen database~\cite{Bergervoet:1990,Stoks:1993} with a $\chi^2/N_{data}\simeq 1$. The more diffuse are the Nijmegen models~\cite{Stoks:1994}, the Argonne models~\cite{Wiringa:1995,Wiringa:2002} and the CD-Bonn~\cite{Machleidt:1996}. Although reproducing the $NN$ scattering data, all these two-nucleon interactions underestimate the triton binding energy, suggesting that the contribution of a three-nucleon ($NNN$) interaction~(TNI) is essential to reproduce the physics of nuclei. The TNI is mainly attributed to the possibility of nucleon excitation in a $\Delta$ resonance and it can be written as different effective three-nucleon interactions which have been fitted on light nuclei~\cite{Carlson:1981,Pieper:2001} and on saturation properties of nuclear matter~\cite{Carlson:1983}. The TNIs typically depend on the choice of the two-body $NN$ potential~\cite{Wiringa:1983}, but the final result with the total Hamiltonian should be independent of the choice.

A different approach to the problem is the realization that low-energy QCD is equivalent to an Effective Field Theory (EFT) which allows for a perturbative expansion that is known as chiral perturbation theory. In the last years modern nucleon-nucleon interaction directly derived from Chiral Effective Field Theory ($\chi$-EFT) have been proposed, at next-to-next-to-next-to-leading order~(N$^3$LO) in the chiral expansion~\cite{Entem:2003,Epelbaum:2005} and recently at optimized next-to-next-to-leading order~(N$^2$LO)~\cite{Ekstrom:2013} (see Ref.~\cite{Machleidt:2011} for a complete review). 
All these potentials are able to reproduce the Nijmegen phase shifts with $\chi/N_{data}^2\simeq1$. TNIs enter naturally at N$^2$LO in this scheme, and they play again a pivotal role in nuclear structure calculations~\cite{Hammer:2013}. The contributions of TNIs at N$^3$LO have also been worked out~\cite{Ishikawa:2007,Bernard:2008,Bernard:2011}. The $\chi$-EFT interactions are typically developed in momentum space, preventing their straightforward application within the Quantum Monte Carlo~(QMC) framework. However, a local version of the $\chi$-EFT potentials in coordinate space up to N$^2$LO has been very recently proposed and employed in QMC calculations~\cite{Gezerlis:2013}.

Nuclear phenomenological Hamiltonians have been widely used to study finite and infinite nuclear systems within different approaches. From now on, we will focus on the Argonne $NN$ potentials and the corresponding TNIs, the Urbana~IX~(UIX)~\cite{Carlson:1983} and the modern Illinois~(ILx)~\cite{Pieper:2001} forms. These potentials have been used to study nuclei, neutron drops, neutron and nuclear matter in Quantum Monte Carlo (QMC) calculations, such as Variational Monte Carlo (VMC)~\cite{Wiringa:1991,Wiringa:1992,Pieper:1992}, Green Function Monte Carlo (GFMC)~\cite{Pudliner:1997,Wiringa:2002,Pieper:2004,Pieper:2005,Schiavilla:2007,Pieper:2008,Lovato:2013,Gandolfi:2011} and Auxiliary Field Diffusion Monte Carlo (AFDMC)~\cite{Sarsa:2003,Pederiva:2004,Gandolfi:2006,Gandolfi:2007,Gandolfi:2011,Gandolfi:2007_SNM,Gandolfi:2009}. Same bare interactions have been also employed in the Fermi Hyper-Netted Chain~(FHNC) approach~\cite{AriasdeSaavedra:2007,Armani:2011}, both for nuclei and nuclear matter. With a projection of the interaction onto the model space, these Hamiltonians are used in Effective Interaction Hyperspherical Harmonics (EIHH)~\cite{Barnea:2001,Barnea:2004} and Non-Symmetrized Hyperspherical Harmonics (NSHH)~\cite{Deflorian:2013} calculations. Finally, same potentials can be also used in Brueckner Hartree Fock~(BHF)~\cite{Li:2006}, Shell-Model (SM)~\cite{Coraggio:2009}, No-Core-Shell-Model (NCSM)~\cite{Navratil:2009} and Coupled Cluster (CC)~\cite{Hagen:2010} calculations by means of appropriate techniques to handle the short-range repulsion of the nucleon-nucleon force, such as Brueckner $G$-matrix approach~\cite{Brueckner:1955,Bethe:2006}, $V_{low-k}$ reduction~\cite{Bogner:2001,Bogner:2002,Bogner:2003}, Unitary Correlation Operator Method~(UCOM)~\cite{Feldmeier:1998} or Similarity Renormalization Group (SRG) evolution~\cite{Bogner:2007,Jurgenson:2011}.
The list of methods that can handle in a successful way the Argonne+TNIs potentials demonstrates the versatility and reliability of this class of phenomenological nuclear Hamiltonians.

Moving from the non-strange nuclear sector, where nucleons are the only baryonic degrees of freedom, to the strange nuclear sector, where also hyperons enter the game, the picture becomes much less clear. There exists only a very limited amount of scattering data from which one could construct high-quality hyperon-nucleon ($YN$) potentials. Data on hypernuclei binding energies and hyperon separation energies are rather scarce and can only partially complete the scheme. 

After the pioneering work reported in Ref.~\cite{Dalitz:1972}, several models have been proposed to describe the $YN$ interaction. The more diffuse are the Nijmegen soft-core models (like NSC89 and NSC97x)~\cite{Nagels:1977,Nagels:1979,Maessen:1989,Rijken:1999,Stoks:1999,Halderson:1999,Halderson:2000} and the J\"ulic potential (J04)~\cite{Holzenkamp:1989,Reuber:1994,Haidenbauer:2005}. A recent review of these interactions, together with Hartree-Fock (HF) calculations have been published by \DH{}apo~\emph{et al.} in Ref.~\cite{Dapo:2008}. In the same framework, extended soft-core Nijmegen potentials for strangeness $S=-2$ have been also developed~\cite{Rijken:2006_I,Rijken:2006_II}. Very recently, the extended soft-core 08 (ESC08) model has been completed, which represents the first unified theoretical framework involving hyperon-nucleon, hyperon-hyperon ($YY$) and also nucleon-nucleon sectors~\cite{Schulze:2013}. This class of interaction has been used in different calculations for hypernuclei~\cite{Hao:1993,HjorthJensen:1996,Vidana:1998,Vidana:2001,Nogga:2002,Dapo:2008,Schulze:2013} and hypermatter~\cite{Dapo:2008,Dapo:2010,Schulze:2011,Vidana:2011} within different methods, but the existing data do not constrain the potentials sufficiently. For example, six different parameterizations of the Nijmegen $YN$ potentials fit equally well the scattering data but produce very different scattering lengths, as reported for instance in Ref.~\cite{Rijken:1999}. In addition, these potentials are not found to yield the correct spectrum of hypernuclear binding energies. For example, the study~\cite{Nogga:2002} of $^4_\Lambda$H and $^4_\Lambda$He that uses Nijmegen models, does not predict all experimental separation energies. Similar conclusions for single- and double-$\Lambda$~hypernuclei have also been drawn in a study employing a different many-body technique~\cite{Vidana:2001}. Even the most recent ESC08 model produces some overbinding of single-$\Lambda$~hypernuclei and a weakly repulsive incremental $\Lambda\Lambda$~energy~\cite{Schulze:2013}, not consistent with the observed weak $\Lambda\Lambda$ attraction in $^{\;\;\,6}_{\Lambda\Lambda}$He.

In analogy with the nucleon-nucleon sector, a $\chi$-EFT approach for the hyperon-nucleon interaction has been also developed. The first attempt was proposed by Polinder and collaborators in 2006~\cite{Polinder:2006}, resulting in a leading order (LO) expansion. Only recently the picture has been improved going to next-to-leading order~(NLO)~\cite{Haidenbauer:2013_HYP2012,Nogga:2013_HYP2012,Haidenbauer:2013}. The $YN$ $\chi$-EFT model is still far away from the theoretical accuracy obtained in the non-strange sector, but it is any case good enough to describe the limited available $YN$ scattering~data.

As an alternative, a cluster model with phenomenological interactions has been proposed by Hiyama and collaborators to study light hypernuclei ~\cite{Hiyama:1997,Hiyama:2001,Hiyama:2002,Hiyama:2009,Hiyama:2010,Hiyama:2013}.
Interesting results on $\Lambda$~hypernuclei have also been obtained within a $\Lambda$-nucleus potential model, in which the need of a functional with a more than linear density dependence was shown, suggesting the importance of a many-body interaction~\cite{Millener:1988}. While studying $s$-shell hypernuclei, the $\Lambda N\rightarrow\Sigma N$ coupling as a three-body $\Lambda NN$ force has been investigated by many authors~\cite{Nogga:2002,Akaishi:2000,Hiyama:2001_conv,Nemura:2002}. Having strong tensor dependence it is found to play an important role, comparable to the TNI effect in non-strange nuclei.

Finally, starting in the 1980s, a class of Argonne-like interactions has been developed by Bodmer, Usmani and Carlson on the grounds of quantum Monte Carlo calculations to describe the $\Lambda$-nucleon force. These phenomenological interactions are written in coordinates space and they include two- and three-body hyperon-nucleon components, mainly coming from two-pion exchange processes and shorter range effects. They have been used in different forms mostly in variational Monte Carlo calculations for single $\Lambda$~hypernuclei ($^3_\Lambda$H~\cite{Bodmer:1988,Shoeb:1999}, $^4_\Lambda$H and $^4_\Lambda$He~\cite{Bodmer:1985,Bodmer:1988,Shoeb:1999,Sinha:2002}, $^5_\Lambda$He~\cite{Bodmer:1988,Shoeb:1999,Usmani:1995_3B,Usmani:1999,Sinha:2002,Usmani:2003,Usmani:2006,Usmani:2008}, $^9_\Lambda$Be~\cite{Bodmer:1984,Shoeb:1998}, $^{13}_{~\Lambda}$C~\cite{Bodmer:1984}, $^{17}_{~\Lambda}$O~\cite{Usmani:1995,Usmani:1995_3B}), double $\Lambda$~hypernuclei ($^{\;\;\,4}_{\Lambda\Lambda}$H, $^{\;\;\,5}_{\Lambda\Lambda}$H, $^{\;\;\,5}_{\Lambda\Lambda}$He~\cite{Shoeb:2004} and $^{\;\;\,6}_{\Lambda\Lambda}$He~\cite{Shoeb:2004,Usmani:2004,Usmani:2006_He6LL}) and in the framework of correlated basis function theory for $\Lambda$~hypernuclei~\cite{AriasdeSaavedra:2001}, typically in connection with the Argonne $NN$ potential.

Within the phenomenological interaction scheme, a generic nuclear system including nucleons and hyperons, can be described by the non relativistic phenomenological Hamiltonian
\begin{equation}
	H=H_{N}+H_{Y}+H_{YN}\;,\label{eq:H}
\end{equation}
where $H_N$ and $H_Y$ are the pure nucleonic and hyperonic Hamiltonians and $H_{YN}$ represents the interaction Hamiltonian connecting the two distinguishable types of baryon:
\begin{align}
	H_{N} &=\frac{\hbar^2}{2m_N}\sum_{i}\nabla_i^2\;+\sum_{i<j}v_{ij}\;\,+\sum_{i<j<k}v_{ijk}\;\;\,+\,\ldots\;,\label{eq:H_N}\\[0.5em]
	H_{Y} &=\frac{\hbar^2}{2m_\Lambda}\sum_{\lambda}\nabla_\lambda^2\;+\sum_{\lambda<\mu}v_{\lambda\mu}\,+\sum_{\lambda<\mu<\nu}v_{\lambda\mu\nu}\;+\,\ldots\;,\label{eq:H_Y}\\[0.5em]
	H_{YN}&=\sum_{\lambda i}v_{\lambda i}\,+\sum_{\lambda,i<j}v_{\lambda ij}\,+\sum_{\lambda<\mu,i}v_{\lambda\mu i}\,+\,\ldots\;.\label{eq:H_YN}
\end{align}
In this context, $A$ is the total number of baryons, $A=\mathcal N_N+\mathcal N_Y$. Latin indices $i,j,k=1,\ldots,\mathcal N_N$ label nucleons and Greek symbols $\lambda,\mu,\nu=1,\ldots,\mathcal N_Y$ are used for the hyperons. The Hamiltonians (\ref{eq:H_N}) and (\ref{eq:H_Y}) contain the kinetic energy operator and two- and three-body interactions for nucleons and hyperons separately. In principles they could include higher order many-body forces that however are expected to be less important. The Hamiltonian (\ref{eq:H_YN}) describes the interaction between nucleons and hyperons, and it involves two-body ($YN$) and three-body ($YNN$ and $YYN$) forces. At present there is no evidence for higher order terms in the hyperon-nucleon sector.

As reported in the previous chapter, experimental data are mainly available for $\Lambda p$ scattering and $\Lambda$~hypernuclei and present experimental efforts are still mostly concentrated in the study of the $S=-1$ hypernuclear sector. Information on heavier hyperon-nucleon scattering and on $\Sigma$ or more exotic hypernuclei are very limited. For these reasons, from now on we will focus on the phenomenological interactions involving just the $\Lambda$~hyperon. We adopt the class of Argonne-like $\Lambda$-nucleon interaction for the strange sector and the nucleon-nucleon Argonne force with the corresponding TNIs (UIX and ILx) for the non-strange sector. An effective $\Lambda\Lambda$ interaction has been also employed.

\section{Interactions: nucleons}
\label{sec:nuc_int}

We report the details of the $NN$ Argonne potential~\cite{Wiringa:1995,Wiringa:2002} and the corresponding TNIs, the Urbana~IX~(UIX)~\cite{Carlson:1983} and the Illinois~(ILx)~\cite{Pieper:2001}. These interactions are written in coordinate space and they include different range components coming from meson (mostly pion) exchange and phenomenological higher order contributions.

\subsection{Two-body $NN$ potential}
\label{subsec:AV18}

The nucleon-nucleon potential Argonne~V18~(AV18)~\cite{Wiringa:1995} contains a complete electromagnetic (EM) interaction and a strong interaction part which is written as a sum of a long-range component $v_{ij}^\pi$ due to one-pion exchange~(OPE) and a phenomenological intermediate- and short-range part $v_{ij}^R$ :
\begin{equation}
	v_{ij}=v_{ij}^\pi+v_{ij}^R \;.
\end{equation}

Ignoring isospin breaking terms, the long-range OPE is given by
\begin{align}
	v_{ij}^\pi=\frac{f_{\pi NN}^2}{4\pi}\frac{m_\pi}{3}\,X_{ij}\,\bm\tau_i\cdot\bm\tau_j \;,
	\label{eq:OPE}
\end{align}
where $\tfrac{f_{\pi NN}^2}{4\pi}=0.075$ is the pion-nucleon coupling constant~\cite{Stoks:1993_pi} and
\begin{align}
	X_{ij}=Y_\pi(r_{ij})\,\bm\sigma_i\cdot\bm\sigma_j+T_\pi(r_{ij})\,S_{ij} \;.
	\label{eq:X_ij}
\end{align}
$\bm \sigma_i$ and $\bm \tau_i$ are Pauli matrices acting on the spin or isospin of nucleons and $S_{ij}$ is the tensor operator
\begin{align}
	S_{ij}=3\left(\bm\sigma_i\cdot\hat{\bm r}_{ij}\right)\left(\bm\sigma_j\cdot\hat{\bm r}_{ij}\right)-\bm\sigma_i\cdot\bm\sigma_j \;.
	\label{eq:S_ij}
\end{align}
The pion radial functions associated with the spin-spin (Yukawa potential) and tensor (OPE tensor potential) parts are
\begin{align}
	Y_\pi(r)&=\frac{\e^{-\mu_\pi r}}{\mu_\pi r}\xi_Y(r) \;, \label{eq:Y_pi} \\[0.5em]
	T_\pi(r)&=\left[1+\frac{3}{\mu_\pi r}+\frac{3}{(\mu_\pi r)^2}\right]\frac{\e^{-\mu_\pi r}}{\mu_\pi r}\xi_T(r) \;, \label{eq:T_pi}
\end{align}
where $\mu_\pi$ is the pion reduced mass
\begin{align}
	\mu_\pi=\frac{m_\pi}{\hbar}=\frac{1}{\hbar}\frac{m_{\pi^0}+2\,m_{\pi^\pm}}{3} \quad\quad \frac{1}{\mu_\pi}\simeq1.4~\text{fm} \;,
	\label{eq:m_pi}
\end{align}
and $\xi_Y(r)$ and $\xi_T(r)$ are the short-range cutoff functions defined by
\begin{align}
	\xi_Y(r)=\xi_T^{1/2}(r)=1-\e^{-cr^2} \quad\quad c=2.1~\text{fm}^{-2}\;.
\end{align}
It is important to note that since $T_\pi(r)\gg Y_\pi(r)$ in the important region where $r\lesssim 2$~fm, the OPE is dominated by the tensor part.

The remaining intermediate- and short-range part of the potential is expressed as a sum of central, $L^2$, tensor, spin-orbit and quadratic spin-orbit terms (respectively labelled as $c$, $l2$, $t$, $ls$, $ls2$) in different $S$, $T$ and $T_z$ states: 
\begin{align}
	\!v_{NN}^R=v_{NN}^c(r)+v_{NN}^{l2}(r)\bm L^2+v_{NN}^t(r)S_{12}+v_{NN}^{ls}(r)\bm L\!\cdot\!\bm S+v_{NN}^{ls2}(r)(\bm L\!\cdot\!\bm S)^2\;,
\end{align}
with the radial functions $v_{NN}^k(r)$ written in the general form
\begin{align}
	v_{NN}^k(r)=I_{NN}^k\,T_\pi^2(r)+\bigg[P_{NN}^k + (\mu_\pi r)\,Q_{NN}^k + (\mu_\pi r)^2\,R_{NN}^k \bigg]\, W(r) \;,
\end{align}
where the $T_\pi^2(r)$ has the range of a two-pion exchange~(TPE) force and $W(r)$ is a Wood-Saxon function which provides the short-range core:
\begin{align}
	W(r)=\Bigl(1+\e^{\frac{r-\bar r}{a}}\Bigr)^{-1} \quad\quad \bar r=0.5~\text{fm},\quad a=0.2~\text{fm}\;.
\end{align}
By imposing a regularization condition at the origin, it is possible to reduce the number of free parameters by one for each $v_{NN}^k(r)$. 
All the parameters in the $\xi(r)$ short-range cutoff functions as well as the other phenomenological constants are fitted on the $NN$ Nijmegen scattering data~\cite{Bergervoet:1990,Stoks:1993}.

The two-body nucleon potential described above can be projected from $S$, $T$, $T_z$ states into an operator format with 18 terms 
\begin{align}
	v_{ij}=\sum_{p=1,18}v_p(r_{ij})\,\mathcal O_{ij}^{\,p} \;.\label{eq:v_ij_Op}
\end{align}

The first 14 operators are charge independent and they are the ones included in the Argonne V14 potential~(AV14):
\begin{align}
	\mathcal O_{ij}^{\,p=1,8} &=\Bigl\{1,\bm\sigma_i\cdot\bm\sigma_j,S_{ij},\bm L_{ij}\cdot\bm S_{ij}\Bigr\}\otimes\Bigl\{1,\bm\tau_i\cdot\bm\tau_j\Bigr\} \;,\\[0.5em]
 	\mathcal O_{ij}^{\,p=9,14}&=\Bigl\{\bm L_{ij}^2,\bm L_{ij}^2\;\bm\sigma_i\cdot\bm\sigma_j,\left(\bm L_{ij}\cdot\bm S_{ij}\right)^2\Bigr\}
 								\otimes\Bigl\{1,\bm\tau_i\cdot\bm\tau_j\Bigr\} \;.
\end{align}
The first eight terms give the higher contribution to the $NN$ interaction and they are the standard ones required to fit $S$ and $P$ wave data in both triplet and singlet isospin states. The first six of them come from the long-range part of OPE and the last two depend on the velocity of nucleons and give the spin-orbit contribution. In the above expressions, $\bm L_{ij}$ is the relative angular momentum of a couple~$ij$ 
\begin{align}
	\bm L_{ij}=\frac{1}{2i}({\bf r}_i-{\bf r}_j)\times(\bm\nabla_i-\bm\nabla_j) \;,\label{eq:LS_ij1}
\end{align}
and $\bm S_{ij}$ the total spin of the pair
\begin{align}
 	\bm S_{ij}=\frac{1}{2}(\bm\sigma_i+\bm\sigma_j) \;.\label{eq:LS_ij2}
\end{align}
Operators from 9 to 14 are included to better describe the Nijmegen higher partial waves phase shifts and the splitting of state with different $J$ values. However, the contribution of these operators is small compared to the total potential energy.

The four last additional operators of the AV18 potential account for the charge symmetry breaking effect, mainly due to the different masses of charged and neutral pions, and they are given by
\begin{align}
	\mathcal O_{ij}^{\,p=15,18}=\Bigl\{T_{ij},(\bm\sigma_i\cdot\bm\sigma_j)\,T_{ij},S_{ij}\,T_{ij},\tau_i^z+\tau_j^z\Bigr\} \;,
\end{align}
where $T_{ij}$ is the isotensor operator defined in analogy with $S_{ij}$ as
\begin{align}
	T_{ij}=3\,\tau_i^z\tau_j^z-\bm\tau_i\cdot\bm\tau_j \;.
\end{align}
The contribution to the total energy given by these four operators is however rather small.

In QMC calculations reduced versions of the original AV18 potential are often employed. The most used one is the Argonne V8'~(AV8')~\cite{Wiringa:2002} that contains only the first eight operators and it is not a simple truncation of AV18 but also a reprojection, which preserves the isoscalar part in all $S$ and $P$ partial waves as well as in the $^3D_1$ wave and its coupling to $^3S_1$. AV8' is about $0.2\div0.3$~MeV per nucleon more attractive than Argonne V18 in light nuclei~\cite{Pieper:2001,Wiringa:2002,Wiringa:2002_url}, but its contribution is very similar to AV18 in neutron drops, where the difference is about 0.06~MeV per neutron~\cite{Pieper:2001}. Other common solutions are the Argonne V6’~(AV6') and V4’~(AV4') potentials~\cite{Wiringa:2002}. AV6' is obtained by deleting the spin-orbit terms from AV8' and adjusting the potential to preserve the deuteron binding. The spin-orbit terms do not contribute to $S$-wave and $^1P_1$ channel of the $NN$ scattering and are the smallest contributors to the energy of $^4$He~\cite{Kamada:2001}, but they are important in differentiating between the $^3P_{0,1,2}$ channels. The AV4' potential eliminates the tensor terms. As a result, the $^1S_0$ and $^1P_1$ potentials are unaffected, but the coupling between $^3S_1$ and $^3D_1$ channels is gone and the $^3P_{0,1,2}$ channels deteriorate further. The Fortran code for the AV18 and AVn' potentials is available at the webpage~\cite{Wiringa:1994}.

\subsection{Three-body $NNN$ potential}
\label{subsec:UIX-ILx}

The Urbana IX three-body force was originally proposed in combination with the Argonne AV18 and AV8'~\cite{Carlson:1983}. Although it slightly underbinds the energy of light nuclei, it has been extensively used to study the equation of state of nuclear and neutron matter~\cite{Akmal:1998,Sarsa:2003,Li:2008,Gandolfi:2009,Gandolfi:2009_gap,Gandolfi:2012}. The Illinois forces~\cite{Pieper:2001}, the most recent of which is the Illinois-7~(IL7)~\cite{Pieper:2008_AIP}, have been introduced to improve the description of both ground- and excited-states of light nuclei, showing an excellent accuracy~\cite{Pieper:2001,Pieper:2005}, but they produce an unphysical overbinding in pure neutron systems~\cite{Maris:2013}.

The three-body Illinois potential consists of two- and three-pion exchange and a phenomenological short-range component (the UIX force does not include the three-pion rings):
\begin{align}
	V_{ijk}=V_{ijk}^{2\pi}+V_{ijk}^{3\pi}+V_{ijk}^R \;.
	\label{eq:V_ijk}
\end{align}

\begin{figure}[ht]
	\centering
	\subfigure[\label{fig:NNN_2pi_p}]{\includegraphics[height=3.7cm]{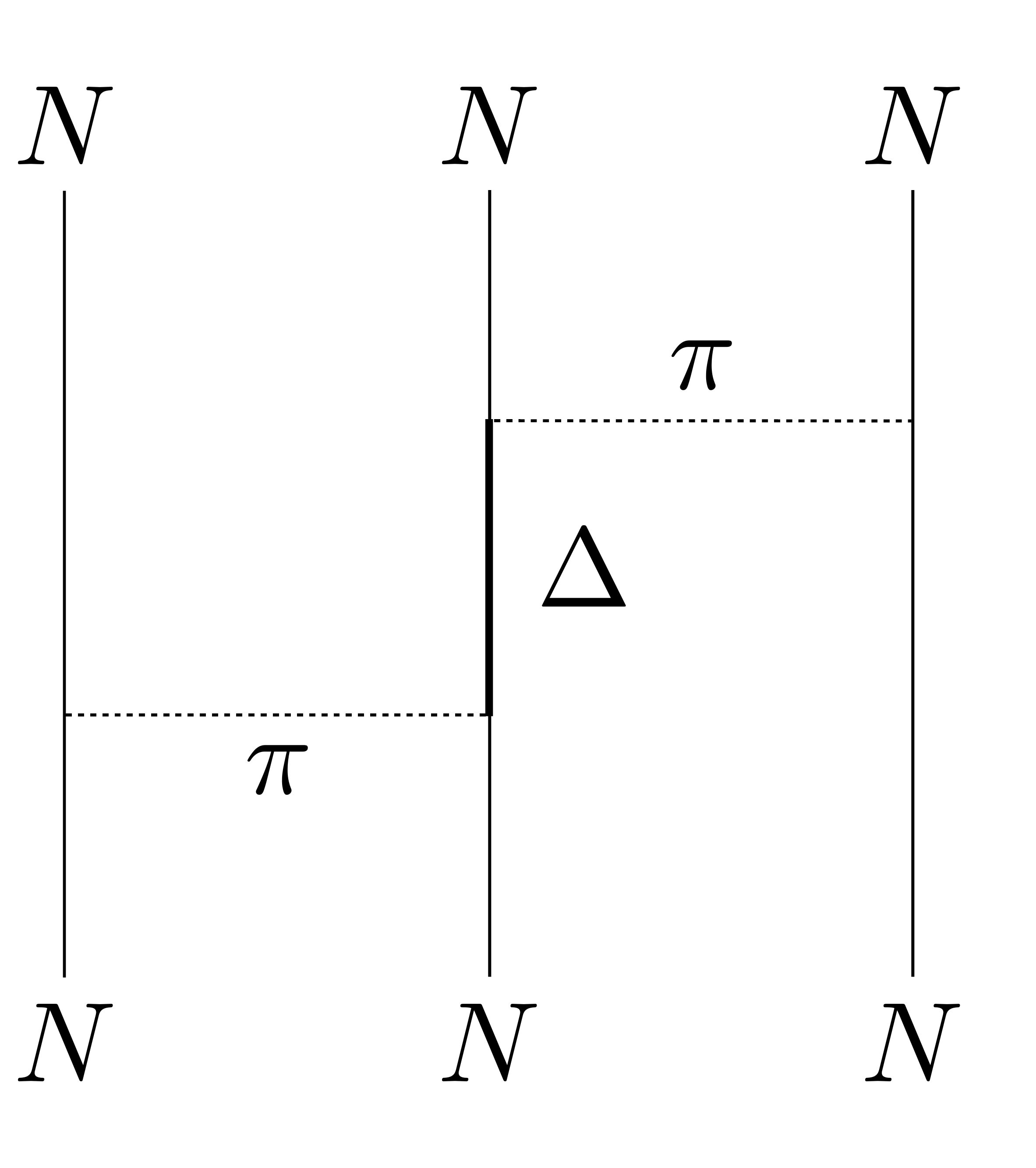}}
	\goodgap\goodgap\goodgap\goodgap\goodgap
	\subfigure[\label{fig:NNN_2pi_s}]{\includegraphics[height=3.7cm]{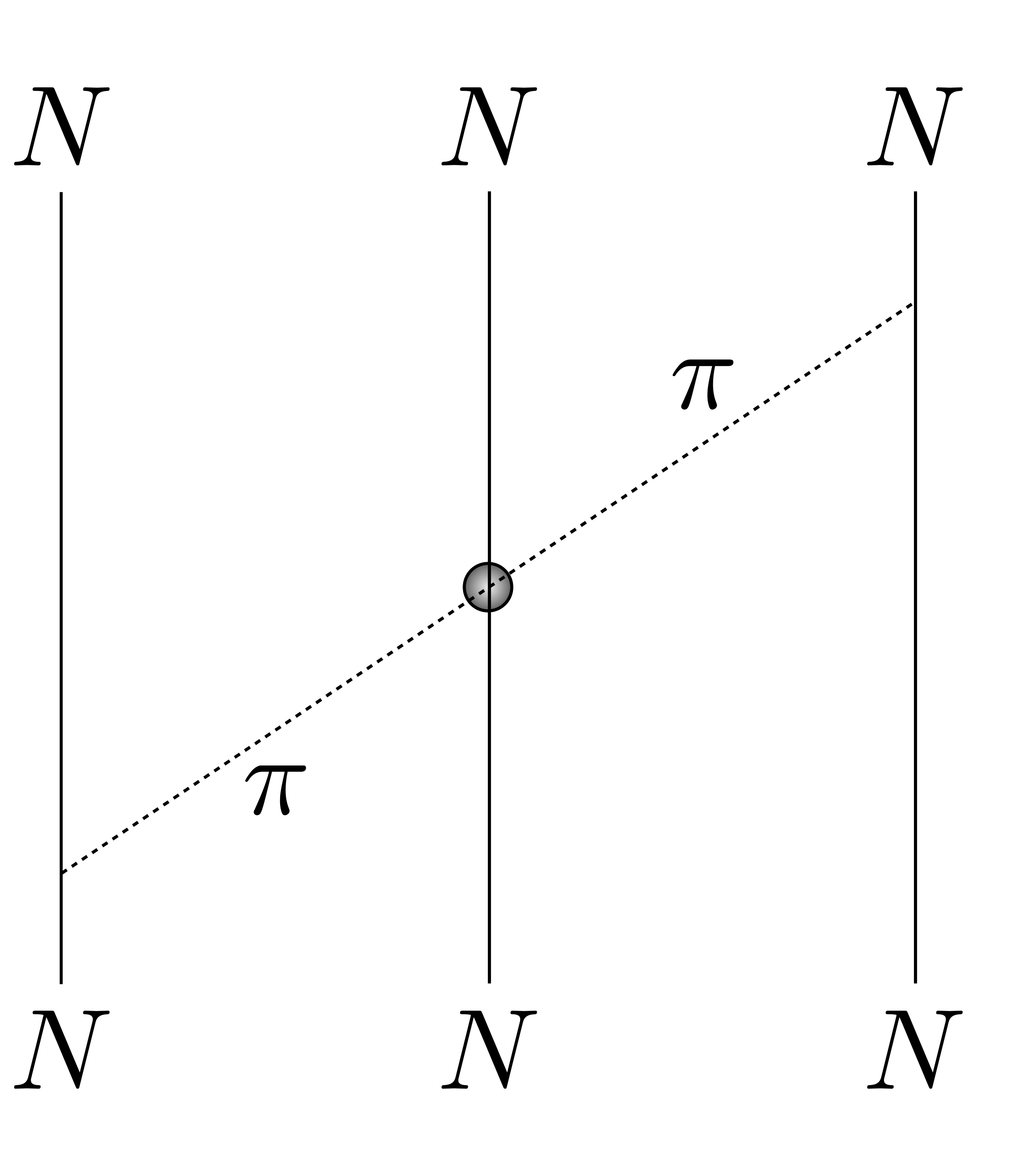}}
	\caption[Two-pion exchange processes in the $NNN$ force]
		{Two-pion exchange processes in the $NNN$ force. \ref{fig:NNN_2pi_p} is the Fujita-Miyazawa $P$-wave term and \ref{fig:NNN_2pi_s} the Tucson-Melbourne $S$-wave term.}
	\label{fig:NNN_2pi}
\end{figure}

The two-pion term, as shown in Fig.~\ref{fig:NNN_2pi}, contains $P$- and $S$-wave $\pi N$ scattering terms (respectively in Fig.~\ref{fig:NNN_2pi_p} and Fig.~\ref{fig:NNN_2pi_s}):
\begin{align}
	V_{ijk}^{2\pi}=V_{ijk}^{2\pi,P}+V_{ijk}^{2\pi,S}\;.
\end{align}

The $P$-wave component, originally introduced by Fujita-Miyazawa~\cite{Fujita:1957}, describes an intermediate excited $\Delta$ resonance produced by the exchange of two pions between nucleons $i$-$j$ and $j$-$k$, as shown in Fig.~\ref{fig:NNN_2pi_p}, and it can be written as
\begin{align}
	V_{ijk}^{2\pi,P}=A_{2\pi}^P\,\mathcal O_{ijk}^{2\pi,P}\;,
\end{align}
where
\begin{subequations}
	\begin{align}
		A_{2\pi}^P &=-\frac{2}{81}\frac{f_{\pi NN}^2}{4\pi}\frac{f_{\pi\Delta N}^2}{4\pi}\frac{m_\pi^2}{m_\Delta-m_N} \;,\\[0.5em]
		\mathcal O_{ijk}^{2\pi,P} &=\sum_{cyclic}\left(\phantom{\frac{1}{4}}\!\!\!\!\Bigl\{X_{ij},X_{jk}\Bigr\}
					 \Bigl\{\bm\tau_i\cdot\bm\tau_j,\bm\tau_j\cdot\bm\tau_k\Bigr\}
		           +\frac{1}{4}\Bigl[X_{ij},X_{jk}\Bigr]\Bigl[\bm\tau_i\cdot\bm\tau_j,\bm\tau_j\cdot\bm\tau_k\Bigr]\right) \;,
	\end{align}
	\label{eq:V_NNN_2pi_P} 
\end{subequations}
and the $X_{ij}$ operator is the same of Eq.~(\ref{eq:X_ij}). The constant $A_{2\pi}^P$ is fitted to reproduce the ground state of light nuclei and properties of nuclear matter.
The $P$-wave TPE term is the longest-ranged nuclear $NNN$ contribution and it is attractive in all nuclei and nuclear matter. However it is very small or even slightly repulsive in pure neutron systems.

The $S$-wave component of TPE three-nucleon force is a simplified form of the original Tucson-Melbourne model~\cite{Coon:1979}, and it involves the $\pi N$ scattering in the $S$-wave as shown in Fig.~\ref{fig:NNN_2pi_s}. It has the following form:
\begin{align}
	V_{ijk}^{2\pi,S}=A_{2\pi}^S\,\mathcal O_{ijk}^{2\pi,S}\;,
\end{align}
where
\begin{subequations}
	\begin{align}
		A_{2\pi}^{S}			  &	= \left(\frac{f_{\pi NN}}{4\pi}\right)^2 a' m_\pi^2 \;,	\\[0.5em]
		\mathcal O_{ijk}^{2\pi,S} &	= \sum_{cyclic}Z_\pi(r_{ij})Z_\pi(r_{jk})\,\bm\sigma_i\cdot\hat{\bm r}_{ij}\,\bm\sigma_k\cdot\hat{\bm r}_{kj}\,\bm\tau_i\cdot\bm\tau_k \;,	
	\end{align}
	\label{eq:V_NNN_2pi_S}
\end{subequations}
and the $Z_\pi(r)$ function is defined as
\begin{align}
	Z_\pi(r)=\frac{\mu_\pi r}{3}\Bigl[Y_\pi(r)-T_\pi(r)\Bigr] \;.
	\label{eq:Z_pi}
\end{align}
The $S$-wave TPE term is required by chiral perturbation theory but in practice its contribution is only 3\%--4\% of $V_{ijk}^{2\pi,P}$ in light nuclei.

\begin{figure}[h]
	\centering
	\subfigure[\label{fig:NNN_3pi_1}]{\includegraphics[height=3.7cm]{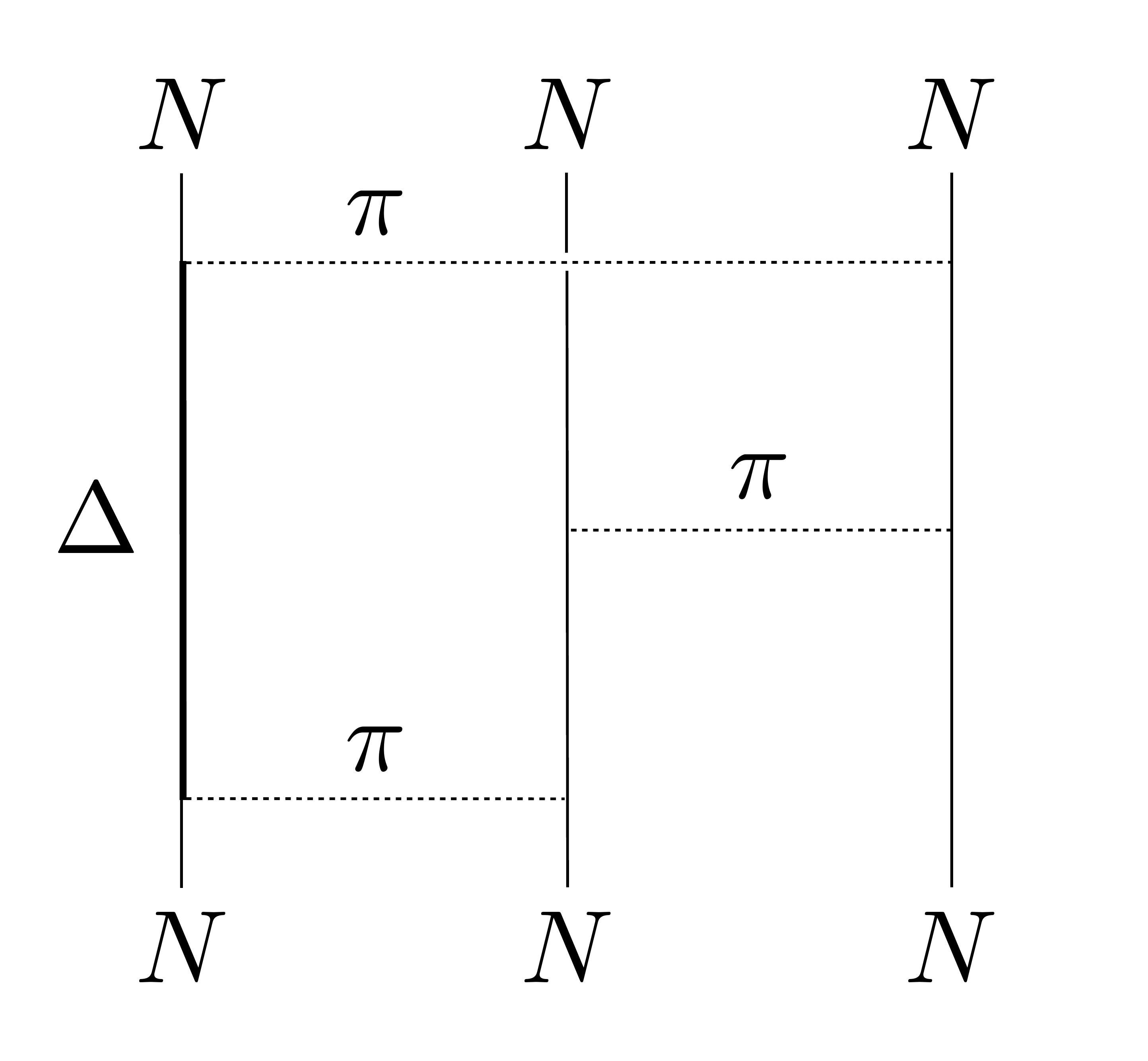}}
	\goodgap\goodgap\goodgap
	\subfigure[\label{fig:NNN_3pi_2}]{\includegraphics[height=3.7cm]{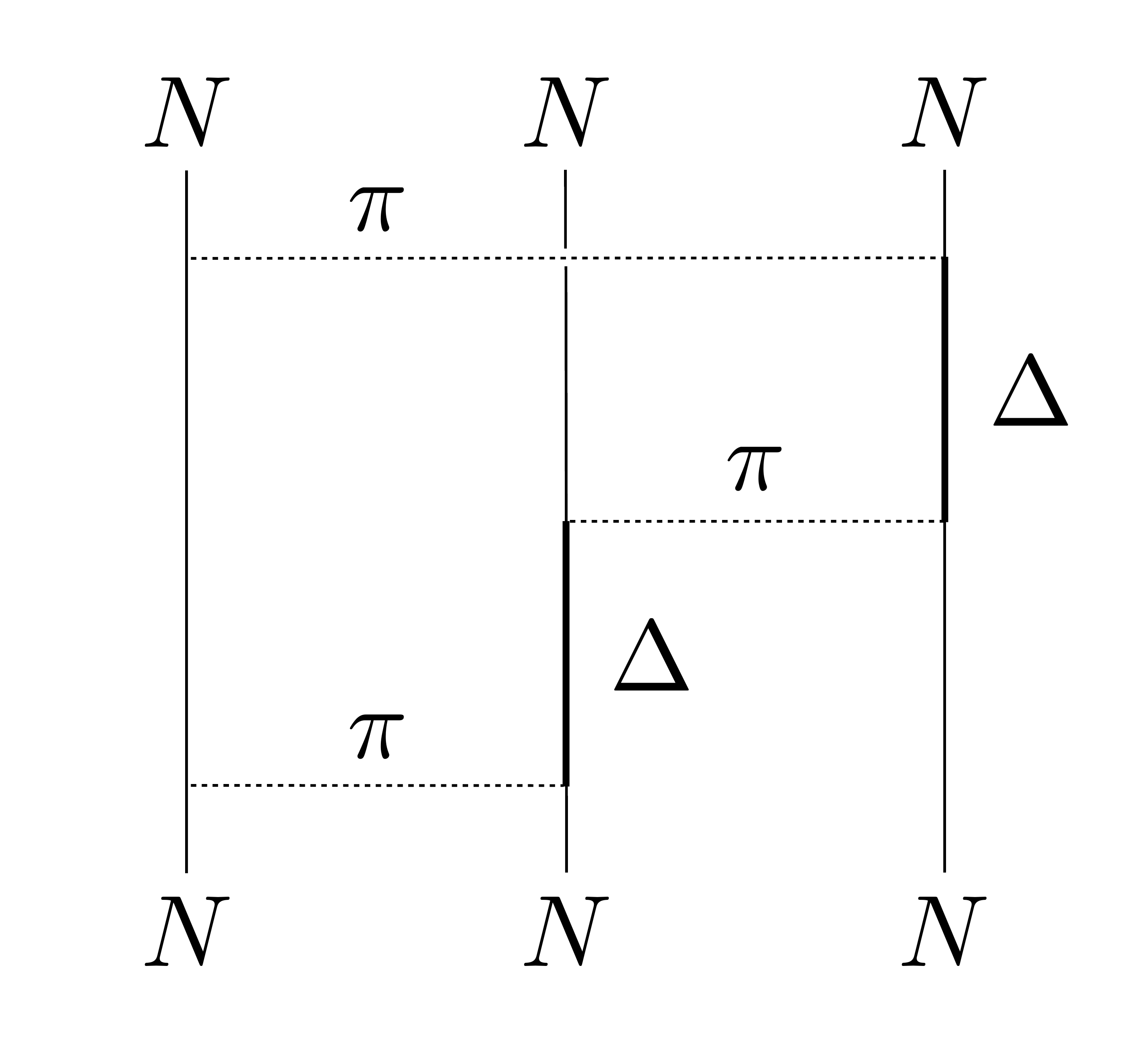}}
	\caption[Three-pion exchange processes in the $NNN$ force]
		{Three-pion exchange processes in the $NNN$ force.}
	\label{fig:NNN_3pi}
\end{figure}

The three-pion term (Fig.~\ref{fig:NNN_3pi}) was introduced in the Illinois potentials. It consists of the subset of three-pion rings that contain only one $\Delta$ mass in the energy denominators. 
As discussed in Ref.~\cite{Pieper:2001}, these diagrams result in a large number of terms, the most important of which are the ones independent of cyclic permutations of $ijk$:
\begin{align}
		V_{ijk}^{3\pi}=A_{3\pi}\,\mathcal O_{ijk}^{3\pi}\;,
\end{align}
where
\begin{subequations}
	\begin{align}
		A_{3\pi} & = \left(\frac{f^2_{\pi NN}}{4\pi}\frac{m_\pi}{3}\right)^3\frac{f^2_{\pi N\Delta}}{f^2_{\pi NN}}\frac{1}{(m_\Delta-m_N)^2} \;,\\[0.5em]
		\mathcal O_{ijk}^{3\pi} &\simeq\frac{50}{3} S_{ijk}^\tau\,S_{ijk}^\sigma+\frac{26}{3} A_{ijk}^\tau\,A_{ijk}^\sigma \;.
	\end{align}
\end{subequations}
The letters $S$ and $A$ denote operators that are symmetric and antisymmetric under the exchange of $j$ with $k$. Superscripts $\tau$ and $\sigma$ label operators containing isospin and spin-space parts, respectively. The isospin operators are
\begin{subequations}
	\begin{align}
		S_{ijk}^\tau &=2+\frac{2}{3}\left(\bm\tau_i\cdot\bm\tau_j+\bm\tau_j\cdot\bm\tau_k+\bm\tau_k\cdot\bm\tau_i\right)=4\,P_{T=3/2}\;,\\[0.5em]
		A_{ijk}^\tau &=\frac{1}{3}\,i\,\bm\tau_i\cdot\bm\tau_j\times\bm\tau_k=-\frac{1}{6}\Bigl[\bm\tau_i\cdot\bm\tau_j,\bm\tau_j\cdot\bm\tau_k\Bigr]\;,
	\end{align}
\end{subequations}
where $S_{ijk}^\tau$ is a projector onto isospin 3/2 triples and $A_{ijk}^\tau$ has the same isospin structure as the commutator part of $V_{ijk}^{2\pi,P}$.
The spin-space operators have many terms and they are listed in the Appendix of Ref.~\cite{Pieper:2001}. An important aspect of this structure is that there is a significant attractive term which acts only in $T=3/2$ triples, so the net effect of $V_{ijk}^{3\pi}$ is slight repulsion in $S$-shell nuclei and larger attraction in $P$-shell nuclei. However, in most light nuclei the contribution of this term is rather small, $\langle V_{ijk}^{3\pi}\rangle\lesssim 0.1 \langle V_{ijk}^{2\pi}\rangle$.

\begin{figure}[h]
	\centering
	\includegraphics[height=3.7cm]{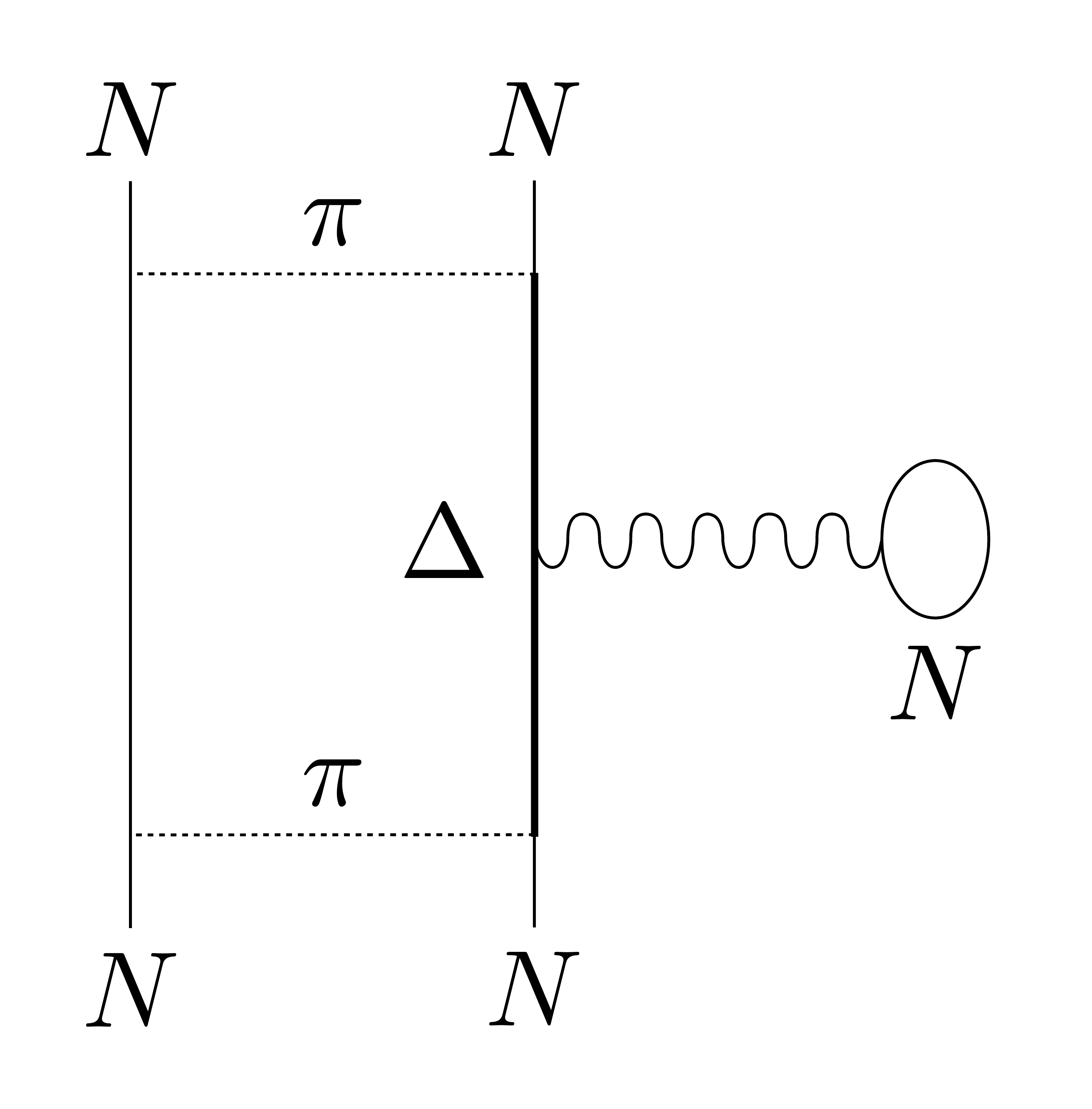}
	\caption[Short-range contribution in the $NNN$ force]{Repulsive short-range contribution included in the $NNN$ force.}
	\label{fig:NNN_2pi_d}
\end{figure}

The last term of Eq.~(\ref{eq:V_ijk}) was introduced to compensate the overbinding in nuclei and the large equilibrium density of nuclear matter given by the previous operators. It is strictly phenomenological and purely central and repulsive, and it describes the modification of the contribution of the TPE $\Delta$-box diagrams to $v_{ij}$ due to the presence of the third nucleon $k$ (Fig.~\ref{fig:NNN_2pi_d}). 
It takes the form:
\begin{align}
	V_{ijk}^R=A_R\,\mathcal O^R_{ijk}=A_R\sum_{cyclic}T_\pi^2(r_{ij})\,T_\pi^2(r_{jk}) \;,
\end{align}
where $T_\pi(r)$ is the OPE tensor potential defined in Eq.~(\ref{eq:T_pi}).

Finally, the Illinois (Urbana IX) TNI can be written as a sum of four different terms:
\begin{align}
	V_{ijk}=A_{2\pi}^P\,\mathcal O^{2\pi,P}_{ijk}+A_{2\pi}^{S}\,\mathcal O^{2\pi,S}_{ijk}+A_{3\pi}\,\mathcal O^{3\pi}_{ijk}+A_R\,\mathcal O^R_{ijk} \;.
\end{align}

\section{Interactions: hyperons and nucleons}
\label{sec:hyper_int}

We present a detailed description of the $\Lambda N$ and $\Lambda NN$ interaction as developed by Bodmer, Usmani and Carlson following the scheme of the Argonne potentials~\cite{Bodmer:1984,Bodmer:1985,Bodmer:1988,Usmani:1995,Usmani:1995_3B,Shoeb:1998,Usmani:1999,Sinha:2002,Usmani:2003,Usmani:2006,Usmani:2008}. The interaction is written in coordinates space and it includes two- and three-body hyperon nucleon components with an explicit hard-core repulsion between baryons and a charge symmetry breaking term. We introduce also an effective $\Lambda\Lambda$ interaction mainly used in variational~\cite{Usmani:2004,Usmani:2006_He6LL} and cluster model~\cite{Hiyama:1997,Hiyama:2002} calculations for double $\Lambda$~hypernuclei.

\subsection{Two-body $\Lambda N$ potential}
\label{subsec:LN}

\subsubsection{$\Lambda N$ charge symmetric potential}
\label{subsec:LN_sym}

\begin{figure}[h]
	\centering
	\subfigure[\label{fig:LN_2pi}]{\includegraphics[height=3.7cm]{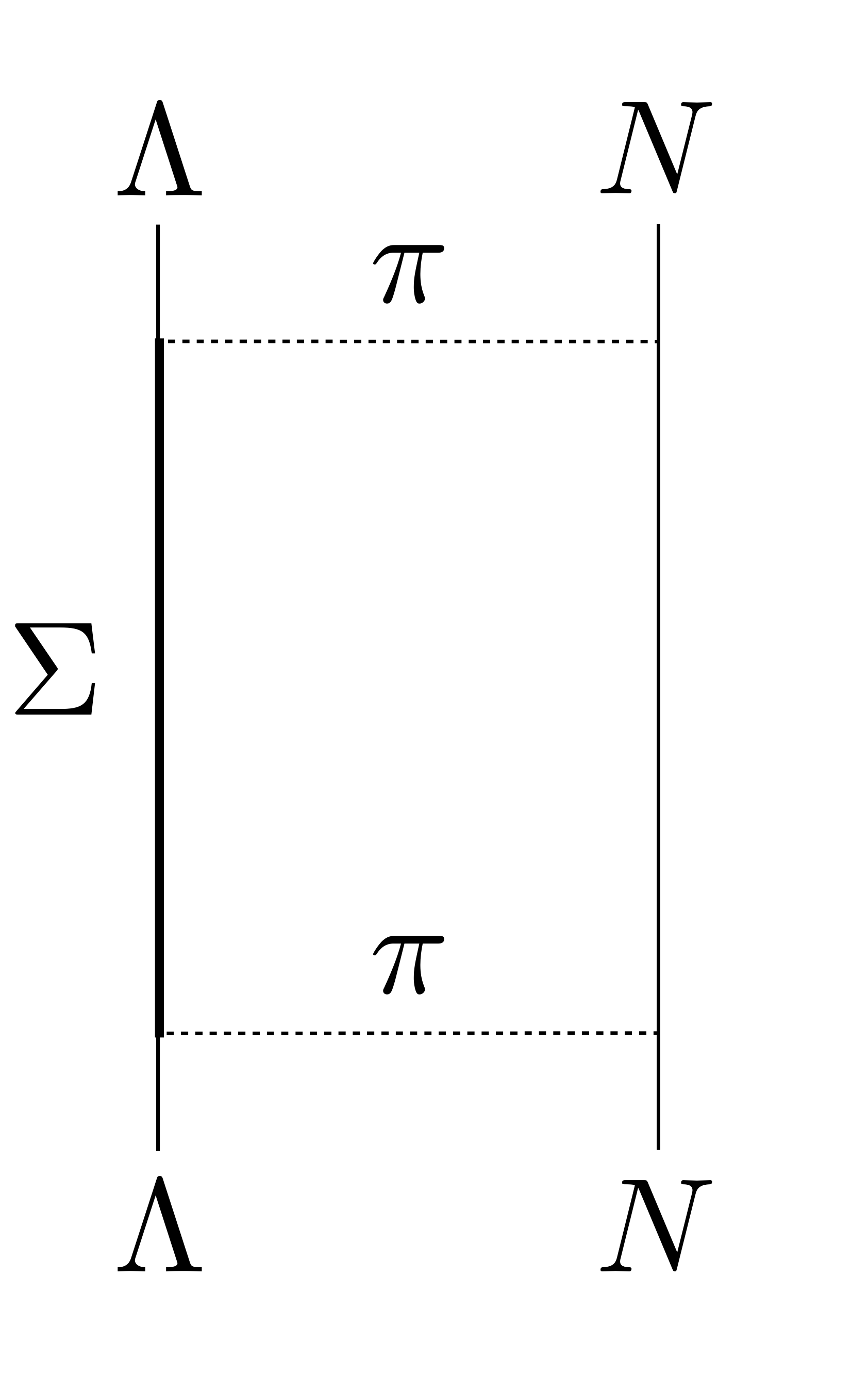}}
	\goodgap\goodgap\goodgap
	\subfigure[\label{fig:LN_2pi_2}]{\includegraphics[height=3.7cm]{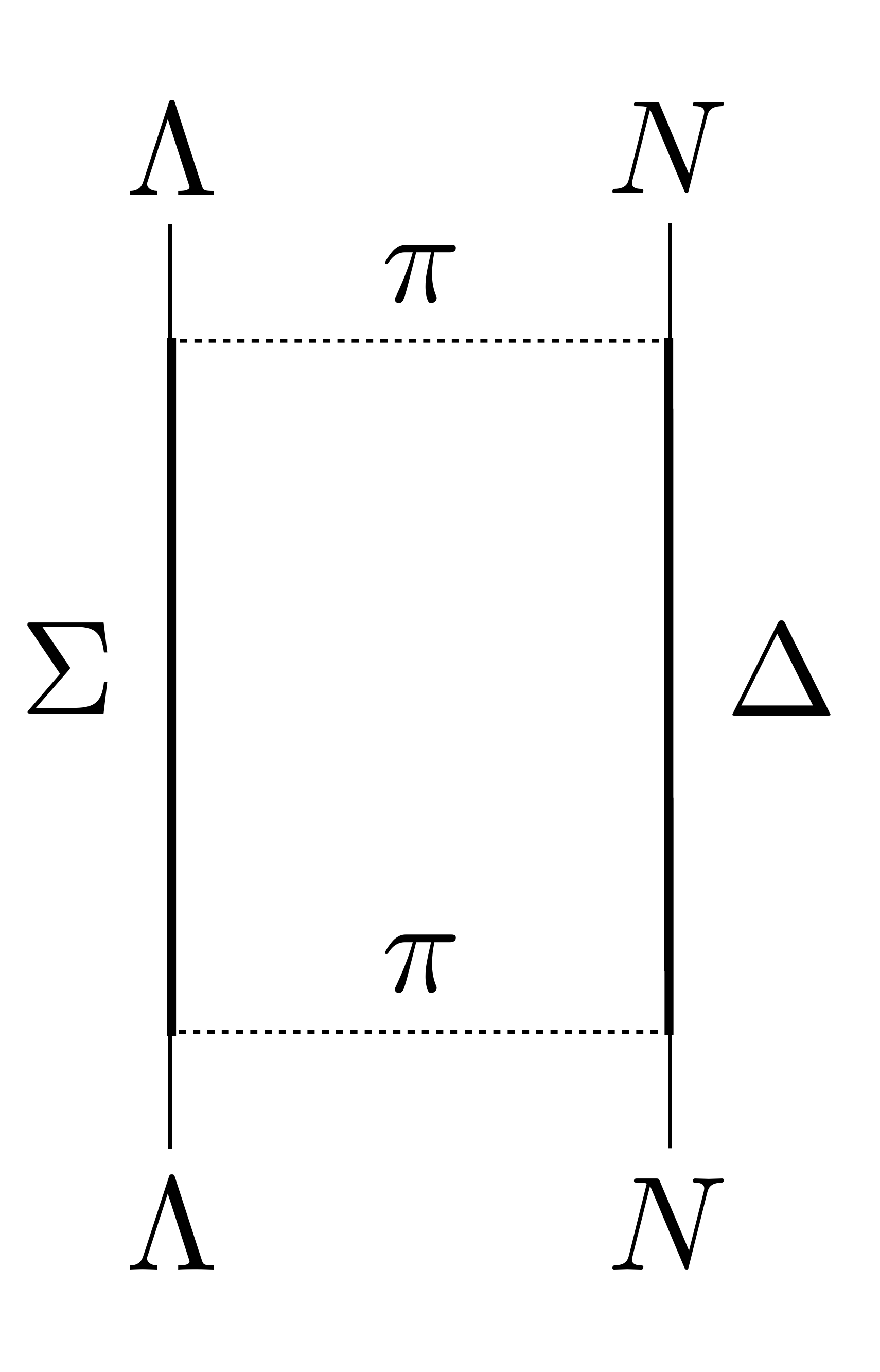}}
	\goodgap\goodgap\goodgap
	\subfigure[\label{fig:LN_K}]{\includegraphics[height=3.7cm]{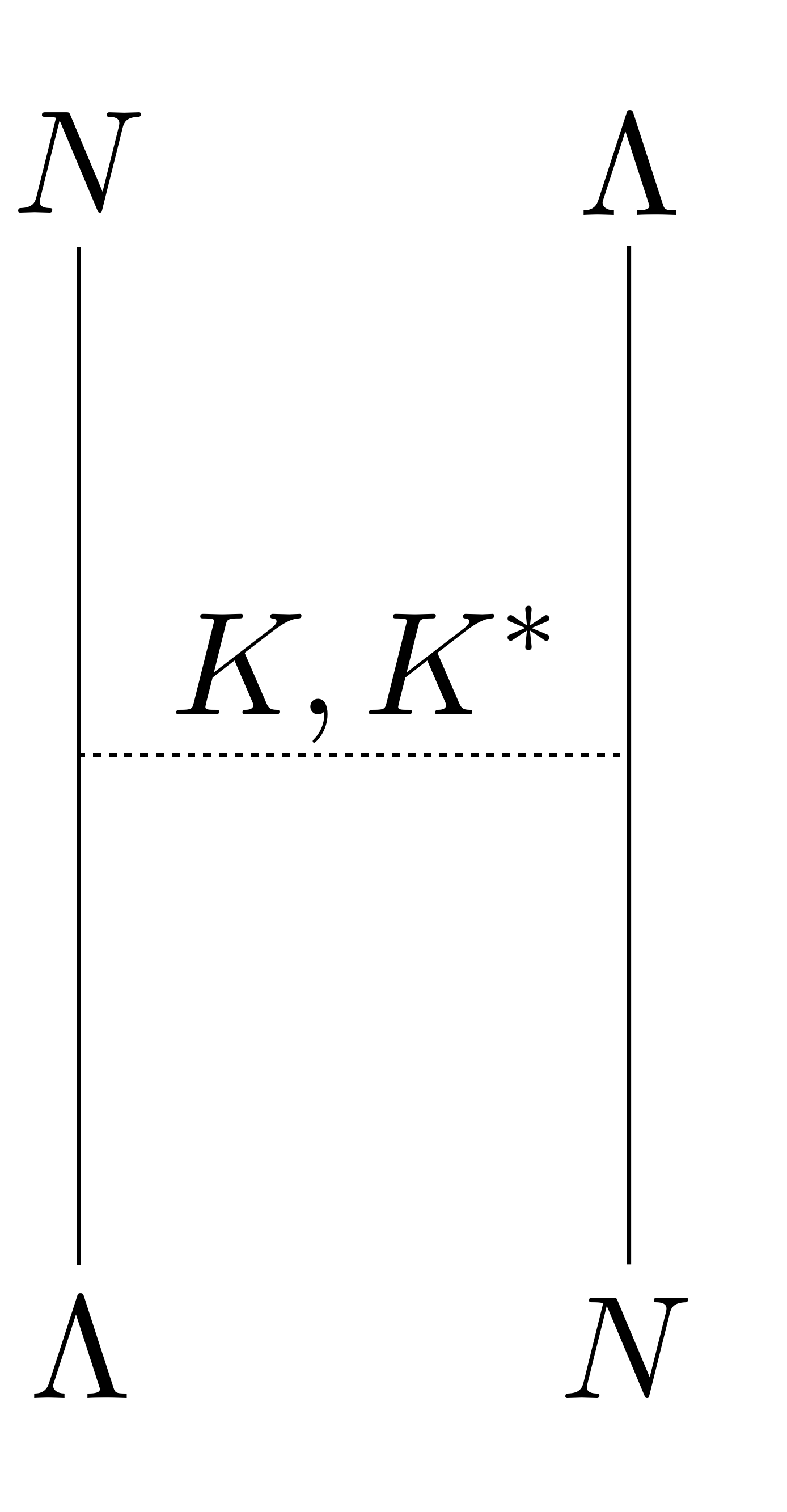}}
	\caption[Meson exchange processes in the $\Lambda N$ force]
		{Meson exchange processes in the $\Lambda N$ force. \ref{fig:LN_2pi} and \ref{fig:LN_2pi_2} are the TPE diagrams. \ref{fig:LN_K} represents the kaon exchange channel.}
	\label{fig:LN} 
\end{figure}

The $\Lambda$~particle has isospin $I=0$, so there is no OPE term, being the strong $\Lambda\Lambda\pi$ vertex forbidden due to isospin conservation. The $\Lambda$~hyperon can thus exchange a pion only with a $\Lambda\pi\Sigma$ vertex. The lowest order $\Lambda N$ coupling must therefore involve the exchange of two pions, with the formation of a virtual $\Sigma$ hyperon, as illustrated in Figs.~\ref{fig:LN_2pi} and \ref{fig:LN_2pi_2}. The TPE interaction is intermediate range with respect to the long range part of $NN$ force. One meson exchange processes can only occur through the exchange of a $K,K^*$ kaon pair, that contributes in exchanging the strangeness between the two baryons, as shown in Fig.~\ref{fig:LN_K}. The $K,K^*$ potential is short-range and contributes to the space-exchange and $\Lambda N$ tensor potential. The latter is expected to be quite weak because the $K$ and $K^*$ tensor contributions have opposite sign~\cite{Shinmura:1984}. 

The $\Lambda N$ interaction has been modeled with an Urbana-type potential~\cite{Lagaris:1981} with spin-spin and space-exchange components and a TPE tail which is consistent with the available $\Lambda p$ scattering data below the $\Sigma$ threshold:
\begin{align}
	v_{\lambda i}=v_0(r_{\lambda i})(1-\varepsilon+\varepsilon\,\mathcal P_x)+\frac{1}{4}v_\sigma T^2_\pi(r_{\lambda i})\,{\bm\sigma}_\lambda\cdot{\bm\sigma}_i \;,
	\label{eq:V_LN}
\end{align}
where
\begin{align}
	v_0(r_{\lambda i})=v_c(r_{\lambda i})-\bar v\,T^2_\pi(r_{\lambda i})\;.
\end{align}
Here,
\begin{align}
	v_c(r)=W_c \Bigl(1+\e^{\frac{r-\bar r}{a}}\Bigr)^{-1}
\end{align}
is a Wood-Saxon repulsive potential introduced, similarly to the Argonne $NN$ interaction, in order to include all the short-range contributions and $T_\pi(r)$ is the regularized OPE tensor operator defined in Eq.~(\ref{eq:T_pi}). The term $\bar v\,T^2_\pi(r_{\lambda i})$ corresponds to a TPE mechanism due to OPE transition potentials 
$\left(\Lambda N\leftrightarrow\Sigma N,\Sigma\Delta\right)$ dominated by their tensor components. The $\Lambda p$ scattering at low energies is well fitted with 
$\bar v=6.15(5)$~MeV. The terms $\bar v=(v_s+3v_t)/4$ and $v_\sigma=v_s-v_t$ are the spin-average and spin-dependent strengths, where $v_s$ and $v_t$ denote singlet- and triplet-state strengths, respectively. $\mathcal P_x$ is the $\Lambda N$ space-exchange operator and $\varepsilon$ the corresponding exchange parameter, which is quite poorly determined from the $\Lambda p$ forward-backward asymmetry to be $\varepsilon\simeq0.1\div0.38$. All the parameters defining the $\Lambda N$ potential can be found in Tab.~\ref{tab:parLN+LNN}.

\subsubsection{$\Lambda N$ charge symmetry breaking potential}
\label{subsubsec:LN_CSB}

The $\Lambda$-nucleon interaction should distinguish between the nucleon isospin channels $\Lambda p$ and $\Lambda n$. 
The mirror pair of hypernuclei $^4_\Lambda$H and $^4_\Lambda$He is the main source of information about the charge symmetry breaking (CSB) $\Lambda N$ interaction.  The experimental data for $A=4$ $\Lambda$~hypernuclei~\cite{Juric:1973}, show indeed a clear difference in the $\Lambda$~separation energies for the $(0^+)$ ground state
\begin{subequations}
	\begin{align}
		B_\Lambda\left(^4_\Lambda\text{H}\right)&=2.04(4)~\text{MeV}\;,\\[0.5em]
		B_\Lambda\left(^4_\Lambda\text{He}\right)&=2.39(3)~\text{MeV}\;,
	\end{align}
\end{subequations}
and for the $(1^+)$ excited state
\begin{subequations}
	\begin{align}
		B_\Lambda^*\left(^4_\Lambda\text{H}\right)&=1.00(6)~\text{MeV}\;,\\[0.5em]
		B_\Lambda^*\left(^4_\Lambda\text{He}\right)&=1.24(6)~\text{MeV}\;.	
	\end{align}
\end{subequations}
The differences in the hyperon separation energies are:
\begin{subequations}
	\begin{align}
		\Delta B_\Lambda&=0.35(6)~\text{MeV}\;,\\[0.5em]
		\Delta B_\Lambda^*&=0.24(6)~\text{MeV}\;.
	\end{align}
\end{subequations}
However, the experimental values $\Delta B_\Lambda$ must be corrected to include the difference $\Delta B_c$ due to the Coulomb interaction in order to obtain the values to be attributed to CSB effects. By means of a variational calculation, Bodmer and Usmani~\cite{Bodmer:1985} estimated the Coulomb contribution to be rather small
\begin{subequations}
	\begin{align}
		|\Delta B_c|&=0.05(2)~\text{MeV}\;,\\[0.5em]
		|\Delta B_c^*|&=0.025(15)~\text{MeV}\;,
	\end{align}
\end{subequations}
and they were able to reproduce the differences in the $\Lambda$ separation energies by means of a phenomenological spin dependent CSB potential. It was found that the CSB interaction is effectively spin independent and can be simply expressed (as subsequently reported in Ref.~\cite{Usmani:1999}) by
\begin{align}
	v_{\lambda i}^{CSB}=C_\tau\,T_\pi^2\left(r_{\lambda i}\right)\tau_i^z \quad\quad C_\tau=-0.050(5)~\text{MeV}\;.
	\label{eq:V_CSB}
\end{align}
Being $C_\tau$ negative, the $\Lambda p$ channel becomes attractive while the $\Lambda n$ channel is repulsive, consistently with the experimental results for $^4_\Lambda$H and $^4_\Lambda$He. The contribution of CSB is expected to be very small in symmetric hypernuclei (if Coulomb is neglected) but could have a significant effect in hypernuclei with an neutron (or proton) excess.

\subsection{Three-body $\Lambda NN$ potential}
\label{subsubsec:LNN}

\begin{figure}[ht]
	\centering
	\subfigure[\label{fig:LNN_pw}]{\includegraphics[height=3.2cm]{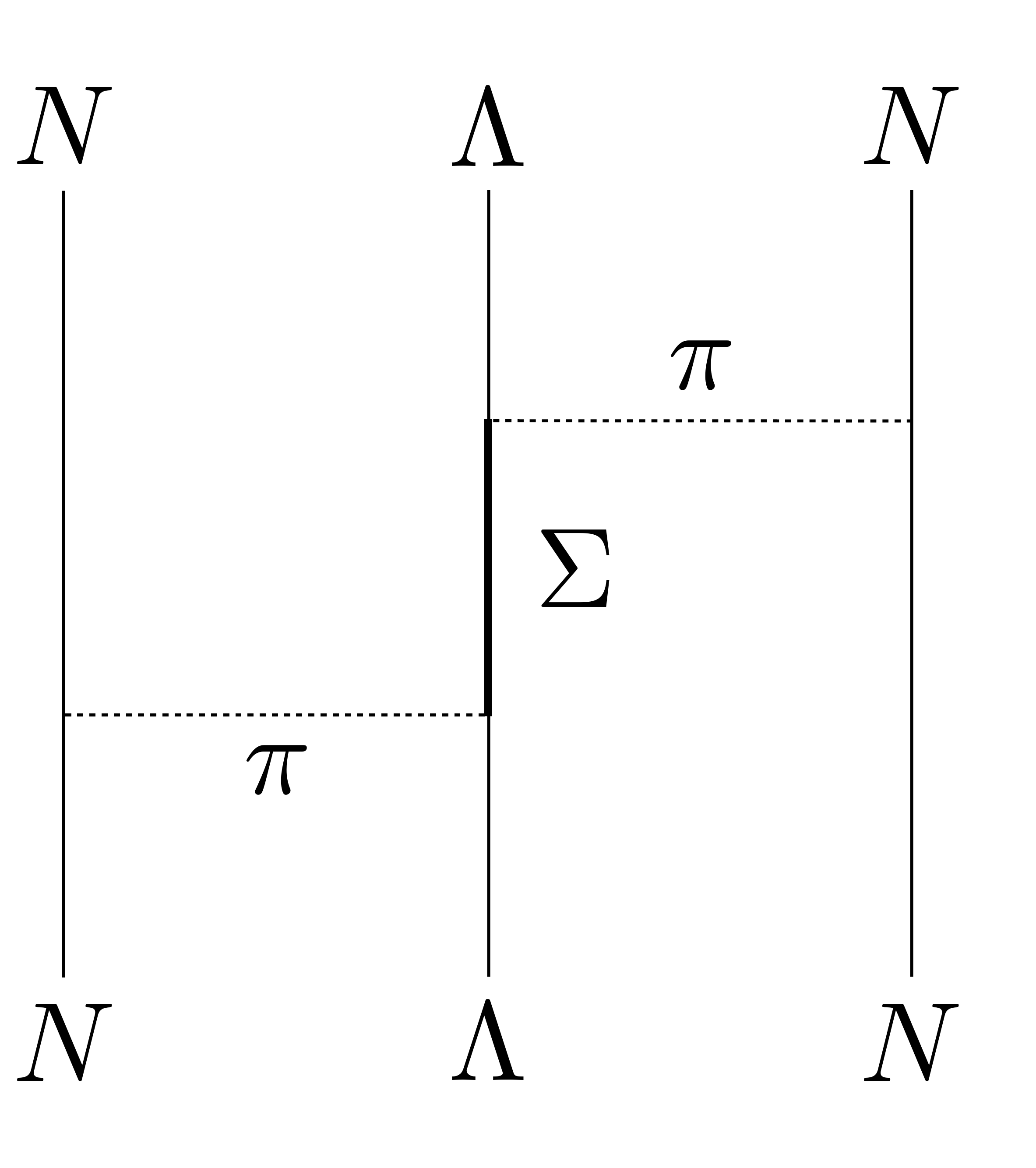}}
	\goodgap\goodgap\goodgap
	\subfigure[\label{fig:LNN_sw}]{\includegraphics[height=3.2cm]{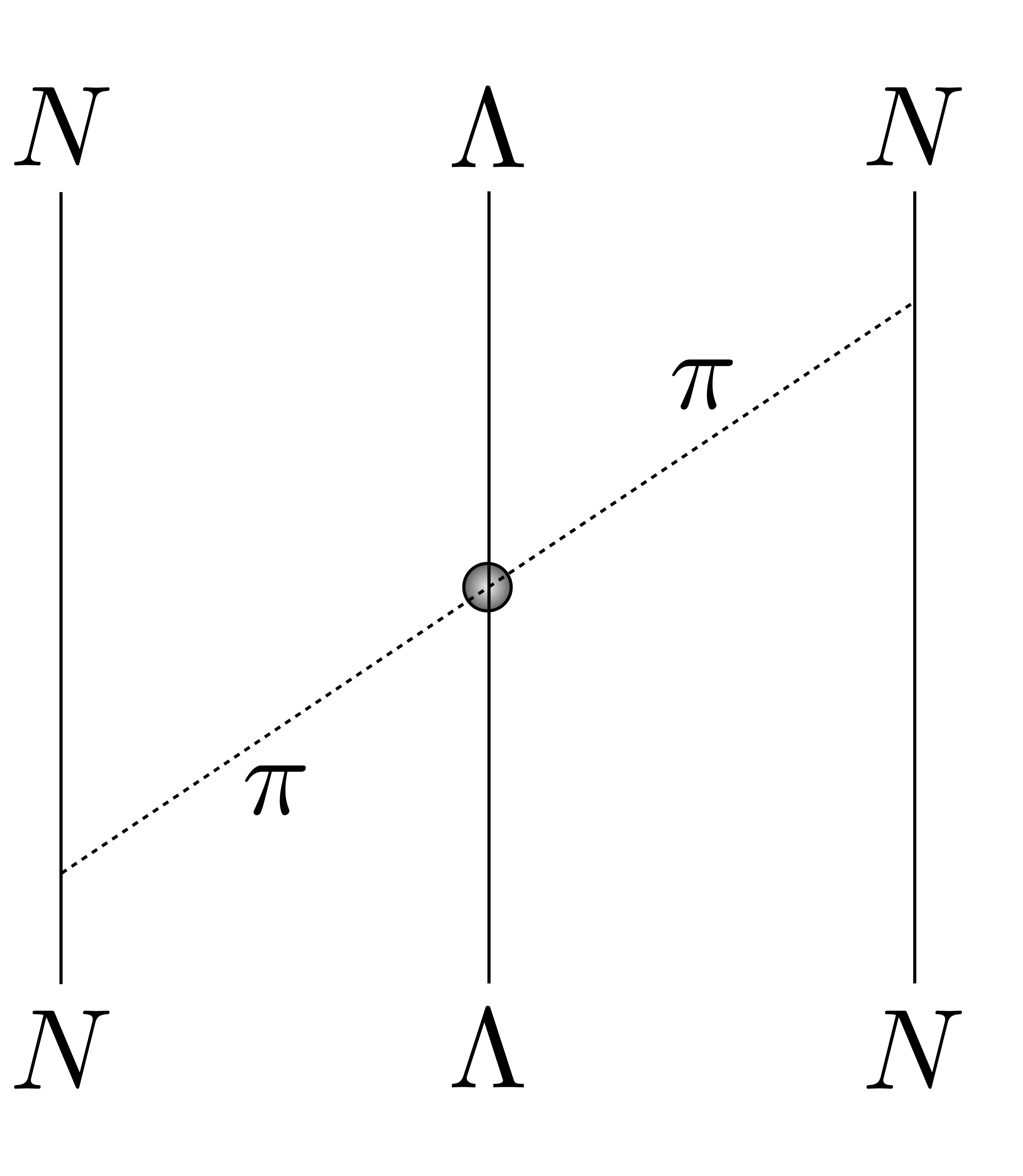}}
	\goodgap\goodgap\goodgap
	\subfigure[\label{fig:LNN_d}]{\includegraphics[height=3.2cm]{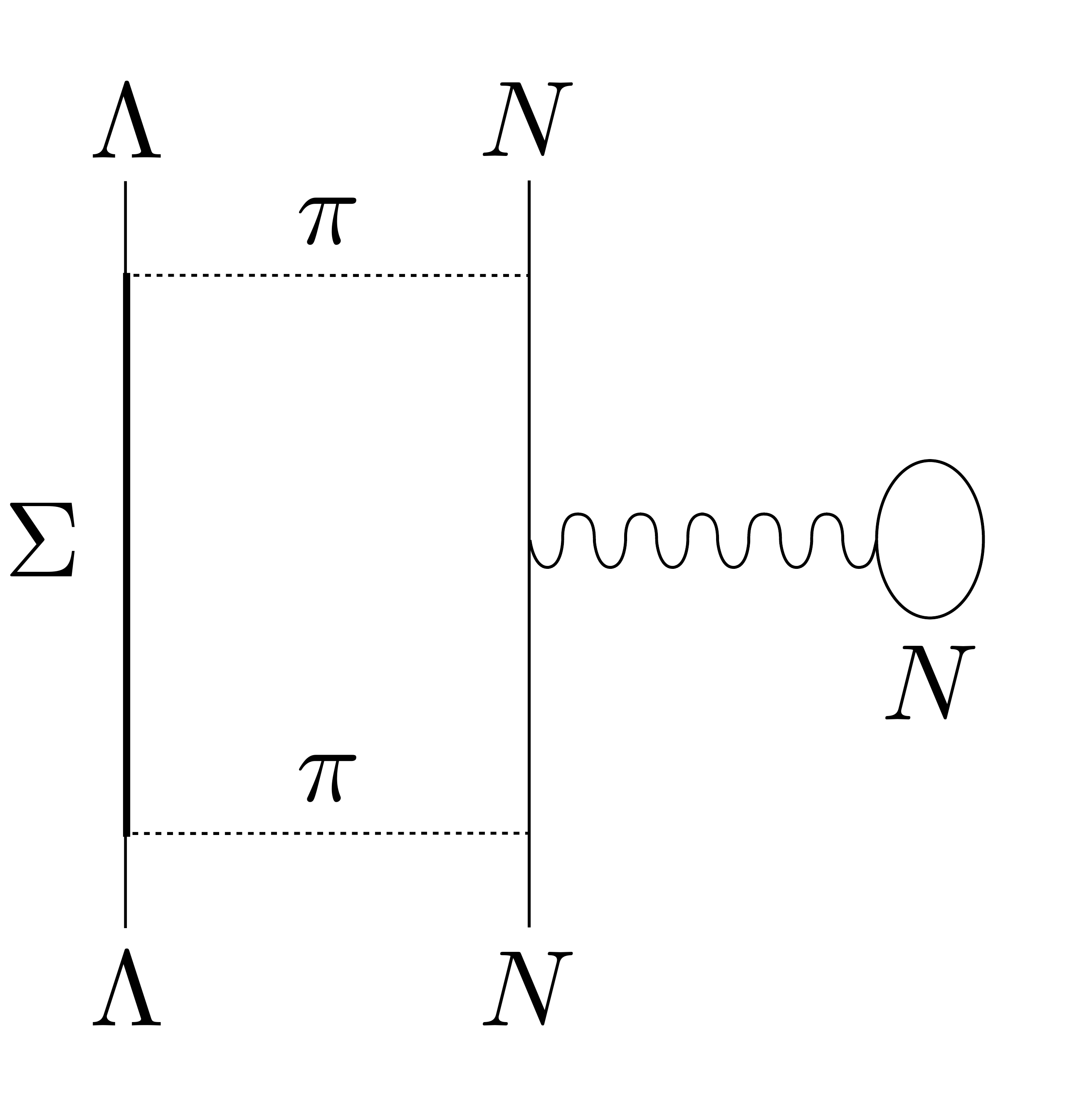}}
	\caption[Two-pion exchange processes in the $\Lambda NN$ force]
		{Two-pion exchange processes in three-body $\Lambda NN$ force. \ref{fig:LNN_pw} and \ref{fig:LNN_sw} are, respectively, the $P$- and $S$-wave TPE contributions. \ref{fig:LNN_d} is the phenomenological dispersive term.}
	\label{fig:LNN} 
\end{figure}

The $\Lambda N$ force as obtained by fitting the $\Lambda p$ scattering does not provide a good account of the experimental binding energies, as in the case of nuclei with the bare $NN$ interaction. A three-body $\Lambda NN$ force is required in this scheme to solve the overbinding. The $\Lambda NN$ potential is at the same TPE order of the $\Lambda N$ force and it includes diagrams involving two nucleons and one hyperon, as reported in Fig.~\ref{fig:LNN}.

The diagrams in Fig.~\ref{fig:LNN_pw} and Fig.~\ref{fig:LNN_sw} correspond respectively to the $P$-wave and $S$-wave TPE 
\begin{align}
	v^{2\pi}_{\lambda ij}=v^{2\pi,P}_{\lambda ij}+v^{2\pi,S}_{\lambda ij}\;,\label{eq:V_LNN_2pi}
\end{align}
that can be written in the following form:
\begin{align}
	v_{\lambda ij}^{2\pi,P}&=\widetilde C_P\,\mathcal O_{\lambda ij}^{2\pi,P}
		&&\hspace{-1.4cm}=-\frac{C_P}{6}\Bigl\{X_{i\lambda}\,,X_{\lambda j}\Bigr\}\,{\bm\tau}_{i}\cdot{\bm\tau}_{j}\;,\label{eq:V_LNN_2pi_P} \\[0.5em]
	v_{\lambda ij}^{2\pi,S}&=C_S\,O_{\lambda ij}^{2\pi,S}
		&&\hspace{-1.4cm}=C_S\,Z\left(r_{\lambda i}\right)Z\left(r_{\lambda j}\right)\,{\bm\sigma}_{i}\cdot\hat{\bm r}_{i\lambda}\,
		{\bm\sigma}_{j}\cdot\hat{\bm r}_{j\lambda}\,{\bm\tau}_{i}\cdot{\bm\tau}_{j}\;.\label{eq:V_LNN_2pi_S}
\end{align}
The structure of $V_{\lambda ij}^{2\pi}$ is very close to the Fujita-Miyazawa $P$-wave term and the Tucson-Melbourne $S$-wave term of the nuclear $V_{ijk}^{2\pi}$ (see Eqs.~(\ref{eq:V_NNN_2pi_P}) and (\ref{eq:V_NNN_2pi_S})). In the hypernuclear sector, however, there are simplifications because only two nucleons at a time enter the picture, so there are no cyclic summations, and the $\Lambda$~particle has isospin zero, thus there is no $\bm\tau_\lambda$ operator involved. As reported in Ref.~\cite{Pieper:2001}, the strength of $V_{ijk}^{2\pi,S}$ is $\left|A_{2\pi}^S\right|\simeq0.8$~MeV. However, in other references it is assumed to have a value of 1.0~MeV. Comparing the Tucson-Melbourne $NNN$ model with Eq.~(\ref{eq:V_LNN_2pi_S}) for the $\Lambda NN$ potential, one may write an identical structure for both $S$-wave $\Lambda NN$ and $NNN$ potentials as follows: 
\begin{align}
	C_S\,\mathcal O_{\lambda ij}^{2\pi,S}=A_S^{2\pi}\,\mathcal O_{ijk}^{2\pi,S}\;.
\end{align}
This directly relates $C_S$ in the strange sector to $A_{2\pi}^S$ in the non-strange sector. Since the $\Sigma$-$\Lambda$ mass difference is small compared to the $\Delta$-$N$ mass difference, the $2\pi$ $\Lambda NN$ potential of $S=-1$ sector is stronger than the non-strange $NNN$ potential of $S=0$ sector. This provides stronger strengths in the case of $\Lambda NN$ potential compared to the $NNN$ potential. It is therefore expected that the value of $C_S$ would be more than 1.0~MeV, and is taken to be 1.5~MeV~\cite{Usmani:2008}. However, the $S$-wave component is expected to be quite weak, at least in spin-zero core hypernuclei, and indeed it has been neglected in variational calculations for $^{17}_{~\Lambda}$O and $^5_\Lambda$He~\cite{Usmani:1995,Usmani:2003,Usmani:2004}.

The last diagram (Fig.~\ref{fig:LNN_d}) represents the dispersive contribution associated with the medium modifications of the intermediate state potentials for the $\Sigma$, $N$, $\Delta$ due to the presence of the second nucleon. This term describes all the short-range contributions and it is expected to be repulsive due to the suppression mechanism associated with the $\Lambda N$-$\Sigma N$ coupling~\cite{Bodmer:1971,Rozynek:1979}. The interaction of the intermediate states $\Sigma$, $N$, $\Delta$ with a nucleon of the medium will be predominantly through a TPE potential, proportional to $T_\pi^2(r)$, with an explicit spin dependence (negligible for spin-zero core hypernuclei):
\begin{align}
	v_{\lambda ij}^D=W_D\,\mathcal O_{\lambda ij}^D=W_D\,T_{\pi}^{2}\left(r_{\lambda i}\right)T^{2}_{\pi}\left(r_{\lambda j}\right)
		\!\!\bigg[1+\frac{1}{6}{\bm\sigma}_\lambda\!\cdot\!\left({\bm\sigma}_{i}+{\bm\sigma}_{j}\right)\bigg]\;.\label{eq:V_LNN_D}
\end{align}
The radial functions $T_\pi(r)$ and $Z_\pi(r)$ are the same of the nuclear potential, see Eq.~(\ref{eq:T_pi}) and Eq.~(\ref{eq:Z_pi}).
The operator $X_{\lambda i}$ is the same of Eq.~(\ref{eq:X_ij}), in which the first nucleon is replaced by the $\Lambda$~particle.

It is important to note that the three-body $\Lambda NN$ interaction have been investigated in variational calculations for $_\Lambda^5$He~\cite{Usmani:1995_3B,Usmani:2003,Usmani:2008}, $_{\Lambda\Lambda}^{\;\;\,6}$He~\cite{Usmani:2004,Usmani:2006_He6LL} and $_{~\Lambda}^{17}$O~\cite{Usmani:1995,Usmani:1995_3B}, resulting in a range of values for the $C_P$ and $W_D$ parameters (see Tab.~\ref{tab:parLN+LNN}) that gives good description of the properties of the studied hypernuclei. A unique set of parameters that reproduces all the available experimental energies for single (and double) $\Lambda$~hypernuclei has not been set yet.

A second crucial observation is that, differently to the nucleon sector, both two- and three-body lambda-nucleon interactions are at the same TPE order. In addition, the mass difference between the $\Lambda$~particle and its excitation $\Sigma$ is much smaller than the mass difference between the nucleon and the $\Delta$ resonance. Thus, the $\Lambda NN$ interaction can not be neglected in this framework but it is a key ingredient in addition to the $\Lambda N$ force for any consistent theoretical calculation involving $\Lambda$~hyperons.

\renewcommand{\arraystretch}{1.4}
\begin{table}[!hb]
	\centering
	\begin{tabular*}{\linewidth}{@{\hspace{5.0em}\extracolsep{\fill}}ccc@{\extracolsep{\fill}\hspace{5.0em}}}
		\toprule
		\toprule
		Constant      & Value            & Unit      \\
		\midrule      
		$W_c$		  & $2137$             & MeV       \\
		$\bar r$	  & $0.5$              & fm        \\
		$a$			  &	$0.2$              & fm        \\
		$v_s$		  &	$6.33, 6.28$       & MeV       \\
		$v_t$		  & $6.09, 6.04$       & MeV       \\
		$\bar v$	  & $6.15(5)$          & MeV       \\
		$v_\sigma$	  & $0.24$             & MeV       \\
		$c$			  &	$2.0$              & fm$^{-2}$ \\
		$\varepsilon$ & $0.1\div0.38$      & ---       \\
		$C_\tau$      & $$-0.050(5)$$      & MeV       \\
		$C_P$         &	$0.5\div2.5$       & MeV       \\
		$C_S$         &	$\simeq 1.5$       & MeV       \\
		$W_D$         &	$0.002\div0.058$   & MeV       \\
		\bottomrule
		\bottomrule
	\end{tabular*}
	\caption[Parameters of the $\Lambda N$ and $\Lambda NN$ interaction]
		{Parameters of the $\Lambda N$ and $\Lambda NN$ interaction~(see~\cite{Usmani:2008} and reference therein). 
		For $C_P$ and $W_D$ the variational allowed range is shown. The value of the charge symmetry breaking parameter $C_\tau$ is from Ref.~\cite{Usmani:1999}.\\}
	\label{tab:parLN+LNN}
\end{table}
\renewcommand{\arraystretch}{1.0}

\subsection{Two-body $\Lambda\Lambda$ potential}
\label{subsec:LL}

Due to the impossibility to collect $\Lambda\Lambda$ scattering data, experimental information about the $\Lambda\Lambda$ interaction can be obtained only from the $\Lambda\Lambda$~separation energy of the observed double $\Lambda$~hypernuclei, $^{\;\;\,6}_{\Lambda\Lambda}$He~\cite{Takahashi:2001,Nakazawa:2010,Ahn:2013}, $^{\;13}_{\Lambda\Lambda}$B~\cite{Nakazawa:2010} and the isotopes of $^{\;10}_{\Lambda\Lambda}$Be ($A=10\div12$)~\cite{Danysz:1963,Nakazawa:2010}. Evidence for the production of $_{\Lambda\Lambda}^{\;\;\,4}$H has been reported in Ref.~\cite{Ahn:2001}, but no information about the $\Lambda\Lambda$~separation energy was found.
On the other hand, there is a theoretical indication for the one-boson exchange (OBE) part of the $\Lambda\Lambda$ interaction coming from the $SU(3)$-invariance of coupling constants, but the $\Lambda\Lambda$ force is still far to be settled.

In the next, we follow the guide line adopted in the three- and four-body cluster models for double $\Lambda$~hypernuclei~\cite{Hiyama:1997,Hiyama:2002}, which was
also used in Faddeev-Yakubovsky calculations for light double $\Lambda$~hypernuclei~\cite{Filikhin:2002} and in variational calculations on $^{\;\;\,4}_{\Lambda\Lambda}$H~\cite{Shoeb:2004,Shoeb:2005}, $^{\;\;\,5}_{\Lambda\Lambda}$H and $^{\;\;\,5}_{\Lambda\Lambda}$He~\cite{Shoeb:2004,Shoeb:2007} and $^{\;\;\,6}_{\Lambda\Lambda}$He~\cite{Usmani:2004,Usmani:2006_He6LL,Shoeb:2004,Shoeb:2007}, with different parametrizations.
The employed OBE-simulating $\Lambda\Lambda$ effective interaction is a low-energy phase equivalent Nijmegen interaction represented by a sum of three Gaussians:
\begin{align}
	&v_{\lambda\mu}=\sum_{k=1}^{3}\left(v_0^{(k)}+v_\sigma^{(k)}\,{\bm\sigma}_\lambda\cdot{\bm\sigma}_\mu\right)\e^{-\mu^{(k)}r_{\lambda\mu}^2}\;. \label{eq:V_LL} 
\end{align}
The most recent parametrization of the potential (see Tab.~\ref{tab:parLL}), was fitted in order to simulate the $\Lambda\Lambda$ sector of the Nijmegen F (NF) interaction~\cite{Nagels:1979,Maessen:1989,Rijken:1999}. The NF is the simplest among the Nijmegen models with a scalar nonet, which seems to be more appropriate than the versions including only a scalar singlet in order to reproduce the weak binding energy indicated by the NAGARA event~\cite{Takahashi:2001}. The components $k=1,2$ of the above Gaussian potential are determined so as to simulate the $\Lambda\Lambda$ sector of NF and the strength of the part for $k=3$ is adjusted so as to reproduce the $^{\;\;\,6}_{\Lambda\Lambda}$He NAGARA experimental double $\Lambda$~separation energy of $7.25\pm 0.19^{+0.18}_{-0.11}$~MeV. In 2010, Nakazawa reported a new, more precise determination of $B_{\Lambda\Lambda}=6.93\pm0.16$~MeV for $^{\;\;\,6}_{\Lambda\Lambda}$He~\cite{Nakazawa:2010}, obtained via the $\Xi^-$ hyperon capture at rest reaction in a hybrid emulsion. This value has been recently revised to $B_{\Lambda\Lambda}=6.91\pm0.16$~MeV by the E373 (KEK-PS) Collaboration~\cite{Ahn:2013}. No references were found about the refitting of the $\Lambda\Lambda$ Gaussian potential on the more recent experimental result, which is in any case compatible with the NAGARA event. We therefore consider the original parametrization of Ref.~\cite{Hiyama:2002}.
\renewcommand{\arraystretch}{1.4}
\begin{table}[!b]
	\centering
	\begin{tabular*}{\linewidth}{@{\hspace{4.0em}\extracolsep{\fill}}cccc@{\extracolsep{\fill}\hspace{4.0em}}}
		\toprule
		\toprule
		$\mu^{(k)}$      & $0.555$  & $1.656$  & $8.163$ \\
		\midrule
		$v_0^{(k)}$      & $-10.67$ & $-93.51$ & $4884$  \\
		$v_\sigma^{(k)}$ & $0.0966$ & $16.08$  & $915.8$ \\
		\bottomrule
		\bottomrule
	\end{tabular*}
	\caption[Parameters of the $\Lambda\Lambda$ interaction]
		{Parameters of the the $\Lambda\Lambda$ interaction. 
		The size parameters $\mu^{(k)}$ are in fm$^{-2}$ and the strengths $v_0^{(k)}$ and $v_\sigma^{(k)}$ are in MeV~\cite{Hiyama:2002}.\\}
	\label{tab:parLL}
\end{table}
\renewcommand{\arraystretch}{1.0}

					% chapter 2: Hamiltonians
	% Chapter 3: Method

\renewcommand{\arraystretch}{1.2}

\chapter{Method}
\label{chap:method}

In nuclear physics, many-body calculations are used to understand the nuclear systems in the non-relativistic regime. When interested in low energy phenomena, a nucleus (or an extensive nucleonic system) can be described as a collection of particles interacting via a potential that depends on positions, momenta, spin and isospin. The properties of the system can be determined by solving a many-body Schr\"odinger equation. Such calculations can study, for example, binding energies, excitation spectra, densities, reactions and many other aspects of nuclei. The equation of state, masses, radii and other properties are obtained by describing astrophysical objects as a nuclear infinite medium. 

The two main problems related to microscopic few- and many-body calculations in nuclear physics are the determination of the Hamiltonian and the method used to accurately solve the Schr\"odinger equation. In the previous chapter, we have already seen how to build a realistic nuclear Hamiltonian, including also strange degrees of freedom. In the next we will focus on the methodological part presenting a class of Quantum Monte Carlo algorithms, the Diffusion Monte Carlo (DMC) and, more in detail, the Auxiliary Field Diffusion Monte Carlo (AFDMC). Such methods are based on evolving a trial wave function in imaginary time to yield the ground state of the system. The DMC method sums explicitly over spin and isospin states and can use very sophisticated wave functions. However, it is limited to small systems. In the AFDMC, in addition to the coordinates, also the spin and isospin degrees of freedom are sampled. It can thus treat larger systems but there are some limitations on the trial wave function and the nuclear potentials that can be handled. 

Strangeness can be included in AFDMC calculations by adding hyperons to the standard nucleons. The interaction between hyperons and nucleons presented in the previous chapter is written in a suitable form to be treated within this algorithm. By extending the AFDMC nuclear wave function to the hyperonic sector, it is possible to study both hypernuclei and hypermatter. A new QMC approach to strange physics is thus now available.

\section{Diffusion Monte Carlo}
\label{sec:DMC}

The Diffusion Monte Carlo method~\cite{Mitas:1998,Pieper:2008,Kalos:2008,Lipparini:2008} projects the ground-state out of a stationary trial wave function $|\psi_T\rangle$ not orthogonal to the true ground state. Consider the many-body time dependent Schr\"odinger equation with its formal solution
\begin{align}
	i\hbar\frac{\partial}{\partial t}|\psi(t)\rangle=(H-E_T)|\psi(t)\rangle\quad\Rightarrow\quad|\psi(t+dt)\rangle=\e^{-\frac{i}{\hbar}(H-E_T)dt}|\psi(t)\rangle\;,
\end{align}
and let move to the imaginary time $\tau=it/\hbar$\footnote{with this definition $\tau$ has the dimensions of the inverse of an energy.}:
\begin{align}
	-\frac{\partial}{\partial\tau}|\psi(\tau)\rangle=(H-E_T)|\psi(\tau)\rangle\quad\Rightarrow\quad|\psi(\tau+d\tau)\rangle=\e^{-(H-E_T)d\tau}|\psi(\tau)\rangle\;.\label{eq:psi_tau}
\end{align}
The stationary states $|\psi(0)\rangle=|\psi_T\rangle$ are the same for both normal and imaginary time Schr\"odinger equations and we can expand them on a complete orthonormal set of eigenvectors $|\varphi_n\rangle$ of the Hamiltonian $H$:
\begin{align}
	|\psi_T\rangle=\sum_{n=0}^\infty c_n|\varphi_n\rangle \;.
\end{align}
Supposing that the $|\psi_T\rangle$ is not orthogonal to the true ground state, i.e. $c_0\ne0$, and that at least the ground state is non degenerate, i.e. $E_n\ge E_{n-1}>E_0$, where $E_n$ are the eigenvalues of $H$ related to $|\varphi_n\rangle$, the imaginary time evolution of $|\psi_T\rangle$ is given by
\begin{align}
	|\psi(\tau)\rangle&=\sum_{n=0}^\infty c_n \e^{-(E_n-E_T)\tau}|\varphi_n\rangle \;,\nonumber\\[0.2em]
	&=c_0\e^{-(E_0-E_T)\tau}|\varphi_0\rangle+\sum_{n=1}^\infty c_n \e^{-(E_n-E_T)\tau}|\varphi_n\rangle \;.\label{eq:psi_0}
\end{align}
If the energy offset $E_T$ is the exact ground state energy $E_0$, in the limit $\tau\rightarrow\infty$ the components of Eq.~(\ref{eq:psi_0}) for $n>0$ vanish and we are left with
\begin{align}
	\lim_{\tau\rightarrow\infty}|\psi(\tau)\rangle=c_0|\varphi_0\rangle\;.\label{eq:tau_limit}
\end{align}
Starting from a generic initial trial wave function $|\psi_T\rangle$ not orthogonal to the ground state, and adjusting the energy offset $E_T$ to be as close as possible to $E_0$, in the limit of infinite imaginary time, one can project out the exact ground state $c_0|\varphi_0\rangle$ giving access to the lowest energy properties of the system. 

Consider the imaginary time propagation of Eq.~(\ref{eq:psi_tau}) and insert a completeness on the orthonormal basis $|R'\rangle$, where $R$ represents a configuration $\{\bm r_1,\ldots,\bm r_\mathcal N\}$ of the $\mathcal N$ particle system with all its degrees of freedom:
\begin{align}
	|\psi(\tau+d\tau)\rangle&=\e^{-(H-E_T)d\tau}|\psi(\tau)\rangle\;,\nonumber\\[0.2em]
	&=\int dR'\e^{-(H-E_T)d\tau}|R'\rangle\langle R'|\psi(\tau)\rangle\;.
\end{align}
Projecting on the coordinates $\langle R|$ leads to
\begin{align}
	\langle R|\psi(\tau+d\tau)\rangle=\int dR'\,\langle R|\e^{-(H-E_T)d\tau}|R'\rangle\langle R'|\psi(\tau)\rangle\;,\label{eq:psi_tau_dtau}
\end{align}
where $\langle R|\e^{-(H-E_T)d\tau}|R'\rangle=G(R,R',d\tau)$ is the Green's function of the operator $(H-E_T)+\frac{\partial}{\partial\tau}$. Recalling that 
$\langle R|\psi(\tau)\rangle=\psi(R,\tau)$, we can write Eq.~(\ref{eq:psi_tau}) as
\begin{align}
	-\frac{\partial}{\partial\tau}\psi(R,\tau)&=(H-E_T)\psi(R,\tau)\;,\label{eq:psi_R_tau}\\[0.5em]
	\psi(R,\tau+d\tau)&=\int dR'\,G(R,R',d\tau)\,\psi(R',\tau)\;.\label{eq:G}
\end{align}

If we consider a non-interacting many-body system, i.e. the Hamiltonian is given by the pure kinetic term
\begin{align}
	H_0=T=-\frac{\hbar^2}{2m}\sum_{i=1}^{\mathcal N}\nabla_i^2\;,
\end{align}
the Schr\"odinger equation~(\ref{eq:psi_R_tau}) becomes a $3\mathcal N$-dimensional diffusion equation. By writing the Green’s function of Eq.~(\ref{eq:G}) in momentum space by means of the Fourier transform, it is possible to show that $G_0$ is a Gaussian with variance proportional to $\tau$
\begin{align}
	G_0(R,R',d\tau)=\left(\frac{1}{4\pi Dd\tau}\right)^{\frac{3\mathcal N}{2}}\!\e^{-\frac{(R-R')^2}{4Dd\tau}}\;,\label{eq:G0}
\end{align}
where $D=\hbar^2/2m$ is the diffusion constant of a set of particles in Brownian motion with a dynamic governed by random collisions. This interpretation can be implemented by representing the wave function $\psi(R,\tau)$ by a set of discrete sampling points, called \emph{walkers}
\begin{align}
	\psi(R,\tau)=\sum_k\delta(R-R_k)\;,
\end{align}
and evolving this discrete distribution for an imaginary time $d\tau$ by means of Eq.~(\ref{eq:G}):
\begin{align}
	\psi(R,\tau+d\tau)=\sum_k G_0(R,R_k,d\tau)\;.
\end{align}
The result is a set of Gaussians that in the infinite imaginary time limit represents a distribution of walkers according to the lowest state of the Hamiltonian, that can be used to calculate the ground state properties of the system.

Let now consider the full Hamiltonian where the interaction is described by a central potential in coordinate space:
\begin{align}
	H=T+V=-\frac{\hbar^2}{2m}\sum_{i=1}^{\mathcal N}\nabla_i^2+V(R)\;.
\end{align}
Because $T$ and $V$ in general do not commute, it is not possible to directly split the propagator in a kinetic and a potential part
\begin{align}
	\e^{-(H-E_T)d\tau}\ne\e^{-Td\tau}\e^{-(V-E_T)d\tau}\;,
\end{align}
and thus the analytic solution of the Green’s function $\langle R|\e^{-(T+V-E_T)d\tau}|R'\rangle$ is not known in most of the cases.
However, by means of the Trotter-Suzuki formula to order $d\tau^3$
\begin{align}
	\e^{-(A+B)d\tau}=\e^{-A\frac{d\tau}{2}}\e^{-B d\tau}\e^{-A\frac{d\tau}{2}}\,+\ord\left(d\tau^3\right)\;,\label{eq:Trotter_3}
\end{align}
which is an improvement of the standard
\begin{align}
	\e^{-(A+B)d\tau}=\e^{-Ad\tau}\e^{-B d\tau}\,+\ord\left(d\tau^2\right)\;,\label{eq:Trotter_2}
\end{align}
in the limit of small imaginary time step $d\tau$ it is possible to write an approximate solution for $\psi(R,\tau+d\tau)$:
\begin{align}
	\psi(R,\tau+d\tau)&\simeq\int dR'\langle R|\e^{-V\frac{d\tau}{2}}\e^{-T d\tau}\e^{-V\frac{d\tau}{2}}\e^{E_Td\tau}|R'\rangle\,\psi(R',\tau)\;,\nonumber\\[0.2em]
	&\simeq\int dR'\underbrace{\langle R|\e^{-T d\tau}|R'\rangle}_{G_0(R,R',d\tau)}
	\underbrace{\phantom{\langle}\!\!\e^{-\left(\frac{V(R)+V(R')}{2}-E_T\right)d\tau}}_{G_V(R,R',d\tau)}\psi(R',\tau)\;,\nonumber\\[0.2em]
	&\simeq\left(\frac{1}{4\pi Dd\tau}\right)^{\frac{3\mathcal N}{2}}\!\!\int dR'\e^{-\frac{(R-R')^2}{4Dd\tau}}\e^{-\left(\frac{V(R)+V(R')}{2}-E_T\right)d\tau} \psi(R',\tau)\;,\label{eq:psi_propag}
\end{align}
which is the same of Eq.~(\ref{eq:G}) with the full Green's function given by
\begin{align}
	G(R,R',d\tau)\simeq G_0(R,R',d\tau)\,G_V(R,R',d\tau)\;.\label{eq:G0-GV}
\end{align}
According to the interacting Hamiltonian, the propagation of $\psi(R,\tau)$ for $d\tau\rightarrow0$ is thus described by the integral~(\ref{eq:psi_propag}) and the long imaginary time evolution necessary to project out the ground state component of the wave function is realized by iteration until convergence is reached. 

The steps of this process, that constitute the Diffusion Monte Carlo algorithm, can be summarized as follows:
\begin{enumerate}
	\item\label{item:DMC1} An initial distribution of walkers $w_i$ with $i=1,\ldots,\mathcal N_w$ is sampled from the trial wave function $\langle R|\psi_T\rangle=\psi_T(R)$ and the starting trial energy $E_T$ is chosen (for instance from a variational calculation or close to the expected result).

	\item\label{item:DMC2} The spacial degrees of freedom are propagated for small imaginary time step $d\tau$ with probability density $G_0(R,R',d\tau)$, i.e. the coordinates of the walkers are diffused by means of a Brownian motion
	\begin{align}
		R=R'+\xi\;,\label{eq:Brown}
	\end{align}
	where $\xi$ is a stochastic variable distributed according to a Gaussian probability density with $\sigma^2=2Dd\tau$ and zero average.

	\item\label{item:DMC3} For each walker, a weight
	\begin{align}
		\omega_i=G_V(R,R',d\tau)=\e^{-\left(\frac{V(R)+V(R')}{2}-E_T\right)d\tau}\;,\label{eq:w}
	\end{align}
	is assigned. The estimator contributions (kinetic energy, potential energy, root mean square radii, densities, \ldots) are evaluated on the imaginary time propagated configurations, weighting the results according to~$\omega_i$.

	\item\label{item:DMC4} The \emph{branching} process is applied to the propagated walkers. $\omega_i$ represents the probability of a configuration to multiply at the next step according to the normalization. This process is realized by generating from each $w_i$ a number of walker copies 
	\begin{align}
		n_i=[\omega_i+\eta_i]\;,
	\end{align}
	where $\eta_i$ is a random number uniformly distributed in the interval $[0,1]$ and $[x]$ means integer part of $x$. In such a way, depending on the potential $V(R)$ and the trial energy $E_T$, some configurations will disappear and some other will replicate, resulting in the evolution of walker population which is now made of $\widetilde{\mathcal N}_w=\sum_{i=1}^{\mathcal N_w}n_i$ walkers. A simple solution in order to control the fluctuations of walker population is to multiply the weight $\omega_i$ by a factor $\mathcal N_w/\widetilde{\mathcal N}_w$, adjusting thus the branching process at each time step. This solution is not efficient if the potential diverges. The corrections applied run-time could generate a lot of copies from just few good parent walkers and the population will be stabilized but not correctly represented. A better sampling technique is described in \S~\ref{subsec:Imp_Samp}.

	\item\label{item:DMC5} Iterate from \ref{item:DMC2} to \ref{item:DMC4} as long as necessary until convergence is reached, i.e. for large enough $\tau$ to reach the infinite limit of Eq.~(\ref{eq:tau_limit}). In this limit, the configurations $\{R\}$ are distributed according to the lowest energy state $\psi_0(R,\tau)$. Therefore, we can compute the ground state expectation values of observables that commute with the Hamiltonian
	\begin{align}
		\!\!\langle\mathcal O\rangle=\frac{\langle\psi_0|\mathcal O|\psi_0\rangle}{\langle\psi_0|\psi_0\rangle}
		=\!\lim_{\tau\rightarrow\infty}\!\frac{\langle\psi_T|\mathcal O|\psi(\tau)\rangle}{\langle\psi_T|\psi(\tau)\rangle}=\!\lim_{\tau\rightarrow\infty}\int\!\!dR\frac{\langle\psi_T|\mathcal O|R\rangle\psi(R,\tau)}{\psi_T(R)\psi(R,\tau)}\;,\label{eq:mixed_int}
	\end{align}
	by means of
	\begin{align}
		\langle\mathcal O\rangle=\frac{\sum_{\{R\}}\langle R|\mathcal O|\psi_T\rangle}{\sum_{\{R\}}\langle R|\psi_T\rangle}
		=\frac{\sum_{\{R\}}\mathcal O\psi_T(R)}{\sum_{\{R\}}\psi_T(R)}\;.\label{eq:mixed}
	\end{align}
	Statistical error bars on expectation values are then estimated by means of block averages and the analysis of auto-correlations on data blocks.
	The direct calculation of the expectation value~(\ref{eq:mixed}) gives an exact result only when $\mathcal O$ is the Hamiltonian $H$ or commutes with $H$, otherwise only ``mixed'' matrix elements $\langle\mathcal O\rangle_m\ne \langle\mathcal O\rangle$ can be obtained. Among the different methods to calculate expectation values for operators that do not commute with $H$, the extrapolation method~\cite{Pieper:2008} is the most widely used. Following this method, one has a better approximation to the ``pure'' (exact) value by means of a linear extrapolation
	\begin{align}
		\langle\mathcal O\rangle_p\simeq2\,\frac{\langle\psi_0|\mathcal O|\psi_T\rangle}{\langle\psi_0|\psi_T\rangle}-\frac{\langle\psi_T|\mathcal O|\psi_T\rangle}{\langle\psi_T|\psi_T\rangle}=2\,\langle\mathcal O\rangle_m-\langle\mathcal O\rangle_v\;,\label{eq:pure1}
	\end{align}
	or, if the operator $\mathcal O$ is positive defined, by means of 
	\begin{align}
		\langle\mathcal O\rangle_p&\simeq\frac{\left(\frac{\langle\psi_0|\mathcal O|\psi_T\rangle}{\langle\psi_0|\psi_T\rangle}\right)^2}{\frac{\langle\psi_T|\mathcal O|\psi_T\rangle}{\langle\psi_T|\psi_T\rangle}}=\frac{\langle\mathcal O\rangle_m^2}{\langle\mathcal O\rangle_v}\;,\label{eq:pure2}
	\end{align}
	where $\langle\mathcal O\rangle_v$ is the variational estimator. The accuracy of the extrapolation method is closely related to the trial wave function used in the variational calculation and on the accuracy of the DMC sampling technique.
\end{enumerate}

For a many-body system, if no constraint is imposed, $H$ has both symmetric and antisymmetric eigenstates with respect to particle exchange. It can be proven~\cite{Courant:1953} that the lowest energy solution, and hence the state projected by imaginary time propagation, is always symmetric. Moreover, in the DMC algorithm, the walkers distribution is sampled through the wave function, that must be positive defined in the whole configuration space for the probabilistic interpretation to be applicable. The projection algorithm described above is thus referred to Boson systems only. The extension for Fermion systems is reported in \S~\ref{subsec:Sign}.

\subsection{Importance Sampling}
\label{subsec:Imp_Samp}

As discussed in the previous section, the basic version of the DMC algorithm is rather inefficient because the weight term of Eq.~(\ref{eq:w}) could suffer of very large fluctuations. Indeed, because the Brownian diffusive process ignores the shape of the potential, there is nothing that prevents two particles from moving very close to each other, even in presence of an hard-core repulsive potential.

The \emph{importance function} techniques~\cite{Mitas:1998,Kalos:2008,Lipparini:2008} mitigates this problem by using an appropriate importance function $\psi_I(R)$ (which is often, but not
necessarily, the same $\psi_T(R)$ used for the projection) to guide the diffusive process. The idea is to multiply Eq.~(\ref{eq:G}) by $\psi_I(R)$
\begin{align}
	\psi_I(R)\psi(R,\tau+d\tau)=\int dR'\,G(R,R',d\tau)\frac{\psi_I(R)}{\psi_I(R')}\psi_I(R')\psi(R',\tau)\;,
\end{align}
and define a new propagator
\begin{align}
	\widetilde G(R,R',d\tau)=G(R,R',d\tau)\frac{\psi_I(R)}{\psi_I(R')}\;,\label{eq:ratio}
\end{align}
and a new function
\begin{align}
	f(R,\tau)=\psi_I(R)\psi(R,\tau)\;,
\end{align}
such that
\begin{align}
	f(R,\tau+d\tau)=\int dR'\,\widetilde G(R,R',d\tau)\,f(R',\tau)\;.\label{eq:f}
\end{align}
$f(R,\tau)$ represents the new probability density from which sample the walker distribution. If $\psi_I(R)$ is suitably chosen, for example to be small in the region where the potential presents the hard-core, then $f(R,\tau)$ contains more information than the original $\psi(R,\tau)$, being correlated to the potential by construction, and thus there is an improvement in the quality of the DMC sampling and a reduction of the fluctuations of the weight~(\ref{eq:w}).

By inserting the new propagator $\widetilde G(R,R',d\tau)$ in Eq.~(\ref{eq:psi_propag}) and expanding near $R'$, it is possible to show (see for instance Refs.~\cite{Lipparini:2008}) that the integration gives an additional drift term in $G_0(R,R',d\tau)$ 
\begin{align}
	G_0(R,R',d\tau)\rightarrow\widetilde G_0(R,R',d\tau)=\left(\frac{1}{4\pi Dd\tau}\right)^{\frac{3\mathcal N}{2}}\e^{-\frac{(R-R'-v_d(R') D d\tau)^2}{4Dd\tau}}\;,\label{eq:G_IS}
\end{align}
where 
\begin{align}
	\bm v_d(R)=2\frac{\bm\nabla\psi_I(R)}{\psi_I(R)}\;,\label{eq:drift}
\end{align}
is a $3\mathcal N$ dimensional \emph{drift velocity} that drives the free diffusion. The branching factor of Eq.~(\ref{eq:w}) modifies in
\begin{align}
	\omega_i\rightarrow\widetilde\omega_i=\e^{-\left(\frac{E_L(R)+E_L(R')}{2}-E_T\right)d\tau}\;,\label{eq:w_IS}
\end{align}
where the potential energy is replaced by the \emph{local energy}
\begin{align}
	E_L(R)=\frac{H\psi_I(R)}{\psi_I(R)}\;.\label{eq:E_L}
\end{align}
If the importance function is sufficiently accurate, the local energy remains close to the ground-state energy throughout the imaginary time evolution and the population of walkers is not subject to large fluctuations.

Going back to the imaginary time dependent Schr\"odinger equation, it is possible to show (details can be found in Refs.~\cite{Lipparini:2008}) that by multiplying Eq.~(\ref{eq:psi_R_tau}) by $\psi_I(R)$ we obtain a non homogenous Fokker-Plank equation for $f(R,\tau)$
\begin{align}
	\!\!-\frac{\partial}{\partial\tau}f(R,\tau)=&-\frac{\hbar^2}{2m}\nabla^2 f(R,\tau)+\frac{\hbar^2}{2m}\bm\nabla\!\cdot\!\Bigl[\bm v_d(R)f(R,\tau)\Bigr]+E_L(R)f(R,\tau)\;,
\end{align}
for which the corresponding Green’s function is given by the two terms of Eqs.~(\ref{eq:G_IS}) and (\ref{eq:w_IS}).

The DMC algorithm including the importance sampling procedure is still the same described in \S~\ref{sec:DMC}, where now the coordinates of the walkers are diffused by the Brownian motion and guided by the drift velocity
\begin{align}
	R=R'+\xi+\bm v_d D d\tau\;,
\end{align}
and the branching process is given by the local energy through the weight (\ref{eq:w_IS}). The expectation values are still calculated by means of Eq.~(\ref{eq:mixed}) but now the sampling function $\psi(R,\tau)$ is replaced by $f(R,\tau)$.

\subsection{Sign Problem}
\label{subsec:Sign}

As discussed in \S~\ref{subsec:Imp_Samp}, the standard DMC algorithm applies to positive defined wave function and the result of the imaginary time projection is a nodeless function. The ground state of a Fermionic system is instead described by an antisymmetric wave function, to which a probability distribution interpretation cannot be given. Moreover, the search for an antisymmetric ground state $|\psi_0^A\rangle$ corresponds to the search for an excited state of the many-body Hamiltonian with eigenvalue
\begin{align}
	E_0^A>E_0^S\;,\label{eq:E_0^A}
\end{align}
where $E_0^S$ and $E_0^A$ are the ground state energies for the Bosonic and the Fermionic system. 

If no constraint is imposed, the Hamiltonian has both eigenstates that are symmetric and antisymmetric with respect to particle exchange. We can thus rewrite Eq.~(\ref{eq:psi_0}) by separating Bosonic and Fermionic components:
\begin{align}
	|\psi(\tau)\rangle=\sum_{n=0}^\infty c_n^S \e^{-(E_n^S-E_T)\tau}|\varphi_n^S\rangle+\sum_{n=0}^\infty c_n^A \e^{-(E_n^A-E_T)\tau}|\varphi_n^A\rangle \;.
\end{align}
If we want to naively apply the standard DMC algorithm to project out the Fermionic ground state, we need to propagate the trial wave function for long imaginary time taking $E_0^A$ as energy reference. If the Fermionic ground state is not degenerate, i.e. $E_n^A\ge E_{n-1}^A>E_0^A$, in the limit $\tau\rightarrow\infty$ we have
\begin{align}
	\lim_{\tau\rightarrow\infty}|\psi(\tau)\rangle=\lim_{\tau\rightarrow\infty}\sum_n c_n^S \e^{-(E_n^S-E_0^A)\tau}|\varphi_n\rangle+c_0^A |\varphi_0^A\rangle \;,	
\end{align}
where at least for $E_0^S$ the Bosonic part diverges due to the condition~(\ref{eq:E_0^A}). However, the exponentially growing component along the symmetric ground state does not affect the expectation of the Hamiltonian. Indeed, during the evaluation of the integral (\ref{eq:mixed_int}) on an antisymmetric trial wave function $\psi_T^A(R)$, the symmetric components of $\psi(R,\tau)$ vanish by orthogonality and in the limit of infinite imaginary time the energy converges to exact eigenvalue $E_0^A$. However, the orthogonality cancellation of the Bosonic terms does not apply to the calculation of the DMC variance for the antisymmetric energy expectation value $\langle E_0^A\rangle$
\begin{align}
	\sigma^2_{E_0^A}=\left|\langle H\rangle_{\psi_T^A}^2-\langle H^2\rangle_{\psi_T^A}\right| \;,
\end{align}
where the second term diverges. We are left thus with an exact eigenvalue affected by an exponentially growing statistical error. The signal to noise ratio exponentially decays. This is the well known \emph{sign problem} and it represents the main limit to the straightforward application of the DMC algorithm to Fermion systems.

In order to extend the DMC method to systems described by antisymmetric wave functions, it is possible to artificially split the configuration space in regions where the trial wave function does not change sign. The multi dimensional surface where the trial wave function vanishes, denoted as \emph{nodal surface}, can be used to constrain the diffusion of the walkers: whenever a walker crosses the nodal surface it is dropped from the calculation. In such a way only the configurations that diffuse according to region of the wave function with definite sign are taken into account. The problem reduces thus to a standard DMC in the subsets of the configuration space delimited by the nodal surface. This approximate algorithm is called \emph{fixed-node}~\cite{Ceperley:1991,Mitas:1998,Mitas:2006} and it can be proven that it always provides an upper bound to the true Fermionic ground state.

The sign problem appears for both real and complex antisymmetric wave functions. The latter is the case of nuclear Hamiltonians. As proposed by Zhang \emph{et al.}~\cite{Zhang:1995,Zhang:1997,Zhang:2003}, the \emph{constrained path} approximation can be used to deal with the sign problem for complex wave functions. 
The general idea is to constraining the path of walkers to regions where the real part of the overlap with the wave function is positive. If we consider a complex importance function $\psi_I(R)$, in order to keep real the coordinates space of the system, the drift term in Eq.~(\ref{eq:G_IS}) must be real. A suitable choice for the drift velocity is thus:
\begin{align}
	\bm v_d(R)=2\frac{\bm\nabla\re\left[\psi_I(R)\right]}{\re\left[\psi_I(R)\right]}\;.
\end{align} 
Consistently, a way to eliminate the decay of the signal to noise ratio consists in requiring that the real part of the overlap of each walker with the importance function must keep the same sign
\begin{align}
	\frac{\re\left[\psi_I(R)\right]}{\re\left[\psi_I(R')\right]}>0\;,
\end{align}
where $R$ and $R'$ denote the coordinates of the system after and before the diffusion of a time step. When this condition is violate, i.e. when the overlap between the importance function and the walker after a diffusive step changes sign, the walker is dropped. In these scheme, the ground state expectation value of an observable $\mathcal O$ (Eq.~(\ref{eq:mixed})) is given by
\begin{align}
	\langle\mathcal O\rangle=\frac{\sum_{\{R\}}\mathcal O\re\left[\psi_T(R)\right]}{\sum_{\{R\}}\re\left[\psi_T(R)\right]}\;.
\end{align} 

Another approach to deal with the complex sign problem is the \emph{fixed phase} approximation, originally introduced for systems whose Hamiltonian contains a magnetic field~\cite{Ortiz:1993}. Let write a complex wave function as
\begin{align}
	\psi(R)=\left|\psi(R)\right|\e^{i\phi(R)} \;,\label{eq:mod_phase}
\end{align}
where $\phi(R)$ is the phase of $\psi(R)$, and rewrite the drift velocity as
\begin{align}
	\bm v_d(R)=2\frac{\bm\nabla\left|\psi_I(R)\right|}{\left|\psi_I(R)\right|}=2\re\left[\frac{\bm\nabla\psi_I(R)}{\psi_I(R)}\right]\;.
\end{align}
With this choice, the weight for the branching process becomes
\begin{align}
	\widetilde\omega_i&=\exp\Bigg\{-\Bigg[\frac{1}{2}\Bigg(-\frac{\hbar^2}{2m}\frac{\nabla^2|\psi_I(R)|}{|\psi_I(R)|}+\frac{V\psi_I(R)}{\psi_I(R)}\nonumber\\[0.2em]
	&\quad-\frac{\hbar^2}{2m}\frac{\nabla^2|\psi_I(R')|}{|\psi_I(R')|}+\frac{V\psi_I(R')}{\psi_I(R')}\Bigg)-E_T\Bigg]d\tau\Bigg\}
	\times\frac{\left|\psi_I(R')\right|}{\left|\psi_I(R)\right|}\frac{\psi_I(R)}{\psi_I(R')}\;,\label{eq:w_PF}
\end{align}
which is the usual importance sampling factor as in Eq.~(\ref{eq:w_IS}) multiplied by an additional factor that corrects for the particular choice of the drift. Using Eq.~(\ref{eq:mod_phase}), the last term of the previous equation can be rewritten as
\begin{align}
	\frac{\left|\psi_I(R')\right|}{\left|\psi_I(R)\right|}\frac{\psi_I(R)}{\psi_I(R')}=\e^{i[\phi_I(R)-\phi_I(R')]} \;.
\end{align}
The so called ``fixed phase'' approximation is then realized by constraining the walkers to have the same phase as the importance function $\psi_I(R)$. It can be applied by keeping the real part of the last expression. In order to preserve the normalization, one has to consider an additional term in the Green’s function due to the phase, that must be added to the weight:
\begin{align}
	\exp\Bigg[-\frac{\hbar^2}{2m}\Bigl(\bm\nabla\phi(R)\Bigr)^2d\tau\Bigg]\;.
\end{align}
This factor can be included directly in $\widetilde\omega_i$ considering the following relation:
\begin{align}
	\re\left[\frac{\nabla^2\psi_I(R)}{\psi_I(R)}\right]=\frac{\nabla^2\left|\psi_I(R)\right|}{\left|\psi_I(R)\right|}-\Bigl(\bm\nabla\phi(R)\Bigr)^2\;.
\end{align}
Thus, by keeping the real part of the kinetic energy in Eq.~(\ref{eq:w_PF}), the additional weight term given by the fixed phase approximation is automatically included. The calculation of expectation values is given now by
\begin{align}
	\langle\mathcal O\rangle=\sum_{\{R\}}\re\left[\frac{\mathcal O\psi_T(R)}{\psi_T(R)}\right]\;, 
\end{align}
i.e. by the evaluation of the real part of a local operator. This is of particular interest for the technical implementation of the DMC algorithm. As we will see in \S~\ref{subsec:Wave}, when dealing with Fermions the wave function can be written as a Slater determinant of single particle states. It can be shown (see Appendix~\ref{app:d_SD}) that the evaluation of local operators acting on Slater determinants can be efficiently implemented by means of the inverse matrix of the determinant. The fixed phase approximation allows thus to deal with the Fermion sign problem and also provides a natural scheme to implement the DMC method. Moreover, the above derivation can be extended to operators other than the kinetic energy. For example, when dealing with nuclear Hamiltonians like~(\ref{eq:H_N}), spin and isospin expectation values can be evaluated by taking the real part of local spin and isospin operators calculated on the Slater determinant. This is actually the standard way to treat the spin-isospin dependent components of the nuclear Hamiltonian in the Auxiliary Field Diffusion Monte Carlo (see \S~\ref{sec:AFDMC}). 

The constrained path and the fixed phase prescriptions are both approximations introduced to deal with the sign problem for complex wave functions. In principle they should yield similar results if the importance function is close enough to the real ground state of the system. Accurate $\psi_I(R)$ are thus needed. An additional important observation is that the DMC algorithm with the constrained path approximation does not necessarily provide an upper bound in the calculation of energy~\cite{Carlson:1999,Wiringa:2002_CP}. Moreover, it has not been proven that the fixed phase approximation gives an upper bound to the real energy. Thus, the extension of the DMC algorithm to Fermion systems described by complex wave functions does not obey to the Rayleigh-Ritz  variational principle. Further details on the fixed node, constrained path and fixed phase approximations can be found in the original papers and an exhaustive discussion is reported in the Ph.D. thesis of Armani~\cite{Armani:2011_thesis}.

\subsection{Spin-isospin degrees of freedom}
\label{subsec:Spin}

If we want to study a nuclear many-body system described by the Hamiltonian~(\ref{eq:H_N}), we need to include also the spin-isospin degrees of freedom in the picture. In order to simplify the notation, in the next with $A$ we will refer to the number of nucleons. Starting from \S~\ref{subsec:Wave} we will restore the convention $A=\mathcal N_N+\mathcal N_\Lambda$. The typical trial many-body wave function used in DMC calculation for nuclear systems takes the form~\cite{Pieper:2008,Wiringa:2002_CP}
\begin{align}
	|\psi_T\rangle=\mathcal S\left[ \prod_{i<j}\left(1+U_{ij}+\sum_k U_{ijk}\right)\right]\prod_{i<j}f_c(r_{ij})|\Phi_A\rangle\;,\label{eq:GFMC_psiT}
\end{align}
where $f_c(r_{ij})$ is a central (mostly short ranged repulsion) correlation, $U_{ij}$ are non commuting two-body correlations induced by $v_{ij}$ (that typically takes the same form of Eq.~(\ref{eq:v_ij_Op}) for $p=2,\ldots,6$) and $U_{ijk}$ is a simplified three-body correlation from $v_{ijk}$. $|\Phi_A\rangle$ is the one-body part of the trial wave function that determines the quantum numbers of the states and it is fully antisymmetric. The central correlation is symmetric with respect to particle exchange and the symmetrization operator $\mathcal S$ acts on the operatorial correlation part of $|\psi_T\rangle$ in order to make the complete trial wave function antisymmetric. The best trial wave function from which (\ref{eq:GFMC_psiT}) is derived, includes also spin-orbit and the full three-body correlations and it is used in VMC calculations. See Refs.~\cite{Arriaga:1995,Wiringa:2002_CP}.

Given $A$ nucleons ($Z$ protons, $A-Z$ neutrons), the trial wave function is a complex vector in spin-isospin space with dimension $\mathcal N_S\times\mathcal N_T$, where $\mathcal N_S$ is the number of spin states and $\mathcal N_T$ the number of isospin states:
\begin{align}
	\mathcal N_S=2^A\quad\quad\quad\mathcal N_T=\left(\begin{array}{c} A\\ Z \end{array}\right)=\frac{A!}{Z!(A-Z)!}\;.
\end{align}
For example, the wave function of an $A=3$ system has 8 spin components and, considering the physical systems for $Z=1$ $\left(^3\text{H}\right)$ or $Z=2$ $\left(^3\text{He}\right)$, 3 isospin states, thus a spin-isospin structure with 24 entries. Using the notation of Ref.~\cite{Pieper:2008}, we can write the spin part of an $A=3$ wave function as a complex 8-vector (ignore antisymmetrization)
\begin{align}
	|\Phi_{A=3}\rangle=\left(
		\begin{array}{c}
			a_{\uparrow\uparrow\uparrow}\\
			a_{\uparrow\uparrow\downarrow}\\
			a_{\uparrow\downarrow\uparrow}\\
			a_{\uparrow\downarrow\downarrow}\\
			a_{\downarrow\uparrow\uparrow}\\
			a_{\downarrow\uparrow\downarrow}\\
			a_{\downarrow\downarrow\uparrow}\\
			a_{\downarrow\downarrow\downarrow}
		\end{array}
	\right)\quad\quad\text{with}\quad a_{\uparrow\downarrow\uparrow}=\langle\uparrow\downarrow\uparrow|\Phi_{A=3}\rangle\;.\label{eq:wave_GFMC}
\end{align}

The potentials ($v_{ij}$, $v_{ijk}$) and correlations ($U_{ij}$, $U_{ijk}$) involve repeated operations on $|\psi_T\rangle$ but the many-body spin-isospin space is closed under the action of the operators contained in the Hamiltonian. As an example, consider the term $\bm\sigma_i\cdot\bm\sigma_j$:
\begin{align}
	\bm\sigma_i\cdot\bm\sigma_j&=2\left(\sigma_i^+\sigma_j^-+\sigma_i^-\sigma_j^+\right)+\sigma_i^z\sigma_j^z\;,\nonumber\\[0.2em]
	&=2\,\mathcal P_{ij}^\sigma-1\;,\nonumber\\[0.2em]
	&=\left(\begin{array}{cccc}
		1 &  0 &  0 & 0 \\
		0 & -1 &  2 & 0 \\
		0 &  2 & -1 & 0 \\
		0 &  0 &  0 & 1
	\end{array}\right)
	\quad\text{acting on}\quad
	\left(\begin{array}{c}
		\uparrow\uparrow   \\
		\uparrow\downarrow \\
		\downarrow\uparrow \\
		\downarrow\downarrow  
	\end{array}\right)\;.
\end{align}
The $\mathcal P_{ij}^\sigma$ exchanges the spin $i$ and $j$, so the operator $\bm\sigma_i\cdot\bm\sigma_j$ does not mix different isospin components and acts on different, non contiguous, 4-element blocks of $|\Phi_{A=3}\rangle$. For $i=2$ and $j=3$ we have for example:
\begin{align}
	\bm\sigma_2\cdot\bm\sigma_3\,|\Phi_{A=3}\rangle=\left(
		\begin{array}{c}
			a_{\uparrow\uparrow\uparrow}\\
			2a_{\uparrow\downarrow\uparrow}-a_{\uparrow\uparrow\downarrow}\\
			2a_{\uparrow\uparrow\downarrow}-a_{\uparrow\downarrow\uparrow}\\
			a_{\uparrow\downarrow\downarrow}\\
			a_{\downarrow\uparrow\uparrow}\\
			2a_{\downarrow\downarrow\uparrow}-a_{\downarrow\uparrow\downarrow}\\
			2a_{\downarrow\uparrow\downarrow}-a_{\downarrow\downarrow\uparrow}\\
			a_{\downarrow\downarrow\downarrow}
		\end{array}
	\right)\;.\label{eq:s2s3psi}
\end{align}
The action of pair operators on $|\psi_T\rangle$, that are the most computationally expensive, results thus in a sparse matrix of (non contiguous) $4\times4$ blocks in the $A$-body problem.

In the Green Function Monte Carlo, which slightly differs from the DMC in the way the propagator is treated, each of the $2^A\frac{A!}{Z!(A-Z)!}$ spin-isospin configurations undergoes to the imaginary time evolution of Eq.~(\ref{eq:G}). The propagation is now acting on the component $a_\alpha$, being $\alpha$ the spin-isospin index,
\begin{align}
	a_\alpha(R,\tau+d\tau)=\sum_\beta\int dR'\,G_{\alpha\beta}(R,R',d\tau)\,a_\beta(R',\tau)\;,\label{eq:a_GFMC}
\end{align}
where the Green's function is a matrix function of $R$ and $R'$ in spin-isospin space, defined as
\begin{align}
	G_{\alpha\beta}(R,R',d\tau)=\langle R,\alpha|\e^{-(H-E_T)d\tau}|R',\beta\rangle\;.\label{eq:G_GFMC}
\end{align}

Due to the the factorial growth in the number of components of the wave function, GFMC cannot deal with systems having a large number of nucleons, like medium-heavy nuclei or nuclear matter.
Standard GFMC calculations are indeed limited up to 12 nucleons~\cite{Pieper:2005,Lusk:2010,Lovato:2013} or 16 neutrons~\cite{Gandolfi:2011}.

\section{Auxiliary Field Diffusion Monte Carlo}
\label{sec:AFDMC}

The AFDMC algorithm was originally introduced by Schmidt and Fantoni~\cite{Schmidt:1999} in order to deal in an efficient way with spin-dependent Hamiltonians. Many details on the AFDMC method can be found in Refs.~\cite{Sarsa:2003,Pederiva:2004,Gandolfi:2007_thesis,Gandolfi:2006,Gandolfi:2007,Gandolfi:2009,Armani:2011_thesis,Lovato:2012_thesis,Lipparini:2008}. The main idea is to move from the many particle wave function of the DMC or GFMC to a single particle wave function. In this representation, going back to the example of the previous section, the spin part of an $A=3$ wave function becomes a tensor product of 3 single particle spin states (ignore antisymmetrization):
\begin{align}
	|\Phi_{A=3}\rangle&=
	\left(\begin{array}{c}
		a_{1\uparrow}\\
		a_{1\downarrow}
	\end{array}\right)_1\!\!\otimes\!
	\left(\begin{array}{c}
		a_{2\uparrow}\\
		a_{2\downarrow}
	\end{array}\right)_2\!\!\otimes\!
	\left(\begin{array}{c}
		a_{3\uparrow}\\
		a_{3\downarrow}
	\end{array}\right)_3\quad\text{with}\quad a_{k\uparrow}=\,_{_k\,}\!\langle\uparrow|\Phi_{A=3}\rangle\;.\label{eq:psi_SP}
\end{align} 
Taking also into account the isospin degrees of freedom, each single particle state becomes a complex 4-vector and the total number of entries for $|\Phi_{A=3}\rangle$ is thus 12, half of the number for the full DMC function of Eq.~(\ref{eq:wave_GFMC}). In the general case, the dimension of the multicomponent vector describing a system with $A$ nucleons scale as $4A$. So, in this picture, the computational cost for the evaluation of the wave function is drastically reduced compared to the DMC-GFMC method when the number of particles becomes large.

The problem of the single particle representation is that it is not closed with respect to the application of quadratic spin (isospin) operators. As done in the previous section (Eq.~(\ref{eq:s2s3psi})), consider the operator $\bm\sigma_2\cdot\bm\sigma_3=2\,\mathcal P_{23}^\sigma-1$ acting on $|\Phi_{A=3}\rangle$:
\begin{align}
	\bm\sigma_2\cdot\bm\sigma_3\,|\Phi_{A=3}\rangle&=
	2\left(\begin{array}{c}
		a_{1\uparrow}\\
		a_{1\downarrow}
	\end{array}\right)_1\!\!\otimes\!
	\left(\begin{array}{c}
		a_{3\uparrow}\\
		a_{3\downarrow}
	\end{array}\right)_2\!\!\otimes\!
	\left(\begin{array}{c}
		a_{2\uparrow}\\
		a_{2\downarrow}
	\end{array}\right)_3\nonumber\\[0.2em]
	&\phantom{=}-\left(\begin{array}{c}
		a_{1\uparrow}\\
		a_{1\downarrow}
	\end{array}\right)_1\!\!\otimes\!
	\left(\begin{array}{c}
		a_{2\uparrow}\\
		a_{2\downarrow}
	\end{array}\right)_2\!\!\otimes\!
	\left(\begin{array}{c}
		a_{3\uparrow}\\
		a_{3\downarrow}
	\end{array}\right)_3\;.
\end{align}
There is no way to express the result as a single particle wave function of the form~(\ref{eq:psi_SP}). At each time step, the straightforward application of the DMC algorithm generates a sum of single particle wave functions. The number of these functions will grows very quickly during the imaginary time evolution, destroying the gain in computational time obtained using a smaller multicomponent trial wave function.

In order to keep the single particle wave function representation and overcome this problem, the AFDMC makes use of the Hubbard-Stratonovich transformation
\begin{align}
	\e^{-\frac{1}{2}\lambda\mathcal O^2 }=\frac{1}{\sqrt{2\pi}}\int\!dx \e^{-\frac{x^2}{2}+\sqrt{-\lambda}\,x\mathcal O}\;,\label{eq:HS}
\end{align}
to linearize the quadratic dependence on the spin-isospin operators by adding the integration over a new variable $x$ called \emph{auxiliary field}. It is indeed possible to show that the single particle wave function is closed with respect to the application of a propagator containing linear spin-isospin operators at most:
\begin{align}
	\e^{-\mathcal O_j d\tau}|\Phi_A\rangle
	&=\e^{-\mathcal O_j d\tau}\bigotimes_i\left(\begin{array}{c} a_{i\uparrow} \\ a_{i\downarrow} \end{array}\right)_i\;,\nonumber \\[0.2em]
	&=\left(\begin{array}{c} a_{1\uparrow} \\ a_{1\downarrow} \end{array}\right)_1
	\!\otimes\cdots\otimes\,\e^{-\mathcal O_j d\tau}\left(\begin{array}{c} a_{j\uparrow} \\ a_{j\downarrow} \end{array}\right)_j
	\!\otimes\cdots\otimes \left(\begin{array}{c} a_{A\uparrow} \\ a_{A\downarrow} \end{array}\right)_A\;,\quad\quad\nonumber\\[0.2em]
	&=\left(\begin{array}{c} a_{1\uparrow} \\ a_{1\downarrow} \end{array}\right)_1
	\!\otimes\cdots\otimes \left(\begin{array}{c} \widetilde a_{j\uparrow} \\ \widetilde a_{j\downarrow} \end{array}\right)_j
	\!\otimes\cdots\otimes \left(\begin{array}{c} a_{A\uparrow} \\ a_{A\downarrow} \end{array}\right)_A\;,
\end{align}
where, working on 2-component spinors, $\mathcal O_j$ can be a $2\times2$ spin or isospin matrix. If we are dealing with the full 4-component spinor, $\mathcal O_j$ can be an extended $4\times4$ spin, isospin or isospin$\,\otimes\,$spin matrix. To get this result we have used the fact that the operator $\mathcal O_j$ is the representation in the $A$-body tensor product space of a one-body operator:
\begin{align}
	\mathcal O_j\equiv\mathbb{I}_1\otimes\cdots\otimes\mathcal O_j\otimes\cdots\otimes\mathbb{I}_A\;.\label{eq:O_j}
\end{align}

Limiting the study to quadratic spin-isospin operators and making use of the Hubbard-Stratonovich transformation, it is thus possible to keep the single particle wave function representation over all the imaginary time evolution. This results in a reduced computational time for the propagation of the wave function compared to GFMC, that allows us to simulate larger systems, from medium-heavy nuclei to the infinite matter. In the next we will see in detail how the AFDMC works on the Argonne V6 like potentials (\S~\ref{subsec:Prop_AV6}), and how it is possible to include also spin-orbit (\S~\ref{subsec:Prop_LS}) and three-body (\S~\ref{subsec:Prop_TNI}) terms for neutron systems. Finally the extension of AFDMC for hypernuclear systems (\S~\ref{subsec:Prop_YN}) is presented.

\subsection{Propagator for nucleons: \texorpdfstring{$\bm\sigma$}{$\sigma$}, \texorpdfstring{$\bm\sigma\cdot\bm\tau$}{$\sigma\cdot\tau$} and \texorpdfstring{$\bm\tau$}{$\tau$} terms}
\label{subsec:Prop_AV6}

Consider the first six components of the Argonne $NN$ potential of Eq.~(\ref{eq:v_ij_Op}). They can be conveniently rewritten as a sum of a spin-isospin independent and a spin-isospin dependent term
\begin{align}
	V_{NN}=\sum_{i<j}\sum_{p=1,6}v_p(r_{ij})\,\mathcal O_{ij}^{\,p}=V_{NN}^{SI}+V_{NN}^{SD} \;,
\end{align}
where
\begin{align}
	V_{NN}^{SI}&=\sum_{i<j}v_1(r_{ij})\;,\label{eq:V_NN_SI}
\end{align}
and
\begin{align}
	V_{NN}^{SD}&=\frac{1}{2}\sum_{i\ne j}\sum_{\alpha\beta}\sigma_{i\alpha}\,A_{i\alpha,j\beta}^{[\sigma]}\,\sigma_{j\beta} 
	+\frac{1}{2}\sum_{i\ne j}\sum_{\alpha\beta\gamma}\tau_{i\gamma}\,\sigma_{i\alpha}\,A_{i\alpha,j\beta}^{[\sigma\tau]}\,\sigma_{j\beta}\,\tau_{j\gamma}\;\nonumber\\[0.2em]
	&\,+\frac{1}{2}\sum_{i\ne j}\sum_\gamma\tau_{i\gamma}\,A_{ij}^{[\tau]}\,\tau_{j\gamma}\;.\label{eq:V_NN_SD}
\end{align}
The $3A\times 3A$ matrices $A^{[\sigma]}$, $A^{[\sigma\tau]}$ and the $A\times A$ matrix $A^{[\tau]}$ are real and symmetric under Cartesian component interchange $\alpha\leftrightarrow\beta$,   under particle exchange $i\leftrightarrow j$ and fully symmetric with respect to the exchange $i\alpha\leftrightarrow j\beta$. They have zero diagonal (no self interaction) and contain proper combinations of the components of AV6 (Latin indices are used for the nucleons, Greek ones refer to the Cartesian components of the operators):
\begin{align}
	A_{ij}^{[\tau]}&=v_2\left(r_{ij}\right)\;,\nonumber\\[0.5em]
	A_{i\alpha,j\beta}^{[\sigma]}&=v_3\left(r_{ij}\right)\delta_{\alpha\beta}
	+v_5\left(r_{ij}\right)\left(3\,\hat r_{ij}^\alpha\,\hat r_{ij}^\beta-\delta_{\alpha\beta}\right)\;,\label{eq:A_NN}\\[0.5em]
	A_{i\alpha,j\beta}^{[\sigma\tau]}&=v_4\left(r_{ij}\right)\delta_{\alpha\beta}
	+v_6\left(r_{ij}\right)\left(3\,\hat r_{ij}^\alpha\,\hat r_{ij}^\beta-\delta_{\alpha\beta}\right)\;,\nonumber
\end{align}
that come from the decomposition of the operators in Cartesian coordinates:
\begin{align}
	\bm\sigma_i\cdot\bm\sigma_j&=\sum_{\alpha\beta}\sigma_{i\alpha}\,\sigma_{j\beta}\,\delta_{\alpha\beta}\;,\label{eq:sigma_dec}\\[0.2em]
	S_{ij}&=\sum_{\alpha\beta}\left(3\,\sigma_{i\alpha}\,\hat r_{ij}^\alpha\,\sigma_{j\beta}\,\hat r_{ij}^\beta-\sigma_{i\alpha}\,\sigma_{j\beta}\,\delta_{\alpha\beta}\right)\;.\label{eq:Sij_dec}
\end{align}
Being real and symmetric, the $A$ matrices have real eigenvalues and real orthogonal eigenstates
\begin{align}
	\sum_{j\beta} A_{i\alpha,j\beta}^{[\sigma]}\,\psi_{n,j\beta}^{[\sigma]}&=\lambda_n^{[\sigma]}\,\psi_{n,i\alpha}^{[\sigma]}\;,\nonumber\\[0.5em]
	\sum_{j\beta} A_{i\alpha,j\beta}^{[\sigma\tau]}\,\psi_{n,j\beta}^{[\sigma\tau]}&=\lambda_n^{[\sigma\tau]}\,\psi_{n,i\alpha}^{[\sigma\tau]}\;,\\[0.5em]
	\sum_{j} A_{ij}^{[\tau]}\,\psi_{n,j}^{[\tau]}&=\lambda_n^{[\tau]}\,\psi_{n,i}^{[\tau]}\;.\nonumber
\end{align}

Let us expand $\sigma_{i\alpha}$ on the complete set of eigenvectors $\psi_{n,i\alpha}^{[\sigma]}$ of the matrix $A_{i\alpha,j\beta}^{[\sigma]}$~:
\begin{align}
	\sigma_{i\alpha}=\sum_{n}c_n^{[\sigma]}\,\psi_{n,i\alpha}^{[\sigma]}=\sum_{n}\left(\sum_{j\beta}\psi_{n,j\beta}^{[\sigma]}\,\sigma_{j\beta}\right)\psi_{n,i\alpha}^{[\sigma]}\;,
	\label{eq:sigma_ia}
\end{align}
where we have used the orthogonality condition
\begin{align}
	\sum_{i\alpha}\psi_{n,i\alpha}^{[\mathcal O]}\,\psi_{m,i\alpha}^{[\mathcal O]}=\delta_{nm}\;.
\end{align}
Using Eq.~(\ref{eq:sigma_ia}) we can recast the first term of Eq.~(\ref{eq:V_NN_SD}) in the following form:
\begin{align}
	&\frac{1}{2}\sum_{i\alpha,j\beta}\sigma_{i\alpha}\,A_{i\alpha,j\beta}^{[\sigma]}\,\sigma_{j\beta}=\nonumber\\[0.2em]
	&\,=\frac{1}{2}\sum_{i\alpha,j\beta}\left\{
	\left[\sum_{n}\left(\sum_{k\gamma}\sigma_{k\gamma}\,\psi_{n,k\gamma}^{[\sigma]}\right)\psi_{n,i\alpha}^{[\sigma]}\right]
	A_{i\alpha,j\beta}^{[\sigma]}
	\left[\sum_{m}\left(\sum_{k\gamma}\sigma_{k\gamma}\,\psi_{m,k\gamma}^{[\sigma]}\right)\psi_{m,j\beta}^{[\sigma]}\right]
	\right\}\;,\nonumber\\[0.2em]
	&\,=\frac{1}{2}\sum_{i\alpha}\left\{
	\left[\sum_{n}\left(\sum_{k\gamma}\sigma_{k\gamma}\,\psi_{n,k\gamma}^{[\sigma]}\right)\psi_{n,i\alpha}^{[\sigma]}\right]
	\left[\sum_{m}\lambda_{m}^{[\sigma]}\left(\sum_{k\gamma}\sigma_{k\gamma}\,\psi_{m,k\gamma}^{[\sigma]}\right)\psi_{m,i\alpha}^{[\sigma]}\right]
	\right\}\;,\nonumber\\[0.2em]
	&\,=\frac{1}{2}\sum_{n}\left(\sum_{k\gamma}\sigma_{k\gamma}\,\psi_{n,k\gamma}^{[\sigma]}\right)^2\!\lambda_n^{[\sigma]}\;.
\end{align}
Similar derivation can be written for the terms $\tau_{i\gamma}\,\sigma_{i\alpha}$ and $\tau_{i\gamma}$ and we can define a new set of operators expressed in terms of the eigenvectors of the matrices~$A$:
\begin{align}
	\mathcal O_n^{[\sigma]}&=\sum_{j\beta}\sigma_{j\beta}\,\psi_{n,j\beta}^{[\sigma]}\;,\nonumber\\[0.5em]
	\mathcal O_{n,\alpha}^{[\sigma\tau]}&=\sum_{j\beta}\tau_{j\alpha}\,\sigma_{j\beta}\,\psi_{n,j\beta}^{[\sigma\tau]}\;,\label{eq:O_n}\\[0.5em]
	\mathcal O_{n,\alpha}^{[\tau]}&=\sum_{j}\tau_{j\alpha}\,\psi_{n,j}^{[\tau]}\;.\nonumber
\end{align}
The spin dependent part of the $NN$ interaction can be thus expressed as follows:
\begin{align}
	\!\!\!\!V_{NN}^{SD}=\frac{1}{2}\sum_{n=1}^{3A}\lambda_n^{[\sigma]}\!\left(\mathcal O_n^{[\sigma]}\right)^2
	\!+\frac{1}{2}\sum_{n=1}^{3A}\sum_{\alpha=1}^3\lambda_n^{[\sigma\tau]}\!\left(\mathcal O_{n\alpha}^{[\sigma\tau]}\right)^2
	\!+\frac{1}{2}\sum_{n=1}^A\sum_{\alpha=1}^3\lambda_n^{[\tau]}\!\left(\mathcal O_{n\alpha}^{[\tau]}\right)^2 \,.\!\label{eq:V_NN_SD_On}
\end{align}

$V_{NN}^{SD}$ is now written in a suitable form for the application of the Hubbard-Stratonovich transformation of Eq.~(\ref{eq:HS}). The propagator $\e^{-V_{NN}^{SD}\,d\tau}$ can be finally recast as:
\begin{align}
	\e^{-\frac{1}{2}\sum_n\lambda_n(\mathcal O_n)^2 d\tau}&=\prod_n\e^{-\frac{1}{2}\lambda_n(\mathcal O_n)^2 d\tau}\,+\ord\left(d\tau^2\right)\;,\nonumber\\[0.2em]
	&\simeq\prod_n\frac{1}{\sqrt{2\pi}}\int\!dx_n\e^{\frac{-x_n^2}{2}+\sqrt{-\lambda_n d\tau}\,x_n\mathcal O_n}\;,\label{eq:HS_applied}
\end{align}
where we have used the compact notation $\mathcal O_n$ to denote the $3A$ $\mathcal O_n^{[\sigma]}$, the $9A$ $O_{n,\alpha}^{[\sigma\tau]}$ and the $3A$ $\mathcal O_{n,\alpha}^{[\tau]}$ operators including the summation over the coordinate index $\alpha$ where needed. The first step of the above equation comes to the fact that in general the operators $\mathcal O_n$ do not commute and so, due to Eq.~(\ref{eq:Trotter_2}), the equality is correct only at order $d\tau^2$.

The standard DMC imaginary time propagation of Eq.~(\ref{eq:G}) needs to be extended to the spin-isospin space, as done in the GFMC algorithm via the projection of Eqs.~(\ref{eq:a_GFMC}) and (\ref{eq:G_GFMC}). In the AFDMC method, spin-isospin coordinates $\{S\}$ are added to the spacial coordinates $\{R\}$, defining a set of walkers which represents the single-particle wave function to be evolved in imaginary time:
\begin{align}
	\psi(R,S,\tau+d\tau)=\int dR'dS'\,G(R,S,R',S',d\tau)\,\psi(R',S',\tau)\;.
\end{align}
Including the integration over the Hubbard-Stratonovich auxiliary fields, the Auxiliary Field DMC Green's function reads (recall Eqs.~(\ref{eq:psi_propag}) and (\ref{eq:G0-GV})):
\begin{align}
	G(R,S,R',S',d\tau)&=\langle R,S|\e^{-(T+V-E_T)d\tau}|R',S'\rangle\;,\nonumber\\[0.2em]
	&\simeq\left(\frac{1}{4\pi Dd\tau}\right)^{\frac{3\mathcal N}{2}}\!\e^{-\frac{(R-R')^2}{4Dd\tau}}
	\e^{-\left(\frac{V_{NN}^{SI}(R)+V_{NN}^{SI}(R')}{2}-E_T\right)d\tau}\times\nonumber\\[0.2em]
	&\quad\,\times\langle S|\prod_{n=1}^{15A}\frac{1}{\sqrt{2\pi}}\int\!dx_n\e^{\frac{-x_n^2}{2}+\sqrt{-\lambda_n d\tau}\,x_n\mathcal O_n}|S'\rangle\;,\label{eq:G_AFDMC}
\end{align}
Each operator $\mathcal O_n$ involves the sum over the particle index $j$, as shown in Eq.~(\ref{eq:O_n}). However, in the $A$-body tensor product space, each $j$ sub-operator is a one-body operator acting on a different single particle spin-isospin states, as in Eq.~(\ref{eq:O_j}). Therefore the $j$-dependent terms commute and we can represent the exponential of the sum over $j$ as a tensor product of exponentials. The result is that the propagation of a spin-isospin state $|S'\rangle$ turns into a product of independent rotations, one for each spin-isospin state. Considering just a spin wave function we have for example: 
\begin{align}
	&\e^{\sqrt{-\lambda_n d\tau}\,x_n\mathcal O_n^{[\sigma]}}|S'\rangle=\nonumber\\[0.2em]
	&\,=\e^{\sqrt{-\lambda_n d\tau}\,x_n\sum_{\beta}\sigma_{1\beta}\,\psi_{n,1\beta}^{[\sigma]}}
	\left(\begin{array}{c} a_{1\uparrow} \\ a_{1\downarrow} \end{array}\right)_1\!\otimes\cdots\otimes 
	\e^{\sqrt{-\lambda_n d\tau}\,x_n\sum_{\beta}\sigma_{A\beta}\,\psi_{n,A\beta}^{[\sigma]}}
	\left(\begin{array}{c} a_{A\uparrow} \\ a_{A\downarrow} \end{array}\right)_A\;,\nonumber\\[0.2em]
	&\,=\left(\begin{array}{c} \widetilde a_{1\uparrow} \\ \widetilde a_{1\downarrow} \end{array}\right)_1\!\otimes\cdots\otimes
	\left(\begin{array}{c} \widetilde a_{A\uparrow} \\ \widetilde a_{A\downarrow} \end{array}\right)_A\;.\label{eq:eO_S}
\end{align}
We can thus propagate spin-isospin dependent wave functions remaining inside the space of single particle states.

The new coefficients $\widetilde a_{j\uparrow}$ and $\widetilde a_{j\downarrow}$ are calculated at each time step for each $\mathcal O_n$ operator. For neutron systems, i.e. for two-component spinors for which only the operator $\mathcal O_n^{[\sigma]}$ is active, there exists an explicit expression for these coefficients. Consider the Landau relations
\begin{align}
	\e^{i\,\vec b\cdot\vec\sigma}&=\cos(|\vec b|)+i\sin(|\vec b|)\frac{\vec b\cdot\vec\sigma}{|\vec b|}\;,\label{eq:Landau_1}\\[0.4em]
	\e^{\vec b\cdot\vec\sigma}&=\cosh(|\vec b|)+\sinh(|\vec b|)\frac{\vec b\cdot\vec\sigma}{|\vec b|}\;,\label{eq:Landau_2}
\end{align}
and identify the $\vec b$ vector with
\begin{align}
	\vec b=\sqrt{|\lambda_n| d\tau}\,x_n\vec\psi_{n,j}^{\,[\sigma]}\quad\quad\quad b_\beta=\sqrt{|\lambda_n| d\tau}\,x_n\psi_{n,j\beta}^{\,[\sigma]}\;.
\end{align}
The following expressions for the coefficients of the rotated spinors can be then written, distinguishing the case $\lambda_n<0$ (Eq.~(\ref{eq:a_lambda1})) and the case $\lambda_n>0$ (Eq.~(\ref{eq:a_lambda2})):
\begin{align}
	&\begin{array}{l}
		\widetilde a_{j\uparrow}\!=\!\!\Biggl[\cosh(|\vec b|)+\sinh(|\vec b|)\sgn\!\left(x_n\right)\frac{\psi_{n,jz}^{[\sigma]}}{|\vec \psi_{n,j}^{\,[\sigma]}|}\Biggr]a_{j\uparrow}\!
		+\sinh(|\vec b|)\sgn\!\left(x_n\right)\!\!\Biggl[\frac{\psi_{n,jx}^{[\sigma]}-i\,\psi_{n,jy}^{[\sigma]}}{|\vec \psi_{n,j}^{\,[\sigma]}|}\Biggr]a_{j\downarrow}\;,\!\!\\[1.3em]
		\widetilde a_{j\downarrow}\!=\!\!\Biggl[\cosh(|\vec b|)-\sinh(|\vec b|)\sgn\!\left(x_n\right)\frac{\psi_{n,jz}^{[\sigma]}}{|\vec \psi_{n,j}^{\,[\sigma]}|}\Biggr]a_{j\downarrow}\!
		+\sinh(|\vec b|)\sgn\!\left(x_n\right)\!\!\Biggl[\frac{\psi_{n,jx}^{[\sigma]}+i\,\psi_{n,jy}^{[\sigma]}}{|\vec \psi_{n,j}^{\,[\sigma]}|}\Biggr]a_{j\uparrow}\;,\!\!	
	\end{array}\label{eq:a_lambda1}\\[0.3em]
	&\begin{array}{l}
		\widetilde a_{j\uparrow}\!=\!\!\Biggl[\cos(|\vec b|)+i\,\sin(|\vec b|)\sgn\!\left(x_n\right)\frac{\psi_{n,jz}^{[\sigma]}}{|\vec \psi_{n,j}^{\,[\sigma]}|}\Biggr]a_{j\uparrow}\!
		+\sin(|\vec b|)\sgn\!\left(x_n\right)\!\!\Biggl[\frac{i\,\psi_{n,jx}^{[\sigma]}+\psi_{n,jy}^{[\sigma]}}{|\vec \psi_{n,j}^{\,[\sigma]}|}\Biggr]a_{j\downarrow}\;,\!\!\\[1.3em]
		\widetilde a_{j\downarrow}\!=\!\!\Biggl[\cos(|\vec b|)+i\,\sin(|\vec b|)\sgn\!\left(x_n\right)\frac{\psi_{n,jz}^{[\sigma]}}{|\vec \psi_{n,j}^{\,[\sigma]}|}\Biggr]a_{j\downarrow}\!
		+\sin(|\vec b|)\sgn\!\left(x_n\right)\!\!\Biggl[\frac{i\,\psi_{n,jx}^{[\sigma]}-\psi_{n,jy}^{[\sigma]}}{|\vec \psi_{n,j}^{\,[\sigma]}|}\Biggr]a_{j\uparrow}\;.\!\!
	\end{array}\label{eq:a_lambda2}
\end{align}

When we are dealing with the full 4-dimension single particle spinors, the four coefficients $\widetilde a$ do not have an explicit expression. The exponential of the $\mathcal O_n$ operators acting on the spinors is calculated via a diagonalization procedure. Consider the general $4\times4$ rotation matrix $B_j$ and its eigenvectors $\Psi_{m,j}\ne0$ and eigenvalues $\mu_{m,j}$:
\begin{align}
	B_j\,\Psi_{m,j}=\mu_{m,j}\,\Psi_{m,j}\quad\Rightarrow\quad\Psi_{m,j}^{-1}\,B_j\,\Psi_{m,j}=\mu_{m,j}\quad\quad m=1,\ldots,4\;. 
\end{align}
Using the formal notation $\vec\Psi_j$ and $\vec\mu_j$ to denote the $4\times4$ matrix of eigenvectors and the 4-dimension vector of eigenvalues, it is possible to write the action of $\e^{B_j}$ on a 4-dimensional single particle spinor $|S'\rangle_j$ as follows:
\begin{align}
	\e^{B_j}|S'\rangle_j&=\vec\Psi_j\vec\Psi_j^{-1}\e^{B_j}\vec\Psi_j\vec\Psi_j^{-1}|S'\rangle_j\;,\nonumber\\[0.2em]
	&=\vec\Psi_j\e^{\vec\Psi_j^{-1}B_j\,\vec\Psi_j}\vec\Psi_j^{-1}|S'\rangle_j\;,\nonumber\\[0.2em]
	&=\vec\Psi_j\e^{\text{diag}\left(\vec\mu_j\right)}\vec\Psi_j^{-1}|S'\rangle_j\;,\nonumber\\[0.2em]
	&=\vec\Psi_j\,\text{diag}\left(\e^{\vec\mu_j}\right)\vec\Psi_j^{-1}|S'\rangle_j\;.
\end{align}
Each component of the rotated spinor $|\widetilde S'\rangle_j$ is thus derived from the eigenvectors and eigenvalues of the rotation matrix $B_j$, which is built starting from the $\mathcal O_n$ operators. Moving from neutrons to nucleons, i.e. adding the isospin degrees of freedom to the system, the computational time spent to rotate each single particle spin-isospin state during the propagation is increased by the time for the diagonalization of the $4\times 4$ Hubbard-Stratonovich rotation matrices. However, the total time for the propagation of the wave function as $A$ becomes large, is dominated by the diagonalization of the potential matrices. Since the cost of this operation goes as the cube of the number of matrix rows (columns), the AFDMC computational time is proportional to $A^3$, which is much slower than the scaling factor $A!$ of GFMC. 

In addition to the diagonalization of the AV6 potential matrices and the spinor rotation matrices, we have to deal with the evaluation of the integral over the auxiliary fields $x_n$. The easiest way, in the spirit of Monte Carlo, is to sample the auxiliary fields from the Gaussian of Eq.~(\ref{eq:G_AFDMC}), which is interpreted as a probability distribution. The sampled values are then used to determine the action of the operators on the spin-isospin part of the wave function as described above. The integral over all the spin-isospin rotations induced by the auxiliary fields eventually recovers the action of the quadratic spin-isospin operators on a trial wave function containing all the possible good spin-isospin states, as the GFMC one. 

In this scheme, the integration over the auxiliary fields is performed jointly with the integration over the coordinates. This generally leads to a large variance. The integral of Eq.~(\ref{eq:G_AFDMC}) should be indeed evaluated for each sampled position and not simply estimated ``on the fly''. A more refined algorithm, in which for each sampled configuration the integral over $x_n$ is calculated by sampling more than one auxiliary variable, has been tested. The energy values at convergence are the same for both approaches. However, in the latter case the variance is much reduced, although the computational time for each move is increased due to the iteration over the newly sampled auxiliary points.

As done in the DMC method, see \S~\ref{subsec:Imp_Samp}, we can introduce an importance function to guide the diffusion in the coordinate space also in the AFDMC algorithm. The drift term (\ref{eq:drift}) is added to the $R-R'$ Gaussian distribution of Eq.~(\ref{eq:G_AFDMC}) and the branching weight $\widetilde\omega_i$ is given by the local energy as in Eq.~(\ref{eq:w_IS}). The idea of the importance sampling can be applied to guide the rotation of the spin-isospin states in the Hubbard-Stratonovich transformation. This can be done by properly shifting the Gaussian over the auxiliary fields of Eq.~(\ref{eq:G_AFDMC}) by means of a drift term~$\bar x_n$:
\begin{align}
	\e^{-\frac{x_n^2}{2}+\sqrt{-\lambda_n d\tau}\,x_n\mathcal O_n}=
	\e^{-\frac{(x_n-\bar x_n)^2}{2}}\e^{\sqrt{-\lambda_n d\tau}\,x_n\mathcal O_n}\e^{-\bar x_n\left(x_n-\frac{\bar x_n}{2}\right)} \;,\label{eq:HS_impsamp}
\end{align}  
where
\begin{align}
	\bar x_n=\re\left[\sqrt{-\lambda_n d\tau}\langle\mathcal O_n\rangle_m\right]\;,
\end{align}
and $\langle\mathcal O_n\rangle_m$ is the mixed expectation value of $\mathcal O_n$ (Eq.~(\ref{eq:mixed})) calculated on the old spin-isospin configurations. The mixed estimator is introduced in order to guide the rotations, by maximizing the overlap between the walker and the trial function, which is not generally picked around $x_n=0$. 

The last factor of Eq.~(\ref{eq:HS_impsamp}) can be interpreted as an additional weight term that has to be included in the total weight. By combining diffusion, rotation and all the additional factors we can derive two different algorithms.
\begin{itemize}
	\item[\emph{v1}]\hypertarget{method:PsiT}{} In the first one, the ratio between the importance functions in the new and old configurations (see Eq.~(\ref{eq:ratio})) is kept explicit. However the drifted Gaussian $\widetilde G_0(R,R',d\tau)$ of Eq.~(\ref{eq:G_IS}) is used for the diffusion in the coordinate space and the drifted Gaussian of Eq.~(\ref{eq:HS_impsamp}) for the sampling of auxiliary fields. The weight for the branching process $\omega_i$ defined in Eq.~(\ref{eq:w}) takes then an overall factor
	\begin{align}
		\frac{\langle\psi_I|RS\rangle}{\langle\psi_I|R'S'\rangle} 
		\e^{-\frac{d(R')\left[d(R')+2(R-R')\right]}{4Dd\tau}} 
		\prod_n\e^{-\bar x_n\left(x_n-\frac{\bar x_n}{2}\right)}\;,
	\end{align}
	due to the counter terms coming from the coordinate drift $d(R)=\bm v_d(R)D d\tau$ added in the original $G_0(R,R',d\tau)$ and from the auxiliary field shift~$\bar x_n$.

	\item[\emph{v2}]\hypertarget{method:Elocal}{} The second algorithm corresponds the local energy scheme described in \S~\ref{subsec:Imp_Samp}. Again the coordinates are diffused via the drifted Gaussian $\widetilde G_0(R,R',d\tau)$ of Eq.~(\ref{eq:G_IS}) and the auxiliary fields are sampled from the shifted Gaussian of Eq.~(\ref{eq:HS_impsamp}). The branching weight $\widetilde w_i$ is instead given by the local energy as in Eq.~(\ref{eq:w_IS}). The counter terms related to $\bar x_n$ are automatically included in the weight because the local energy $E_L(R,S)=\frac{H\psi_I(R,S)}{\psi_I(R,S)}$ takes now contributions from all the spin-isospin operators of the full potential $V_{NN}$. Actually, the term $\e^{-\bar x_n x_n}$ vanishes during the auxiliary field integration because $x_n$ can take positive and negative values. The term $\frac{\bar x_n^2}{2}$ is nothing but the $-\frac{1}{2}\lambda_n \langle\mathcal O_n\rangle_m^2 d\tau$ contribution already included in the weight via $E_L(R,S)$.
\end{itemize}
Given the same choice for the drift term, that depends, for example, on the constraint applied to deal with the sign problem, the two algorithms are equivalent and should sample the same Green’s function. 

In both versions, the steps that constitute the AFDMC algorithm are almost the same of the DMC one, reported in \S~\ref{sec:DMC}. The starting point is the initial distribution of walkers, step~\ref{item:DMC1}. In step~\ref{item:DMC2} the diffusion of the coordinates is performed including the drift factor. Now also the spin-isospin degrees of freedom are propagated, by means of the Hubbard-Stratonovich rotations and the integral over the auxiliary fields. As in step~\ref{item:DMC3}, a weight is assigned to each walker, choosing one of the two equivalent solutions proposed above (explicit $\psi_I$~ratio or local energy). Both propagation and weight depend on the prescription adopted in order to keep under control the sign problem. Usually the fixed phase approximation (see \S~\ref{subsec:Sign}) is applied with the evaluation of local operators. The branching process follows then the DMC version described in step~\ref{item:DMC4} and the procedure is iterated in the same way with the computation of expectation values at convergence.

\subsection{Propagator for neutrons: spin-orbit terms}
\label{subsec:Prop_LS}

In the previous section we have seen how to deal in an efficient way with a propagator containing the first six components of the Argonne two-body potential. Next terms in Eq.~(\ref{eq:v_ij_Op}) are the spin-orbit contributions for $p=7,8$. Although an attempt to treat the spin-orbit terms for nucleon systems has been reported by Armani in his Ph.D. thesis~\cite{Armani:2011_thesis} (together with a possible $\bm L_{ij}^2$ inclusion for $p=9$), at present the $\bm L_{ij}\cdot\bm S_{ij}$ operator is consistently employed in the AFDMC algorithm only for neutron systems. No other terms of the $NN$ interaction are included in the full propagator, neither for nucleons nor for neutrons, although a perturbative treatment of the remaining terms of AV18 is also possible~\cite{Pieper:2008}. The full derivation of the neutron spin-orbit propagator is reported in Ref.~\cite{Sarsa:2003}. Here we want just to sketch the idea behind the treatment of this non local term for which the corresponding Green's function is not trivial to be derived.

Consider the spin-orbit potential for neutrons:
\begin{align}
	v_{ij}^{LS}=v_{LS}(r_{ij})\,\bm L_{ij}\cdot\bm S_{ij}=v_{LS}(r_{ij})\left(\bm L\cdot\bm S\right)_{ij}\;,
\end{align}
where
\begin{align}
	 v_{LS}(r_{ij})=v_7(r_{ij})+v_8(r_{ij})\;,
\end{align}
and $\bm L_{ij}$ and $\bm S_{ij}$ are defined respectively by Eqs.~(\ref{eq:LS_ij1}) and (\ref{eq:LS_ij2}). As reported in Ref.~\cite{Pieper:1998}, one way to evaluate the propagator for $\bm L\cdot\bm S$ is to consider the expansion at first order in $d\tau$
\begin{align}
	\e^{-v_{LS}(r_{ij})\left(\bm L\cdot\bm S\right)_{ij}d\tau}\simeq\left[1-v_{LS}(r_{ij})\left(\bm L\cdot\bm S\right)_{ij}d\tau\right]\;,\label{eq:e_SP}
\end{align}
acting on the free propagator $G_0(R,R',d\tau)$ of Eq.~(\ref{eq:G0}). The derivatives terms of the above expression give
\begin{align}
	\left(\bm\nabla_i-\bm\nabla_j\right)G_0(R,R',d\tau)=-\frac{1}{2Dd\tau}\left(\Delta\bm r_i-\Delta\bm r_j\right)G_0(R,R',d\tau)\;,
\end{align}
where $\Delta\bm r_i=\bm r_i-\bm r'_i$ . We can then write:
\begin{align}
	&\left(\bm L\cdot\bm S\right)_{ij}G_0(R,R',d\tau)=\nonumber\\[0.2em]
	&\quad\quad=-\frac{1}{4i}\frac{1}{2Dd\tau}\left(\bm r_i-\bm r_j\right)\times\left(\Delta\bm r_i-\Delta\bm r_j\right)\cdot\left(\bm\sigma_i+\bm\sigma_j\right)G_0(R,R',d\tau)\;,\nonumber\\[0.4em]
	&\quad\quad=-\frac{1}{4i}\frac{1}{2Dd\tau}\left(\bm\Sigma_{ij}\times\bm r_{ij}\right)\cdot\left(\Delta\bm r_i-\Delta\bm r_j\right)G_0(R,R',d\tau)\;,
\end{align}
where $\bm\Sigma_{ij}=\bm\sigma_i+\bm\sigma_j$ and $\bm r_{ij}=\bm r_i-\bm r_j$, and the relation $\bm a\cdot\left(\bm b\times\bm c\right)=\bm c\cdot\left(\bm a\times\bm b\right)$ has been used.

By inserting the last expression in Eq.~(\ref{eq:e_SP}) and re-exponentiating, including also the omitted sum over particle indices $i$ and $j$, the following propagator is obtained:
\begin{align}
	\mathcal P_{LS}\simeq\e^{\sum_{i\ne j}\frac{v_{LS}(r_{ij})}{8iD}\left(\bm\Sigma_{ij}\times\bm r_{ij}\right)\cdot\left(\Delta\bm r_i-\Delta\bm r_j\right)}\;.
\end{align}
The effect of $\mathcal P_{LS}$ can be studied starting from the formal solution
\begin{align}
	\psi(R,S,\tau+d\tau)\stackrel{LS}{\simeq}\int dR'dS'\, G_0(R,R',d\tau)\,\mathcal P_{LS}\,\psi(R',S',\tau)\;,
\end{align}
and expanding the propagator to the second order and the wave function $\psi(R',S',\tau)$ to the first order in $R-R'$. It is possible to show (see Ref.~\cite{Sarsa:2003} for the details) that the spin-orbit contribution of the propagator takes a simple form, but two- and three-body extra corrections appear. However, in the case of neutrons these additional terms contain quadratic spin operators and so they can be handled by the Hubbard-Stratonovich transformation and the rotations over new auxiliary fields.

\subsection{Propagator for neutrons: three-body terms}
\label{subsec:Prop_TNI}

As reported in \S~\ref{subsec:UIX-ILx}, the Illinois (Urbana IX) TNI can be written as a sum of four different terms:
\begin{align}
	V_{ijk}=A_{2\pi}^P\,\mathcal O^{2\pi,P}_{ijk}+A_{2\pi}^{S}\,\mathcal O^{2\pi,S}_{ijk}+A_{3\pi}\,\mathcal O^{3\pi}_{ijk}+A_R\,\mathcal O^R_{ijk} \;.
\end{align}
For neutron systems, being $\bm\tau_i\cdot\bm\tau_j=1$, the operator structure simplify in such a way that $V_{ijk}$ can be recast as a sum of two-body terms only~\cite{Sarsa:2003,Pederiva:2004}. We can therefore handle also the TNI in the AFDMC propagator by means of the Hubbard-Stratonovich transformation. Let analyze how each term of the above relation can be conveniently rewritten for neutron systems.

\begin{itemize}
	\item $\mathcal O^{2\pi,P}_{ijk}$~\emph{term}. The $P$-wave 2$\pi$ exchange term (and also the 3$\pi$ exchange one) of Eq.~(\ref{eq:V_NNN_2pi_P}) includes the OPE operator $X_{ij}$, defined in Eq.~(\ref{eq:X_ij}). $X_{ij}$ involves the $\bm\sigma_i\cdot\bm\sigma_j$ and the $S_{ij}$ operators that can be decomposed via Eqs.~(\ref{eq:sigma_dec}) and (\ref{eq:Sij_dec}) in order to define a $3A\times 3A$ matrix $X_{i\alpha,j\beta}$ analogous to the $A_{i\alpha,j\beta}^{[\sigma]}$ of Eq.~(\ref{eq:A_NN}), where $v_3(r_{ij})\!\rightarrow Y_\pi(r_{ij})$ and $v_5(r_{ij})\!\rightarrow T_\pi(r_{ij})$. The OPE operator can be thus expressed as
	\begin{align}
		X_{ij}=\sigma_{i\alpha}\,X_{i\alpha,j\beta}\,\sigma_{j\beta}\;,
	\end{align}
	where the matrix $X_{i\alpha,j\beta}$ is real with zero diagonal and has the same symmetries of $A_{i\alpha,j\beta}^{[\sigma]}$. The commutator over the $\bm\tau_i$ operators vanishes, while the anticommutator gives simply a factor 2. Recalling that $X_{ij}=X_{ji}$ we can derive the following relation:
	\begin{align}
		\sum_{i<j<k}\mathcal O^{2\pi,P}_{ijk}
		&=\frac{1}{3!}\sum_{i\ne j\ne k}\sum_{cyclic}2\phantom{\frac{1}{4}}\!\!\!\!\Bigl\{X_{ij},X_{jk}\Bigr\}\;,\nonumber\\[0.2em]
		&=2\sum_{i\ne j\ne k}X_{ik}X_{kj}\;,\nonumber\\[-0.4em]
		&=2\sum_{i\ne j}\sum_{\alpha\beta}\sigma_{i\alpha}\Biggl(\sum_{k\gamma} X_{i\alpha,k\gamma}\,X_{k\gamma,j\beta}\Biggr)\sigma_{j\beta}\;,\nonumber\\[0.2em]
		&=2\sum_{i\ne j}\sum_{\alpha\beta}\sigma_{i\alpha}\,X^2_{i\alpha,j\beta}\,\sigma_{j\beta}\;.
	\end{align} 

	\item $\mathcal O^{2\pi,S}_{ijk}$~\emph{term}. In the $S$-wave TPE term the isospin operators do not contribute and we are left with
	\begin{align}
		\sum_{i<j<k}\mathcal O_{ijk}^{2\pi,S}
		&=\frac{1}{3!}\sum_{i\ne j\ne k}\sum_{cyclic}Z_\pi(r_{ij})Z_\pi(r_{jk})\,\bm\sigma_i\cdot\hat{\bm r}_{ij}\,\bm\sigma_k\cdot\hat{\bm r}_{kj}\;,\nonumber\\[0.2em]
		&=\frac{1}{2}\sum_{i\ne j}\sum_{\alpha\beta}\sigma_{i\alpha}\Biggl[\sum_k Z_\pi(r_{ik})\,\hat r_{ik}^\alpha\,Z_\pi(r_{jk})\,\hat r_{jk}^\beta \Biggr]\sigma_{j\beta}\;,\nonumber\\[0.2em]
		&=\frac{1}{2}\sum_{i\ne j}\sum_{\alpha\beta}\sigma_{i\alpha}\,Z_{i\alpha,j\beta}\,\sigma_{j\beta}\;.
	\end{align}	

	\item $\mathcal O^{3\pi}_{ijk}$~\emph{term}. The 3$\pi$ exchange term, even with the isospin reduction for neutrons, still keeps a very complicated operator structure. As reported in Ref.~\cite{Pederiva:2004}, this factor can be conveniently written as a sum of a spin independent and a spin dependent components
	\begin{align}
		\sum_{i<j<k}\mathcal O_{ijk}^{3\pi}=V_c^{3\pi}+V_\sigma^{3\pi}\;,
	\end{align}
	with
	\begin{align}
		V_c^{3\pi}&=\frac{400}{18}\sum_{i\ne j}X_{i\alpha,j\beta}^2\,X_{i\alpha,j\beta}\;,\\[0.5em]
		V_\sigma^{3\pi}&=\frac{200}{54}\sum_{i\ne j}\sum_{\alpha\beta}\sigma_{i\alpha}\Biggl(\sum_{\gamma\delta\mu\nu}
		X_{i\gamma,j\mu}^2\,X_{i\delta,j\nu}\,\varepsilon_{\alpha\gamma\delta}\,\varepsilon_{\beta\mu\nu}\Biggr)\sigma_{j\beta}\;,\nonumber\\[0.2em]
		&=\frac{200}{54}\sum_{i\ne j}\sum_{\alpha\beta}\sigma_{i\alpha}\,W_{i\alpha,j\beta}\,\sigma_{j\beta}\;,
	\end{align}
	where $\varepsilon_{\alpha\beta\gamma}$ is the full antisymmetric tensor.	 

	\item $\mathcal O^R_{ijk}$~\emph{term}. The last spin independent term can be recast as a two body operator as follows
	\begin{align}
		\sum_{i<j<k}\mathcal O^R_{ijk}=G_0^R+\frac{1}{2}\sum_i\left(G_i^R\right)^2\;,
	\end{align}
	with
	\begin{align}
		G_0^R&=-\sum_{i<j}T_\pi^4(r_{ij})\;,\\[0.5em]
		G_i^R&=\sum_{k\ne i}T_\pi^2(r_{ik})\;.
	\end{align}
\end{itemize}

Finally, for neutron systems we can still write the spin dependent part of the $NN$ potential in the form of Eq.~(\ref{eq:V_NN_SD}), with the inclusion of TNI contributions:
\begin{align}
	V_{NN}^{SD}&=\frac{1}{2}\sum_{i\ne j}\sum_{\alpha\beta}\sigma_{i\alpha}\,A_{i\alpha,j\beta}^{[\sigma]}\,\sigma_{j\beta} \;,
\end{align}
where now 
\begin{align}
	A_{i\alpha,j\beta}^{[\sigma]}\longrightarrow
	A_{i\alpha,j\beta}^{[\sigma]}+2A_{2\pi}^P\,X^2_{i\alpha,j\beta}+\frac{1}{2}A_{2\pi}^{S}\,Z_{i\alpha,j\beta}+\frac{200}{54}\,A_{3\pi}\,W_{i\alpha,j\beta}\;.
\end{align}
The central term of the two-body potential of Eq.~(\ref{eq:V_NN_SI}) keeps also contributions from the TNI  3$\pi$ exchange term and from the phenomenological term, and it reads now:
\begin{align}
	V_{NN}^{SI}\longrightarrow V_{NN}^{SI}+A_{3\pi}V_c^{3\pi}+A_R\left[G_0^R+\frac{1}{2}\sum_i\left(G_i^R\right)^2\right]\;.
\end{align}

\subsection{Wave functions}
\label{subsec:Wave}

In this section the trial wave functions used in AFDMC calculations for nuclear and hypernuclear systems will be presented, distinguishing between the case of finite and infinite systems. Restoring the convention of Chapter~\ref{chap:hamiltonians}, which is commonly used in the literature for hypernuclear systems, $A$ will refer to the total number of baryons, $\mathcal N_N$ nucleons plus $\mathcal N_\Lambda$ lambda particles. Latin indices will be used for the nucleons, Greek $\lambda$, $\mu$ and $\nu$ indices for the lambda particles. Finally, the first letters of the Greek alphabet ($\alpha,\beta,\gamma,\delta,\ldots$) used as indices will refer to the Cartesian components of the operators.

\subsubsection{Non strange finite and infinite systems}
\label{subsubsubsec:Wave_non_strange}

As already sketched in \S~\ref{sec:AFDMC}, the AFDMC wave function is written in the single particle state representation. The trial wave function for nuclear systems, which is used both as projection and importance function $|\psi_I\rangle\equiv|\psi_T\rangle$, is assumed of the form~\cite{Gandolfi:2007,Gandolfi:2009}
\begin{align}
	\psi_T^N(R_N,S_N)=\prod_{i<j}f_c^{NN}(r_{ij})\,\Phi_N(R_N,S_N)\;,\label{eq:psi_N}
\end{align}
where $R_N=\{\bm r_1,\ldots,\bm r_{\mathcal N_N}\}$ are the Cartesian coordinates and $S_N=\{s_1,\ldots,s_{\mathcal N_N}\}$ the spin-isospin coordinates, represented as complex 4- or 2-component vectors:
\begin{align}
	\text{nucleons:}&\quad s_i=\left(\begin{array}{c} 
			a_i \\ b_i \\ c_i \\ d_i
		\end{array}\right)_i
		\!=a_i|p\uparrow\rangle_i+b_i|p\downarrow\rangle_i+c_i|n\uparrow\rangle_i+d_i|n\downarrow\rangle_i \;,\\[0.5em]
	\text{neutrons:}&\quad s_i=\left(\begin{array}{c} 
			a_i \\ b_i
		\end{array}\right)_i
		\!=a_i|n\uparrow\rangle_i+b_i|n\downarrow\rangle_i\;,
\end{align}
with $\left\{|p\uparrow\rangle,|p\downarrow\rangle,|n\uparrow\rangle,|n\downarrow\rangle\right\}$ the proton-neutron-up-down basis.

The function $f_c^{NN}(r)$ is a symmetric and spin independent Jastrow correlation function, solution of the Schr\"odinger-like equation for $f_c^{NN}(r<d)$
\begin{align}
	-\frac{\hbar^2}{2\mu_{NN}}\nabla^2 f_c^{NN}(r)+\eta\,v_c^{NN}(r)f_c^{NN}(r)=\xi f_c^{NN}(r)\;,\label{eq:Jastrow}
\end{align}
where $v_c^{NN}(r)$ is the spin independent part of the two-body $NN$ interaction, $\mu_{NN}=m_N/2$ the reduced mass of the nucleon pair and $\eta$ and the healing distance $d$ are variational parameters. For distances $r\ge d$ we impose $f_c^{NN}(r)=1$. The role of the Jastrow function is to include the short-range correlations in the trial wave function. In the AFDMC algorithm the effect is simply a reduction of the overlap between pairs of particles, with the reduction of the energy variance. Since there is no change in the phase of the wave function, the $f_c^{NN}$ function does not influence the computed energy value in the long imaginary time projection.

The antisymmetric part $\Phi_N(R_N,S_N)$ of the trial wave function depends on the system to be studied (finite or infinite). As already seen, it is generally built starting from single particle states $\varphi_\epsilon^N(\bm r_i,s_i)$, where $\epsilon$ is the set of quantum numbers describing the state and $\bm r_i$, $s_i$ the single particle nucleon coordinates. The antisymmetry property is then realized by taking the Slater determinant of the $\varphi_\epsilon^N$:
\begin{align}
	\Phi_N(R_N,S_N)=\mathcal A\Bigg[\prod_{i=1}^{\mathcal N_N}\varphi_\epsilon^N(\bm r_i,s_i)\Bigg]=\det\Bigl\{\varphi_\epsilon^N(\bm r_i,s_i)\Bigr\}\;.\label{eq:Phi_N}
\end{align}

For nuclei and neutron drops~\cite{Gandolfi:2007} a good set of quantum number is given by $\epsilon=\{n,j,m_j\}$. The single particle states are written as:
\begin{align}
	\varphi_\epsilon^N(\bm r_i,s_i)=R_{n,j}^N(r_i)\Bigl[Y_{l,m_l}^N(\Omega)\,\chi_{s,m_s}^N(s_i)\Bigr]_{j,m_j}\;,
\end{align}
where $R_{n,j}^N$ is a radial function, $Y_{l,m_l}^N$ the spherical harmonics depending on the solid angle $\Omega$ and $\chi_{s,m_s}^N$ the spinors in the proton-neutron-up-down basis. The angular functions are coupled to the spinors using the Clebsh-Gordan coefficients to have orbitals in the $\{n,j,m_j\}$ basis according to the usual shell model classification of the nuclear single particle spectrum. For finite systems, in order to make the wave function translationally invariant, the single particle orbitals have to be defined with respect to the center of mass (CM) of the system. We have thus:
\begin{align}
	\varphi_\epsilon^N(\bm r_i,s_i)\longrightarrow\varphi_\epsilon^N(\bm r_i-\bm r_{CM},s_i)\quad\quad\text{with}\quad \bm r_{CM}=\frac{1}{\mathcal N_N}\sum_{i=1}^{\mathcal N_N} \bm r_i\;.
\end{align}
In order to deal with new shifted coordinates, we need to correct all the first and second derivatives of trial wave function with respect to $\bm r_i$. The derivation of such corrections is reported in Appendix~\ref{app:Wave}. The choice of the radial functions $R_{n,j}^N$ depends on the system studied and, typically, solutions of the self-consistent Hartree-Fock problem with Skyrme interactions are adopted. For nuclei the Skyrme effective interactions of Ref.~\cite{Bai:1997} are commonly used. For neutron drops, the Skyrme SKM force of Ref.~\cite{Pethick:1995} has been considered. 

An additional aspect to take care when dealing with finite systems, is the symmetry of the wave function. Because the AFDMC projects out the lowest energy state not orthogonal to the starting trial wave function, it is possible to study a state with given symmetry imposing to the trial wave function the total angular momentum $J$ experimentally observed. This can be achieved by taking a sum over a different set of determinants 
\begin{align}
	\det\Bigl\{\varphi_\epsilon^N(\bm r_i,s_i)\Bigr\}\longrightarrow\left[\sum_\kappa c_\kappa\,\text{det}_\kappa\Bigl\{\varphi_{\epsilon_\kappa}^N(\bm r_i,s_i)\Bigr\}\right]_{J,M_J}\;,
\end{align}
where the $c_\kappa$ coefficients are determined in order to have the eigenstate of total angular momentum $J=j_1+\ldots+j_{\mathcal N_N}$.

For nuclear and neutron matter~\cite{Gandolfi:2009}, the antisymmetric part of the wave function is given by the ground state of the Fermi gas, built from a set of plane waves. The infinite uniform system at a given density is simulated with $\mathcal N_N$ nucleons in a cubic box of volume $L^3$ replicated into the space by means of periodic boundary conditions (PBC):
\begin{align}
	\varphi_\epsilon^N(\bm r_1+L\hat{\bm r},\bm r_2,\ldots,s_i)=\varphi_\epsilon^N(\bm r_1,\bm r_2,\ldots,s_i)\;.\label{eq:PBC}
\end{align}
Working in a discrete space, the momentum vectors are quantized and can be expressed as
\begin{align}
	\bm k_\epsilon=\frac{2\pi}{L}\left(n_x,n_y,n_z\right)_\epsilon\;,\label{eq:k_vec}
\end{align}
where $\epsilon$ labels the quantum state and $n_x$, $n_y$ and $n_z$ are integer numbers labelling the momentum shell. The single particle states are then given by
\begin{align}
	\varphi_\epsilon^N(\bm r_i,s_i)=\e^{-i\bm k_\epsilon\cdot\bm r_i}\chi_{s,m_s}^N(s_i)\;.
\end{align}

In order to meet the requirement of homogeneity and isotropy, the shell structure of the system must be closed. The total number of Fermions in a particular spin-isospin configuration that can be correctly simulated in a box corresponds to the closure of one of the $\left(n_x,n_y,n_z\right)_\epsilon$ shells. The list of the first closure numbers is 
\begin{align}
	\mathcal N_c=1,7,19,27,33,57,81,93\ldots\;.\label{eq:n_c}
\end{align}
Given a closure number $\mathcal N_c^N$, we can thus simulate an infinite system by means of a periodic box with $2\,\mathcal N_c^N$ neutrons (up and down spin) or $4\,\mathcal N_c^N$ nucleons (up and down spin and isospin). Although the use of PBC should reduce the finite-size effects, in general there are still sizable errors in the kinetic energy arising from shell effects in filling the plane wave orbitals, even at the closed shell filling in momentum space. However, in the thermodynamical limit $\mathcal N_c^N\rightarrow\infty$, exact results should be obtained. For symmetric nuclear matter~(SNM), 28, 76 and 108 nucleons have been used~\cite{Gandolfi:2007_SNM}, resulting in comparable results for the energy per particle at a given density. In the case of pure neutron matter~(PNM), finite-size effects are much more evident~\cite{Gandolfi:2009} and the thermodynamical limit is not reached monotonically. Typically, PNM is simulated using 66 neutrons, which was found to give the closest kinetic energy compared to the Fermi gas in the range of $\mathcal N_c^N$ corresponding to feasible computational times.

As reported in Ref.~\cite{Lin:2001}, twist-averaged boundary conditions (TABC) can be imposed on the trial wave function to reduce the finite-size effects. One can allow particles to pick up a phase $\theta$ when they wrap around the periodic boundaries:
\begin{align}
	\varphi_\epsilon^N(\bm r_1+L\hat{\bm r},\bm r_2,\ldots,s_i)=\e^{i\theta}\varphi_\epsilon^N(\bm r_1,\bm r_2,\ldots,s_i)\;.\label{eq:TABC}
\end{align}
The boundary condition $\theta=0$ corresponds to the PBC, $\theta\ne 0$ to the TABC. It has been shown that if the twist phase is integrated over, the finite size effects are substantially reduced. TABC has been used in PNM calculations~\cite{Gandolfi:2009}, showing a small discrepancy in the energy per particle for 38, 45, 66 and 80 neutrons at fixed density. A remarkable result is that the PNM energy for 66 neutrons using PBC is very close to the extrapolated result obtained employing the TABC, validating then the standard AFDMC calculation for 66 particles. Compare to PBC, employing the TABC results in a more computational time and they have not been used in this work.

\subsubsection{Strange finite and infinite systems}
\label{subsubsubsec:Wave_strange}

The $\Lambda$~hyperon, having isospin zero, does not participate to the isospin doublet of nucleons. Referring to hypernuclear systems, we can therefore treat the additional strange baryons as distinguishable particles writing a trial wave function of the form
\begin{align}
	\psi_T(R,S)=\prod_{\lambda i}f_c^{\Lambda N}(r_{\lambda i})\,\psi_T^N(R_N,S_N)\,\psi_T^\Lambda(R_\Lambda,S_\Lambda)\;,\label{eq:Psi_T}
\end{align}
where $R=\{\bm r_1,\ldots,\bm r_{\mathcal N_N},\bm r_1,\ldots,\bm r_{\mathcal N_\Lambda}\}$ and $S=\{s_1,\ldots,s_{\mathcal N_N},s_1,\ldots,s_{\mathcal N_\Lambda}\}$ refer to the coordinates of all the baryons and $\psi_T^N(R_N,S_N)$ is the nucleon single particle wave function of Eq.~(\ref{eq:psi_N}). $\psi_T^\Lambda(R_\Lambda,S_\Lambda)$ is the lambda single particle wave function that takes the same structure of the nucleon one:
\begin{align}
	\psi_T^\Lambda(R_\Lambda,S_\Lambda)=\prod_{\lambda<\mu}f_c^{\Lambda\Lambda}(r_{\lambda\mu})\,\Phi_\Lambda(R_\Lambda,S_\Lambda)\;,\label{eq:psi_L}
\end{align}
with
\begin{align}
	\Phi_\Lambda(R_\Lambda,S_\Lambda)=\mathcal A\Bigg[\prod_{\lambda=1}^{\mathcal N_\Lambda}\varphi_\epsilon^\Lambda(\bm r_\lambda,s_\lambda)\Bigg]=\det\Bigl\{\varphi_\epsilon^\Lambda(\bm r_\lambda,s_\lambda)\Bigr\}\;.
\end{align}
$R_\Lambda=\{\bm r_1,\ldots,\bm r_{\mathcal N_\Lambda}\}$ are the hyperon Cartesian coordinates and $S_\Lambda=\{s_1,\ldots,s_{\mathcal N_\Lambda}\}$ the hyperon spin coordinates, represented by the 2-dimension spinor in the lambda-up-down basis:
\begin{align}
	s_\lambda=\left(\begin{array}{c} 
		u_\lambda \\ v_\lambda
	\end{array}\right)_\lambda
	\!=u_\lambda|\Lambda\uparrow\rangle_\lambda+v_\lambda|\Lambda\downarrow\rangle_\lambda\;.
\end{align}

The $\Lambda\Lambda$ Jastrow correlation function $f_c^{\Lambda\Lambda}(r)$ is calculated by means of Eq.~(\ref{eq:Jastrow}) for the hyperon-hyperon pair using the central channel of the $\Lambda\Lambda$ potential of Eq.~(\ref{eq:V_LL}). Eq.~(\ref{eq:Jastrow}) is also used to calculate the hyperon-nucleon correlation function $f_c^{\Lambda N}(r)$ of the hypernuclear wave function~(\ref{eq:Psi_T}) by considering the pure central term of the $\Lambda N$ potential of Eq.~(\ref{eq:V_LN}) and using the reduced mass
\begin{align}
	\mu_{\Lambda N}=\frac{m_\Lambda\,m_N}{m_\Lambda+m_N}\;.
\end{align}

For $\Lambda$~hypernuclei (and $\Lambda$~neutron drops) the hyperon single particle states take the same structure as the nuclear case, and they read:
\begin{align}
	\varphi_\epsilon^\Lambda(\bm r_\lambda,s_\lambda)=R_{n,j}^\Lambda(r_\lambda)\Bigl[Y_{l,m_l}^\Lambda(\Omega)\,\chi_{s,m_s}^\Lambda(s_\lambda)\Bigr]_{j,m_j}\;.\label{eq:varphi_L}
\end{align}
Although the AFDMC code for hypernuclei is set up for an arbitrary number of hyperons, we focused on single and double $\Lambda$~hypernuclei. Having just two hyperons to deal with, only one radial function $R_{n,j}^\Lambda$ is needed. Being the mass difference between the neutron and the $\Lambda$~particle small, we used the neutron $1s_{1/2}$ radial function also for the hyperon. 

Dealing with finite systems, the coordinates of all the baryons must be related to the CM, that now is given by the coordinates of particles with different mass. Nucleon and hyperon single particle orbitals are thus defined as:
\begin{align}
	\begin{array}{rcl}
		\varphi_\epsilon^N(\bm r_i,s_i)\!\!\!&\longrightarrow&\!\!\varphi_\epsilon^N(\bm r_i-\bm r_{CM},s_i)\\[0.5em]
		\varphi_\epsilon^\Lambda(\bm r_\lambda,s_\lambda)\!\!\!&\longrightarrow&\!\!\varphi_\epsilon^\Lambda(\bm r_\lambda-\bm r_{CM},s_\lambda)
	\end{array}
\end{align}
where
\begin{align}
	\bm r_{CM}=\frac{1}{M}\left(m_N\sum_{i=1}^{\mathcal N_N}\bm r_i+m_\Lambda\sum_{\lambda=1}^{\mathcal N_\Lambda}\bm r_\lambda\right)
	\quad\text{with}\quad M=\mathcal N_N\,m_N+\mathcal N_\Lambda\,m_\Lambda\;.
\end{align}
As in the nuclear case, the use of relative coordinates introduces corrections in the calculation of the derivatives of the trial wave function. For hypernuclei such corrections, and in general the evaluation of derivatives, are more complicated than for nuclei. This is because we have to deal with two set of spacial coordinates ($R_N$ and $R_\Lambda$) and the Jastrow function $f_c^{\Lambda N}$ depends on both. The derivatives of the trial wave function including CM corrections are reported in Appendix~\ref{app:Wave}.

For $\Lambda$~neutron (nuclear) matter the antisymmetric part of the hyperon wave function is given by the ground state of the Fermi gas, as for nucleons. We are thus dealing with two Slater determinants of plane waves with $\bm k_\epsilon$~vectors quantized in the same $L^3$ box (see Eq.~(\ref{eq:k_vec})). The dimension of the simulation box, and thus the quantization of the $\bm k_\epsilon$ vectors, is given by the total numeric baryon density 
\begin{align}
	\rho_b=\frac{\mathcal N_b}{L^3}=\frac{\mathcal N_N+\mathcal N_\Lambda}{L^3}=\rho_N+\rho_\Lambda\;,\label{eq:rho_b}
\end{align}
and the number of nucleons and lambda particles. The hyperon single particle states correspond then to
\begin{align}
	\varphi_\epsilon^\Lambda(\bm r_\lambda,s_\lambda)=\e^{-i\bm k_\epsilon\cdot\bm r_\lambda}\chi_{s,m_s}^\Lambda(s_\lambda)\;,
\end{align}
where the $\bm k_\epsilon$ structure derived from $\rho_b$ is used also for the the nuclear part. The requirements of homogeneity and isotropy discussed in the previous section still apply and so the lambda plane waves have to close their own momentum shell structure. Given the list of closure numbers~(\ref{eq:n_c}), we can add $2\,\mathcal N_c^\Lambda$ hyperons (up and down spin) to the $2\,\mathcal N_c^N$ neutrons or $4\,\mathcal N_c^N$ nucleons in the periodic box.

The wave functions described so far are appropriate only if we consider nucleons and hyperons as distinct particles. In this way, it is not possible to include the $\Lambda N$ exchange term of Eq.~(\ref{eq:V_LN}) directly in the propagator, because it mixes hyperon and nucleon states. The complete treatment of this factor would require a drastic change in the AFDMC code and/or a different kind of nuclear-hypernuclear interactions, as briefly discussed in Appendix~\ref{app:Px}. A perturbative analysis of the $v_0(r_{\lambda i})\,\varepsilon\,\mathcal P_x$ term is however possible and it is reported in the next section.

\subsection{Propagator for hypernuclear systems}
\label{subsec:Prop_YN}

Consider a many-body system composed by nucleons and hyperons, interacting via the full Hamiltonian~(\ref{eq:H}) and described by the trial wave function~(\ref{eq:Psi_T}). Suppose to switch off all the spin-isospin interactions in all the channels and keep only the central terms:
\begin{align}
	H=T_N+T_\Lambda+V_{NN}^c+V_{\Lambda\Lambda}^c+V_{\Lambda N}^c\;,
\end{align}
where also the central contributions from the three-body interactions are included. Neglecting the spin-isospin structure of the trial wave function we can follow the idea of the standard DMC described in \S~\ref{sec:DMC} and write the analog of Eq.~(\ref{eq:psi_propag}):
\begin{align}
	&\psi(R_N,R_\Lambda,\tau+d\tau)\simeq\int dR'_N\,dR'_\Lambda\langle R_N,R_\Lambda|
	\e^{-\left(V_{NN}^c+V_{\Lambda\Lambda}^c+V_{\Lambda N}^c\right)\frac{d\tau}{2}}
	\e^{-T_N d\tau}\e^{-T_\Lambda d\tau}\times\nonumber\\[0.5em]
	&\hspace{0.5cm}\times\e^{-\left(V_{NN}^c+V_{\Lambda\Lambda}^c+V_{\Lambda N}^c\right)\frac{d\tau}{2}}\e^{E_Td\tau}
	|R'_N,R'_\Lambda\rangle\,\psi(R'_N,R'_\Lambda,\tau)\;,\nonumber\\[0.8em]
	&\simeq\int dR'_N\,dR'_\Lambda\underbrace{\langle R_N|\e^{-T_N d\tau}|R'_N\rangle}_{G_0^N(R_N,R'_N,d\tau)}
	\underbrace{\langle R_\Lambda|\e^{-T_\Lambda d\tau}|R'_\Lambda\rangle}_{G_0^\Lambda(R_\Lambda,R'_\Lambda,d\tau)}\times\nonumber\\[0.5em]
	&\hspace{0.5cm}\times\underbrace{\phantom{\langle}\!\!
	\e^{-\left(\widetilde V_{NN}^c(R_N,R'_N)+\widetilde V_{\Lambda\Lambda}^c(R_\Lambda,R'_\Lambda)+\widetilde V_{\Lambda N}^c(R_N,R_\Lambda,R'_N,R'_\Lambda)-E_T\right)d\tau}}
	_{G_V(R_N,R_\Lambda,R'_N,R'_\Lambda,d\tau)}\psi(R'_N,R'_\Lambda,\tau)\;,\nonumber\\[0.8em]
	&\simeq\left(\frac{1}{4\pi D_N d\tau}\right)^{\frac{3\mathcal N_N}{2}}\!\!\left(\frac{1}{4\pi D_\Lambda d\tau}\right)^{\frac{3\mathcal N_\Lambda}{2}}\!\!
	\int dR'_N\,dR'_\Lambda\e^{-\frac{(R_N-R'_N)^2}{4D_N d\tau}}\e^{-\frac{(R_\Lambda-R'_\Lambda)^2}{4D_\Lambda d\tau}}\times\nonumber\\[0.5em]
	&\hspace{0.5cm}\times\e^{-\left(\widetilde V_{NN}^c(R_N,R'_N)+\widetilde V_{\Lambda\Lambda}^c(R_\Lambda,R'_\Lambda)
	+\widetilde V_{\Lambda N}^c(R_N,R_\Lambda,R'_N,R'_\Lambda)-E_T\right)d\tau}\psi(R'_N,R'_\Lambda,\tau)\;,
	\label{eq:psi_hyp_propag}
\end{align}
where 
\begin{align}
	\widetilde V_{NN}^c(R_N,R'_N)&=\frac{1}{2}\Bigl[V_{NN}^c(R_N)+V_{NN}^c(R'_N)\Bigr]\;,\nonumber\\[0.5em]
	\widetilde V_{\Lambda\Lambda}^c(R_\Lambda,R'_\Lambda)&=\frac{1}{2}\Bigl[V_{\Lambda\Lambda}^c(R_\Lambda)+V_{\Lambda\Lambda}^c(R'_\Lambda)\Bigr]\;,\label{eq:V_tilde}\\[0.5em]
	\widetilde V_{\Lambda N}^c(R_N,R_\Lambda,R'_N,R'_\Lambda)&=\frac{1}{2}\Bigl[V_{\Lambda N}^c(R_N,R_\Lambda)+V_{\Lambda N}^c(R'_N,R'_\Lambda)\Bigr]\;,\nonumber
\end{align}
and $D_N=\hbar^2/2m_N$ and $D_\Lambda=\hbar^2/2m_\Lambda$ are the diffusion constants of the Brownian motion of nucleons and lambda particles. 

The evolution in imaginary time is thus performed in the same way of the standard DMC algorithm. A set of walkers, which now contains nucleon and hyperon coordinates, is diffused according to $G_0^N(R_N,R'_N,d\tau)$ and $G_0^\Lambda(R_\Lambda,R'_\Lambda,d\tau)$. A weight $\omega_i=G_V(R_N,R_\Lambda,R'_N,R'_\Lambda,d\tau)$ is assigned to each waker and it is used for the estimator contributions and the branching process. We can also apply the importance function technique, the result of which is the inclusion of a drift term in the diffusion of each type of baryon and the use of the local energy for the branching weight. The drift velocities take the same form of Eq.~(\ref{eq:drift}), but now the derivatives are calculated with respect to nucleon or hyperon coordinates, including all the possible CM (for finite systems) and Jastrow corrections, as reported in Appendix~\ref{app:Wave}.

Reintroduce now the spin-isospin structure in the wave function and consider then the spin-isospin dependent interactions. For the nuclear part, we can still deal with AV6 like potentials for nucleon systems by means of the Hubbard-Stratonovich transformation, as discussed in \S~\ref{subsec:Prop_AV6}. In the case of pure neutron systems, we can also include spin-orbit and three-body contributions as reported in \S~\ref{subsec:Prop_LS} and \S~\ref{subsec:Prop_TNI}. In the next we will discuss how to deal with the spin-isospin dependent part of the hypernuclear potentials, in both two- and three-body channels.

\subsubsection{Two-body terms}
\label{subsubsec:Prop_LN}

Consider the full two-body $\Lambda N$ interaction described in the previous chapter:
\begin{align}
	V_{\Lambda N}&=\sum_{\lambda i}\left(v_{\lambda i}+v_{\lambda i}^{CSB}\right)\;,\nonumber\\[0.5em]
	&=\sum_{\lambda i}v_0(r_{\lambda i})(1-\varepsilon)+\sum_{\lambda i}v_0(r_{\lambda i})\,\varepsilon\,\mathcal P_x
	+\sum_{\lambda i}\frac{1}{4}v_\sigma T^2_\pi(r_{\lambda i})\,\bm\sigma_\lambda\cdot\bm\sigma_i\nonumber \\[0.2em]
	&\quad+\sum_{\lambda i}C_\tau\,T_\pi^2\left(r_{\lambda i}\right)\tau_i^z\nonumber\;,\\[0.5em]
	&=\sum_{\lambda i}v_0(r_{\lambda i})(1-\varepsilon)+\sum_{\lambda i}B_{\lambda i}^{[\mathcal P_x]}\,\mathcal P_x
	+\sum_{\lambda i}\sum_\alpha\sigma_{\lambda\alpha}\,B_{\lambda i}^{[\sigma]}\,\sigma_{i\alpha}+\sum_i B_i^{[\tau]}\,\tau_i^z\;,\label{eq:V_LN_prop}
\end{align}
where
\begin{align}
	B_{\lambda i}^{[\mathcal P_x]}=v_0(r_{\lambda i})\,\varepsilon\quad\quad
	B_{\lambda i}^{[\sigma]}=\frac{1}{4}v_\sigma T^2_\pi(r_{\lambda i})\quad\quad
	B_i^{[\tau]}=\sum_\lambda C_\tau\,T_\pi^2\left(r_{\lambda i}\right)\;.
\end{align}
The first term of Eq.~(\ref{eq:V_LN_prop}) is simply a spin independent factor and can be included in the $V_{\Lambda N}^c$ contribution of Eq.~(\ref{eq:psi_hyp_propag}). The remaining terms involve operators acting on nucleons and hyperons and need to be discussed separately.
\begin{itemize}
	\item $\bm\sigma_\lambda\cdot\bm\sigma_i$~\emph{term}. The quadratic spin-spin term of the $\Lambda N$ interaction is written in same form of the nucleon-nucleon one of Eq.~(\ref{eq:V_NN_SD}). However, in general the matrix $B_{\lambda i}^{[\sigma]}$ is not a square matrix ($\dim B_{\lambda i}^{[\sigma]}=\mathcal N_\Lambda\times\mathcal N_N$) and so we can not follow the derivation of \S~\ref{subsec:Prop_AV6}. Recalling that we are working with single particle wave functions and that each spin-isospin operator is the representation in the $A$-body tensor product space of a one-body operator as in Eq.~(\ref{eq:O_j}), we can write
	\begin{align}
		\!\!\!\!\sum_\alpha\sigma_{\lambda\alpha}\otimes\sigma_{i\alpha}=\frac{1}{2}\sum_\alpha\left[\left(\sigma_{\lambda\alpha}\oplus\sigma_{i\alpha}\right)^2
		-\left(\sigma_{\lambda\alpha}\otimes\mathbb I_{i\alpha}\right)^2-\left(\mathbb I_{\lambda\alpha}\otimes\sigma_{i\alpha}\right)^2\right]\;.\!
	\end{align}
	The square of the Pauli matrices of the last two terms gives the identity with respect to the single particle state $\lambda$ or $i$, so that they can be simply written as a spin independent contribution
	\begin{align}
		\sum_\alpha\sigma_{\lambda\alpha}\otimes\sigma_{i\alpha}=-3+\frac{1}{2}\sum_\alpha\left(\mathcal O_{\lambda i,\alpha}^{[\sigma_{\Lambda N}]}\right)^2\;,
	\end{align}
	where we have defined a new spin-spin operator 
	\begin{align}
		\mathcal O_{\lambda i,\alpha}^{[\sigma_{\Lambda N}]}=\sigma_{\lambda\alpha}\oplus\sigma_{i\alpha}\;.\label{eq:O_LN}
	\end{align}
	We can now write the the $\bm\sigma_\lambda\cdot\bm\sigma_i$ term as follows
	\begin{align}
		V_{\Lambda N}^{\sigma\sigma}&=\sum_{\lambda i}\sum_\alpha\sigma_{\lambda\alpha}\,B_{\lambda i}^{[\sigma]}\,\sigma_{i\alpha}\;,\nonumber\\[0.2em]
		&=-3\sum_{\lambda i}B_{\lambda i}^{[\sigma]}+\frac{1}{2}\sum_{\lambda i}\sum_\alpha B_{\lambda i}^{[\sigma]}\left(\mathcal O_{\lambda i,\alpha}^{[\sigma_{\Lambda N}]}\right)^2
	\end{align}
	The first term is a central factor that can be included in $V_{\Lambda N}^c$. The second term is written in the same way of the spin-isospin dependent part of the nuclear interaction of Eq.~(\ref{eq:V_NN_SD_On}). With a little abuse of notation $n=\{\lambda, i\}={1,\ldots,\mathcal N_N\,\mathcal N_\Lambda}$, the spin dependent part of the propagator for the $\bm\sigma_\lambda\cdot\bm\sigma_i$ takes a suitable form for the application of the Hubbard-Stratonovich transformation:
	\begin{align}
		\e^{-\sum_{\lambda i}\sum_\alpha\sigma_{\lambda\alpha}\,B_{\lambda i}^{[\sigma]}\,\sigma_{i\alpha}\,d\tau}
		&=\e^{-3\sum_n B_n^{[\sigma]}
		-\frac{1}{2}\sum_{n\alpha} B_n^{[\sigma]}\left(\mathcal O_{n,\alpha}^{[\sigma_{\Lambda N}]}\right)^2d\tau}\;,\nonumber\\[0.2em]
		&=\e^{-V_{\Lambda N}^{c\,[\sigma]}}\prod_{n\alpha}\e^{-\frac{1}{2}B_n^{[\sigma]}\left(\mathcal O_{n\alpha}^{[\sigma_{\Lambda N}]}\right)^2d\tau}\,+\ord\!\left(d\tau^2\right)\;,\nonumber\\[0.2em]
		&\simeq\e^{-V_{\Lambda N}^{c\,[\sigma]}}\prod_{n\alpha}\frac{1}{\sqrt{2\pi}}\int\!dx_{n\alpha}\e^{\frac{-x_{n\alpha}^2}{2}+\sqrt{-B_n^{[\sigma]} d\tau}\,x_{n\alpha}\mathcal O_{n\alpha}^{[\sigma_{\Lambda N}]}}\;.
	\end{align}
	Recalling Eq.~(\ref{eq:G_AFDMC}), we can write the hyperon spin dependent part of the AFDMC propagator for hypernuclear systems as
	\begin{align}
		\langle S_N S_\Lambda|\prod_{n\alpha=1}^{3\mathcal N_N\mathcal N_\Lambda}\!\frac{1}{\sqrt{2\pi}}\int\!dx_{n\alpha}\e^{\frac{-x_{n\alpha}^2}{2}+\sqrt{-B_n^{[\sigma]} d\tau}\,x_{n\alpha}\mathcal O_{n\alpha}^{[\sigma_{\Lambda N}]}}|S'_N S'_\Lambda\rangle\;.		
	\end{align}
	By the definition of Eq.~(\ref{eq:O_LN}), it comes out that the action of the operator $\mathcal O_{n\alpha}^{[\sigma_{\Lambda N}]}$ on the spinor $|S'_N,S'_\Lambda\rangle$ factorizes in a $\sigma_{i\alpha}$ rotation for the nucleon spinor $|S_N\rangle$ and a $\sigma_{\lambda\alpha}$ rotation for the $\Lambda$ spinor $|S_\Lambda\rangle$, coupled by the same coefficient $\sqrt{-B_n^{[\sigma]} d\tau}\,x_{n\alpha}$. 
	
	\item $\tau_i^z$~\emph{term}. As already seen in \S~\ref{sec:AFDMC}, the single particle wave function is closed with respect to the application of a propagator containing linear spin-isospin operators. The action of the CSB potential corresponds to the propagator
	\begin{align}
		\e^{-\sum_iB_i^{[\tau]}\,\tau_i^z\,d\tau}=\prod_i\e^{-B_i^{[\tau]}\,\tau_i^z\,d\tau}\,+\ord\left(d\tau^2\right)\;,
	\end{align}
	that, acting on the trial wave function, simply produces a rotation of the nucleon spinors, as in Eq.~(\ref{eq:eO_S}). Being the CSB term already linear in $\tau_i^z$, there is no need for Hubbard-Stratonovich transformation. The $\tau_i^z$ rotations can be applied after the integration over auxiliary fields on the spinors modified by the Hubbard-Stratonovich rotations. In the $\psi_I$~ratio AFDMC algorithm~(\hyperlink{method:PsiT}{\emph{v1}}) there are no additional terms in the weight coming from the CSB rotations. If we use the local energy version of the algorithm~(\hyperlink{method:Elocal}{\emph{v2}}), we need to subtract the CSB contribution to $E_L(R)$ (Eq.~(\ref{eq:E_L})) because there are no counter terms coming from the importance sampling on auxiliary fields. Note that in neutron systems, $\tau_i^z$ gives simply a factor $-1$, so the CSB becomes a positive central contribution ($C_\tau<0$) to be added in $V_{\Lambda N}^c$.
	
	\item $\mathcal P_x$~\emph{term}. As discussed in \S~\ref{subsubsubsec:Wave_strange}, the structure of our trial wave function for hypernuclear systems prevents the straightforward inclusion of the $\Lambda N$ space exchange operator in the AFDMC propagator. We can try to treat this contribution perturbatively: $\mathcal P_x$ is not included in the propagator but its effect is calculated as a standard estimator on the propagated wave function. The action of $\mathcal P_x$ is to exchange the position of one nucleon and one hyperon, modifying thus the CM of the whole system due to the mass difference between the baryons. To compute this potential contribution we have thus to sum over all the hyperon-nucleon pairs. For each exchanged pair, all the positions are referred to the new CM and the wave function is evaluated and accumulated. Then, particles are moved back to the original positions and a new pair is processed. At the end of the sum the contribution $\sum_{\lambda i}\mathcal P_x\,\psi$ is obtained. As reported in Refs.~\cite{Shoeb:1998,Usmani:2006,Usmani:2006_He6LL,Usmani:2008}, the $\Lambda N$ space exchange operator induces strong correlations and thus a perturbative approach might not be appropriate. A possible non perturbative extension of the AFDMC code for the space exchange operator is outlined in Appendix~\ref{app:Px}.
\end{itemize}

In the two body hypernuclear sector a $\Lambda\Lambda$ interaction is also employed, as reported in \S~\ref{subsec:LL}. The potential described in Eq.~(\ref{eq:V_LL}) can be recast as
\begin{align}
	V_{\Lambda\Lambda}&=\sum_{\lambda<\mu}\sum_{k=1}^{3}\left(v_0^{(k)}+v_\sigma^{(k)}\,{\bm\sigma}_\lambda\cdot{\bm\sigma}_\mu\right)\e^{-\mu^{(k)}r_{\lambda\mu}^2}\;,\nonumber\\[0.2em]
	&=\sum_{\lambda<\mu}\sum_{k=1}^{3}v_0^{(k)}\e^{-\mu^{(k)}r_{\lambda\mu}^2}+\frac{1}{2}\sum_{\lambda\ne\mu}\sum_{\alpha}\sigma_{\lambda\alpha}\,C_{\lambda\mu}^{[\sigma]}\,\sigma_{\mu\alpha}\;,
\end{align}
where
\begin{align}
	C_{\lambda\mu}^{[\sigma]}=\sum_{k=1}^{3}v_\sigma^{(k)}\e^{-\mu^{(k)}r_{\lambda\mu}^2}\;.
\end{align}
The first term of $V_{\Lambda\Lambda}$ is a pure central factor to be included in $V_{\Lambda\Lambda}^c$, while the second part has exactly the same form of the isospin component of Eq.~(\ref{eq:V_NN_SD}). We can thus diagonalize the $C$ matrix and define a new operator $\mathcal O_{n,\alpha}^{[\sigma_\Lambda]}$ starting from the eigenvectors $\psi_{n,\lambda}^{[\sigma_\Lambda]}$:
\begin{align}
	\mathcal O_{n,\alpha}^{[\sigma_\Lambda]}&=\sum_{\lambda}\sigma_{\lambda\alpha}\,\psi_{n,\lambda}^{[\sigma_\Lambda]}\;.
\end{align}
In this way, the spin dependent part of the hyperon-hyperon interaction becomes
\begin{align}
	V_{\Lambda\Lambda}^{SD}=\frac{1}{2}\sum_{n=1}^{\mathcal N_\Lambda}\sum_{\alpha=1}^3\lambda_n^{[\sigma_\Lambda]}\!\left(\mathcal O_{n\alpha}^{[\sigma_\Lambda]}\right)^2 \;,
\end{align}
and we can apply the Hubbard-Stratonovich transformation to linearize the square dependence of $\mathcal O_{n\alpha}^{[\sigma_\Lambda]}$ introducing the integration over $3\,\mathcal N_\Lambda$ new auxiliary fields and the relative $|S'_\Lambda\rangle$ rotations.

At the end, using the diagonalization of the potential matrices and the derivation reported in this section, the spin-isospin dependent part of the nuclear and hypernuclear two-body potentials (but spin-orbit term for simplicity) reads:
\begin{align}
	V_{NN}^{SD}+V_{\Lambda N}^{SD}&=
	   \frac{1}{2}\sum_{n=1}^{3\mathcal N_N}\lambda_n^{[\sigma]}\!\left(\mathcal O_n^{[\sigma]}\right)^2
	\!+\frac{1}{2}\sum_{n=1}^{3\mathcal N_N}\sum_{\alpha=1}^3\lambda_n^{[\sigma\tau]}\!\left(\mathcal O_{n\alpha}^{[\sigma\tau]}\right)^2
	\!+\frac{1}{2}\sum_{n=1}^{\mathcal N_N}\sum_{\alpha=1}^3\lambda_n^{[\tau]}\!\left(\mathcal O_{n\alpha}^{[\tau]}\right)^2\nonumber\\[0.4em]
	&\,+\frac{1}{2}\sum_{n=1}^{\mathcal N_\Lambda}\sum_{\alpha=1}^3\lambda_n^{[\sigma_\Lambda]}\!\left(\mathcal O_{n\alpha}^{[\sigma_\Lambda]}\right)^2
	\!+\frac{1}{2}\sum_{n=1}^{\mathcal N_N\mathcal N_\Lambda}\!\sum_{\alpha=1}^3 B_n^{[\sigma]}\!\left(\mathcal O_{n\alpha}^{[\sigma_{\Lambda N}]}\right)^2
	\!+\sum_{i=1}^{\mathcal N_N} B_i^{[\tau]}\,\tau_i^z\;.\label{eq:SD_Op}
\end{align}
Using a compact notation, the AFDMC propagator for hypernuclear systems of Eq.~(\ref{eq:psi_hyp_propag}) with the inclusion of spin-isospin degrees of freedom becomes:
\begin{align}
	&\hspace{-0.5em}G(R,S,R',S',d\tau)=\langle R,S|\e^{-(T_N+T_\Lambda+V_{NN}+V_{\Lambda\Lambda}+V_{\Lambda N}-E_T)d\tau}|R',S'\rangle\;,\nonumber\\[0.8em]
	&\hspace{0.4em}\simeq\left(\frac{1}{4\pi D_N d\tau}\right)^{\frac{3\mathcal N_N}{2}}\!\!\left(\frac{1}{4\pi D_\Lambda d\tau}\right)^{\frac{3\mathcal N_\Lambda}{2}}\!\!
	\e^{-\frac{(R_N-R'_N)^2}{4D_N d\tau}}\e^{-\frac{(R_\Lambda-R'_\Lambda)^2}{4D_\Lambda d\tau}}\times\nonumber\\[0.7em]
	&\hspace{0.6cm}\times\e^{-\left(\widetilde V_{NN}^c(R_N,R'_N)+\widetilde V_{\Lambda\Lambda}^c(R_\Lambda,R'_\Lambda)
	+\widetilde V_{\Lambda N}^c(R_N,R_\Lambda,R'_N,R'_\Lambda)-E_T\right)d\tau}\nonumber\\[0.2em]
	&\hspace{0.6cm}\times\langle S_N,S_\Lambda|\prod_{i=1}^{\mathcal N_N}\e^{-B_i^{[\tau]}\,\tau_i^z\,d\tau}
	\prod_{n=1}^{\mathcal M}\frac{1}{\sqrt{2\pi}}\int\!dx_n\e^{\frac{-x_n^2}{2}+\sqrt{-\lambda_n d\tau}\,x_n\mathcal O_n}|S'_N,S'_\Lambda\rangle\;,\label{eq:Prop_full}
\end{align}
where $|R,S\rangle\equiv|R_N,R_\Lambda,S_N,S_\Lambda\rangle$ is the state containing all the coordinates of the baryons and $\widetilde V_{NN}^c$, $\widetilde V_{\Lambda\Lambda}^c$ and $\widetilde V_{\Lambda N}^c$ defined in Eqs.~(\ref{eq:V_tilde}) contain all the possible central factors. Formally, $\mathcal M=15\,\mathcal N_N+3\,\mathcal N_\Lambda+3\,\mathcal N_N\mathcal N_\Lambda$ and $\mathcal O_n$ stays for the various operators of Eq.~(\ref{eq:SD_Op}), which have a different action on the spinors $|S'_N,S'_\Lambda\rangle$. The $\mathcal O_n^{[\sigma]}$, $\mathcal O_{n\alpha}^{[\sigma\tau]}$ and $\mathcal O_{n\alpha}^{[\tau]}$ act on the nucleon spinor $|S'_N\rangle$. The $\mathcal O_{n\alpha}^{[\sigma_\Lambda]}$ rotates the lambda spinor $|S'_\Lambda\rangle$. $\mathcal O_{n\alpha}^{[\sigma_{\Lambda N}]}$ acts on both baryon spinors with a separate rotation for nucleons and hyperons coupled by the same coefficient $(-B_n^{[\sigma]} d\tau)^{1/2}\,x_n$ (recall Eq.~(\ref{eq:O_LN})). The algorithm follows then the nuclear version (\S~\ref{subsec:Prop_AV6}) with the sampling of the nucleon and hyperon coordinates and of the auxiliary fields, one for each linearized operator. The application of the propagator of Eq.~(\ref{eq:Prop_full}) has the effect to rotate the spinors of the baryons. The weight for each walker is then calculate starting from the central part of the interaction with possible counter terms coming from the importance sampling on spacial coordinates and on auxiliary fields (algorithm~\hyperlink{method:PsiT}{\emph{v1}}), or by means of the local energy (algorithm~\hyperlink{method:Elocal}{\emph{v2}}). Fixed phase approximation, branching process and expectation values are the same discussed in \S~\ref{sec:DMC}.

\subsubsection{Three-body terms}
\label{subsubsec:Prop_LNN}

We have already shown in \S~\ref{subsec:Prop_TNI} that for neutron systems the three-body nucleon force can be recast as a sum of two-body terms only. In the case of the three-body $\Lambda NN$ interaction it is possible to verify that the same reduction applies both for nucleon and neutron systems. Let consider the full potential of Eqs.~(\ref{eq:V_LNN_2pi}) and~(\ref{eq:V_LNN_D})
\begin{align}
	V_{\Lambda NN}=\sum_{\lambda,i<j}\left(v_{\lambda ij}^{2\pi,P}+v_{\lambda ij}^{2\pi,S}+v_{\lambda ij}^{D}\right)\;,
\end{align}
and assume the notations:
\begin{align}
	T_{\lambda i}=T_{\pi}(r_{\lambda i})\quad\quad
	Y_{\lambda i}=Y_{\pi}(r_{\lambda i})\quad\quad
	Q_{\lambda i}=Y_{\lambda i}-T_{\lambda i}\;.
\end{align}
By expanding the operators over the Cartesian components as done in Eqs.~(\ref{eq:sigma_dec}) and (\ref{eq:Sij_dec}), it is possible to derive the following relations:
\begin{align}
	V_{\Lambda NN}^{2\pi,S}&=\frac{1}{2}\sum_{i\ne j}\sum_{\alpha\beta\gamma}\tau_{i\gamma}\,\sigma_{i\alpha}\left(-\frac{C_P}{3}\sum_\lambda\sum_\delta\Theta_{\lambda i}^{\alpha\delta}\,\Theta_{\lambda j}^{\beta\delta}\right)\sigma_{j\beta}\,\tau_{j\gamma}\;,\\[0.5em]
	V_{\Lambda NN}^{2\pi,P}&=\frac{1}{2}\sum_{i\ne j}\sum_{\alpha\beta\gamma}\tau_{i\gamma}\,\sigma_{i\alpha}\,\Xi_{i\alpha,j\beta}\,\sigma_{j\beta}\,\tau_{j\gamma}\;,\\[0.5em]
	V_{\Lambda NN}^{D}&=W_D\!\sum_{\lambda,i<j}T^2_{\lambda i}\,T^2_{\lambda j}+\frac{1}{2}\sum_{\lambda i}\sum_{\alpha}\sigma_{\lambda\alpha}\,D_{\lambda i}^{[\sigma]}\,\sigma_{i\alpha}\;,
\end{align}
where
\begin{align}
	\Theta_{\lambda i}^{\alpha\beta}&=Q_{\lambda i}\,\delta^{\alpha\beta}+3\,T_{\lambda i}\hat r_{\lambda i}^\alpha\,\hat r_{\lambda i}^\beta\;,\label{eq:Theta}\\[0.5em]
	\Xi_{i\alpha,j\beta}&=\frac{1}{9}C_S\,\mu_\pi^2 \sum_\lambda\,Q_{i\lambda}\,Q_{\lambda j}\,|r_{i\lambda}||r_{j\lambda}|\,\hat r_{i\lambda}^\alpha\,\hat r_{j\lambda}^\beta\;,\label{eq:Xi}\\[0.5em]
	D_{\lambda i}^{[\sigma]}&=\frac{1}{3} W_D\!\sum_{j,j\ne i} T^2_{\lambda i}\,T^2_{\lambda j}\;.
\end{align}
By combining then Eq.~(\ref{eq:Theta}) and Eq.~(\ref{eq:Xi}), the TPE term of the three-body hyperon-nucleon interaction can be recast as:
\begin{align}
	V_{\Lambda NN}^{2\pi}=\frac{1}{2}\sum_{i\ne j}\sum_{\alpha\beta\gamma}\tau_{i\gamma}\,\sigma_{i\alpha}\,D_{i\alpha,j\beta}^{[\sigma\tau]}\,\sigma_{j\beta}\,\tau_{j\gamma}\;,
\end{align}
where
\begin{align}
	D_{i\alpha,j\beta}^{[\sigma\tau]}&=\sum_\lambda\Bigg\{\!-\frac{1}{3} C_P Q_{\lambda i}Q_{\lambda j}\delta_{\alpha\beta}
	-C_P Q_{\lambda i}T_{\lambda j}\,\hat r_{j\lambda}^{\,\alpha}\,\hat r_{j\lambda}^{\,\beta}
	-C_P Q_{\lambda j}T_{\lambda i}\,\hat r_{i\lambda}^{\,\alpha}\,\hat r_{i\lambda}^{\,\beta}\nonumber\\[0.2em]
	&\quad+\left[-3\,C_P T_{\lambda i}T_{\lambda j } \left({\sum_\delta}\,\hat r_{i\lambda}^{\,\delta}\,\hat r_{j\lambda}^{\,\delta}\right)
	+\frac{1}{9} C_S \mu_\pi^2 Q_{\lambda i} Q_{\lambda j}\,|r_{i\lambda}||r_{j\lambda}|\right]\,\hat r_{i\lambda}^{\,\alpha}\,\hat r_{j\lambda}^{\,\beta}\Bigg\}\;.
\end{align}
Finally, the $\Lambda NN$ interaction takes the following form:
\begin{align}
	V_{\Lambda NN}&=W_D\!\sum_{\lambda,i<j}T^2_{\lambda i}\,T^2_{\lambda j}
	+\frac{1}{2}\sum_{\lambda i}\sum_{\alpha}\sigma_{\lambda\alpha}\,D_{\lambda i}^{[\sigma]}\,\sigma_{i\alpha}\nonumber\\[0.2em]
	&\quad\,+\frac{1}{2}\sum_{i\ne j}\sum_{\alpha\beta\gamma}\tau_{i\gamma}\,\sigma_{i\alpha}\,D_{i\alpha,j\beta}^{[\sigma\tau]}\,\sigma_{j\beta}\,\tau_{j\gamma}\;.
	\label{eq:V_LNN_prop}
\end{align}
The first term is a pure central factor that can be included in $V_{\Lambda N}^c$. The second factor is analogous to the $\bm\sigma_\lambda\cdot\bm\sigma_i$ term~(\ref{eq:V_LN_prop}) of the two body hyperon-nucleon interaction. The last term acts only on the spin-isospin of the two nucleons $i$ and $j$ and has the same structure of the nuclear $\bm\sigma\cdot\bm\tau$ contribution described by the matrix $A_{i\alpha,j\beta}^{[\sigma\tau]}$. The three-body hyperon-nucleon interaction is then written as a sum of two-body operators only, of the same form of the ones already discussed for the $NN$ and $\Lambda N$ potentials. We can therefore include also these contributions in the AFDMC propagator of Eq.~(\ref{eq:Prop_full}) by simply redefining the following matrices:
\begin{align}
	B_{\lambda i}^{[\sigma]}&\longrightarrow B_{\lambda i}^{[\sigma]}+D_{\lambda i}^{[\sigma]}\;,\\[0.5em]
	A_{i\alpha,j\beta}^{[\sigma\tau]}&\longrightarrow A_{i\alpha,j\beta}^{[\sigma\tau]}+D_{i\alpha,j\beta}^{[\sigma\tau]}\;.
\end{align}
The algorithm follows then the steps already discussed in the previous section. Note that in the case of pure neutron systems, the last term of Eq.~(\ref{eq:V_LNN_prop}) simply reduces to a $\bm\sigma_i\cdot\bm\sigma_j$ contribution that is included in the propagator by redefining the nuclear matrix $A_{i\alpha,j\beta}^{[\sigma]}$.

With the AFDMC method extended to the hypernuclear sector, we can study finite and infinite lambda-nucleon and lambda-neutron systems. In the first case we can 
treat Hamiltonians that include the full hyperon-nucleon, hyperon-nucleon-nucleon and hyperon-hyperon interaction of Chapter~\ref{chap:hamiltonians}, but we are limited to the Argonne V6 like potentials for the nuclear sector. However it has been shown that this approach gives good results for finite nuclei~\cite{Gandolfi:2007} and nuclear matter~\cite{Gandolfi:2007_SNM,Gandolfi:2010}. In the latter case, instead, we can also add the nucleon spin-orbit contribution, so AV8 like potentials, and the three-neutron force. The neutron version of the AFDMC code has been extensively and successfully applied to study the energy differences of oxygen~\cite{Gandolfi:2006} and calcium~\cite{Gandolfi:2008} isotopes, the properties of neutron drops~\cite{Pederiva:2004,Gandolfi:2011,Maris:2013} and the properties of neutron matter in connection with astrophysical observables~\cite{Sarsa:2003,Gandolfi:2009,Gandolfi:2009_gap,Gandolfi:2012}. Very recently, the AFDMC algorithm has been also used to perform calculations for neutron matter using chiral effective field theory interactions~\cite{Gezerlis:2013}.

\renewcommand{\arraystretch}{1.0}

					% chapter 3: Method
	% Chapter 4: Results: finite systems

\chapter{Results: finite systems}
\label{chap:results_finite}

This chapter reports on the analysis of finite systems, nuclei and hypernuclei. For single $\Lambda$~hypernuclei a direct comparison of energy calculations with experimental results is given for the $\Lambda$~separation energy, defined as:
\begin{align}
	B_{\Lambda}\left(\,^A_\Lambda\text{Z}\,\right)=E\left(\,^{A-1}\text{Z}\,\right)-E\left(\,^A_\Lambda\text{Z}\,\right)\;,\label{eq:B_L}
\end{align}
where, using the notation of the previous chapters, $^A_\Lambda\text{Z}$ refers to the hypernucleus and $^{A-1}\text{Z}$ to the corresponding nucleus. $E$ is the binding energy of the system, i.e. the expectation value of the Hamiltonian on the ground state wave function
\begin{align}
	E(\kappa)=\frac{\langle\psi_{0,\kappa}|H_\kappa|\psi_{0,\kappa}\rangle}{\langle\psi_{0,\kappa}|\psi_{0,\kappa}\rangle}\;,\quad\quad\kappa=\text{nuc},\text{hyp}\;,
\end{align}
that we can compute by means of the AFDMC method. In the case of double $\Lambda$~hypernuclei, the interesting experimental observables we can have access are the double $\Lambda$~separation energy
\begin{align}
	&B_{\Lambda\Lambda}\left(\,^{~\,A}_{\Lambda\Lambda}\text{Z}\,\right)=E\left(\,^{A-2}\text{Z}\,\right)-E\left(\,^{~\,A}_{\Lambda\Lambda}\text{Z}\,\right)\;,\label{eq:B_LL}
\end{align}
and the incremental $\Lambda\Lambda$~energy
\begin{align}
	&\Delta B_{\Lambda\Lambda}\left(\,^{~\,A}_{\Lambda\Lambda}\text{Z}\,\right)=B_{\Lambda\Lambda}\left(\,^{~\,A}_{\Lambda\Lambda}\text{Z}\,\right)
	-2 B_\Lambda\left(\,^{A-1}_{\quad\;\Lambda}\text{Z}\,\right)\;.\label{eq:dB_LL}
\end{align}
The calculation of these quantities proceeds thus with the computation of the binding energies for both strange and non strange systems. Moreover it is interesting to compare other observables among the systems with strangeness $0$, $-1$ and $-2$, such as the single particle densities. By looking at the densities in the original nucleus and in the one modified by the addition of the lambda particles, information about the hyperon-nucleon interaction can be deduced.

As reported in Ref.~\cite{Gandolfi:2007}, the ground state energies of $^4$He, $^8$He, $^{16}$O and $^{40}$Ca have been computed using the AV6' potential (\S~\ref{subsec:AV18}). Due to the limitations in the potential used, the results cannot reproduce the experimental energies and all the nuclei result less bound than expected. However, given the same simplified interaction, the published AFDMC energies are close to the GFMC results, where available. AFDMC has also been used to compute the energy differences between oxygen~\cite{Gandolfi:2006} and calcium~\cite{Gandolfi:2008} isotopes, by studying the external neutrons with respect to a nuclear core obtained from Hartree-Fock calculations using Skyrme forces. In this case the results are close to the experimental ones.

The idea behind the AFDMC analysis of $\Lambda$~hypernuclei follows in some sense the one assumed in the study of oxygen and calcium isotopes by the analysis of energy differences. The two-body nucleon interaction employed is limited to the first six operators of AV18. However, if we use the same potential for the nucleus and the core of the corresponding hypernucleus, and take the difference between the binding energies of the two systems, the uncertainties in the $NN$ interaction largely cancel out. We shall see that this assumption, already used in other works~\cite{Dalitz:1972,Bodmer:1988}, is indeed consistent with our results, thereby confirming that the specific choice of the nucleon Hamiltonian does not significantly affect the results on $B_\Lambda$. On the grounds of this observation, we can focus on the interaction between hyperons and nucleons, performing QMC simulations with microscopic interactions in a wide mass range.

\section{Nuclei}
\label{sec:nuc}

Let us start with the AFDMC study of finite nuclei. In the previous chapter, we have seen that two versions of the AFDMC algorithm, that should give the same results, are available (\hyperlink{method:PsiT}{\emph{v1}} and \hyperlink{method:Elocal}{\emph{v2}}). Before including the strange degrees of freedom, we decided to test the stability and accuracy of the two algorithms, within the fixed phase approximation, by performing some test simulations on $^4$He. The result of $-27.13(10)$~MeV for the AV6' potential reported in Ref.~\cite{Gandolfi:2007}, was obtained employing the algorithm~\hyperlink{method:Elocal}{\emph{v2}} using single particle Skyrme orbitals and a particular choice of the parameters for the solution of the Jastrow correlation equation (see \S~\ref{subsubsubsec:Wave_non_strange}). In order to check the AFDMC projection process, we tried to modify the starting trial wave function:
\begin{itemize}
	\item we changed the healing distance $d$ and the quencher parameter $\eta$ for the Jastrow function $f_c^{NN}$;
	\item we used a different set of radial functions, labelled as HF-B1~\cite{Co:2011_comm}, coming from Hartree-Fock calculations for the effective interaction B1 of Ref.~\cite{Brink:1967}. The B1 is a phenomenological two-body nucleon-nucleon potential fitted in order to reproduce the binding energies and densities of various light nuclei and of nuclear matter in the HF approximation.
\end{itemize}

Although a central correlation function should not affect the computed energy value, in the version \hyperlink{method:Elocal}{\emph{v2}} of the algorithm an unpleasant dependence on $f_c^{NN}$ was found, and in particular as a function of the quencher parameter $\eta$. This dependence is active for both the AV4' and the AV6' potentials, regardless of the choice of the single particle orbitals. The time step extrapolation ($d\tau\rightarrow 0$) of the energy does not solve the issue. Energy differences are still more than 1~MeV among different setups for the trial wave function. By varying the parameter $\eta$ from zero (no Jastrow at all) to one (full central channel of the $NN$ potential used for the solution of Eq.~(\ref{eq:Jastrow})), the energies increase almost linearly. For example, in the case of AV4' for the Skyrme orbital functions, the energy of $^4$He goes from $-31.3(2)$~MeV for $\eta=0$, to $-27.2(2)$~MeV for $\eta=1$. Same effect is found for the HF-B1 orbitals with energies going from $-32.5(2)$~MeV to $-28.4(2)$~MeV. The inclusion of the pure central Jastrow introduces thus strong biases in the evaluation of the total energy. Moreover, there is also a dependence on the choice of the single particle orbitals, as shown from the results for AV4'. Same conclusions follow for the AV6' potential.

On the grounds of these observations we moved from the AFDMC local energy scheme to the importance function ratio scheme (version \hyperlink{method:PsiT}{\emph{v1}}), with no importance sampling on auxiliary fields. In this case the bias introduced by the Jastrow correlation function is still present but reduced to $0.3\div0.4$~MeV for AV4' and $0.1\div0.2$~MeV for AV6'. In spite of the improvement with respect to the previous case, we decided to remove this source of uncertainty from the trial wave function and proceed with the test of the \hyperlink{method:PsiT}{\emph{v1}} algorithm with no Jastrow. It has to be mentioned that a new sampling procedure, for both coordinates and auxiliary fields, capable to reduce the dependence on central correlations is being studied.

\begin{figure}[ht]
	\centering
	\includegraphics[width=\linewidth]{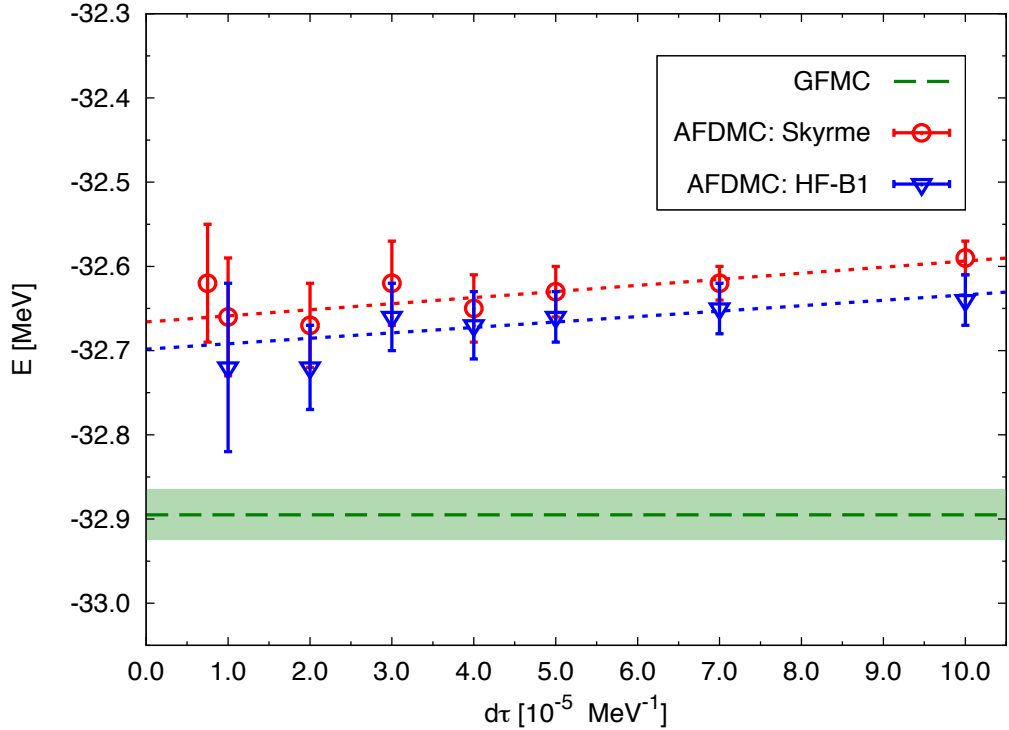}
	\caption[Binding energies: $E$ vs. $d\tau$ for $^4$He, Argonnne V4']
		{Binding energy of $^4$He as a function of the Monte Carlo imaginary time step. Results are obtained using the AV4' $NN$ potential. 
		Red dots are the AFDMC results for the Skyrme radial orbitals. Blue triangles the ones for the HF-B1 orbitals.
		For comparison, the GFMC result of Ref.~\cite{Wiringa:2002_url}, corrected by the Coulomb contribution (see text for details), is reported with the green band.}
	\label{fig:E_He4_V4p}
\end{figure}

\begin{figure}[ht]
	\centering
	\includegraphics[width=\linewidth]{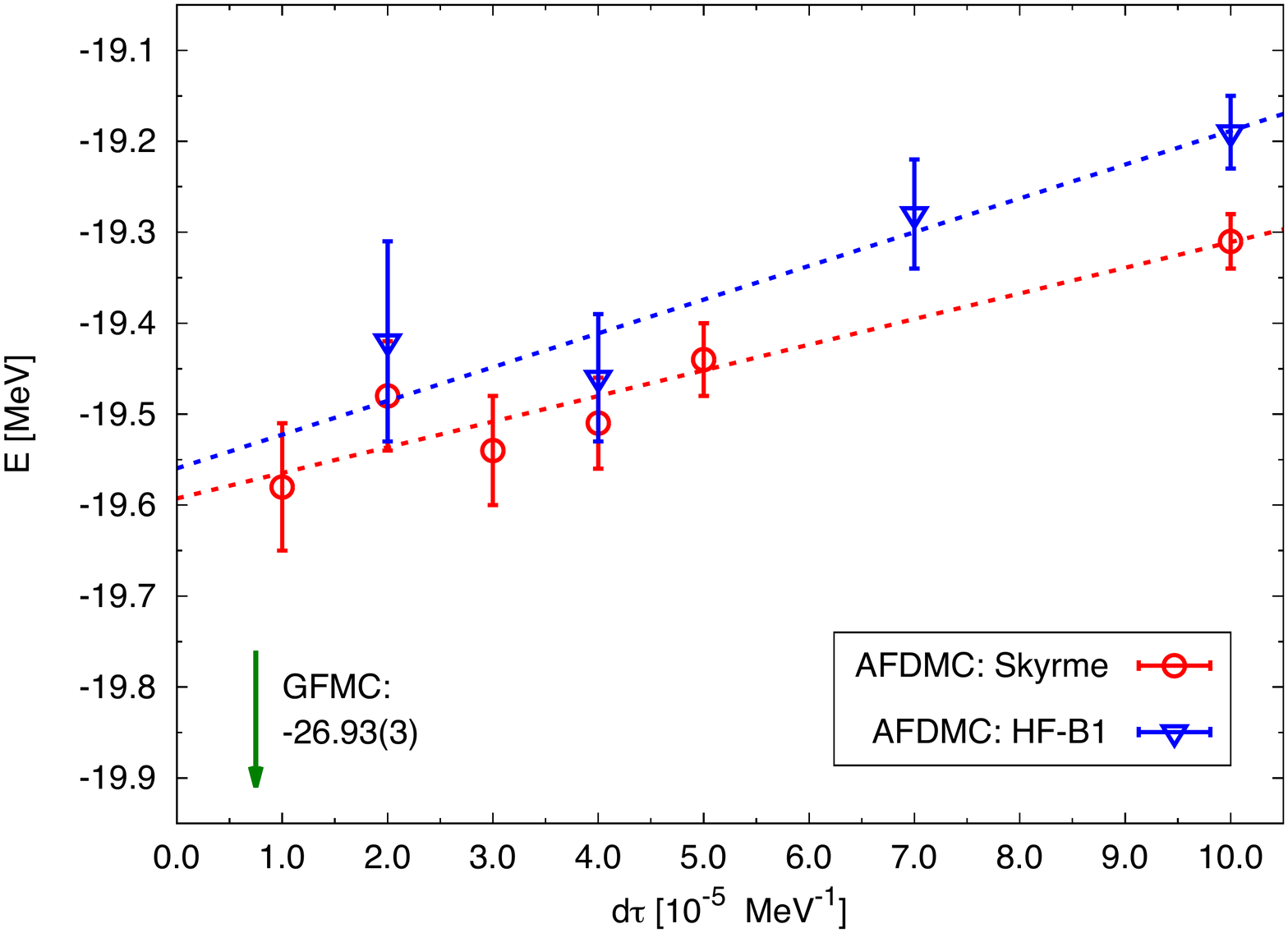}
	\caption[Binding energies: $E$ vs. $d\tau$ for $^4$He, Argonnne V6']
		{Binding energy of $^4$He as a function of the Monte Carlo imaginary time step. Results are obtained using the AV6' $NN$ potential. 
		As in Fig.~\ref{fig:E_He4_V4p}, red dots refers to the AFDMC results for the Skyrme radial functions and blue triangles for the HF-B1 orbitals.
		The green arrow points to the GFMC result.}
	\label{fig:E_He4_V6p}
\end{figure}

As shown in Figs.~\ref{fig:E_He4_V4p} and \ref{fig:E_He4_V6p}, the \hyperlink{method:PsiT}{\emph{v1}} extrapolated energies obtained using different single particle orbitals are consistent within the Monte Carlo statistical errors, both for the AV4' and the AV6' potentials.
For $\mathcal N_N=4$ we can compare the AFDMC results with the GFMC ones. In our calculations the Coulomb interaction is not included. A precise VMC estimate, that should be representative also for the GFMC estimate, of the Coulomb expectation value for $^4$He is $0.785(2)$~MeV~\cite{Wiringa:2012_comm}. The AFDMC values of $-32.67(8)$~MeV (Skyrme) and $-32.7(1)$~MeV (HF-B1) for $^4$He with AV4' are thus very close to the GFMC $-32.11(2)$~MeV of Ref.~\cite{Wiringa:2002_url} for the same potential once the Coulomb contribution is subtracted. Our results are still $\sim0.1\div0.2$~MeV above the GFMC one, most likely due to the removal of the sign problem constraint applied at the end of the GFMC runs (release node procedure~\cite{Ceperly:1980}).

Although AFDMC and GFMC energies for $^4$He described by the AV4' potential are consistent, a clear problem appears using the AV6' interaction (Fig.~\ref{fig:E_He4_V6p}). With the two sets of radial functions, the energies are $-19.59(8)$~MeV (Skyrme) and $-19.53(13)$~MeV (HF-B1) and thus the AFDMC actually projects out the same ground state. However, the GFMC estimate is $-26.15(2)$~MeV minus the Coulomb contribution. This large difference in the energies cannot be attributed to the GFMC release node procedure. The difference in using AV4' and AV6' is the inclusion of the tensor term $S_{ij}$ of Eq.~(\ref{eq:S_ij}). The Hamiltonian moves then from real to complex and this might result in a phase problem during the imaginary time propagation. There might be some issues with the fixed phase approximation or with the too poor trial wave function (or both), which does not include operatorial correlations. This is still an unsolved question but many ideas are being tested. According to the lack of control on the AFDMC simulations for the AV6' potential, from now on we will limit the study to AV4'. As we shall see, this choice does not affect the result on energy differences as the hyperon separation energy, which is the main observable of this study for finite systems.

In order to complete the check of the accuracy of the algorithm \hyperlink{method:PsiT}{\emph{v1}} for $^4$He, we performed simulations using the Minnesota potential of Ref.~\cite{Thompson:1977}. This two-nucleon interaction has the same operator structure of AV4' but much softer cores. Our AFDMC result for the energy is $-30.69(7)$~MeV. It has to be compared with the $-29.937$~MeV ($-30.722$~MeV with the Coulomb subtraction) obtained with the Stochastic Variational Method (SVM)~\cite{Varga:1995}, that has been proven to give consistent results with the GFMC algorithm for $^4$He~\cite{Kamada:2001}. The agreement of the results is remarkable.

Moreover, we tested the consistency of the \hyperlink{method:PsiT}{\emph{v1}} algorithm for the AV4' potential by studying the deuteron, tritium and oxygen nuclei.
\begin{itemize}
	\item The AFDMC binding energy for $^2$H is $-2.22(5)$~MeV, in agreement with the experimental $-2.225$~MeV. The result is significant because, although the Argonne V4' was exactly fitted in order to reproduce the deuteron energy, our starting trial wave function is just a Slater determinant of single particle orbitals, with no correlations.
	\item The result for $^3$H is $-8.74(4)$~MeV, close to the GFMC $-8.99(1)$~MeV of Ref.~\cite{Wiringa:2002_url}. As for $^4$He, the small difference in the energies is probably due to the release node procedure in GFMC. Without the Coulomb contribution, we obtained the same energy $-8.75(4)$~MeV also for $^3$He. In AV4' there are no charge symmetry breaking terms. Therefore, this result can be seen as a consistency test on the correct treatment of the spin-isospin operators acting on the wave function during the Hubbard-Stratonovich rotations. 
	\item For $^{16}$O we found the energy values of $-176.8(5)$~MeV for the Skyrme orbitals and $-174.3(8)$~MeV for the HF-B1 radial functions. The energy difference is of order 1\% even for a medium mass nucleus. The projection mechanism is working accurately regardless the starting trial function. GFMC results are limited to 12 nucleons~\cite{Pieper:2005,Lusk:2010,Lovato:2013}, so we cannot compare the two methods for $\mathcal N_N=16$. The binding energy cannot be compared with the experimental data due to the poor employed Hamiltonian. However the AFDMC results are consistent with the overbinding predicted by the available GFMC energies for AV4'~\cite{Wiringa:2002_url} and the nucleus results stable under alpha particle break down, as expected.
\end{itemize}
On the grounds of the results of these consistency checks, in the present work we adopt the version \hyperlink{method:PsiT}{\emph{v1}} of the AFDMC algorithm employing the nuclear potential AV4’ for both nuclei and hypernuclei. 

\renewcommand{\arraystretch}{1.4}
\begin{table}[ht]
	\centering
	\begin{tabular*}{\linewidth}{@{\hspace{1.0em}\extracolsep{\fill}}lccccc@{\extracolsep{\fill}\hspace{1.0em}}}
		\toprule
		\toprule
		System & $E_{\text{AFDMC}}$ & $E_{\text{GFMC}}$ & $E_{\text{exp}}$ & $E_{\text{AFDMC}}/\mathcal N_N$ & $E_{\text{exp}}/\mathcal N_N$\\
		\midrule
		\hspace{0.6em}$^2$H     &    -2.22(5)  &     ---     &   -2.225 &  -1.11 & -1.11 \\
		\hspace{0.6em}$^3$H     &    -8.74(4)  &   -8.99(1)  &   -8.482 &  -2.91 & -2.83 \\
		\hspace{0.6em}$^3$He    &    -8.75(4)  &     ---     &   -7.718 &  -2.92 & -2.57 \\
		\hspace{0.6em}$^4$He    &   -32.67(8)  &  -32.90(3)  &  -28.296 &  -8.17 & -7.07 \\
		\hspace{0.6em}$^5$He    &   -27.96(13) &  -31.26(4)  &  -27.406 &  -5.59 & -5.48 \\
		\hspace{0.6em}$^6$He    &   -29.87(14) &  -33.00(5)  &  -29.271 &  -4.98 & -4.88 \\
		\hspace{0.3em}$^{12}$C  &  -77.31(25)* &     ---     &  -92.162 &  -6.44 & -7.68 \\
		\hspace{0.3em}$^{15}$O  &  -144.9(4)   &     ---     & -111.955 &  -9.66 & -7.46 \\
		\hspace{0.3em}$^{16}$O  &  -176.8(5)   &     ---     & -127.619 & -11.05 & -7.98 \\
		\hspace{0.3em}$^{17}$O  &  -177.0(6)   &     ---     & -131.762 & -10.41 & -7.75 \\
		\hspace{0.3em}$^{40}$Ca &  -597(3)     &     ---     & -342.052 & -14.93 & -8.55 \\
		\hspace{0.3em}$^{48}$Ca &  -645(3)     &     ---     & -416.001 & -13.44 & -8.67 \\
		\hspace{0.3em}$^{90}$Zr & -1457(6)     &     ---     & -783.899 & -16.19 & -8.71 \\
		\bottomrule
		\bottomrule
	\end{tabular*}
	\caption[Binding energies: nuclei, $2\le A-1\le 90$]
		{Binding energies (in MeV) for different nuclei. AFDMC and GFMC results are obtained using the the AV4' $NN$ potential. 
		The GFMC data are from Ref.~\cite{Wiringa:2002_url} corrected by the Coulomb contribution (see text for details).
		In the fourth column the experimental results are from Ref.~\cite{Zagrebaev:1999}. Errors are less than 0.1~KeV.
		In the last two columns the calculated and experimental binding energies per particle. For the note * on $^{12}$C see the text.}
	\label{tab:E_nuc}
\end{table}
\renewcommand{\arraystretch}{1.0}

\begin{figure}[!hb]
	\centering
	\includegraphics[width=\linewidth]{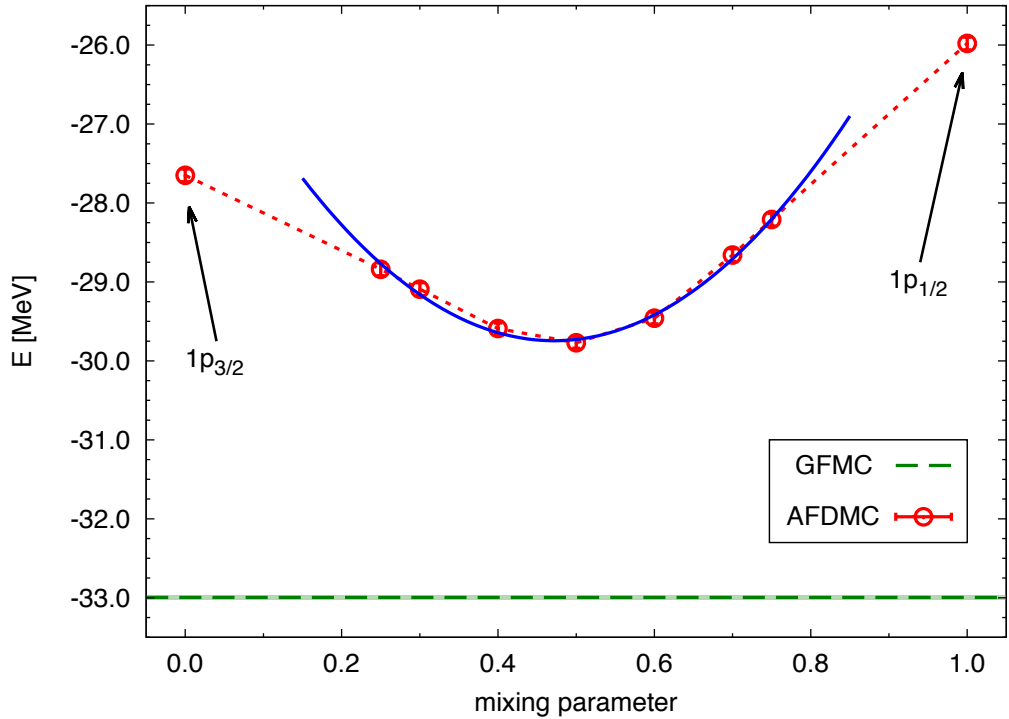}
	\caption[Binding energies: $E$ vs. mixing parameter for $^6$He]
		{$^6$He binding energy as a function of the mixing parameter $c$ of Eq.~(\ref{eq:He6_mix}). 
		The arrows point to the results for the pure $1p_{3/2}$ ($-27.65(8)$~MeV)
		and $1p_{1/2}$ ($-25.98(8)$~MeV) 
		configurations used for the two external neutrons. 
		The green line is the GFMC result of Ref.~\cite{Wiringa:2002_url} corrected by the VMC Coulomb expectation contribution
		$0.776(2)$~MeV~\cite{Wiringa:2012_comm}.}
	\label{fig:E_He6}
\end{figure}

As reported in Tab.~\ref{tab:E_nuc}, the resulting absolute binding energies using AV4' are not comparable with experimental ones, as expected, due to the lack of information about the nucleon interaction in the Hamiltonian. With the increase of the number of particles, the simulated nuclei become more an more bound until the limit case of $^{90}$Zr, for which the estimated binding energy is almost twice the experimental one. Looking at the results for helium isotopes, we can see that for $\mathcal N_N=3$ and $4$ the energies are compatible with GFMC calculations, once the Coulomb contribution is removed. For $^5$He and $^6$He instead, we obtained discrepancies between the results for the two methods. However this is an expected result. When moving to open shell systems, as $^5$He and $^6$He with one or two neutrons out of the first $s$ shell, the structure of the wave function becomes more complicated and results are more dependent on the employed $\psi_T$. For example, in the case of $^6$He, in order to have total angular momentum zero, the two external neutrons can occupy the $1p_{3/2}$ or the $1p_{1/2}$ orbitals of the nuclear shell model classification. By using just one of the two $p$~shells, one gets the unphysical result $E(^5\text{He})<E(^6\text{He})$. The reported binding energy has been instead obtained by considering the linear combination of the Slater determinants giving $J=0$ 
\begin{align}
	\Phi(R_N,S_N)=(1-c)\det\Bigl\{\varphi_\epsilon^N(\bm r_i,s_i)\Bigr\}_{1p_{3/2}}\!\!+c\,\det\Bigl\{\varphi_\epsilon^N(\bm r_i,s_i)\Bigr\}_{1p_{1/2}}\;,\label{eq:He6_mix}
\end{align}
and minimizing the energy with respect to the mixing parameter $c$, as shown in Fig.~\ref{fig:E_He6}. However the final result is still far from the GFMC data. This is a clear indication that better wave functions are needed for open shell systems. A confirmation of that is the non physical result obtained for the $^{12}$C nucleus (marked in Tab.~\ref{tab:E_nuc} with *), which is even less bound than expected, although the employed AV4' potential, resulting thus unstable under $\alpha$ break down. In the case of $\mathcal N_N=12$ indeed, the 8 additional neutrons and protons to the alpha core have just been placed in the $1p_{3/2}$ shell without any linear combination of the other possible setups giving zero total angular momentum. This result will be useful in the hyperon separation energy estimate anyway. In fact, we shall see in the next section that regardless the total binding energies, by using the same nucleon potential to describe nuclei and the core of hypernuclei, the obtained hyperon separation energy is in any case realistic.

Last comment on a technical detail regarding the computation of AFDMC observables. As shown in Fig.~\ref{fig:E_He4_V4p} and \ref{fig:E_He4_V6p}, the extrapolation of the energy values in the limit $d\tau\rightarrow0$ is linear. This is consistent with the application of the Trotter-Suzuki formula of Eq.~(\ref{eq:Trotter_2}) in the Hubbard-Stratonovich transformation~(\ref{eq:HS_applied}), that is thus correct at order $\sqrt{d\tau}^{\,2}$. Focusing on the AV4' case, for $^4$He the time step extrapolation is almost flat. The differences between the final results and the energies computed at large $d\tau$ are less than $0.5\%$ and almost within the statistical errors of the Monte Carlo run. The situation dramatically changes with the increase of the particle number. For $\mathcal N_N=16$ this difference is around $2\%$. For 40 and 48 particles, large time step values and the extrapolated ones are, respectively, $6\%$ and $8.5\%$ different. Therefore, the binding energies must always be carefully studied by varying the time step of the AFDMC run. The same behavior has been found for observables other than the total energy (single particle densities and radii). Each reported result in this chapter has been thus obtained by means of a computationally expensive procedure of imaginary time extrapolation.

\section{Single $\Lambda$~hypernuclei}
\label{sec:l_hyp}

When a single $\Lambda$~particle is added to a core nucleus, the wave function of Eq.~(\ref{eq:Psi_T}) is given by
\begin{align}
	\psi_T(R,S)=\prod_{i}f_c^{\Lambda N}(r_{\Lambda i})\,\psi_T^N(R_N,S_N)\,\varphi_\epsilon^\Lambda(\bm r_\Lambda,s_\Lambda)\;.\label{eq:psi_singleL}
\end{align}
The structure of the nucleon trial wave function is the same of Eq.~(\ref{eq:psi_N}), used in the AFDMC calculations for nuclei. The hyperon Slater determinant is simply replaced by the single particle state $\varphi_\epsilon^\Lambda$ of Eq.~(\ref{eq:varphi_L}), assumed to be the neutron $1s_{1/2}$ radial function, as already described in the previous chapter. In order to be consistent with the calculations for nuclei, we neglected the Jastrow $\Lambda N$ correlation function which was found to produce a similar but smaller bias on the total energy. As radial functions we used the same Skyrme set employed in the calculations for the nuclei of Tab.~\ref{tab:E_nuc}.

The $\Lambda$ separation energies defined in Eq.~(\ref{eq:B_L}), are calculated by taking the difference between the nuclei binding energies presented in the previous section, and the AFDMC energies for hypernuclei, given the same nucleon potential. By looking at energy differences, we studied the contribution of the $\Lambda N$ and $\Lambda NN$ terms defined in Chapter~\ref{chap:hamiltonians}. By comparing AFDMC results with the expected hyperon separation energies, information about the hyperon-nucleon interaction are deduced. Some qualitative properties have been also obtained by studying the nucleon and hyperon single particle densities and the root mean square radii.

\subsection{Hyperon separation energies}
\label{subsec:E_l}

We begin the study of $\Lambda$~hypernuclei with the analysis of closed shell hypernuclei, in particular $^5_\Lambda$He and $^{17}_{~\Lambda}$O. We have seen in the previous section that the AFDMC algorithm is most accurate in describing closed shell nuclei. Results for $^4$He and $^{16}$O with the AV4' potential are indeed consistent and under control. This give us the possibility to realistically describe the hyperon separation energy for such systems and deduce some general properties of the employed hyperon-nucleon force.

The step zero of this study was the inclusion in the Hamiltonian of the $NN$ AV4' interaction and the two-body $\Lambda N$ charge symmetric potential of Eq.~(\ref{eq:V_LN}). The employed parameters $\bar v$ and $v_\sigma$ are reported in Tab~\ref{tab:parLN+LNN}. The exchange parameter $\varepsilon$ has been initially set to zero due to the impossibility of including the space exchange operator directly in the AFDMC propagator (see \S~\ref{subsubsubsec:Wave_strange}). As reported in Tab.~\ref{tab:BL_He5L-O17L_I}, the AV4' $\Lambda$~separation energy for $^5_\Lambda$He is more than twice the expected value. For the heavier $^{17}_{~\Lambda}$O the discrepancy is even larger. Actually, this is an expected result. As firstly pointed out by Dalitz~\cite{Dalitz:1972}, $\Lambda N$ potentials, parameterized to account for the low-energy $\Lambda N$ scattering data and the binding energy of the $A=3,4$ hypernuclei, overbind $^5_\Lambda$He by $2\div3$~MeV. That is, the calculated $A=5$ $\Lambda$~separation energy is about a factor of 2 too large. This fact is usually reported as \emph{$A=5$~anomaly}. With only a $\Lambda N$ potential fitted to $\Lambda p$ scattering, the heavier hypernuclei result then strongly overbound.

\renewcommand{\arraystretch}{1.4}
\begin{table}[t]
	\centering
	\begin{tabular*}{\linewidth}{@{\hspace{1.0em}\extracolsep{\fill}}lcccc@{\extracolsep{\fill}\hspace{1.0em}}}
		\toprule
		\toprule
		\multirow{2}{*}{$NN$ potential} & \multicolumn{2}{c}{$^5_\Lambda$He}  & \multicolumn{2}{c}{$^{17}_{~\Lambda}$O} \\
		\cmidrule(l){2-3}\cmidrule(l){4-5}
		& \hspace{1em}$V_{\Lambda N}$ & $V_{\Lambda N}$+$V_{\Lambda NN}$ & \hspace{1em}$V_{\Lambda N}$ & $V_{\Lambda N}$+$V_{\Lambda NN}$ \\
		\midrule
		Argonne V4'\ &     \hspace{1em} 7.1(1)     &             5.1(1)               &   \hspace{1em} 43(1)        &        19(1) \\
		Argonne V6'\ &     \hspace{1em} 6.3(1)     &             5.2(1)               &   \hspace{1em} 34(1)        &        21(1) \\
		Minnesota\   &     \hspace{1em} 7.4(1)     &             5.2(1)               &   \hspace{1em} 50(1)        &        17(2) \\
		\midrule
		Expt.        & \multicolumn{2}{c}{3.12(2)} & \multicolumn{2}{c}{13.0(4)} \\
		\bottomrule
		\bottomrule
	\end{tabular*}
	\caption[$\Lambda$~separation energies: $\Lambda N+\Lambda NN$ set (I) for $^5_\Lambda$He and $^{17}_{~\Lambda}$O]
		{$\Lambda$~separation energies (in MeV) for $^5_\Lambda$He and $^{17}_{~\Lambda}$O obtained using different nucleon potentials (AV4', AV6', Minnesota)
		and different hyperon-nucleon interaction (two-body alone and two-body plus three-body, 
		set of parameters~(\hyperlink{par_I}{I}))~\cite{Lonardoni:2013_PRC(R)}. 
		In the last line the experimental $B_\Lambda$ for $^5_\Lambda$He is from Ref.~\cite{Juric:1973}.
		Since no experimental data for $^{17}_{~\Lambda}$O exists, the reference separation energy is the semiempirical value reported in Ref.~\cite{Usmani:1995}.}
	\label{tab:BL_He5L-O17L_I}
\end{table}
\renewcommand{\arraystretch}{1.0}

As suggested by the same Dalitz~\cite{Dalitz:1972} and successively by Bodmer and Usmani~\cite{Bodmer:1988}, the inclusion of a $\Lambda$-nucleon-nucleon potential may solve the overbinding problem. This is indeed the case, as reported for instance in Refs.~\cite{Usmani:1995,Sinha:2002,Usmani:2008}. Therefore, in our AFDMC calculations we included the three-body $\Lambda NN$ interaction developed by Bodmer, Usmani and Carlson and described in \S~\ref{subsubsec:LNN}. Among the available parametrizations coming from different VMC studies of light hypernuclei, the set of parameters for the $\Lambda NN$ potential has been originally taken from Ref.~\cite{Usmani:1995_3B}, being the choice that made the variational $B_\Lambda$ for $_\Lambda^5$He and $^{17}_{~\Lambda}$O compatible with the expected results. It reads:
\begin{equation*}
	\hypertarget{par_I}{(\text{I})}\phantom{I}\quad
	\left\{
	\begin{array}{rcll}
		C_P&\!=\!&0.60& \!\text{MeV}\\
		C_S&\!=\!&0.00& \!\text{MeV}\\
		W_D&\!=\!&0.015&\!\text{MeV}
	\end{array}
	\right.
\end{equation*}
The inclusion of the $\Lambda NN$ force reduces the overbinding and thus the hyperon separation energies, as reported in Tab.~\ref{tab:BL_He5L-O17L_I}. Although the results are still not compatible with the experimental ones, the gain in energy due to the inclusion of the three-body hypernuclear force is considerable. 

It has to be pointed out that this result might in principle depend on the particular choice of the $NN$ interaction used to describe both nucleus and hypernucleus. One of the main mechanisms that might generate this dependence might be due to the different environment experienced by the hyperon in the hypernucleus because of the different nucleon densities and correlations generated by each $NN$ potential. To discuss this possible dependence, we performed calculations with different $NN$ interactions having very different saturation properties. As it can be seen from Tab.~\ref{tab:BL_He5L-O17L_I}, for $^5_\Lambda$He the extrapolated $B_\Lambda$ values with the two-body $\Lambda N$ interaction alone are about 10\% off and well outside statistical errors. In contrast, the inclusion of the three-body $\Lambda NN$ force gives a similar $\Lambda$~binding energy independently to the choice of the $NN$ force. On the grounds of this observation, we feel confident that the use of AV4', for which AFDMC calculations for nuclei are under control, will in any case return realistic estimates of $B_\Lambda$ for larger masses when including the $\Lambda NN$ interaction. We checked this assumption performing simulations in $^{17}_{~\Lambda}$O, where the discrepancy between the $\Lambda$~separation energy computed using the different $NN$ interactions and the full $\Lambda N$+$\Lambda NN$ force is less than few per cent (last column of Tab.\ref{tab:BL_He5L-O17L_I}). The various $NN$ forces considered here are quite different. The AV6' includes a tensor force, while AV4' and Minnesota have a simpler structure with a similar operator structure but very different intermediate- and short-range correlations. The fact that the inclusion of the $\Lambda NN$ force does not depend too much on the nuclear Hamiltonian is quite remarkable, because the different $NN$ forces produce a quite different saturation point for the nuclear matter EoS, suggesting that our results are pretty robust. 

\begin{figure}[!htb]
	\centering
	\includegraphics[width=\linewidth]{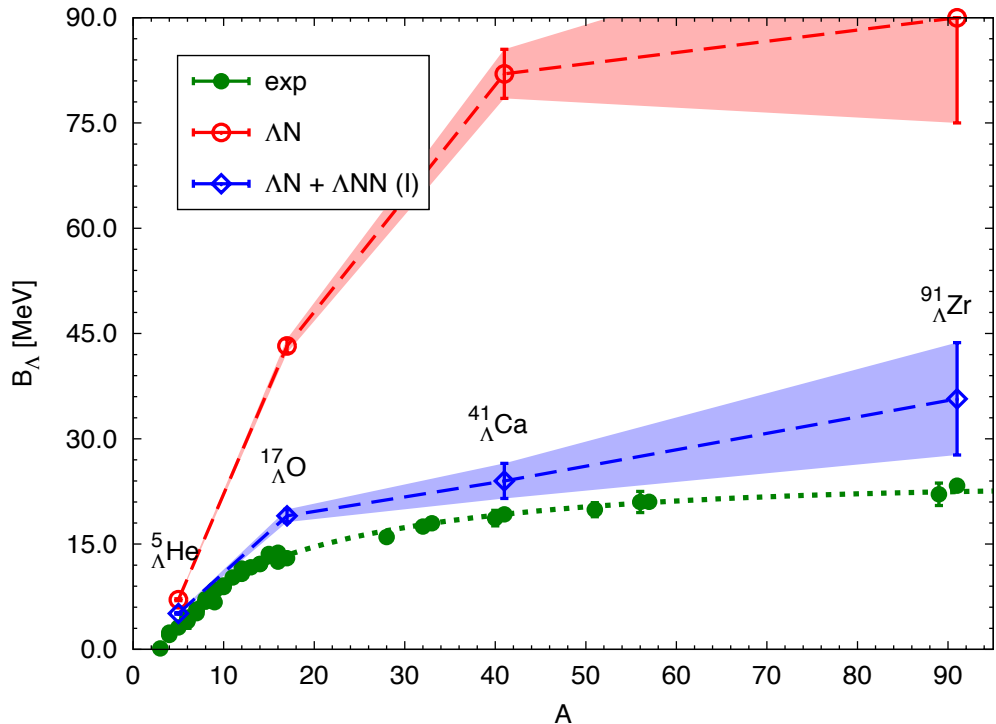}
	\caption[$\Lambda$ separation energy vs. $A$: closed shell hypernuclei]
		{$\Lambda$ separation energy as a function of $A$ for closed shell hypernuclei, adapted from Ref.~\cite{Lonardoni:2013_PRC(R)}. 
		Solid green dots~(dashed curve) are the available $B_\Lambda$ experimental or semiempirical values. Empty red dots~(upper banded curve) refer to 
		the AFDMC results for the two-body $\Lambda N$ interaction alone. Empty blue diamonds~(middle banded curve) are the results with the inclusion also of the 
		three-body hyperon-nucleon force in the parametrization (\hyperlink{par_I}{I}).}
	\label{fig:BL-A_cshell}
\end{figure}

For $^5_\Lambda$He the hyperon separation energy with the inclusion of the $\Lambda NN$ force with the set of parameters~(\hyperlink{par_I}{I}) reduces of a factor $\sim1.4$. For $^{17}_{~\Lambda}$O the variation is around $40\div50\%$. In order to check the effect of the three-body force with increasing the particle number, we performed simulations for the next heavier closed or semi-closed shell $\Lambda$~hypernuclei, $^{41}_{~\Lambda}$Ca and $^{91}_{~\Lambda}$Zr. The $\Lambda$~separation energies for all the studied closed shell hypernuclei are shown in Fig.~\ref{fig:BL-A_cshell}. While the results for lighter hypernuclei might be inconclusive in terms of the physical consistency of the $\Lambda NN$ contribution to the hyperon binding energy in AFDMC calculations, the computations for $^{41}_{~\Lambda}$Ca and $^{91}_{~\Lambda}$Zr reveal a completely different picture. The saturation binding energy provided by the $\Lambda N$ force alone is completely unrealistic, while the inclusion of the $\Lambda NN$ force gives results that are much closer to the experimental behavior. Therefore, the $\Lambda$-nucleon-nucleon force gives a very important repulsive contribution towards a realistic description of the saturation of medium-heavy hypernuclei~\cite{Lonardoni:2013_PRC(R)}. However, with the given parametrization, only a qualitative agreement wiht the expected separation energies is reproduced. A refitting procedure for the three-body hyperon-nucleon interaction might thus improve the quality of the results.

As already discussed in \S~\ref{subsubsec:LNN}, the $C_S$ parameter can be estimated by comparing the $S$-wave term of the $\Lambda NN$ force with the Tucson-Melbourne component of the $NNN$ interaction. We take the suggested $C_S=1.50$~MeV value~\cite{Usmani:2008}, in order to reduce the number of fitting parameters. This choice is justified because the $S$-wave component of the three-body $\Lambda NN$ interaction is sub-leading. We indeed verified that a change in the $C_S$ value yields a variation of the total energy within statistical error bars, and definitely much smaller than the variation in energy due to a change of the $W_D$ parameter.

\begin{figure}[!htb]
	\centering
	\includegraphics[width=\linewidth]{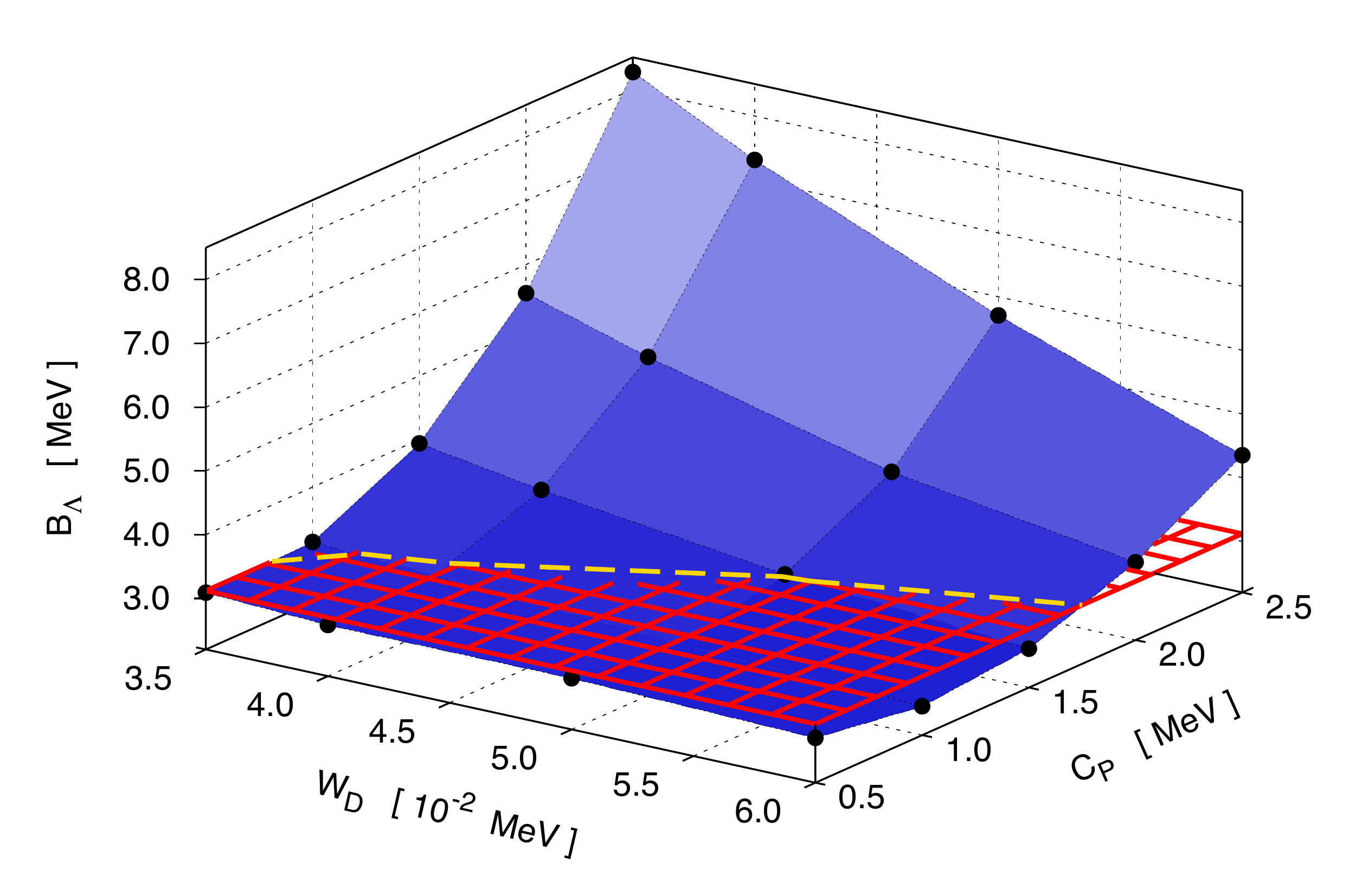}
	\caption[$\Lambda$~separation energy for $^5_\Lambda$He vs. $W_D-C_P$: 3D plot]
		{$\Lambda$~separation energy for $^5_\Lambda$He as a function of strengths $W_D$ and $C_P$ of the three-body $\Lambda NN$ 
		interaction~\cite{Lonardoni:2013_PRC}. The red grid represents the experimental $B_\Lambda=3.12(2)$~MeV~\cite{Juric:1973}. 
		The dashed yellow curve is the interception between the expected result and the $B_\Lambda$ surface in the $W_D-C_P$ parameter space.
		Statistical error bars on AFDMC results (solid black dots) are of the order of $0.10\div0.15$~MeV.}
	\label{fig:Wd-Cp_3D}
\end{figure}

\begin{figure}[!htb]
	\centering
	\includegraphics[width=\linewidth]{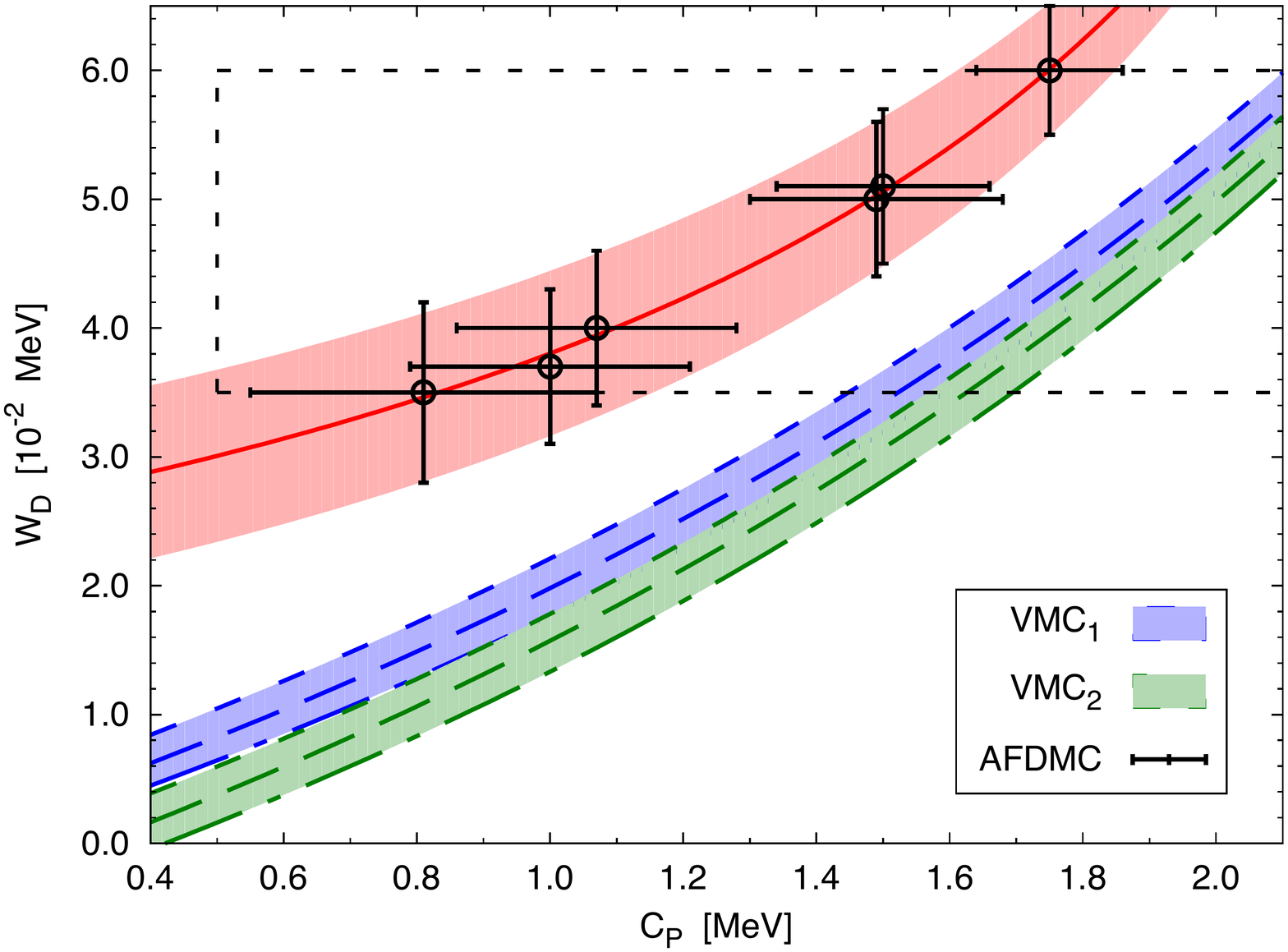}
	\caption[$\Lambda$~separation energy for $^5_\Lambda$He vs. $W_D-C_P$: 2D plot]
		{Projection of Fig.~\ref{fig:Wd-Cp_3D} on the $W_D-C_P$ plane~\cite{Lonardoni:2013_PRC}. Error bars come from a realistic conservative estimate of 
		the uncertainty in the determination of the parameters due to the statistical errors of the Monte Carlo calculations. 
		Blue and green dashed, long-dashed and dot-dashed lines (lower curves) are the variational results of Ref.~\cite{Usmani:2008} for different 
		$\varepsilon$ and $\bar v$ (two-body $\Lambda N$ potential). The dashed box corresponds to the parameter domain of Fig.~\ref{fig:Wd-Cp_3D}. 
		Black empty dots and the red band (upper curve) are the projected interception describing the possible set of parameters reproducing the
		experimental~$B_\Lambda$.}
	\label{fig:Wd-Cp_2D}
\end{figure}

In Fig.~\ref{fig:Wd-Cp_3D} we report the systematic study of the $\Lambda$~separation energy of $_\Lambda^5$He as a function of both $W_D$ and $C_P$. Solid black dots are the AFDMC results. The red grid represents the experimental $B_\Lambda=3.12(2)$~MeV~\cite{Juric:1973}. The dashed yellow curve follows the set of parameters reproducing the expected $\Lambda$~separation energy. The same curve is also reported in Fig.~\ref{fig:Wd-Cp_2D} (red banded curve with black empty dots and error bars), that is a projection of Fig.~\ref{fig:Wd-Cp_3D} on the $W_D-C_P$ plane. The dashed box represents the $W_D$ and $C_P$ domain of the previous picture. For comparison, also the variational results of Ref.~\cite{Usmani:2008} are reported. Green curves are the results for $\bar v=6.15$~MeV and $v_\sigma=0.24$~MeV, blue ones for $\bar v=6.10$~MeV and $v_\sigma=0.24$~MeV. Dashed, long-dashed and dot-dashed lines correspond respectively to $\varepsilon=0.1$, $0.2$ and $0.3$. In our calculations we have not considered different combinations for the parameters of the two-body $\Lambda N$ interaction, focusing on the three-body part. We have thus kept fixed $\bar v$ and $v_\sigma$ to the same values of the green curves of Fig.~\ref{fig:Wd-Cp_2D} which are the same reported in Tab.~\ref{tab:parLN+LNN}. Moreover, we have set $\varepsilon=0$ for all the hypernuclei studied due to the impossibility of exactly including the $\mathcal P_x$ exchange operator in the propagator. A perturbative analysis of the effect of the $v_0(r)\varepsilon(\mathcal P_x-1)$ term on the hyperon separation energy is reported in~\S~\ref{subsubsec:Px}.

As it can be seen from Fig.~\ref{fig:Wd-Cp_3D}, $B_\Lambda$ significantly increases with the increase in $C_P$, while it decreases with $W_D$. This result is consistent with the attractive nature of $V_{\Lambda NN}^{2\pi,P}$ and the repulsion effect induced by $V_{\Lambda NN}^D$. It is also in agreement with all the variational estimates on $^5_\Lambda$He (see for instance Refs.~\cite{Sinha:2002,Usmani:2008}). Starting from the analysis of the results in the $W_D-C_P$ space for $_\Lambda^5$He, we performed simulations for the next closed shell hypernucleus $^{17}_{~\Lambda}$O. Using the parameters in the red band of Fig.~\ref{fig:Wd-Cp_2D} we identified a parametrization able to reproduce the experimental $B_\Lambda$ for both $_\Lambda^5$He and $^{17}_{~\Lambda}$O at the same time within the AFDMC framework:
\begin{equation*}
	\hypertarget{par_II}{(\text{II})}\quad
	\left\{
	\begin{array}{rcll}
		C_P&\!=\!&1.00& \!\text{MeV}\\
		C_S&\!=\!&1.50& \!\text{MeV}\\
		W_D&\!=\!&0.035&\!\text{MeV}
	\end{array}
	\right.
\end{equation*}

\begin{figure}[!htb]
	\centering
	\includegraphics[width=\linewidth]{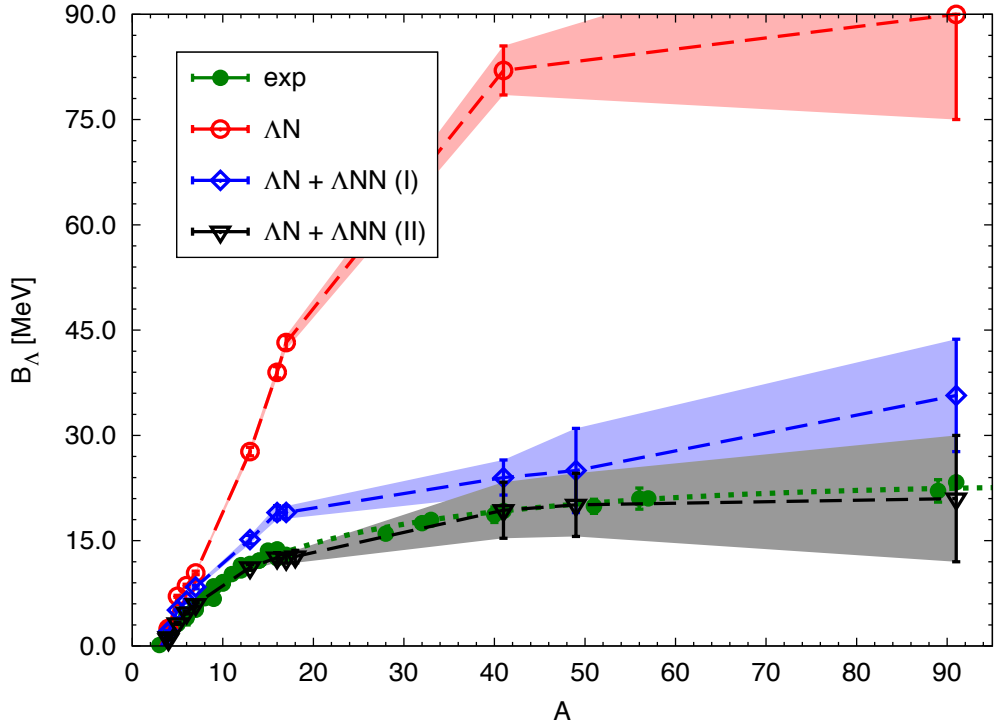}
	\caption[$\Lambda$ separation energy vs. $A$]
		{$\Lambda$ separation energy as a function of $A$. 
		Solid green dots~(dashed curve) are the available $B_\Lambda$ experimental or semiempirical values. Empty red dots~(upper banded curve) refer to 
		the AFDMC results for the two-body $\Lambda N$ interaction alone. Empty blue diamonds~(middle banded curve) and empty black triangles~(lower banded curve)
		are the results with the inclusion also of the three-body hyperon-nucleon force, respectively for the parametrizations (\hyperlink{par_I}{I}) and
		(\hyperlink{par_II}{II}).}
	\label{fig:BL-A}
\end{figure}

\begin{figure}[!htb]
	\centering
	\includegraphics[width=\linewidth]{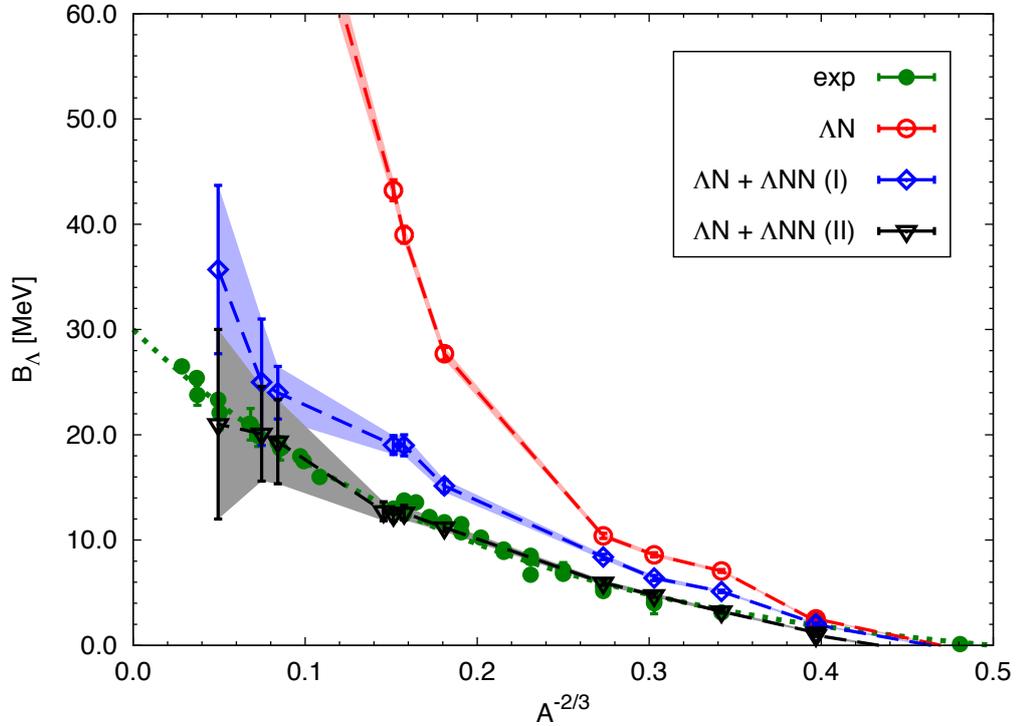}
	\caption[$\Lambda$~separation energy vs. $A^{-2/3}$]
		{$\Lambda$~separation energy as a function of $A^{-2/3}$, adapted from Ref.~\cite{Lonardoni:2013_PRC}.
		The key is the same of Fig.~\ref{fig:BL-A}.}
	\label{fig:BL-A23}
\end{figure}

Given the set (\hyperlink{par_II}{II}), the $\Lambda$~separation energy of the closed shell hypernuclei reported in Fig.~\ref{fig:BL-A_cshell} has been re-calculated. We have seen that $B_\Lambda$ is not sensitive neither to the details of the $NN$ interaction, nor to the total binding energies of nuclei and hypernuclei, as verified by the good results in Tab.~\ref{tab:BL_He5L-O17L_I} even for the problematic case of AV6' (see \S~\ref{sec:nuc}). On the grounds of this observation, we tried to simulate also open shell hypernuclei, using the $\Lambda N$, $\Lambda NN$ set (\hyperlink{par_I}{I}) and $\Lambda NN$ set (\hyperlink{par_II}{II}) potentials. The binding energies for these systems might not be accurate, as in the case of the corresponding nuclei. The hyperon separation energy is expected to be in any case realistic. All the results obtained so far in the mass range $3 \leq A \leq 91$ are summarized in Fig.~\ref{fig:BL-A} and Fig.~\ref{fig:BL-A23}. 

We report $B_\Lambda$ as a function of $A$ and $A^{-2/3}$, which is an approximation of the $A$ dependence of the kinetic term of the Hamiltonian. Solid green dots are the available experimental data, empty symbols the AFDMC results. The red curve is obtained using only the two-body hyperon-nucleon interaction in addition to the nuclear AV4' potential. The blue curve refers to the results for the same systems when also the three-body $\Lambda NN$ interaction with the old set of parameters~(\hyperlink{par_I}{I}) is included. The black lower curve shows the results obtained by including the three-body hyperon-nucleon interaction described by the new parametrization~(\hyperlink{par_II}{II}). A detailed comparison between numerical and experimental results for the hyperon-separation energy is given in Tab.~\ref{tab:BL}.

\renewcommand{\arraystretch}{1.4}
\begin{table}[!htb]
	\centering
	\begin{tabular*}{\linewidth}{@{\hspace{3.0em}\extracolsep{\fill}}lccc@{\extracolsep{\fill}\hspace{3.0em}}}
		\toprule
		\toprule
		System & $E$ & $B_\Lambda$ & Expt. $B_\Lambda$ \\
		\midrule
		\hspace{0.6em}$^3_\Lambda$H        &   -1.00(14) & -1.22(15)  &  0.13(5)  \hspace{1.2em}\cite{Juric:1973}    \\
		\hspace{0.6em}$^4_\Lambda$H        &   -9.69(8)  &  0.95(9)   &  2.04(4)  \hspace{1.2em}\cite{Juric:1973}    \\
		\hspace{0.6em}$^4_\Lambda$He       &   -9.97(8)  &  1.22(9)   &  2.39(3)  \hspace{1.2em}\cite{Juric:1973}    \\
		\hspace{0.6em}$^5_\Lambda$He       &  -35.89(12) &  3.22(14)  &  3.12(2)  \hspace{1.2em}\cite{Juric:1973}    \\
		\hspace{0.6em}$^6_\Lambda$He       &  -32.72(15) &  4.76(20)  &  4.25(10) \hspace{0.7em}\cite{Juric:1973}    \\
		\hspace{0.6em}$^7_\Lambda$He       &  -35.82(15) &  5.95(25)  &  5.68(28) \hspace{0.7em}\cite{Nakamura:2013} \\
		\hspace{0.3em}$^{13}_{~\Lambda}$C  &  -88.5(26)* & 11.2(4)    & 11.69(12) \hspace{0.2em}\cite{Cantwell:1974} \\
		\hspace{0.3em}$^{16}_{~\Lambda}$O  & -157.5(6)   & 12.6(7)    & 12.50(35) \hspace{0.2em}\cite{Pile:1991}     \\
		\hspace{0.3em}$^{17}_{~\Lambda}$O  & -189.2(4)   & 12.4(6)    & 13.0(4)   \hspace{0.8em}\cite{Usmani:1995}   \\
		\hspace{0.3em}$^{18}_{~\Lambda}$O  & -189.7(6)   & 12.7(9)    & \hspace{-3.0em}---                           \\
		\hspace{0.3em}$^{41}_{~\Lambda}$Ca & -616(3)     & 19(4)      & 19.24(0)  \hspace{0.3em}\cite{Bodmer:1988}   \\
		\hspace{0.3em}$^{49}_{~\Lambda}$Ca & -665(4)     & 20(5)      & \hspace{-3.0em}---                           \\
		\hspace{0.3em}$^{91}_{~\Lambda}$Zr & -1478(7)    & 21(9)      & 23.33(0)  \hspace{0.3em}\cite{Bodmer:1988}   \\
		\bottomrule
		\bottomrule
	\end{tabular*}
	\caption[$\Lambda$~separation energies: $\Lambda$~hypernuclei, $3\le A\le 91$]
		{Binding energies and $\Lambda$~separation energies (in MeV) obtained using the two-body plus three-body hyperon-nucleon interaction with the 
		set of parameters~(\hyperlink{par_II}{II})~\cite{Lonardoni:2013_PRC}. The results already include the CSB contribution. 
		The effect is evident only for light systems, as discussed in the next section.
		In the last column, the expected $B_\Lambda$ values. 
		Since no experimental data for $A=17,41,91$ exists, the reference separation energies are semiempirical values.}
	\label{tab:BL}
\end{table}
\renewcommand{\arraystretch}{1.0}

From Fig.~\ref{fig:BL-A} and Fig.~\ref{fig:BL-A23} we can see that the new parametrization for the three-body hyperon-nucleon interaction correctly reproduces the experimental saturation property of the $\Lambda$~separation energy. All the separation energies for $A\ge 5$ are compatible or very close to the expected results, where available, as reported in Tab.~\ref{tab:BL}. Since for $^{18}_{~\Lambda}$O and $^{49}_{~\Lambda}$Ca no experimental data have been found, the values of $12.7(9)$~MeV and $20(5)$~MeV are AFDMC predictions, that follows the general trend of the experimental curve. Although for $A\ge 41$ the Monte Carlo statistical error bars become rather large, the extrapolation of the $\Lambda$~binding energy for $A\rightarrow\infty$ points to the correct region for the expected value $D_\Lambda\sim30$~MeV of $s_\Lambda$ states in nuclear matter. 

We can find the same problems discussed in the case of nuclei (\S~\ref{sec:nuc}) in the analysis of the total hypernuclear binding energies for $A\ge 5$. For instance, the binding energy of $^{13}_{~\Lambda}$C is non physical, as for the energy of the core nucleus $^{12}$C. However, the energy difference is consistent with the expected result. Moreover, for the core wave function of $^7_\Lambda$He we have used the same mixing parameter adopted in the description of $^6$He (see Eq.~(\ref{eq:He6_mix}) and Fig.~\ref{fig:E_He6}), in order to have at least the correct ordering in the hypernuclear energy spectrum. However, the same hyperon separation energy can be found by just using the $1p_{3/2}$ shell for the outer neutrons for both strange and non strange nucleus. Our working hypothesis regarding the computation of the hyperon separation energy is thus correct, at least for medium-heavy hypernuclei.

For $A<5$ our results are more than 1~MeV off from experimental data. For $^3_\Lambda$H, the $\Lambda$~separation energy is even negative, meaning that the hypernucleus is less bound than the corresponding nucleus $^2$H. We can ascribe this discrepancy to the lack of accuracy of our wave function for few-body systems. Since the $\Lambda$~hyperon does not suffer from Pauli blocking by the other nucleons, it can penetrate into the nuclear interior and form deeply bound hypernuclear states. For heavy systems the $\Lambda$~particle can be seen as an impurity that does not drastically alter the nuclear wave function. Therefore, the trial wave function of Eq.~(\ref{eq:psi_singleL}) with the single particle state $\varphi_\epsilon^\Lambda$ described by the $1s_{1/2}$ neutron orbital, is accurate enough as starting point for the imaginary time propagation. For very light hypernuclei, for which the first nucleonic $s$ shell is not closed, this might not be the case. In order to have a correct projection onto the ground state, the single particle orbitals of both nucleons and lambda might need to be changed when the hyperon is added to the nucleus. Moreover, in very light hypernuclei, the neglected nucleon-nucleon and hyperon-nucleon correlations, might result in non negligible contributions to the $\Lambda$~binding energy. A study of these systems within a few-body method or a different projection algorithm like the GFMC, might solve this issue.

\subsubsection{Effect of the charge symmetry breaking term}
\label{subsubsec:CSB}

The effect of the CSB potential has been studied for the $A=4$ mirror hypernuclei. As reported in Tab.~\ref{tab:CSB_A4}, without the CSB term there is no difference in the $\Lambda$ binding energy of $^4_\Lambda$H and $^4_\Lambda$He. When CSB is active, a splitting appears due to the different behavior of the $\Lambda p$ and $\Lambda n$ channels. The strength of the difference $\Delta B_\Lambda^{CSB}$ is independent on the parameters of the three-body $\Lambda NN$ interaction and it is compatible with the experimental result~\cite{Juric:1973}, although the $\Lambda$~separation energies are not accurate.

\renewcommand{\arraystretch}{1.4}
\begin{table}[!ht]
	\centering
	\begin{tabular*}{\linewidth}{@{\hspace{2.0em}\extracolsep{\fill}}ccccc@{\extracolsep{\fill}\hspace{2.0em}}}
		\toprule
		\toprule
		Parameters & System & $B_\Lambda^{sym}$ & $B_\Lambda^{CSB}$ & {$\Delta B_\Lambda^{CSB}$} \\
		\midrule
		\multirow{2}{*}{Set (\hyperlink{par_I}{I})}   
		& $^4_\Lambda$H\hspace{0.4em} & 1.97(11) &  1.89(9) & \multirow{2}{*}{0.24(12)} \\       
		& $^4_\Lambda$He              & 2.02(10) &  2.13(8) &                           \\[0.8em]
		\multirow{2}{*}{Set (\hyperlink{par_II}{II})} 
		& $^4_\Lambda$H\hspace{0.4em} & 1.07(8)  &  0.95(9) & \multirow{2}{*}{0.27(13)} \\       
		& $^4_\Lambda$He              & 1.07(9)  &  1.22(9) &                           \\
		\midrule
		\multirow{2}{*}{Expt.~\cite{Juric:1973}} & $^4_\Lambda$H\hspace{0.4em} &  {---}   &  2.04(4) & \multirow{2}{*}{0.35(5)\phantom{0}}  \\
		& $^4_\Lambda$He              &  {---}   &  2.39(3) &                           \\
		\bottomrule
		\bottomrule
	\end{tabular*}
	\caption[$\Lambda$~separation energies: $A=4$ mirror hypernuclei]
		{$\Lambda$~separation energies (in MeV) for the $A=4$ mirror $\Lambda$~hypernuclei with (fourth column) and without (third column)
		the inclusion of the charge symmetry breaking term~\cite{Lonardoni:2013_PRC}. In the last column the difference in the separation energy induced by
		the CSB interaction. First and second rows refer to different set of parameters for the $\Lambda NN$ interaction, while the last row is the
		experimental result.}
	\label{tab:CSB_A4}
\end{table}
\renewcommand{\arraystretch}{1.0}

\renewcommand{\arraystretch}{1.4}
\begin{table}[!htb]
	\centering
	\begin{tabular*}{\linewidth}{@{\hspace{3.0em}\extracolsep{\fill}}ccccc@{\extracolsep{\fill}\hspace{3.0em}}}
		\toprule
		\toprule
		System                             & $p$ & $n$ &   $\Delta_{np}$    & $\Delta B_\Lambda$ \\
		\midrule                                                    
		\hspace{-0.5em}$^4_\Lambda$H       & 1 & 2 &       $+1$         & $-0.12(8)$ \\[0.5em]     
		$^4_\Lambda$He                     & 2 & 1 &       $-1$         & $+0.15(9)$ \\       
		$^5_\Lambda$He                     & 2 & 2 & \hspace{0.78em}$0$ & $+0.02(9)$ \\       
		$^6_\Lambda$He                     & 2 & 3 &       $+1$         & $-0.06(8)$ \\       
		$^7_\Lambda$He                     & 2 & 4 &       $+2$         & $-0.18(8)$ \\[0.5em]
		\hspace{-0.7em}$^{16}_{~\Lambda}$O & 8 & 7 &       $-1$         & $+0.27(35)$ \\      
		\hspace{-0.7em}$^{17}_{~\Lambda}$O & 8 & 8 & \hspace{0.78em}$0$ & $+0.15(35)$ \\      
		\hspace{-0.7em}$^{18}_{~\Lambda}$O & 8 & 9 &       $+1$         & $-0.74(49)$ \\      
		\bottomrule
		\bottomrule
	\end{tabular*}
	\caption[$\Lambda$ separation energies: effect of the CSB potential]
		{Difference (in MeV) in the hyperon separation energies induced by the CSB term (Eq.~(\ref{eq:V_CSB})) for different
		hypernuclei~\cite{Lonardoni:2013_PRC}. The fourth column reports the difference between the number of neutrons and protons. 
		Results are obtained with the full two- plus three-body (set (\hyperlink{par_II}{II})) hyperon-nucleon interaction.
		In order to reduce the errors, $\Delta B_\Lambda$ has been calculated by taking the difference between total hypernuclear binding energies, 
		instead of the hyperon separation energies.}
	\label{tab:CSB}
\end{table}
\renewcommand{\arraystretch}{1.0}

The same CSB potential of Eq.~(\ref{eq:V_CSB}) has been included in the study of hypernuclei for $A>4$. In Tab.~\ref{tab:CSB} the difference in the hyperon separation energies $\Delta B_\Lambda=B_\Lambda^{CSB}-B_\Lambda^{sym}$ is reported for different hypernuclei up to $A=18$. The fourth column shows the difference between the number of neutrons and protons $\Delta_{np}=\mathcal N_n-\mathcal N_p$. For the symmetric hypernuclei $^5_\Lambda$He and $^{17}_{~\Lambda}$O the CSB interaction has no effect, being this difference zero. In the systems with neutron excess ($\Delta_{np}>0$), the effect of the CSB consists in decreasing the hyperon separation energy compared to the charge symmetric case. When $\Delta_{np}$ becomes negative, $\Delta B_\Lambda>0$ due to the attraction induced by the CSB potential in the $\Lambda p$ channel, that produces more bound hypernuclei. Being $\Delta_{np}$ small, these effects are in any case rather small and they become almost negligible compared to the statistical errors on $B_\Lambda$ when the number of baryons becomes large enough ($A>16$). However, in the case of $\Lambda$~neutron matter, the CSB term might have a relevant effect for large enough $\Lambda$ fraction.

\subsubsection{Effect of the hyperon-nucleon space-exchange term}
\label{subsubsec:Px}

As already mentioned in the previous chapter, the inclusion of the $\Lambda N$ space exchange operator of Eq.~(\ref{eq:V_LN}) in the AFDMC propagator is not yet possible. In \S~\ref{subsubsec:Prop_LN} we presented a possible perturbative approach for the treatment of such term. In Tab.~\ref{tab:Px} we report the results of this analysis.

All the results for $^{41}_{~\Lambda}$Ca are consistent within the statistical errors. On the contrary, for lighter systems the $\Lambda$~separation energy seems rather sensitive to the value of the exchange parameter $\varepsilon$. Considering larger values for $\varepsilon$, $B_\Lambda$ generally increases. This trend is opposite to what is found for instance in Ref.~\cite{Usmani:2006}. We recall that only the computation of the Hamiltonian expectation value by means of Eq.~(\ref{eq:mixed}) gives exact results. For other operators, like the space exchange $\mathcal P_x$, the pure estimators have to be calculated with the extrapolation method via the two relations (\ref{eq:pure1}) or (\ref{eq:pure2}). The variational estimate $\langle\mathcal P_x\rangle_v$ is thus needed. In the mentioned reference, the importance of space exchange correlations for variational estimates is discussed. Being these correlations neglected in this work, our perturbative treatment of the $\mathcal P_x$ contribution might not be accurate. Moreover, the evidence of the importance of space exchange correlations might invalid the perturbative approach itself. An effective but more consistent treatment of this term could consist in a slight change in the strength of the central $\Lambda N$ potential. However, due to the very limited information about the space exchange parameter and its effect on single $\Lambda$~hypernuclei heavier than $^5_\Lambda$He, this approach has not been considered in the present work. Recent calculations of many hadron systems within an EFT treatment at NLO for the full $SU(3)$ hadronic spectrum confirmed indeed that exchange terms are sub-leading~\cite{Haidenbauer:2013}.

\renewcommand{\arraystretch}{1.4}
\begin{table}[!h]
	\centering			
	\begin{tabular*}{\linewidth}{@{\hspace{3.0em}\extracolsep{\fill}}cccc@{\extracolsep{\fill}\hspace{3.0em}}}
		\toprule
		\toprule
		System                             & $\varepsilon=0.0$ & $\varepsilon=0.1$ & $\varepsilon=0.3$ \\
		\midrule                                              
		\hspace{0.7em}$^5_\Lambda$He       &       3.22(14)    &        3.89(15)   &       4.67(25)    \\
		$^{17}_{~\Lambda}$O                &      12.4(6)      &       12.9(9)     &      14.0(9)      \\ 
		\hspace{0.3em}$^{41}_{~\Lambda}$Ca &       19(4)       &        21(5)      &       25(7)       \\
		\bottomrule
		\bottomrule
	\end{tabular*}
	\caption[$\Lambda$ separation energies: effect of the $\Lambda N$ exchange potential]
		{Variation of the $\Lambda$ separation energy as a consequence of the exchange potential $v_0(r)\varepsilon(\mathcal P_x-1)$ 
		in the $\Lambda N$ interaction of Eq.~(\ref{eq:V_LN}). The contribution of $\mathcal P_x$ is treated perturbatively 
		for different value of the parameter $\varepsilon$. The interaction used is the full AV4'+$\Lambda N$+$\Lambda NN$ 
		set~(\hyperlink{par_II}{II}). Results are expressed in MeV.}
	\label{tab:Px}
\end{table}
\renewcommand{\arraystretch}{1.0}

\subsection{Single particle densities and radii}
\label{subsec:dens_l}

Single particle densities can be easily computed in Monte Carlo calculations by considering the expectation value of the density operator
\begin{align}
	\hat\rho_\kappa(r)=\sum_{i}\delta(r-r_i)\quad\quad \kappa=N,\Lambda\;,
\end{align}
where $i$ is the single particle index running over nucleons for $\rho_N=\langle\hat\rho_N\rangle$ or hyperons for $\rho_\Lambda=\langle\hat\rho_\Lambda\rangle$. The normalization is given by 
\begin{align}
	\int dr 4\pi r^2 \rho_\kappa(r)=1\;.
\end{align}
Root mean square radii $\langle r_\kappa^2\rangle^{1/2}$ are simply calculated starting from the Cartesian coordinates of nucleons and hyperons. A consistency check between AFDMC densities and radii is then taken by verifying the relation
\begin{align}
	\langle r_\kappa^2\rangle=\int dr 4\pi r^4 \rho_\kappa(r)\;.\label{eq:r2}
\end{align}

Before reporting the results we recall that also for densities and radii the AFDMC calculation can only lead to mixed estimators. The pure estimators are thus approximated by using Eq.~(\ref{eq:pure1}) or Eq.~(\ref{eq:pure2}). The two relations should lead to consistent results. This is the case for the nucleon and hyperon radii. In computing the densities instead, the low statistics for $r\rightarrow0$ generates differences in the two approaches. For nucleons these discrepancies are almost within the statistical errors. For hyperons, the much reduced statistics (1 over $A-1$ for single $\Lambda$~hypernuclei) and the fact that typically the $\Lambda$ density is not peaked in $r=0$, create some uncertainties in the region for small $r$, in particular for the first estimator. We therefore chose to adopt the pure estimator of Eq.~(\ref{eq:pure2}) to have at least a positive definite estimate. Finally, it has to be pointed out that the pure extrapolated results are sensitive to the quality of the variational wave function and the accuracy of the projection sampling technique. Although we successfully tested the AFDMC propagation, we are limited in the choice of the VMC wave function. In order to be consistent with the mixed estimators coming from AFDMC calculations, we considered the same trial wave functions also for the variational runs. This might introduce some biases in the evaluation of pure estimators. Therefore, the results presented in the following have to be considered as a qualitative study on the general effect of the hypernuclear forces on the nucleon and hyperon distributions.

\begin{figure}[!htb]
	\centering
	\includegraphics[width=\linewidth]{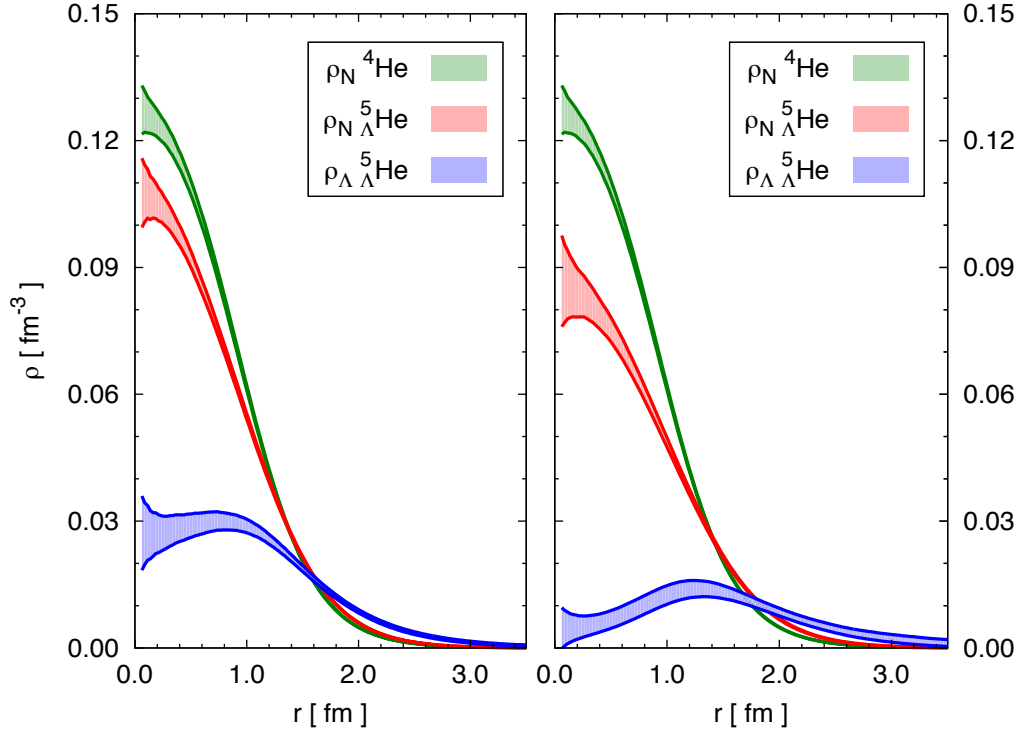}
	\caption[Single particle densities: $N$ and $\Lambda$ in $^4$He and $^5_\Lambda$He]
		{Single particle densities for nucleons in $^4$He [green, upper banded curve] and for nucleons [red, middle banded curve] and the lambda particle 
		[blue, lower banded curve] in $^5_\Lambda$He~\cite{Lonardoni:2013_PRC}. In the left panel the results for the two-body $\Lambda N$ interaction alone.
		In the right panel the results with the inclusion also of the three-body hyperon-nucleon force in the parametrization~(\hyperlink{par_II}{II}).
		The AV4' potential has been used for the nuclear core.}
	\label{fig:rho_He5L}
\end{figure}

In Fig.~\ref{fig:rho_He5L} we report the results for the single particle densities for $^4$He and $^5_\Lambda$He. The green curves are the densities of nucleons in the nucleus, while the red and blue curves are, respectively, the density of nucleons and of the lambda particle in the hypernucleus. In the left panel the results are obtained using AV4' for the nuclear part and the two-body $\Lambda N$ interaction alone for the hypernuclear component. In the right panel the densities are calculated with the full two- plus three-body (set (\hyperlink{par_II}{II})) hyperon-nucleon interaction.

\begin{figure}[p]
	\centering
	\includegraphics[width=\linewidth]{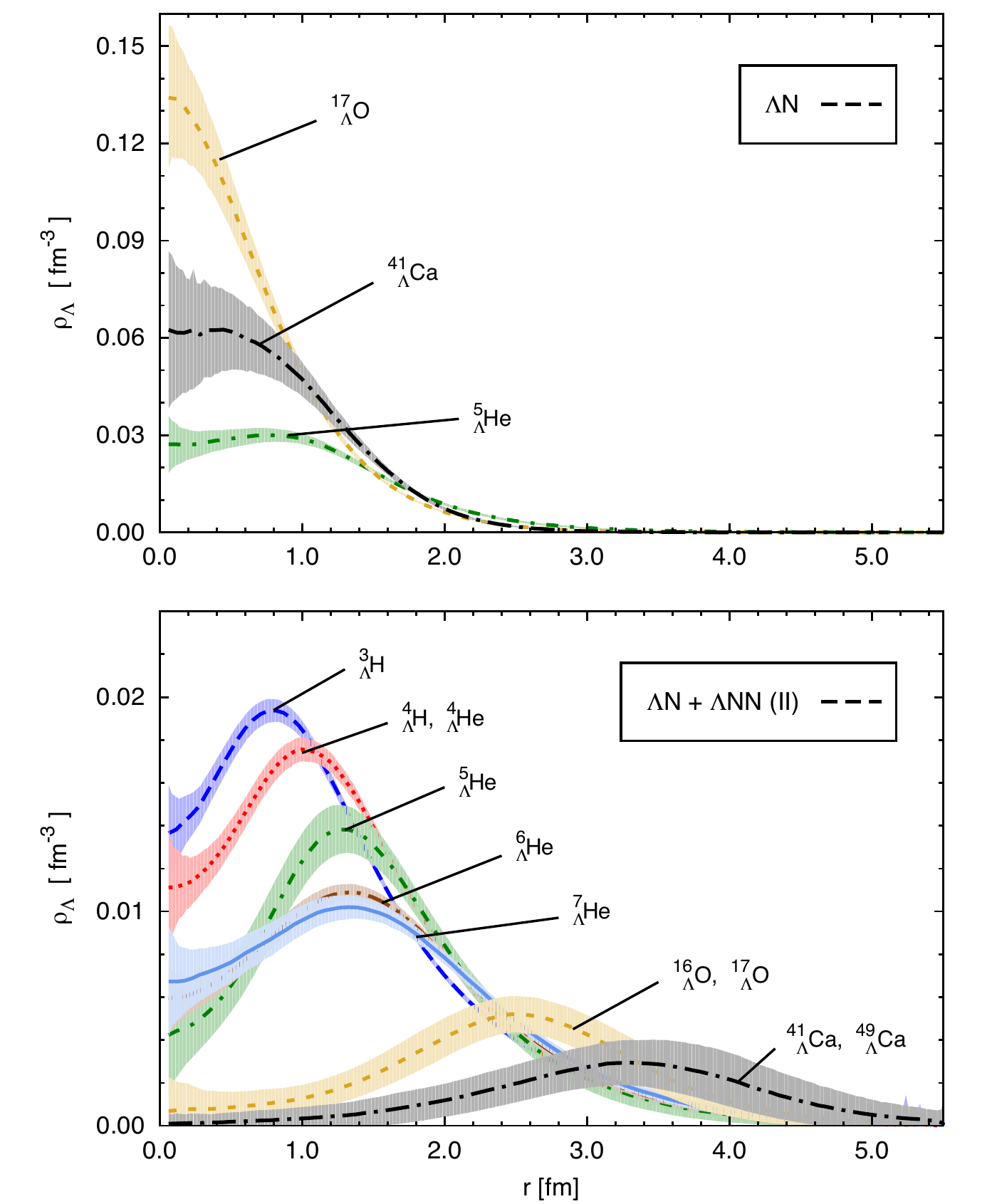}
	\caption[Single particle densities: $\Lambda$ in hypernuclei for $3\le A\le 91$]
		{Single particle densities for the $\Lambda$~particle in different hypernuclei~\cite{Lonardoni:2013_PRC}. Top panel reports the results for the
		two-body $\Lambda N$ interaction alone. Bottom panel shows the results when the three-body hyperon-nucleon interaction with the set of
		parameters~(\hyperlink{par_II}{II}) is also included. The nuclear core is described by the AV4' potential.}
	\label{fig:L_rho}
\end{figure}

The addition of the $\Lambda$~particle to the nuclear core of $^4$He has the effect to slightly reduce the nucleon density in the center. The $\Lambda$~particle tries to localize close to $r=0$, enlarging therefore the nucleon distribution. When the three-body $\Lambda NN$ interaction is turned on (right panel of Fig.~\ref{fig:rho_He5L}), the repulsion moves the nucleons to large distances but the main effect is that the hyperon is pushed away from the center of the system. As can be seen from Fig.~\ref{fig:L_rho}, this effect is much more evident for large $A$. When the hypernucleus is described by the $\Lambda N$ interaction alone, the $\Lambda$~particle is localized near the center, in the range $r<2$~fm (left panel of Fig.~\ref{fig:L_rho}). The inclusion of the three-body $\Lambda NN$ potential forces the hyperon to move from the center, in a region that roughly correspond to the skin of nucleons (see Tab.~\ref{tab:radii}). Although these densities are strictly dependent to the nuclear interaction, by using the AV6' potential we found the same qualitative effects on the $\Lambda$~particle, confirming the importance of the three-body hyperon-nucleon interaction and its repulsive nature. Due to the limitations discussed above and the use of too simplified interactions for the nucleon-nucleon force, the comparison with the available VMC density profiles~\cite{Usmani:2003,Usmani:1995} is difficult.

\renewcommand{\arraystretch}{1.4}
\begin{table}[ht]
	\centering
	\begin{tabular*}{\linewidth}{@{\hspace{2.0em}\extracolsep{\fill}}ccccc@{\extracolsep{\fill}\hspace{2.0em}}}
		\toprule
		\toprule
		\multirow{2}{*}{System} & \multicolumn{2}{c}{nucleus} & \multicolumn{2}{c}{hypernucleus} \\
		\cmidrule(l){2-3}\cmidrule(l){4-5}
		                                   & \hspace{1.3em}$r_N^{\text{exp}}$ & $r_N$   & \hspace{1.3em}$r_N$    & $r_\Lambda$ \\
		\midrule                             
		    $^2$H~~-~~$^3_\Lambda$H        & \hspace{1.3em}2.142 & 1.48(8)  & \hspace{1.3em}1.9(1)   & 2.00(16) \\
		    $^3$H~~-~~$^4_\Lambda$H        & \hspace{1.3em}1.759 & 1.5(1)   & \hspace{1.3em}1.77(9)  & 2.12(15) \\
		   $^3$He~~-~~$^4_\Lambda$He       & \hspace{1.3em}1.966 & 1.5(1)   & \hspace{1.3em}1.77(9)  & 2.10(14) \\
		   $^4$He~~-~~$^5_\Lambda$He       & \hspace{1.3em}1.676 & 1.57(9)  & \hspace{1.3em}1.58(7)  & 2.2(2)   \\
		   $^5$He~~-~~$^6_\Lambda$He       & \hspace{1.1em} ---  & 2.02(16) & \hspace{1.3em}2.16(17) & 2.43(17) \\
		   $^6$He~~-~~$^7_\Lambda$He       & \hspace{1.3em}2.065 & 2.3(2)   & \hspace{1.3em}2.4(2)   & 2.5(2)   \\
		 $^{15}$O~~-~~$^{16}_{~\Lambda}$O  & \hspace{1.1em} ---  & 2.20(12) & \hspace{1.3em}2.3(1)   & 3.2(3)   \\
		 $^{16}$O~~-~~$^{17}_{~\Lambda}$O  & \hspace{1.3em}2.699 & 2.16(12) & \hspace{1.3em}2.23(11) & 3.3(3)   \\
		 $^{17}$O~~-~~$^{18}_{~\Lambda}$O  & \hspace{1.3em}2.693 & 2.26(13) & \hspace{1.3em}2.32(14) & 3.3(3)   \\
		$^{40}$Ca~~-~~$^{41}_{~\Lambda}$Ca & \hspace{1.3em}3.478 & 2.8(2)   & \hspace{1.3em}2.8(2)   & 4.2(5)   \\
		$^{48}$Ca~~-~~$^{49}_{~\Lambda}$Ca & \hspace{1.3em}3.477 & 3.1(2)   & \hspace{1.3em}3.1(2)   & 4.3(5)   \\
		\bottomrule
		\bottomrule
	\end{tabular*}
	\caption[Nucleon and hyperon radii in hypernuclei for $3\le A\le49$]
		{Nucleon and hyperon root mean square radii (in fm) for nuclei and corresponding $\Lambda$~hypernuclei. The employed nucleon-nucleon potential is AV4'.
		For the strange sector we used the full two- plus three-body hyperon-nucleon force in the parametrization~(\hyperlink{par_II}{II}).
		The experimental nuclear charge radii are from Ref.~\cite{Angeli:2013}. Errors are on the fourth digit.}
	\label{tab:radii}
\end{table}
\renewcommand{\arraystretch}{1.0}

In Tab.~\ref{tab:radii} we report the nucleon and hyperon root mean square radii for nuclei and hypernuclei. The experimental nuclear charge radii are reported as a reference. AFDMC $r_N$, that do not distinguish among protons and neutrons, are typically smaller than the corresponding experimental results. This can be understood as a consequence of the employed AV4' $NN$ interaction that overbinds nuclei. The main qualitative information is that the hyperon radii are systematically larger than the nucleon ones, as expected by looking at the single particle densities. Starting from $A=5$, the nucleon radii in the nucleus and the corresponding hypernucleus do not change, although the differences in the nucleon densities for $r\rightarrow0$. This is due to the small contribution to the integral (\ref{eq:r2}) given by the density for $r$ close to zero. For the hypernuclei with $A<5$, AFDMC calculations predict larger $r_N$ when the hyperon is added to the core nucleus. This is inconsistent with the results of Ref.~\cite{Sinha:2002}, where a shrinking of the core nuclei due to the presence of the $\Lambda$~particle in $A\le 5$ hypernuclei is found. We need to emphasize once more that the results presented in this section are most likely strictly connected to the employed nucleon-nucleon potential. For instance, the shrinkage of hypernuclei has been investigated experimentally by $\gamma$-ray spectroscopy~\cite{Hashimoto:2006,Tanida:2001}. In the experiment of Ref.~\cite{Tanida:2001}, by looking at the electric quadrupole transition probability from the excited $5/2^+$ state to the ground state in $^7_\Lambda$Li, a $19\%$ shrinkage of the intercluster distance was inferred, assuming the two-body cluster structure core+deuteron. Therefore, the AFDMC study of densities and radii, differently from the analysis of $\Lambda$~separation energies, cannot lead to accurate results at this level. It has to be considered as a first explorative attempt to get hypernuclear structure information from Diffusion Monte Carlo simulations.

\section{Double $\Lambda$~hypernuclei}
\label{sec:ll_hyp}

In the single particle wave function representation, two $\Lambda$~particles with antiparallel spin can be added to a core nucleus filling the first hyperon $s$ shell, assumed to be the neutron $1s_{1/2}$ Skyrme radial function as in the case of single $\Lambda$~hypernuclei. The complete hypernuclear wave function is given by Eq.~(\ref{eq:Psi_T}), where the nucleon trial wave function is the same used in the AFDMC calculations for nuclei and in this case also the hyperon Slater determinant is employed. Although the effect on the total energy introduced by a $\Lambda\Lambda$ correlation function is found to be negligible, for consistency with the calculations for nuclei and single $\Lambda$~hypernuclei we neglected the central Jastrow correlations. 

The double $\Lambda$~separation energy and the incremental $\Lambda\Lambda$~energy of Eqs.~(\ref{eq:B_LL}) and (\ref{eq:dB_LL}) are calculated starting from the energy of the nucleus and the corresponding single and double $\Lambda$~hypernuclei described by the same $NN$ AV4' potential. Due to the difficulties in treating open shell nuclei and the limited amount of data about double $\Lambda$~hypernuclei, we performed the AFDMC study for just the lightest $\Lambda\Lambda$~hypernucleus for which energy experimental information are available, $^{\;\;\,6}_{\Lambda\Lambda}$He.

\subsection{Hyperon separation energies}
\label{subsec:E_ll}

In Tab.~\ref{tab:BLL} we report the total binding energies for $^4$He, $^5_\Lambda$He and $^{\;\;\,6}_{\Lambda\Lambda}$He in the second column, the single or double hyperon separation energies in the third and the incremental binding energy in the last column. The value of $B_{\Lambda\Lambda}$ confirms the weak attractive nature of the $\Lambda\Lambda$ interaction~\cite{Hiyama:2002,Nagels:1979,Maessen:1989,Rijken:1999}. Starting from $^4$He and adding two hyperons with $B_\Lambda=3.22(14)$~MeV, the energy of $^{\;\;\,6}_{\Lambda\Lambda}$He would be $1.0\div1.5$~MeV less than the actual AFDMC result. Therefore the $\Lambda\Lambda$ potential of Eq.~(\ref{eq:V_LL}) induces a net attraction between hyperons, at least at this density.

\renewcommand{\arraystretch}{1.4}
\begin{table}[!ht]
	\centering
	\begin{tabular*}{\linewidth}{@{\hspace{3.0em}\extracolsep{\fill}}cccc@{\extracolsep{\fill}\hspace{3.0em}}}
		\toprule
		\toprule
		System                          &   {$E$}   & {$B_{\Lambda(\Lambda)}$} & $\Delta B_{\Lambda\Lambda}$ \\
		\midrule
		\hspace{0.7em}$^4$He            & -32.67(8)  &            ---           &          ---                \\
		\hspace{0.6em}$^5_\Lambda$He    & -35.89(12) &          3.22(14)        &          ---                \\
		$^{\;\;\,6}_{\Lambda\Lambda}$He & -40.6(3)   &          7.9(3)          &         1.5(4)              \\
		\midrule
		$^{\;\;\,6}_{\Lambda\Lambda}$He & {Expt.~\cite{Takahashi:2001}} & {$7.25\pm 0.19^{+0.18}_{-0.11}$} & {$1.01\pm 0.20^{+0.18}_{-0.11}$} \\
		\bottomrule
		\bottomrule
	\end{tabular*}
	\caption[$\Lambda$~separation energies: $^{\;\;\,6}_{\Lambda\Lambda}$He]
		{Comparison between $^4$He and the corresponding single and double $\Lambda$~hypernuclei~\cite{Lonardoni:2013_PRC}. In the second column the total
		binding energies are reported. The third column shows the single or double $\Lambda$~separation energies. In the last column the incremental binding energy
		$\Delta B_{\Lambda\Lambda}$ is reported. All the results are obtained using the complete two- plus three-body 
		(set~(\hyperlink{par_II}{II})) hyperon-nucleon interaction with the addition of the $\Lambda\Lambda$ force of Eq.~(\ref{eq:V_LL}). 
		The results are expressed~in~MeV.}
	\label{tab:BLL}
\end{table}
\renewcommand{\arraystretch}{1.0}

Our $B_{\Lambda\Lambda}$ and $\Delta B_{\Lambda\Lambda}$ are very close to the expected results for which the potential has originally been fitted within the cluster model. The latest data $B_{\Lambda\Lambda}=6.91(0.16)$~MeV and $\Delta B_{\Lambda\Lambda}=0.67(0.17)$~MeV of Ref.~\cite{Ahn:2013} suggest a weaker attractive force between the two hyperons. A refit of the interaction of the form proposed in Eq.~(\ref{eq:V_LL}) would be required. It would be interesting to study more double $\Lambda$~hypernuclei within the AFDMC framework with the $\Lambda N$, $\Lambda NN$ and $\Lambda\Lambda$ interaction proposed. Some experimental data are available in the range $A=7\div13$, but there are uncertainties in the identification of the produced double $\Lambda$~hypernuclei, reflecting in inconsistencies about the sign of the $\Lambda\Lambda$ interaction~\cite{Dover:1991,Yamamoto:1991}. An ab-initio analysis of these systems might put some constraints on the hyperon-hyperon force, which at present is still poorly known, and give information on its density dependence. Also the inclusion of the $\Lambda\Lambda N$ force would be important.

\subsection{Single particle densities and radii}
\label{subsec:dens_ll}

\begin{figure}[p]
	\centering
	\includegraphics[width=\linewidth]{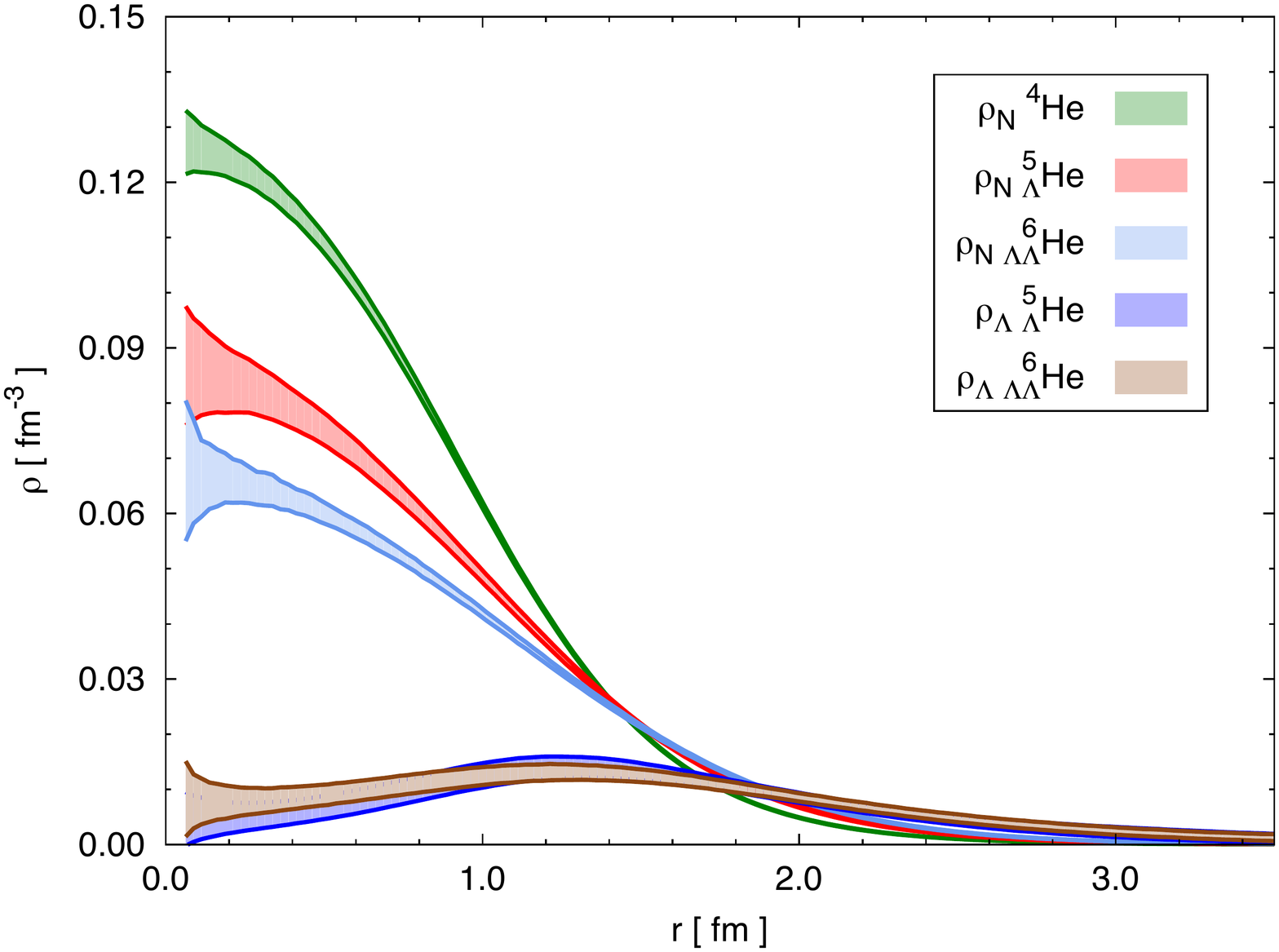}
	\caption[Single particle densities: $N$ and $\Lambda$ in $^4$He, $^5_\Lambda$He and $^{\;\;\,6}_{\Lambda\Lambda}$He]
		{Single particle densities for nucleons in $^4$He [green banded curve], $^5_\Lambda$He [red banded curve] and 
		$^{\;\;\,6}_{\Lambda\Lambda}$He [light blue banded curve], 
		and for the $\Lambda$~particle in $^5_\Lambda$He [blue banded curve] and $^{\;\;\,6}_{\Lambda\Lambda}$He [brown banded curve].
		The results are obtained using the AV4' potential for nucleons and the two- plus three-body hyperon-nucleon force~(\hyperlink{par_II}{II}). 
		In the case of $^{\;\;\,6}_{\Lambda\Lambda}$He, the $\Lambda\Lambda$ interaction of Eq.~(\ref{eq:V_LL}) is also employed.}
	\label{fig:rho_He6LL}
\end{figure}

For the sake of completeness, we also report the results for the single particle densities (Fig.~\ref{fig:rho_He6LL}) and root mean square radii (Tab.~\ref{tab:radii_He6LL}) for the double $\Lambda$~hypernucleus $^{\;\;\,6}_{\Lambda\Lambda}$He. By looking at the densities profiles, when a second hyperon is added to $^5_\Lambda$He, the nucleon density at the center reduces further. The hyperon density, instead, seems to move a bit toward $r=0$ consistently with weak attractive behavior of the employed $\Lambda\Lambda$ interaction. However, the nucleon and hyperon radii are almost the same of $^5_\Lambda$He. These conclusions are thus rather speculative, particularly recalling the discussion on single particle densities of \S~\ref{subsec:dens_l}.

\renewcommand{\arraystretch}{1.4}
\begin{table}[!hb]
	\centering
	\begin{tabular*}{\linewidth}{@{\hspace{5.0em}\extracolsep{\fill}}ccc@{\extracolsep{\fill}\hspace{5.0em}}}
		\toprule
		\toprule
		System                             &  $r_N$  & $r_\Lambda$ \\
		\midrule                             
		   \hspace{0.7em}$^4$He                      & 1.57(9) & ---    \\
		   \hspace{0.6em}$^5_\Lambda$He              & 1.58(7) & 2.2(2) \\
		   $^{\;\;\,6}_{\Lambda\Lambda}$He & 1.7(2)  & 2.3(2) \\
		\bottomrule
		\bottomrule
	\end{tabular*}
	\caption[Nucleon and hyperon radii for $^{\;\;\,6}_{\Lambda\Lambda}$He]
		{Nucleon and hyperon root mean square radii (in fm) for $^4$He and the corresponding single and double $\Lambda$~hypernuclei. 
		The employed interactions are the $NN$ AV4' plus the full two- and three-body hyperon-nucleon force (set~(\hyperlink{par_II}{II})).}
	\label{tab:radii_He6LL}
\end{table}
\renewcommand{\arraystretch}{1.0}

\newpage
\phantom{Empty page}

					% chapter 4: Results: finite systems
	% Chapter 5: Results: infinite systems

\chapter{Results: infinite systems}
\label{chap:results_infinite}

Neutron matter has been deeply investigated in previous works using the Auxiliary Field DMC algorithm. The EoS at zero temperature has been derived in both constrained path~\cite{Sarsa:2003} and fixed phase~\cite{Gandolfi:2009} approximations. In the low density regime, the $^{1\!}S_0$ superfluid energy gap has also been studied~\cite{Gandolfi:2009_gap}. In the high density regime, the connections between three-body forces, nuclear symmetry energy and the neutron star maximum mass are extensively discussed in Refs.~\cite{Gandolfi:2012,Gandolfi:2013}.

In this chapter we will review some details of the AFDMC simulations for pure neutron matter (PNM). They will be useful to extend the calculations for the inclusion of strange degrees of freedom. We will then focus on the hyperon neutron matter (YNM), firstly with the test of the AFDMC algorithm extended to the strange sector in connection with the developed hyperon-nucleon interactions. Starting from the derivation of the threshold density for the appearance of $\Lambda$~hyperons, a first attempt to construct a realistic EoS for YNM will be presented. The corresponding limit for the maximum mass will be finally discussed.

\section{Neutron matter}
\label{sec:nmatt_eos}

As already described in Chapter~\ref{chap:method}, due to the simplification in the potentials for neutron only systems, PNM can investigated by means of AFDMC calculations using the Argonne V8' two-body potential and including three-body forces. The contribution of terms in the Argonne potential beyond spin-orbit are usually very small in nuclei and in low density nuclear and neutron matter. It may become significative only for very large densities~\cite{Gandolfi:2009}. Predicted maximum masses of a NS for the two Argonne potentials are very close and both below $1.8M_\odot$, as a consequence of the softness of the corresponding EoS~\cite{Akmal:1998,Gandolfi:2012}. Being the present observational limit for $M_{\max}$ around $2M_\odot$~\cite{Demorest:2010,Antoniadis:2013}, three-neutron forces must be repulsive at high densities. As reported in Ref.~\cite{Maris:2013}, the Illinois~7 TNI is attractive and produces a too soft EoS. The Urbana~IX interaction instead provides a strong repulsive contribution to the total energy. The inclusion of the UIX force in addition to the two-body AV8' interaction in AFDMC calculations for PNM generates a rather stiff EoS. The predicted maximum mass is around $2.4M_\odot$~\cite{Gandolfi:2012}, in agreement with the result coming from the AV18+UIX calculation of Akmal, Pandharipande and Ravenhall~\cite{Akmal:1998}. It follows that the AFDMC method to solve the AV8'+UIX nuclear Hamiltonian is a valuable tool for the investigation of neutron matter properties and neutron stars observables. This is the starting point for the study of $\Lambda$~neutron matter.

All the AFDMC results for PNM have been obtained using the version \hyperlink{method:Elocal}{\emph{v2}} of the algorithm. Simulations are typically performed at fixed imaginary time step $d\tau=2\cdot10^{-5}~\text{MeV}^{-1}$, that should be small enough to provide a good approximation of the extrapolated result~\cite{Sarsa:2003}. The wave function of Eq.~(\ref{eq:psi_N}) includes a Jastrow correlation function among neutrons and a Slater determinant of plane waves coupled with two-component spinors. For infinite neutron systems, AFDMC calculations do not depend on the Jastrow functions. Moreover by changing the algorithm to version \hyperlink{method:PsiT}{\emph{v1}}, results are less than 1\% different. This is because the employed trial wave function is already a good approximation of the real ground state wave function. Moreover the interaction is simplified with respect to the case of finite nucleon systems due to absence of the $\bm\tau_i\cdot\bm\tau_j$ contributions.

In Chapter~\ref{chap:method} we have seen that finite size effects appear because of the dependence of the Fermi gas kinetic energy to the number of particles. The kinetic energy oscillations of $\mathcal N_F$ free Fermions imply that the energy of $\mathcal N_F=38$ is lower than either $\mathcal N_F=14$ or $\mathcal N_F=66$. This is reflected in the energy of PNM for different number of neutrons with PBC conditions (Eq.~(\ref{eq:PBC})). At each density it follows that $E(38)<E(14)<E(66)$~\cite{Gandolfi:2009}. However, as already discussed in~\S~\ref{subsec:Wave}, the results for 66 neutrons are remarkably close to the extrapolated TABC energy. 66 is thus the typical number of particle employed in AFDMC calculations for PNM.

Finite size effects could appear also from the potential, in particular at high density, depending on the range of the interaction. Monte Carlo calculations are generally performed in a finite periodic box with size $L$ and all inter-particle distances are truncated within the sphere of radius $L/2$. Usually, tail corrections due to this truncation are estimated with an integration of the two-body interaction from $L/2$ up to infinity. However, this is possible only for spin independent terms. As originally reported in Ref.~\cite{Sarsa:2003}, in order to correctly treat all the tail corrections to the potential, it is possible to include the contributions given by neighboring cells to the simulation box. Each two-body contribution to the potential is given~by
\begin{align}
	v_p(r)\equiv v_p(|x,y,z|)\longrightarrow\sum_{i_x,i_y,i_z}v_p\Bigl(\big|(x+i_xL)\hat x+(y+i_yL)\hat y+(z+i_zL)\hat z\big|\Bigr)\;,
\end{align}
where $v_p(r)$ are the potential functions of Eq.~(\ref{eq:v_ij_Op}) and $i_x,i_y,i_z$ are $0,\pm1,\pm2,\ldots$ depending on the number of the boxes considered. The inclusion of the first 26 additional neighbor cells, that corresponds to $i_x,i_y,i_z$ taking the values $-1$, $0$ and $1$, is enough to extend the calculation for inter-particle distances larger than the range of the potential~\cite{Sarsa:2003,Gandolfi:2009}. Finite-size corrections due to three-body forces can be included in the same way as for the nucleon-nucleon interaction, although their contribution is very small compared to the potential energy. Their effect is appreciable only for a small number of particles and at large density, i.e., if the size of the simulation box is small. We will see that these corrections are actually non negligible for the correct computation of energy differences in $\Lambda$~neutron matter. By looking at the results reported in the mentioned references, for PNM we can estimate that the finite-size errors in AFDMC calculations, due to both kinetic and potential energies, do not exceed 2\% of the asymptotic value of the energy calculated by using TABC. 

It was found~\cite{Gandolfi:2009,Gandolfi:2012} that the EoS of PNM can be accurately parametrized using the following polytrope functional form:
\begin{align}
	E(\rho_n)=a\left(\frac{\rho_n}{\rho_0}\right)^\alpha+b\left(\frac{\rho_n}{\rho_0}\right)^\beta\;,\label{eq:poly}
\end{align}
where $E(\rho_n)$ is the energy per neutron as a function of the neutron density $\rho_n$, and the parameters $a$, $\alpha$, $b$, and $\beta$ are obtained by fitting the QMC results. $\rho_0=0.16~\text{fm}^{-3}$ is the nuclear saturation density. AFDMC energies per particle as a function of the neutron density, together with the fitted parameters for both AV8' and the full AV8'+UIX Hamiltonians, are reported in Tab.~\ref{tab:E_nmatt}. The plots of the EoS are shown in the next section, Fig.~\ref{fig:eos_Lfrac}.

\renewcommand{\arraystretch}{1.0}
\begin{table}[!ht]
	\centering
	\begin{tabular*}{\linewidth}{@{\hspace{4.0em}\extracolsep{\fill}}ccc@{\extracolsep{\fill}\hspace{4.0em}}}
		\toprule
		\toprule
		$\rho_n$ & AV8' & AV8'+UIX     \\
		\midrule                             
		0.08  &   9.47(1)  &  10.49(1)  \\
		0.16  &  14.47(2)  &  19.10(2)  \\
		0.24  &  19.98(3)  &  31.85(3)  \\
		0.32  &  26.45(3)  &  49.86(5)  \\
		0.40  &  34.06(5)  &  74.19(5)  \\
		0.48  &  42.99(8)  & 105.9(1)   \\
		0.56  &    ---     & 145.3(1)   \\
		0.60  &  58.24(8)  & 168.1(2)   \\
		0.70  &  73.3(1)   &    ---     \\
		\midrule  
		$\begin{aligned}      
			&\phantom{a=2.04(7)} \\
			&\text{polytrope}    \\ 
			&\text{parameters}   \\ 
			&\phantom{\beta=0.47(1)}
		\end{aligned}$ &  
		$\begin{aligned}      
			a&=2.04(7)      \\
			\alpha&=2.15(2) \\ 
			b&=12.47(47)    \\ 
			\beta&=0.47(1)   
		\end{aligned}$ &    
		$\begin{aligned}
			a&=5.66(3)      \\
			\alpha&=2.44(1) \\
			b&=13.47(3)     \\
			\beta&=0.51(1)    
		\end{aligned}$ \\
		\bottomrule
		\bottomrule
	\end{tabular*}
	\caption[Energy per particle: neutron matter]
		{Energy per particle in neutron matter for selected densities~\cite{Maris:2013,Gandolfi:2013}. 
		$a$, $\alpha$, $b$ and $\beta$ are the fitted polytrope coefficients of Eq.~(\ref{eq:poly}).}
	\label{tab:E_nmatt}
\end{table}
\renewcommand{\arraystretch}{1.0}

\section{$\Lambda$~neutron matter}
\label{sec:Lnmatt_eos_xl}

The study of $\Lambda$~neutron matter follows straightforwardly from PNM calculations with the extension of the wave function (Eq.~(\ref{eq:Psi_T})) and the inclusion of the strange part of the Hamiltonian (Eqs.~(\ref{eq:H_Y}) and (\ref{eq:H_YN})), in analogy with the simulations for finite strange systems. In addition to the Slater determinant of plane waves for neutrons, there is now the determinant for the $\Lambda$~particles. Both sets of plane waves have quantized $\bm k_\epsilon$ vectors given by Eq.~(\ref{eq:k_vec}), and each type of baryon fills its own momentum shell. As discussed in \S~\ref{subsubsubsec:Wave_non_strange}, the requirement of homogeneity and isotropy implies the closure of the momentum shell structure, both for neutrons and hyperons. The consequence is that in AFDMC calculations we are limited in the possible choices for the $\Lambda$~fraction, defined as
\begin{align}
	x_\Lambda=\frac{\rho_\Lambda}{\rho_b}=\frac{\mathcal N_\Lambda}{\mathcal N_n+\mathcal N_\Lambda}\;,
\end{align}
where $\rho_\Lambda$ is the hyperon density and $\rho_b$ the total baryon density of Eq.~(\ref{eq:rho_b}). Employing the TABC (Eq.~(\ref{eq:TABC})) would allow to consider a number of particles corresponding to open shells, providing more freedom in the choice of $x_\Lambda$. However, this approach has not been investigate in this work.

As soon as the hyperons appear in the bulk of neutrons, i.e. above a $\Lambda$~threshold density $\rho_\Lambda^{th}$, the EoS becomes a function of both baryon density and $\Lambda$~fraction, which are connected by the equilibrium condition $\mu_\Lambda=\mu_n$ (see \S~\ref{sec:ns}). The $\Lambda$~threshold density and the function $x_\Lambda(\rho_b)$ are key ingredients to understand the high density properties of hypermatter and thus to predict the maximum mass. We will start the discussion with the test analysis of $\Lambda$~neutron matter at a fixed $\Lambda$~fraction. We will then move to the realistic case of variable $x_\Lambda$.

\subsection{Test: fixed $\Lambda$ fraction}
\label{subsec:x_lambda}

In order to test the feasibility of AFDMC calculations for hypermatter, we considered the limiting case of small $\Lambda$~fraction, in order to look at the hyperon as a small perturbation in the neutron medium. We filled the simulation box with 66 neutrons and just one $\Lambda$~particle, i.e. $x_\Lambda=0.0149$. Although the first momentum shell for the strange baryons is not completely filled (for $\mathcal N_c=1$ the occupation number is 2, spin up and spin down $\Lambda$~particles), the requirement of homogeneity and isotropy is still verified. The first $\bm k_\epsilon$ vector, indeed, is $\frac{2\pi}{L}(0,0,0)$ and thus the corresponding plane wave is just a constant, giving no contribution to the kinetic energy. In order to keep the $\Lambda$~fraction small we are allowed to use one or two hyperons in the box (next close shell is for 14 particles) and, possibly, change the number of neutrons, as we will see. Using just one lambda hyperon there is no need to include the $\Lambda\Lambda$ interaction. The closest hyperon will be in the next neighboring cell at distances larger than the range of the hyperon-hyperon force, at least for non extremely high densities. Therefore, we proceeded with the inclusion of the AV8'+UIX potentials for neutrons, adding the $\Lambda N$+$\Lambda NN$ interactions in both parametrizations~(\hyperlink{par_I}{I})~and~(\hyperlink{par_II}{II}).

\begin{figure}[p]
	\centering
	\includegraphics[width=\linewidth]{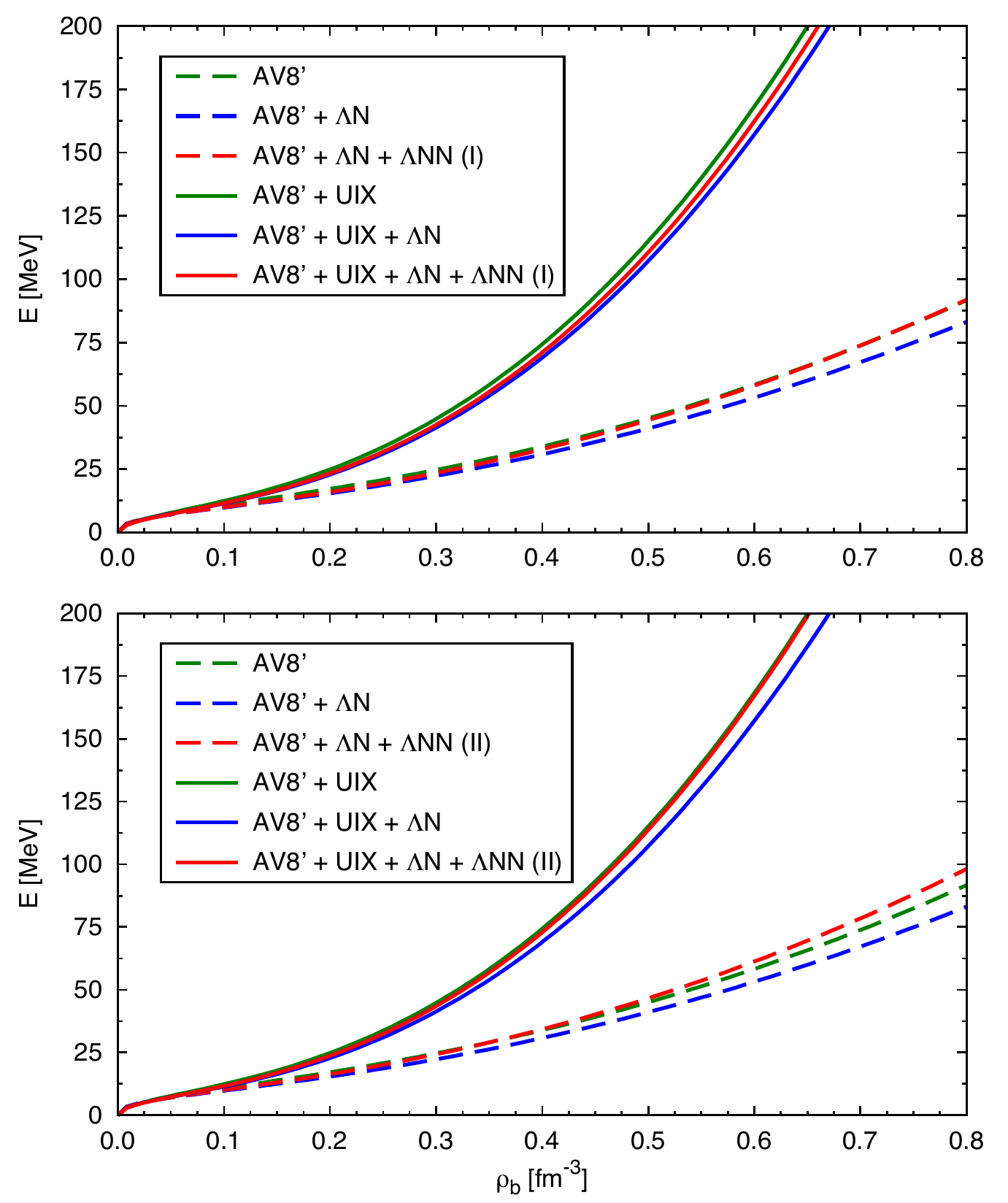}
	\caption[Energy per particle vs. baryon density at fixed $\Lambda$ fraction]
		{Energy per particle as a function of the baryon density for $\Lambda$~neutron matter at fixed $\Lambda$~fraction $x_\Lambda=0.0149$.
		Green curves refer to the PNM EoS, blue and red to the YNM EoS with the inclusion of the two-body and two- plus three-body hyperon nucleon force.
		In the upper panel the results are for the $\Lambda NN$ parametrization~(\hyperlink{par_I}{I}). In the lower panel the set~(\hyperlink{par_II}{II})
		has been used. Dashed lines are obtained using the AV8' nucleon-nucleon potential. Solid lines represent the results with the inclusion of 
		the $NNN$ Urbana~IX potential.}
	\label{fig:eos_Lfrac}
\end{figure}

In Tab.~\ref{tab:E_Lnmatt} we report the energy as a function of the baryon density for different combinations of the employed potentials. The parameters of the polytrope function of Eq.~(\ref{eq:poly}) that fits the AFDMC results are also shown. The plot of the fits, for both PNM and YNM are reported in Fig.~\ref{fig:eos_Lfrac}.

By looking at the dashed lines, corresponding to calculations without the neutron TNI, it is evident the softness of the PNM EoS (green) discussed in the previous section. The addition of the hyperon-nucleon two-body interaction (blue) implies, as expected (see \S~\ref{sec:ns}), a further reduction of the energy per particle, even for the small and constant $\Lambda$~fraction. The inclusion of the three-body $\Lambda NN$ interaction (red), instead, makes the EoS stiffer at high density, even stiffer than the PNM one for the set of parameters~(\hyperlink{par_II}{II}). This result is rather interesting because it means that the hyperon-nucleon force used has a strong repulsive component that is effective also at densities larger than nuclear saturation density, where the interaction was originally fitted on medium-heavy hypernuclei.

When the Urbana~IX TNI is employed (solid lines), the PNM EoS (green) becomes stiff. As in the previous case, the inclusion of the two-body $\Lambda N$ interaction softens the EoS (blue), although the effect is not dramatic for the small $x_\Lambda$ considered. The three-body hyperon-nucleon force gives a repulsive contribution to the total energy (red). The effect is more evident for the parametrization~(\hyperlink{par_II}{II}), for which the PNM and YNM EoS are almost on top of each other. The small constant fraction of hyperons in the neutron medium induces very small modifications in the energy per particle. This is due to the repulsive contribution of the $\Lambda NN$ interaction still active at high densities. 

These results do not describe the realistic EoS for $\Lambda$~neutron matter, because they are computed at a fixed $\Lambda$~fraction for each baryon density. However, the high density part of the curves gives us some indication about the behavior of the hyperon-nucleon interaction in the infinite medium. The fundamental observation is that the $\Lambda NN$ force is repulsive, confirming our expectations. By varying the $\Lambda$~fraction, for example considering two hyperons over 66 neutrons, the qualitative picture drawn in Fig.~\ref{fig:eos_Lfrac} is the same, but a small reasonable increase in the softening of the EoS is found. This is consistent with the theoretical prediction related to the appearance of strange baryons in NS matter and gives us the possibility to quantitatively predict the entity of the softening in a Quantum Monte Carlo framework.

\renewcommand{\arraystretch}{1.0}
\begin{table}[p]
	\centering
	\begin{tabular*}{\linewidth}{@{\hspace{1.5em}\extracolsep{\fill}}cccc@{\extracolsep{\fill}\hspace{1.5em}}}
		\toprule
		\toprule
		\multirow{2}{*}{$\rho_b$} & AV8' & AV8' & AV8' \\
		         & $\Lambda N$ & $\Lambda N$+$\Lambda NN$~(\hyperlink{par_I}{I}) & $\Lambda N$+$\Lambda NN$~(\hyperlink{par_II}{II}) \\
		\midrule 
		0.08  &  8.71(1)   &  8.84(1)   &  8.92(1)   \\   
		0.16  &  13.11(3)  &  13.44(2)  &  13.76(1)  \\   
		0.24  &  17.96(2)  &  18.71(2)  &  19.31(3)  \\   
		0.32  &  23.81(4)  &  25.02(4)  &  26.09(3)  \\   
		0.40  &  30.72(4)  &  32.75(6)  &  34.20(6)  \\   
		0.48  &  38.84(6)  &  42.03(6)  &  43.99(4)  \\   
		0.56  &  48.37(7)  &  52.30(8)  &  55.18(8)  \\   
		0.60  &  53.24(7)  &  57.9(1)   &  61.42(7)  \\   
		0.70  &  67.1(1)   &  74.0(1)   &  78.7(1)   \\       
		0.80  &  83.1(1)   &  91.7(1)   &  98.0(1)   \\
		\midrule  
		$\begin{aligned}      
			&\phantom{a=2.54(13)} \\
			&\text{polytrope}    \\ 
			&\text{parameters}   \\ 
			&\phantom{\beta=0.38(2)}
		\end{aligned}$ &  
		$\begin{aligned}      
			a&=2.54(13)      \\
			\alpha&=2.00(3) \\ 
			b&=10.52(15)    \\ 
			\beta&=0.38(2)   
		\end{aligned}$ &  
		$\begin{aligned}
			a&=2.80(13)      \\
			\alpha&=2.02(3) \\
			b&=10.60(16)     \\
			\beta&=0.38(2)    
		\end{aligned}$ &
		$\begin{aligned}
			a&=2.75(9)      \\
			\alpha&=2.07(2) \\
			b&=10.98(11)     \\
			\beta&=0.41(2)    
		\end{aligned}$ \\  
		\bottomrule
		\bottomrule\\
		\toprule
		\toprule
		\multirow{2}{*}{$\rho_b$} & AV8'+UIX & AV8'+UIX & AV8'+UIX \\
		         & $\Lambda N$ & $\Lambda N$+$\Lambda NN$~(\hyperlink{par_I}{I}) & $\Lambda N$+$\Lambda NN$~(\hyperlink{par_II}{II}) \\
		\midrule 
		0.08  &  9.72(2)   &  9.77(1)    &  9.87(1)   \\   
		0.16  &  17.53(2)  &  17.88(2)   &  18.16(1)  \\   
		0.24  &  29.29(5)  &  29.93(2)   &  30.57(2)  \\   
		0.32  &  46.17(7)  &  47.38(5)   &  48.55(4)  \\   
		0.40  &  68.86(8)  &  71.08(7)   &  72.87(7)  \\   
		0.48  &  98.71(8)  &  101.7(1)   &  104.68(9) \\   
		0.56  &  135.9(1)  &  140.19(9)  &  144.(1)   \\   
		0.60  &  157.0(1)  &  162.3(1)   &  167.0(1)  \\
		\midrule  
		$\begin{aligned}      
			&\phantom{a=5.48(12)} \\
			&\text{polytrope}    \\ 
			&\text{parameters}   \\ 
			&\phantom{\beta=0.47(1)}
		\end{aligned}$ &  
		$\begin{aligned}      
			a&=5.48(12)      \\
			\alpha&=2.42(1) \\ 
			b&=12.06(14)    \\ 
			\beta&=0.47(1)   
		\end{aligned}$ &
		$\begin{aligned}
			a&=5.55(5)      \\
			\alpha&=2.44(1) \\
			b&=12.32(6)     \\
			\beta&=0.49(1)    
		\end{aligned}$ &
		$\begin{aligned}
			a&=5.76(7)      \\
			\alpha&=2.43(1) \\
			b&=12.39(8)     \\
			\beta&=0.49(1)    
		\end{aligned}$ \\ 
		\bottomrule
		\bottomrule
	\end{tabular*}
	\caption[Energy per particle: $\Lambda$~neutron matter]
		{Energy per particle in $\Lambda$~neutron matter as a function of the baryon density. The $\Lambda$ fraction is fixed at $x_\Lambda=0.0149$.
		Different columns correspond to different nucleon-nucleon and hyperon-nucleon potentials.
		$a$, $\alpha$, $b$ and $\beta$ are the fitted polytrope coefficients (Eq.~(\ref{eq:poly})).
		The curves are reported in Fig.~\ref{fig:eos_Lfrac}.}
	\label{tab:E_Lnmatt}
\end{table}
\renewcommand{\arraystretch}{1.0}

\begin{figure}[p]
	\centering
	\includegraphics[width=\linewidth]{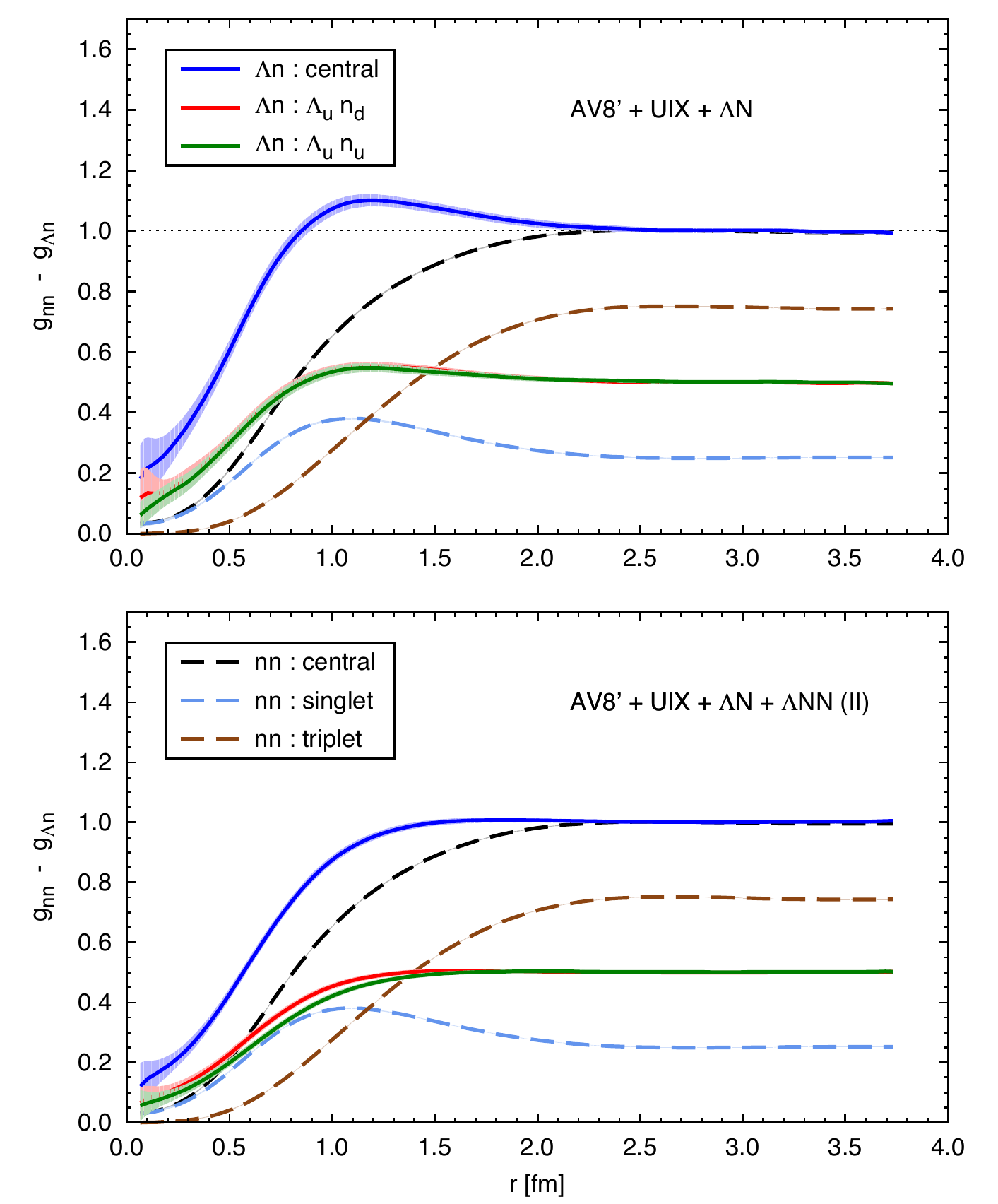}
	\caption[Pair correlation functions at fixed $\Lambda$ fraction: $\rho_b=0.16~\text{fm}^{-3}$]
		{$nn$ (dashed lines) and $\Lambda n$ (solid lines) pair correlation functions in $\Lambda$~neutron matter 
		for $\rho_b=0.16~\text{fm}^{-3}$ and $x_\Lambda=0.0149$.
		The nucelon-nucleon potential is AV8'+UIX. In the upper panel only the two-body hyperon-nucleon potential has been used. 
		In the lower panel also the three body $\Lambda NN$ force in the parametrization~(\hyperlink{par_II}{II}) has been considered.
		The subscript $u$ ($d$) refers to the neutron or lambda spin up (down).}
	\label{fig:gofr_r0.16_V8p+UIX+LN(LNN)}
\end{figure}

\begin{figure}[p]
	\centering
	\includegraphics[width=\linewidth]{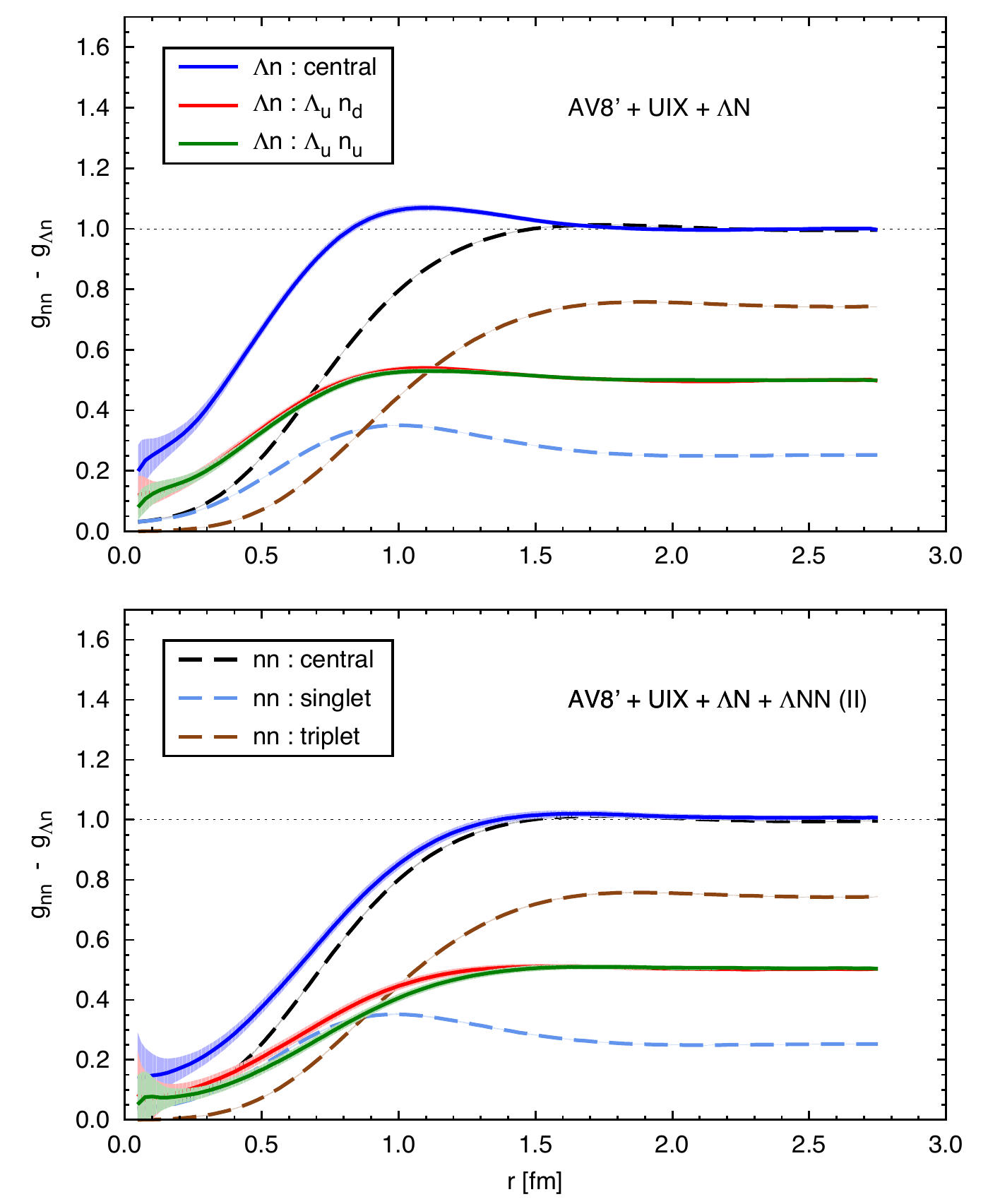}
	\caption[Pair correlation functions at fixed $\Lambda$ fraction: $\rho_b=0.40~\text{fm}^{-3}$]
		{Same of Fig.~\ref{fig:gofr_r0.16_V8p+UIX+LN(LNN)} but for the baryon density $\rho_b=0.40~\text{fm}^{-3}$.}
	\label{fig:gofr_r0.40_V8p+UIX+LN(LNN)}
\end{figure}

Before moving to the derivation of the $\Lambda$~threshold density and the hypermatter EoS, let us analyze the pair correlation functions calculated for $\Lambda$~neutron matter at fixed $\Lambda$~fraction $x_\Lambda=0.0149$. Figs.~\ref{fig:gofr_r0.16_V8p+UIX+LN(LNN)} and \ref{fig:gofr_r0.40_V8p+UIX+LN(LNN)} report the neutron-neutron and lambda-neutron pair correlation functions $g(r)$ for different baryon density, $\rho_b=\rho_0$ and $\rho_b=0.40~\text{fm}^{-3}$. Dashed lines refer to $g_{nn}(r)$ in the central (black), spin singlet (light blue) and spin triplet (brown) channels. Solid lines to $g_{\Lambda n}(r)$ in the central (blue), $\Lambda$~spin~up - $n$~spin~down (red) and $\Lambda$~spin~up - $n$~spin~up (green) channels respectively. In the upper panels results obtained using the two-body $\Lambda N$ interaction only are displayed. In the lower panels the three-body $\Lambda NN$ force in the parametrization~(\hyperlink{par_II}{II}) is also included. 

The main information we can obtain from the plots is the non negligible effect on inter-particle distances due to the inclusion of the three-body $\Lambda NN$ force. Without TNI among hyperons and neutrons, the central $\Lambda n$ correlation function presents a maximum around $1.0\div1.2$~fm, depending on the density. This is a consequence of the attractive $\Lambda N$ force that tends to create a shell of neutrons surrounding the hyperon impurity. The effect is also visible at high density, although reduced. When the $\Lambda NN$ is considered, the shell effect disappears and the $g_{\Lambda n}(r)$ resembles the neutron-neutron one, particularly at high density. The inclusion of the repulsive three-body force avoids the clustering of $\Lambda$~particles in favor of a more homogenous lambda-neutron medium. The use of a $\Lambda n$ central correlation, has the only effect of reducing the value of $g_{\Lambda n}(r)$ in the origin, moving the central functions close to the PNM ones. For the small $\Lambda$~fraction considered here, the neutron-neutron correlation functions are not sensitive to the presence of the hyperon. Indeed, similar results can be obtained for PNM.

It is interesting to observe the projection of the pair correlation functions in the spin channels. For neutrons the Pauli principle tends to suppress the presence of close pairs of particles with parallel spin. For the $\Lambda$-$n$ pair, theoretically there is no Pauli effect because the two particles belong to different isospin spaces. However, the employed hyperon-nucleon interaction involves a $\bm\sigma_\lambda\cdot\bm\sigma_i$ contribution (recall Eqs.~(\ref{eq:V_LN}) and (\ref{eq:V_LNN_D})). This is almost negligible in the case of the $\Lambda N$ potential alone (upper panels of Figs.~\ref{fig:gofr_r0.16_V8p+UIX+LN(LNN)} and \ref{fig:gofr_r0.40_V8p+UIX+LN(LNN)}). It has instead a sizable effect in the dominant three-body force, for which the channel $\Lambda$~spin~up - $n$~spin~down separates from the $\Lambda$~spin~up - $n$~spin~up, revealing a (weak) net repulsion between parallel configurations. Same effect can be found for $\Lambda$ reversed spin.

\subsection{$\Lambda$~threshold density and the equation of state}
\label{subsec:Lnmatt_eos}

In order to address the problem of $\Lambda$~neutron matter, we make use of a formal analogy with the study of two components Fermi gas used in the analysis of asymmetric nuclear matter. When protons are added to the bulk of neutrons, the energy per baryon can be expressed in terms of the isospin asymmetry 
\begin{align}
	\delta_I=\frac{\rho_n-\rho_p}{\rho_n+\rho_p}=1-2x_p\quad\quad x_p=\frac{\rho_p}{\rho_b}\;,
\end{align}
as a sum of even powers of $x_p$
\begin{align}
	E_{pn}(\rho_b,x_p)=E_{pn}(\rho_b,1/2)+S_{pn}^{(2)}(\rho_b)(1-2x_p)^2+S_{pn}^{(4)}(\rho_b)(1-2x_p)^4+\ldots\;,\label{eq:E_pn}
\end{align}
where $x_p$ is the proton fraction and $S_{pn}^{(2i)}(\rho_b)$ with $i=1,2,\ldots$ are the nuclear symmetry energies. Typically, higher order corrections for $i>1$ are ignored. The nuclear symmetry energy $S_{pn}(\rho_b)\equiv S_{pn}^{(2)}(\rho_b)$ is then defined as the difference between the energy per baryon of PNM $E_{\text{PNM}}(\rho_b)=E_{pn}(\rho_b,0)$ and the energy per baryon of symmetric nuclear matter~(SNM) $E_{\text{SNM}}(\rho_b)=E_{pn}(\rho_b,1/2)$.

$E_{pn}(\rho_b,x_p)$ can be rewritten in terms of the PNM energy:
\begin{align}
	E_{pn}(\rho_b,x_p)&=E_{\text{SNM}}(\rho_b)+S_{pn}(\rho_b)\Bigl(1-2x_p\Bigr)^2\;,\nonumber\\[0.2em]
	&=E_{\text{SNM}}(\rho_b)+\Bigl[E_{\text{PNM}}(\rho_b)-E_{\text{SNM}}(\rho_b)\Bigr]\Bigl(1-2x_p\Bigr)^2\;,\nonumber\\[0.2em]
	&=E_{\text{PNM}}(\rho_b)+S_{pn}(\rho_b)\Bigl(-4x_p+4x_p^2\Bigr)\;.\label{eq:E_asym}
\end{align}
In AFDMC calculations the Coulomb interaction is typically neglected. The difference between PNM and asymmetric nuclear matter is thus related to the isospin dependent terms of the nucleon-nucleon interactions. The effect of these components of the potential is parametrized by means of a function of the proton fraction and a function of the baryon density.

We can try to make an analogy between asymmetric nuclear matter and hypermatter, by replacing the protons with the $\Lambda$~particles. In this case the difference with the PNM case is given by the ``strangeness asymmetry''
\begin{align}
	\delta_S=\frac{\rho_n-\rho_\Lambda}{\rho_n+\rho_\Lambda}=1-2x_\Lambda\;,
\end{align}
and the effect on the energy per particle is related to the hyperon-nucleon interactions and the difference in mass between neutron and $\Lambda$. 
In the case of $\Lambda$~neutron matter, the analog of Eq.~(\ref{eq:E_pn}) should contain also odd powers of $\delta_S$. These contributions are negligible for asymmetric nuclear matter due to the smallness of the charge symmetry breaking in $NN$ interaction. Being the $\Lambda$~particles distinguishable from neutrons, there are no theoretical arguments to neglect the linear term in $(1-2x_\Lambda)$. However, we can try to express the energy per particle of $\Lambda$~neutron matter as an expansion over the $\Lambda$~fraction, by introducing an ``hyperon symmetry energy'' $S_{\Lambda n}(\rho_b)$ such that 
\begin{align}
	E_{\Lambda n}(\rho_b,x_\Lambda)=E_{\text{PNM}}(\rho_b)+S_{\Lambda n}(\rho_b)\Bigl(-x_\Lambda+x_\Lambda^2\Bigr)\;.\label{eq:E_YNM}
\end{align}

The expression for the energy difference directly follows from Eq.~(\ref{eq:E_YNM}):
\begin{align}
	\Delta E_{\Lambda n}(\rho_b,x_\Lambda)=E_{\Lambda n}(\rho_b,x_\Lambda)-E_{\text{PNM}}(\rho_b)=S_{\Lambda n}(\rho_b)\Bigl(-x_\Lambda+x_\Lambda^2\Bigr)\;.
	\label{eq:DeltaE}
\end{align}
The idea is then to perform simulations for different $\Lambda$~fraction in order to fit the hyperon symmetry energy $S_{\Lambda n}(\rho_b)$. The main problem in this procedure is the limitation in the values of the hyperon fraction we can consider. In order to keep $x_\Lambda$ small we can use up to 2 lambdas in the first momentum shell and try to vary the number of neutrons from 66 to 14, as reported in Tab.~\ref{tab:L_frac}. In fact, moving to the next $\Lambda$ shell implies a total of 14 strange baryons and a number of neutrons that is computationally demanding. Moreover, we cannot neglect the $\Lambda\Lambda$ interaction for 14 hyperons in a box, even at low density. The inclusion of the hyperon-hyperon force would lead to additional uncertainties in the calculation and it has not been taken into account at this point.

\renewcommand{\arraystretch}{1.4}
\begin{table}[!ht]
	\centering
	\begin{tabular*}{\linewidth}{@{\hspace{4.0em}\extracolsep{\fill}}ccccc@{\extracolsep{\fill}\hspace{4.0em}}}
		\toprule
		\toprule
		$\mathcal N_n$ & $\mathcal N_\Lambda$ & $\mathcal N_b$ & $x_\Lambda$ & $x_\Lambda~\%$ \\
		\midrule                             
		66  &  0  &  66  & 0.0000  & 0.0\% \\
		66  &  1  &  67  & 0.0149  & 1.5\% \\
		54  &  1  &  55  & 0.0182  & 1.8\% \\
		38  &  1  &  39  & 0.0256  & 2.6\% \\
		66  &  2  &  68  & 0.0294  & 2.9\% \\
		54  &  2  &  56  & 0.0357  & 3.6\% \\
		38  &  2  &  40  & 0.0500  & 5.0\% \\
		14  &  1  &  15  & 0.0667  & 6.7\% \\
		\bottomrule
		\bottomrule
	\end{tabular*}
	\caption[Baryon number and $\Lambda$~fraction]
		{Neutron, lambda and total baryon number with the corresponding $\Lambda$~fraction for $\Lambda$~matter calculations.}
	\label{tab:L_frac}
\end{table}
\renewcommand{\arraystretch}{1.0}

Because of finite size effects, we have to be careful in calculating the difference $\Delta E_{\Lambda n}$. Being the $\Lambda$ fraction small, we can suppose that these effects on the total energy are mainly due to neutrons. By taking the difference between YNM and PNM energies for the same number of neutrons, the finite size effects should cancel out. We can see the problem from a different equivalent point of view. The starting point is the energy of PNM obtained with 66 neutrons in the box. If we consider the $\Lambda$~matter described by $66n+1\Lambda$ or $66n+2\Lambda$ there are no problems in evaluating $\Delta E_{\Lambda n}$. When moving to a different $\Lambda$ fraction, the number of neutrons $\mathcal M$ in the strange matter has to be changed. In order to take care of the modified neutron shell, a reasonable approach is to correct the YNM energy by the contribution given by the PNM ``core'' computed with 66 and $\mathcal M$ neutrons:
\begin{align}
	E_{\Lambda n}^{corr}(\rho_b,x_\Lambda)&=E_{\Lambda n}^{\mathcal M}(\rho_b,x_\Lambda)
	+\Bigl[E_{\text{PNM}}^{66}(\rho_b)-E_{\text{PNM}}^{\mathcal M}(\rho_b)\Bigr]\nonumber\\[0.5em]
	&=E_{\text{PNM}}^{66}(\rho_b)+S_{\Lambda n}(\rho_b)\Bigl(-x_\Lambda+x_\Lambda^2\Bigr)\;.
\end{align} 
In this way we obtain
\begin{align}
	\Delta E_{\Lambda n}(\rho_b,x_\Lambda)=S_{\Lambda n}(\rho_b)\Bigl(-x_\Lambda+x_\Lambda^2\Bigr)
	=E_{\Lambda n}^{\mathcal M}(\rho_b,x_\Lambda)-E_{\text{PNM}}^{\mathcal M}(\rho_b)\;,
\end{align}
that exactly corresponds to the result of Eq.~(\ref{eq:DeltaE}).

We verified that energy oscillations for different number of particles keep the same ordering and relative magnitude around the value for 66 neutrons when the density is changed. Actually this is true only when finite size effects due to the truncation of the interaction are also considered. The effect of tail corrections due to the potential is indeed severe, because it depends on both the number of particles and the density, getting worst for few particles and at high densities. In order to control these effects, we performed simulations for PNM and YNM with different number of neutrons including tail corrections for the $NN$ potential and also for the $NNN$, $\Lambda N$ and $\Lambda NN$ forces which are all at the same TPE order and thus have similar interaction range. The result is that, once all the finite size effects are correctly taken into account, the $\Delta E_{\Lambda n}$ values for different densities and number of particles, thus hyperon fraction, can actually be compared. 

\begin{figure}[!hbt]
	\centering
	\includegraphics[width=\linewidth]{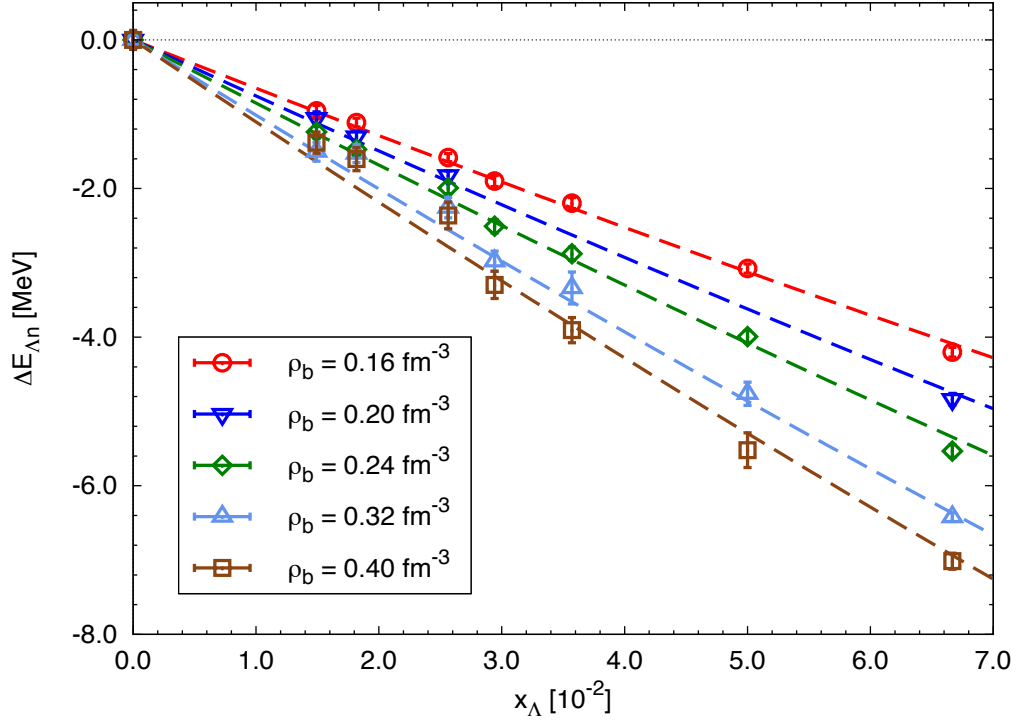}
	\caption[YNM and PNM energy difference vs. $\Lambda$ fraction]
		{YNM and PNM energy difference as a function of the $\Lambda$~fraction for different baryon densities. 
		The employed potential is the full AV8'+UIX+$\Lambda N$+$\Lambda NN$ parametrization~(\hyperlink{par_II}{II}).
		Dashed lines correspond to the quadratic fit $\Delta E_{\Lambda n}(x_\Lambda)=S_{\Lambda n}(-x_\Lambda+x_\Lambda^2)$. 
		In the range of $\Lambda$ fraction shown, $\Delta E_{\Lambda n}$ is essentially given by the linear term in $x_\Lambda$.}
	\label{fig:E_x}
\end{figure}

The result of this analysis is reported in Fig.~\ref{fig:E_x}. The values of the difference $\Delta E_{\Lambda n}$ are shown as a function of the $\Lambda$~fraction for different baryon densities up to $\rho_b=0.40~\text{fm}^{-3}$. As expected, the energy difference is almost linear in $x_\Lambda$, at least for the range of $\Lambda$~fraction that has been possible to investigate. For $x_\Lambda=0.0294,0.0357,0.05$ two hyperons are involved in the calculation. For these cases, we also tried to include the hyperon-hyperon interaction in addition to the AV8'+UIX+$\Lambda N$+$\Lambda NN$ potentials. The $\Lambda\Lambda$ contribution is negligible up to $\rho_b\sim2.5\rho_0$, where some very small effects are found, although compatible with the previous results within the statistical error bars. For densities higher than $\rho_b=0.40~\text{fm}^{-3}$, finite size effects become harder to correct. Although the distribution of energy values generally follows the trend of the lower density data, the approximations used to compute $\Delta E_{\Lambda n}$ might not be accurate enough. A more refined procedure to reduce the dependence on shell closure, for example involving the twist-averaged boundary conditions, it is possibly needed.

\begin{figure}[!ht]
	\centering
	\includegraphics[width=\linewidth]{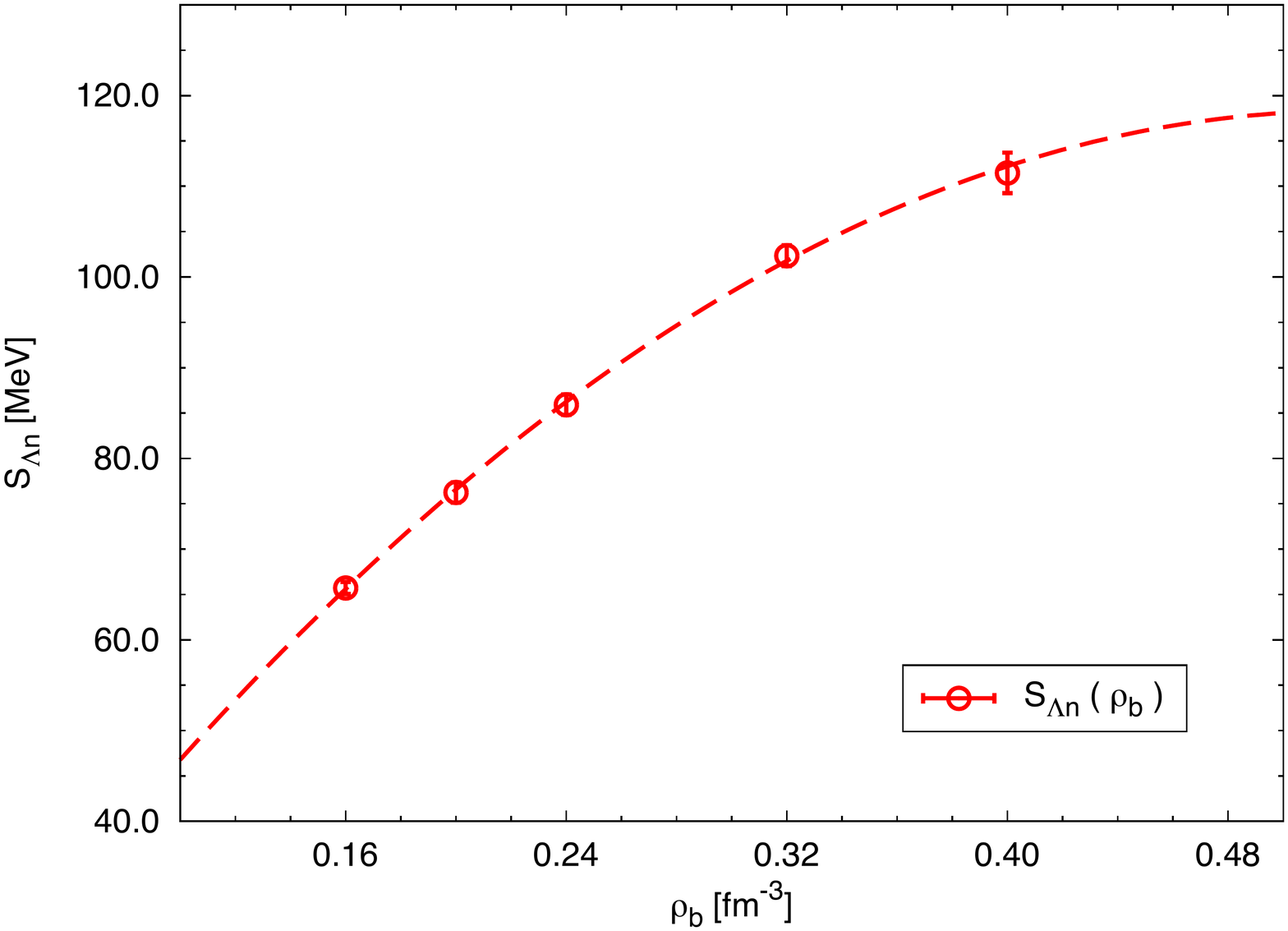}
	\caption[Hyperon symmetry energy vs. baryon density]
		{Hyperon symmetry energy as a function of the baryon density. Red dots are the points obtained by the the quadratic fit 
		$\Delta E_{\Lambda n}(x_\Lambda)=S_{\Lambda n}(-x_\Lambda+x_\Lambda^2)$ on the data of Fig.~\ref{fig:E_x}. 
		The dashed line is the $S_{\Lambda n}(\rho_b)$ fitted curve of Eq.~(\ref{eq:S_rho}).}
	\label{fig:S_rho}
\end{figure}

We used the quadratic function $\Delta E_{\Lambda n}(x_\Lambda)=S_{\Lambda n}(-x_\Lambda+x_\Lambda^2)$ to fit the $\Delta E_{\Lambda n}$ values of Fig.~\ref{fig:E_x}. For each density the coefficient $S_{\Lambda n}$ has been plotted as a function of the baryon density, as shown in Fig.~\ref{fig:S_rho}.
In the case of asymmetric nuclear matter, close to the saturation density the nuclear symmetry energy is parametrized with a linear function of the density~\cite{Gandolfi:2012}. The data in Fig.~\ref{fig:S_rho} actually manifest a linear behavior for $\rho_b\sim\rho_0$ but the trend deviates for large density. We can try to fit the $S_{\Lambda n}$ points including the second order term in the expansion over $\rho_b-\rho_0$:
\begin{align}
	S_{\Lambda n}(\rho_b)=S_{\Lambda n}^{(0)}+S_{\Lambda n}^{(1)}\left(\frac{\rho_b-\rho_0}{\rho_0}\right)
	+S_{\Lambda n}^{(2)}\left(\frac{\rho_b-\rho_0}{\rho_0}\right)^2\;. \label{eq:S_rho}
\end{align}
The results are shown in Fig.~\ref{fig:S_rho} with the dashed line. The three parameters of the $S_{\Lambda n}(\rho_b)$ function are reported in Tab.~\ref{tab:S_rho}.

\renewcommand{\arraystretch}{1.4}
\begin{table}[!hb]
	\centering
	\begin{tabular*}{\linewidth}{@{\hspace{5.0em}\extracolsep{\fill}}ccc@{\extracolsep{\fill}\hspace{5.0em}}}
		\toprule
		\toprule
		$S_{\Lambda n}^{(0)}$ & $S_{\Lambda n}^{(1)}$ & $S_{\Lambda n}^{(2)}$ \\
		\midrule              
		65.6(3) & 46.4(1.6) & -10.2(1.3) \\
		\bottomrule
		\bottomrule
	\end{tabular*}
	\caption[Coefficients of the hyperon symmetry energy fit]
		{Coefficients (in MeV) of the hyperon symmetry energy function of Eq.~(\ref{eq:S_rho}). 
		The parameters are obtained from the quadratic fit on the $\Delta E_{\Lambda n}$ results reported in Fig.~\ref{fig:E_x}.}
	\label{tab:S_rho}
\end{table}
\renewcommand{\arraystretch}{1.0}

After fitting the hyperon symmetry energy we have a complete parametrization for the EoS of $\Lambda$~neutron matter depending on both baryon density and $\Lambda$~fraction (Eq.~(\ref{eq:E_YNM})). For $x_\Lambda=0$ the relation reduces to the EoS of PNM parametrized by the polytrope of Eq.~(\ref{eq:poly}) whose coefficients are reported in Tab.~\ref{tab:E_nmatt}. For $x_\Lambda>0$ the presence of hyperons modifies the PNM EoS through the hyperon symmetry energy and the quadratic term in $x_\Lambda$. The derivation of $S_{\Lambda n}$ has been performed for small $x_\Lambda$ ($\sim10\%$), corresponding to a baryon density up to $\sim3\rho_0$. However, this should be enough to derive at least the $\Lambda$~threshold density by imposing the chemical potentials equilibrium condition $\mu_\Lambda=\mu_n$.

Let us start defining the energy density $\mathcal E$ for the $\Lambda$~neutron matter as
\begin{align}
	\mathcal E_{\Lambda n}(\rho_b,x_\Lambda)&=\rho_b E_{\Lambda n}(\rho_b,x_\Lambda)+\rho_n m_n+\rho_\Lambda m_\Lambda\;,\nonumber\\[0.2em]
	&=\rho_b\Bigl[E_{\Lambda n}(\rho_b,x_\Lambda)+m_n+x_\Lambda\Delta m \Bigr]\;,
\end{align}
where 
\begin{align}
	\rho_n=(1-x_\Lambda)\rho_b\quad\quad\quad\rho_\Lambda=x_\Lambda \rho_b\;,\label{eq:rho_nl}
\end{align}
and $\Delta m=m_\Lambda-m_n$. For $x_\Lambda=0$ the relation corresponds to the PNM case. The chemical potential is generally defined as the derivative of the energy density with respect to the number density, evaluated at fixed volume:
\begin{align}
	\mu=\frac{\partial\mathcal E}{\partial\rho}\Bigg|_V\;.\label{eq:mu}
\end{align}
In AFDMC calculations, because of the requirement of the momentum shell closure, the number of particles has to be fixed. The density is increased by changing the volume, i.e. reducing the size of the simulation box. Therefore, Eq.~(\ref{eq:mu}) must include a volume correction of the form
\begin{align}
	\mu=\frac{\partial\mathcal E}{\partial\rho}+\rho\frac{\partial E}{\partial\rho}\;.
\end{align}
Our chemical potentials are thus given by
\begin{align}
	\mu_\kappa(\rho_b,x_\Lambda)=\frac{\partial\mathcal E_{\Lambda n}(\rho_b,x_\Lambda)}{\partial\rho_\kappa}
	+\rho_\kappa\frac{\partial E_{\Lambda n}(\rho_b,x_\Lambda)}{\partial\rho_\kappa}\;,
\end{align}
where $\kappa=n,\Lambda$ and the derivatives of the energy per particle and energy density must be calculated with respect to $\rho_b$ and $x_\Lambda$:
\begin{align}
	\frac{\partial\mathcal F_{\Lambda n}(\rho_b,x_\Lambda)}{\partial\rho_\kappa}
	&=\frac{\partial\mathcal F_{\Lambda n}(\rho_b,x_\Lambda)}{\partial\rho_b}\frac{\partial\rho_b}{\partial\rho_\kappa}
	+\frac{\partial\mathcal F_{\Lambda n}(\rho_b,x_\Lambda)}{\partial x_\Lambda}\frac{\partial x_\Lambda}{\partial\rho_\kappa}\;.
\end{align}
Recalling Eq.~(\ref{eq:rho_nl}) we have
\begin{align}
	\frac{\partial\rho_b}{\partial\rho_n}=1 \quad\quad \frac{\partial\rho_b}{\partial\rho_\Lambda}=1 \quad\quad
	\frac{\partial x_\Lambda}{\partial\rho_n}=-\frac{x_\Lambda}{\rho_b} \quad\quad \frac{\partial x_\Lambda}{\partial\rho_\Lambda}=\frac{1-x_\Lambda}{\rho_b}\;,
\end{align}
and thus the neutron and lambda chemical potentials take the form:
\begin{align}
	\mu_n(\rho_b,x_\Lambda)&=\frac{\partial\mathcal E_{\Lambda n}}{\partial\rho_b}
	-\frac{x_\Lambda}{\rho_b}\frac{\partial\mathcal E_{\Lambda n}}{\partial x_\Lambda}
	+(1-x_\Lambda)\rho_b\frac{\partial E_{\Lambda n}}{\partial\rho_b}
	-x_\Lambda(1-x_\Lambda)\frac{\partial E_{\Lambda n}}{\partial x_\Lambda}\;,\\[0.5em]
	\mu_\Lambda(\rho_b,x_\Lambda)&=\frac{\partial\mathcal E_{\Lambda n}}{\partial\rho_b}
	+\frac{1-x_\Lambda}{\rho_b}\frac{\partial\mathcal E_{\Lambda n}}{\partial x_\Lambda}
	+x_\Lambda\rho_b\frac{\partial E_{\Lambda n}}{\partial\rho_b}
	+x_\Lambda(1-x_\Lambda)\frac{\partial E_{\Lambda n}}{\partial x_\Lambda}\;.
\end{align}

\begin{figure}[!ht]
	\centering
	\includegraphics[width=\linewidth]{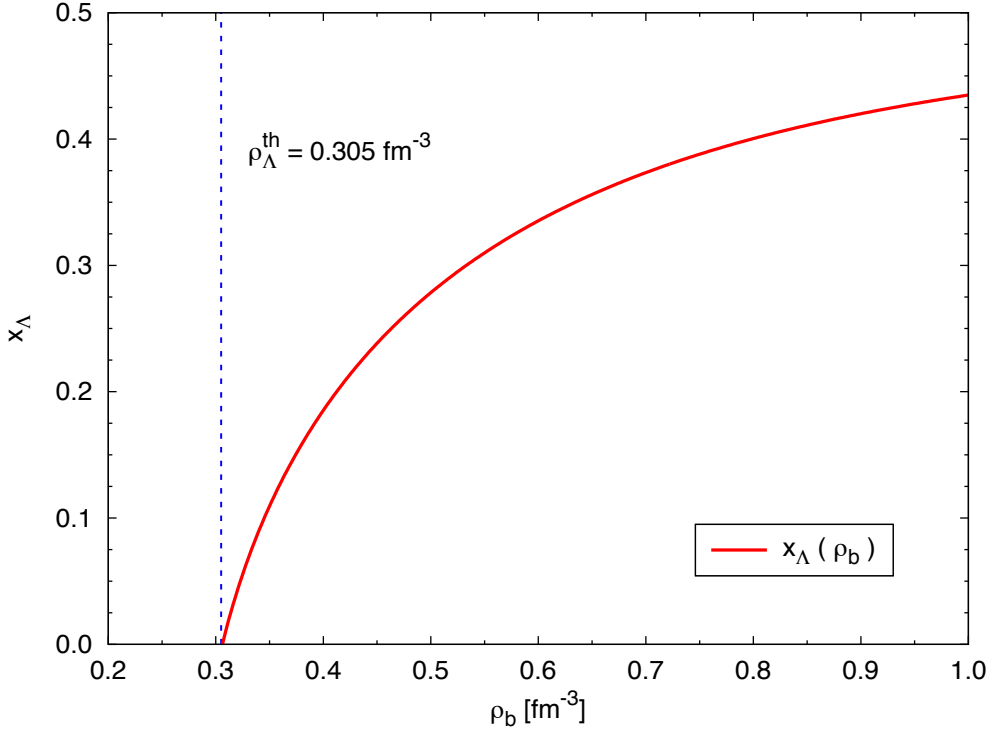}
	\caption[$x_\Lambda(\rho_b)$ function and $\Lambda$~threshold density]
		{$\Lambda$~fraction as a function of the baryon density. The curve describes the equilibrium condition $\mu_\Lambda=\mu_n$.
		The red line is the result for the quadratic fit on the $\Delta E_{\Lambda n}$ data of Fig.~\ref{fig:E_x}.
		The blue dotted vertical line indicates the $\Lambda$~threshold densities $\rho_\Lambda^{th}$ such that $x_\Lambda(\rho_\Lambda^{th})=0$.}
	\label{fig:x_rho}
\end{figure}

The two $\mu_n$ and $\mu_\Lambda$ surfaces in the $\rho_b-x_\Lambda$ space cross each other defining the curve $x_\Lambda(\rho_b)$ reported in Fig.~\ref{fig:x_rho}. This curve describes the equilibrium condition $\mu_\Lambda=\mu_n$. It thus defines the $\Lambda$~threshold density $x_\Lambda(\rho_\Lambda^{th})=0$ and provides the equilibrium $\Lambda$~fraction for each density. For the given parametrization of the hyperon symmetry energy, the threshold density is placed around $1.9\rho_0$, which is consistent with the theoretical indication about the onset of strange baryons in the core of a NS. Once the $\Lambda$~particles appear, the hyperon fraction rapidly increases due to the decrease of the energy and pressure that favors the $n\rightarrow\Lambda$ transition (see \S~\ref{sec:ns}). However, there is a saturation effect induced by the repulsive nature of the hyperon-nucleon interaction that slows down the production of $\Lambda$~particle at higher density.

\begin{figure}[!hb]
	\centering
	\includegraphics[width=\linewidth]{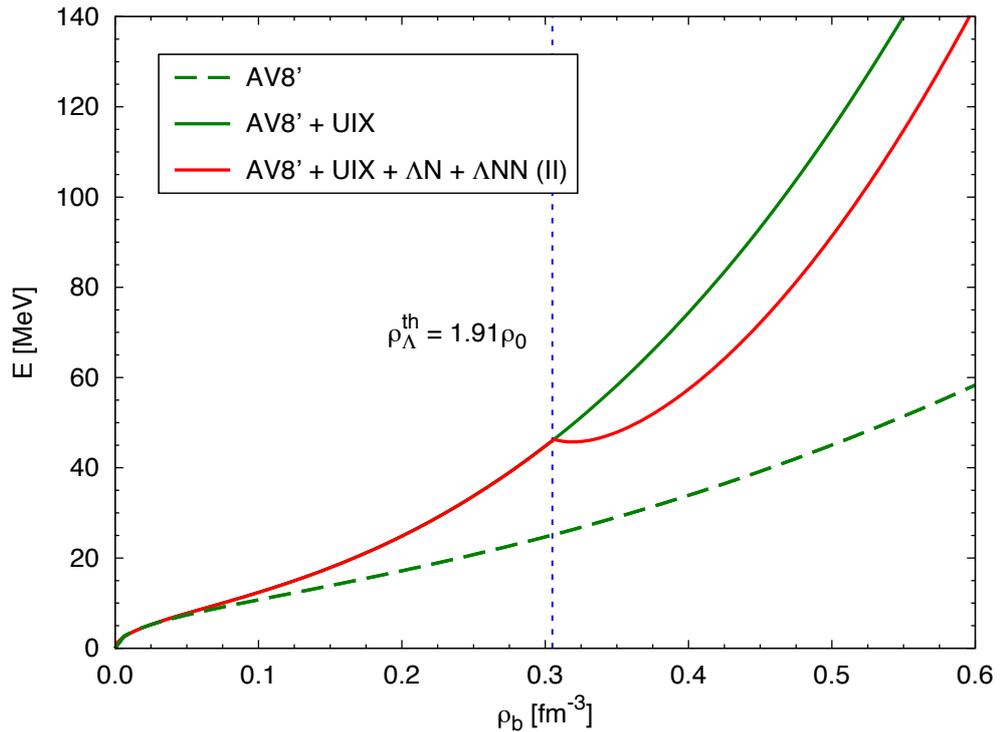}
	\caption[YNM equation of state]
		{Equation of state for $\Lambda$~neutron matter. Green solid (dashed) curves refer to the PNM EoS calculated with the AV8'+UIX (AV8') potential.
		Red line is the EoS for YNM corresponding to the quadratic fit on the $\Delta E$ data of Fig.~\ref{fig:E_x}.
		The employed hyperon-nucleon potential is the full two- plus three-body in the parametrization~(\hyperlink{par_II}{II}).
		The $\Lambda$~threshold density is displayed with the blue dotted vertical line.}
	\label{fig:eos_real}
\end{figure}

By using the $\Lambda$~threshold density $\rho_\Lambda^{th}$ and the equilibrium $\Lambda$~fraction values $x_\Lambda(\rho_b)$ in Eq.~(\ref{eq:E_YNM}), we can finally address the $\Lambda$~neutron matter EoS. The result is reported in Fig.~\ref{fig:eos_real}. The green dashed line is the PNM EoS for AV8', the green solid line the one for AV8'+UIX. Red curve is instead the YNM EoS coming from the AV8'+UIX+$\Lambda N$+$\Lambda NN$~(\hyperlink{par_II}{II}) potentials. At the threshold density there is a strong softening of the EoS induced by the rapid production of hyperons. However the EoS becomes soon almost as stiff as the PNM EoS due to hyperon saturation and the effect of the repulsion among hyperons and neutrons. In $\rho_b=\rho_\Lambda^{th}$ there is a phase transition between PNM and YNM. For densities close to the threshold density, the pressure becomes negative. This is a non physical finite size effect due to the small number of particles considered in the simulations, not large enough for the correct description of a phase transition. However, in the thermodynamical limit the effect should disappear. We could mitigate this effect by using a Maxwell construction between the PNM and the YNM EoS. The details of the density dependence of the energy per baryon at the hyperon threshold are however not relevant for the derivation of the maximum mass.

The derived model for the EoS of $\Lambda$~neutron matter should be a good approximation up to $\rho_b\sim3\rho_0$. The behavior of the energy per baryon after this limit depends on density and $\Lambda$~fraction to which we do not have controlled access with the present AFDMC calculations. Moreover, starting from $\rho_b>0.6~\text{fm}^{-3}$, $\Sigma^0$ hyperons could be formed, as shown in Fig.~\ref{fig:chemicalpot}. The behavior of the energy curve should thus be different. However, there are already strong indications for a weak softening of the EoS induced by the presence of hyperons in the neutron bulk when the hyperon-nucleon potentials employed for hypernuclei are used.

\subsection{Mass-radius relation and the maximum mass}
\label{subsec:Lnmatt_Mmax}

In Chapter~\ref{chap:strangeness} we have seen that, given the EoS, the mass-radius relation and the predicted maximum mass are univocally determined. The $M(R)$ curves are the solutions of the TOV equations~(\ref{eq:TOV}), which involve the energy density $\mathcal E$ and the pressure $P$. For YNM the energy density is given by Eq.~(\ref{eq:E_YNM}) supplemented by the hyperon threshold density and the $x_\Lambda(\rho_b)$ curve. For the pressure we can simply use the relation
\begin{align}
	P_{\Lambda n}(\rho_b,x_\Lambda)=\rho_b^2\frac{\partial E_{\Lambda n}(\rho_b,x_\Lambda)}{\partial\rho_b}\;,
\end{align}
where the additional term due to density dependence of the $\Lambda$ fraction vanishes once the equilibrium condition $\mu_\Lambda=\mu_n$ is given. 

\begin{figure}[!htb]
	\centering
	\includegraphics[width=\linewidth]{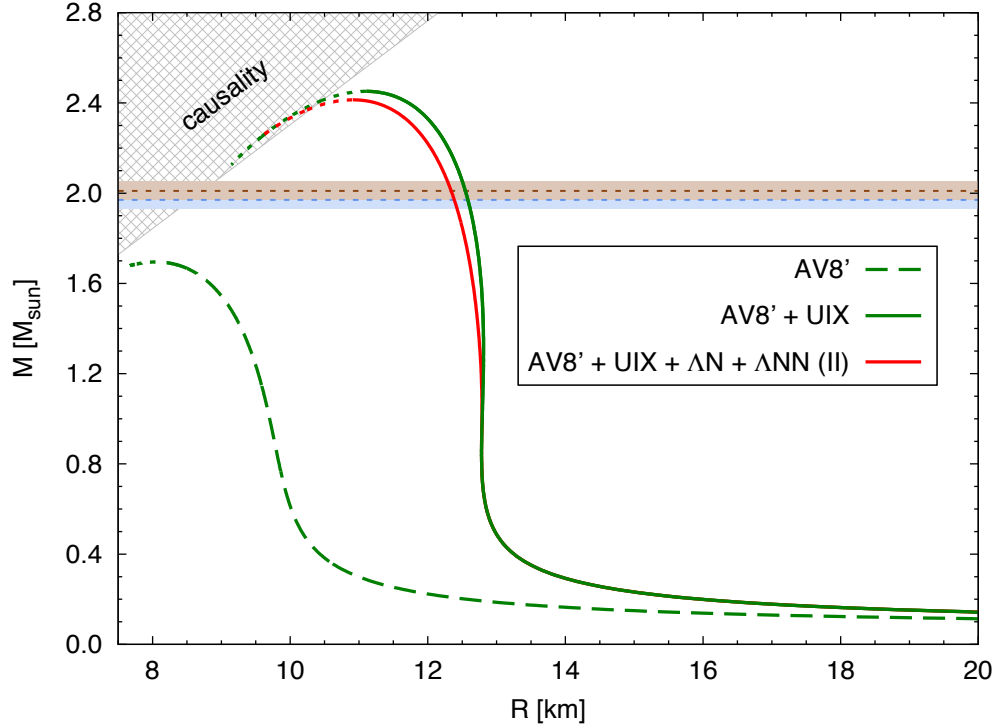}
	\caption[YNM mass-radius relation]
		{Mass-radius relation for $\Lambda$~neutron matter. Green solid (dashed) curves refer to the PNM calculation with the AV8'+UIX (AV8') potential.
		Red line is the result for the YNM corresponding to the quadratic fit on the $\Delta E_{\Lambda n}$ data of Fig.~\ref{fig:E_x}.
		The light blue and brown bands correspond to the masses of the millisecond pulsars PSR J1614-2230 ($1.97(4)M_\odot$)~\cite{Demorest:2010} and
		PSR J1903+0327 ($2.01(4)M_\odot$)~\cite{Antoniadis:2013}. The gray shaded region is the excluded part of the plot according to causality.}
	\label{fig:mofr}
\end{figure}

Fig.~\ref{fig:mofr} reports the $M(R)$ curves solution of the TOV equations for the EoS reported in Fig.~\ref{fig:eos_real}. Green curves are the PNM relations for AV8' (dashed) and AV8'+UIX (solid). Red one is the result for the $\Lambda$~neutron matter described by the full nucleon-nucleon and hyperon-nucleon interaction in the parametrization~(\hyperlink{par_II}{II}). The shaded region corresponds to the excluded region by the causality condition~\cite{Steiner:2010}
\begin{align}
	M\lesssim \beta\frac{c^2}{G} R \quad\quad\quad \beta=\frac{1}{2.94}\;,
\end{align}
where $G$ is the gravitational constant and $c$ the speed of light. The curves with the inclusion of the TNI partially enter the forbidden region. This is due to the behavior of our EoS that evaluated for very high densities becomes superluminal. A connection to the maximally stiff EoS given by the condition $P<1/3\,\mathcal E$ should be needed. However, we can estimate the effect on the maximum mass to be rather small, not changing the general picture.

The maximum mass for PNM obtained using the Argonne~V8' and Urbana~IX potentials is reduced from $\sim2.45M_\odot$ to $\sim2.40M_\odot$ by the inclusion of $\Lambda$~hyperons. This small reduction follows by the stiffness of the YNM EoS for densities larger than $\rho_b\sim3\rho_0$, up to which our model gives a good description of the strange system. However, by limiting the construction of the $M(R)$ relation in the range of validity of the employed YNM model, the mass of the star is already at $\sim1.81M_\odot$ around $R=12.5$~km, and at $\sim1.98M_\odot$ if we extend the range up to $\rho_b=0.55~\text{fm}^{-3}$. These values are larger than the predicted maximum mass for hypermatter in all (B)HF calculations (see \S~\ref{sec:ns}). 

Regardless of the details of the real behavior of the EoS for $\rho_b>3\rho_0$, we can speculate that a maximum mass of $2M_\odot$ can be supported by the $\Lambda$~neutron matter described by means of the realistic AV8'+UIX potentials plus the here developed two- and three-body hyperon-nucleon interactions. The key ingredient of the picture is the inclusion of the repulsive $\Lambda NN$ force that has been proven to give a fundamental contribution in the realistic description of $\Lambda$~hypernuclei. Although very preliminary, our first AFDMC calculations for hypermatter suggest that a $2M_\odot$ neutron star including hyperons can actually exist.

\begin{figure}[p]
	\centering
	\includegraphics[width=\linewidth]{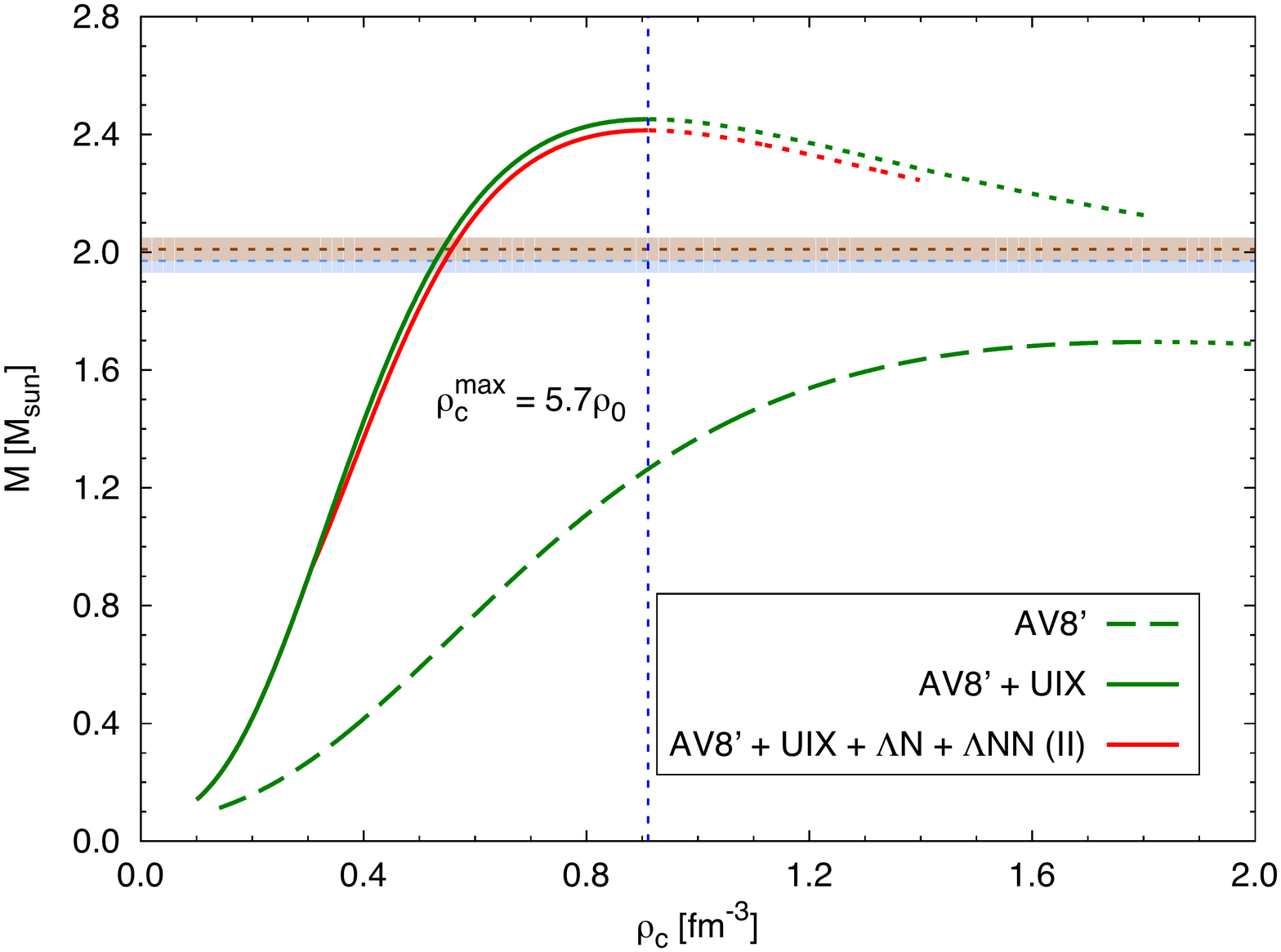}
	\caption[YNM mass-central density relation]
		{Stellar mass versus central density for $\Lambda$~neutron matter. The key is the same of Fig.~\ref{fig:mofr}. 
		The vertical blue dotted line represents the maximum central density for the stability of the star when TNI forces are considered.}
	\label{fig:mofrho}
\end{figure}

The solution of the TOV equations provides additional information on the central density $\rho_c$ of the star. The behavior of the star mass as a function of the central density determines the stability condition of the NS trough the relation $dM(\rho_c)/d\rho_c>0$. For non rotating neutron stars, configurations that violate this condition are unstable and will collapse into black holes~\cite{Haensel:2006}. As can be seen from Fig.~\ref{fig:mofrho} where the mass-central density relation is reported, the maximum mass also determines the maximum central density for stable NSs. Within our model, $\rho_c^{\max}$ is around $5.7\rho_0$ for both PNM and YNM when the three-nucleon force is considered in the calculation. Given the fact the inter-particle distance scale as $\rho_c^{-1/3}$, we can estimate that for the given $\rho_c^{\max}$ baryons are not extremely packed. The baryon-baryon distances are of the order of few fermi, comparable to the range of the hard core of the nucleon-nucleon and hyperon-nucleon interactions considered. Therefore, in this framework there is no evidence for the appearance of exotic phases like quark matter. Our YNM EoS is stiff enough to realistically describe the infinite medium supporting a $2M_\odot$ NS without requiring other additional degrees of freedom for the inner core.

\newpage
\phantom{Empty page}

					% chapter 5: Results: infinite systems
	% Chapter 6: Conclusions

\chapter{Conclusions}
\label{chap:conclusion}

\fancyhead[LO]{\emph{Conclusions}}					% redefinition of left and right header fields for this chapter

In this work the recent developments in Quantum Monte Carlo calculations for nuclear systems including strange degrees of freedom have been reported. The Auxiliary Field Diffusion Monte Carlo algorithm has been extended to the strange sector by the inclusion of the lightest among the hyperons, the $\Lambda$~particle. This gave us the chance to perform detailed calculations for $\Lambda$~hypernuclei, providing a microscopic framework for the study of the hyperon-nucleon interaction in connection with the available experimental information. The extension of the method for strange neutron matter, put the basis for the first Diffusion Monte Carlo analysis of the hypernuclear medium, with the derivation of neutron star observables of great astrophysical interest.

The main outcome of the study of $\Lambda$~hypernuclei, is that, within the employed phenomenological model for hyperon-nucleon forces, the inclusion of a three-body $\Lambda NN$ interaction is fundamental to reproduce the ground state physics of medium-heavy hypernuclei, in particular the observed saturation property of the hyperon binding energy. By accurately refitting the three-body hyperon-nucleon interaction, we obtain a substantial agreement with the experimental separation energies, that are strongly overestimated by the use of a bare $\Lambda N$ interaction. The result is of particular interest because with the employed algorithm, heavy hypernuclei up to 91 particles have been investigated within the same theoretical framework, providing a realistic description able to reproduce the extrapolation of the hyperon binding energy in the infinite medium. By employing an effective hyperon-hyperon interaction, first steps in the study of $S=-2$ $\Lambda$~hypernuclei have also been taken. The interest in these systems is motivated by the controversial results coming from both theoretical and experimental studies.

Preliminary AFDMC results on hypermatter indicate that the hyperon-nucleon interaction fitted on finite strange nuclei leads to a stiff equation of state for the strange infinite medium. Within our model, $\Lambda$~particles start to appear in the neutron bulk around twice the saturation density, consistently with different theoretical previsions. However, the predicted softening of the equation of state seems not to be dramatic, due to the strongly repulsive nature of the employed three-body hyperon-nucleon force. This fact helps to understand how the necessary appearance of hyperons at some value of the nucleon density in the inner core of a neutron star might eventually be compatible with the observed neutron star masses of order $2M_\odot$. 

Both works on hypernuclei and hypermatter represent the first Diffusion Monte Carlo study of finite and infinite strange nuclear systems, and thus are subject to further improvements. The algorithm for (hyper)nuclei should be refined in order to become more independent from the starting trial wave function that should include also correlations other than the pure central. Together with the accurate treatment of the tensor (and spin-orbit) potential term and, possibly, with the inclusion of the density dependent nucleon-nucleon interaction developed in the framework of correlated basis function~\cite{Lovato:2011}, the algorithm might become a powerful tool for the precise investigation not only of energy differences but also of other structural ground state properties such as density and radii. From the methodological point of view, the algorithm for infinite strange systems could benefit from the inclusion of twist-averaged boundary conditions, that would allow for a more refined study of the equation of state of the hypernuclear medium and thus the derivation of the maximum mass.

It would be interesting to perform benchmark calculations with the employed hyperon-nucleon force by means of few-body methods. This would reduce the uncertainties on the fitted interaction, providing more insight on the structure of the phenomenological potential for light hypernuclei. On the other hand, by projecting the three-body interaction on the triplet and singlet isospin channels, it would be possible to fit the experimental data for large hypernuclei in order to better capture the features of the interaction that are relevant for the neutron star physics without significantly change the compatibility of the results with the lighter strange nuclei. This could definitely determine a stiff equation of state for the hyperon neutron matter supporting a $2M_\odot$ star.

In the same contest, the study of asymmetric nuclear matter with the inclusion of hyperon degrees of freedom is very welcome. At present this project has not started yet and so the goal is far to be achieved. However this is one of the more promising direction in order to describe the properties of stellar matter at high densities by means of accurate microscopic calculations with realistic interactions.

The very recent indication of a bound $\Lambda nn$ three-body system~\cite{Rappold:2013_PRC(R)}, might motivate the AFDMC investigation of hyper neutron drops. Weakly bound systems are typically not easily accessible by means of standard AFDMC method for finite systems. The study of neutron systems confined by an external potential with the inclusion of one or more hyperons, could give fundamental information about the hyperon-neutron and hyperon-hyperon interaction in connection with the experimental evidence of light neutron rich hypernuclei, such as $^6_\Lambda$H~\cite{Agnello:2012_H6L}, or the theoretical speculation of exotic neutron systems, as the bound $\Lambda\Lambda nn$ system.

					% chapter 6: Conclusions
	% Appendix

\renewcommand{\arraystretch}{1.2}
\titleformat{\chapter}[display]{\huge\bfseries}{\huge Appendix \huge\thechapter}{0.1em}{\titlerule\vspace{1em}}[\vspace{0.1em}]	% redefinition of appendices title
												
\fancyhead[LO]{\emph{\nouppercase{\rightmark}}}	% back to main page style preferences for left and right header fields
											
\appendix

\chapter{AFDMC wave functions}
\label{app:Wave}

\section{Derivatives of the wave function: CM corrections}
\label{app:CM}

As seen in \S~\ref{subsec:Wave}, for finite systems the single particle orbitals must be referred to the CM of the system: $\bm r_p\rightarrow\bm r_p-\bm r_{CM}$.
Each derivative with respect to nucleon or hyperon coordinates has thus to be calculated including CM corrections. Let Call $\bm\rho_i$ the relative coordinates and $\bm r_i$ the absolute ones for nucleons, and $\bm\rho_\Lambda$, $\bm r_\lambda$ the analogues for the hyperons. Then 
\begin{align}
	\bm\rho_i=\bm r_i-\bm\rho_{CM} \qquad \bm\rho_\lambda=\bm r_\lambda-\bm\rho_{CM} \;,
\end{align}
with
\begin{align}
	\bm\rho_{CM}=\frac{1}{M}\left(m_N\sum_k \bm r_k+m_\Lambda\sum_\nu\bm r_\nu\right) \qquad M=\mathcal N_N\,m_N+\mathcal N_\Lambda\,m_Nm_\Lambda\;.
\end{align}
In order to simplify the notation, in the next we will use $r_p$ instead of $\bm r_p$. The equations for the first derivatives will be valid for the Cartesian component of the position vectors. In the relations for the second derivatives implicit sums over Cartesian components will be involved.

Consider a function of the relative nucleon and hyperon coordinates:
\begin{align}
	f(\rho_N,\rho_\Lambda)\equiv f(\rho_1,\ldots,\rho_{\mathcal N_N},\rho_1,\ldots,\rho_{\mathcal N_\Lambda})\;,
\end{align}
In order to calculate the derivatives of $f(\rho_N,\rho_\Lambda)$ with respect to $r_p$, we need to change variable. Recalling that now all the coordinates (nucleons and hyperons) are connected together via the CM, we have
\begin{align}
	\frac{\partial}{\partial r_i}f(\rho_N,\rho_\Lambda)
	&=\sum_j\frac{\partial\rho_j}{\partial r_i}\frac{\partial}{\partial\rho_j}f(\rho_N,\rho_\Lambda)
	+\sum_\mu\frac{\partial\rho_\mu}{\partial r_i}\frac{\partial}{\partial\rho_\mu}f(\rho_N,\rho_\Lambda)\;,\\[0.2em]
	\frac{\partial}{\partial r_\lambda}f(\rho_N,\rho_\Lambda)
	&=\sum_\mu \frac{\partial\rho_\mu}{\partial r_\lambda}\frac{\partial}{\partial \rho_\mu}f(\rho_N,\rho_\Lambda)
	+\sum_j \frac{\partial\rho_j}{\partial r_\lambda}\frac{\partial}{\partial \rho_j}f(\rho_N,\rho_\Lambda)\;,
\end{align}
where
\begin{align}
	\frac{\partial\rho_j}{\partial r_i}=\delta_{ij}-\frac{m_N}{M}\,,\quad\;
	\frac{\partial\rho_\mu}{\partial r_i}=-\frac{m_N}{M}\,, \quad\;
	\frac{\partial\rho_\mu}{\partial r_\lambda}=\delta_{\lambda\mu}-\frac{m_\Lambda}{M}\,, \quad\;
	\frac{\partial\rho_j}{\partial r_\lambda}=-\frac{m_\Lambda}{M} \;.
\end{align}
The CM corrected first derivates take then the form:
\begin{align}
	\frac{\partial}{\partial r_i}f(\rho_N,\rho_\Lambda)
	&=\left[\frac{\partial}{\partial\rho_i}-\frac{m_N}{M}\left(\sum_j\frac{\partial}{\partial\rho_j}
	+\sum_\mu\frac{\partial}{\partial\rho_\mu}\right)\right]f(\rho_N,\rho_\Lambda) \;,\label{eq:d_CM_N} \\[0.2em]
	\frac{\partial}{\partial r_\lambda}f(\rho_N,\rho_\Lambda)&=\left[\frac{\partial}{\partial\rho_\lambda}
	-\frac{m_\Lambda}{M}\left(\sum_j\frac{\partial}{\partial\rho_j}
	+\sum_\mu\frac{\partial}{\partial\rho_\mu}\right)\right]f(\rho_N,\rho_\Lambda) \;.\label{eq:d_CM_L}
\end{align}
For the second derivatives we have:
\begin{align}
	\frac{\partial^2}{\partial r_i^2}f(\rho_N,\rho_\Lambda)
	&=\left[\frac{\partial^2}{\partial\rho_i^2}-2\frac{m_N}{M}\left(\sum_j\frac{\partial^2}{\partial\rho_i\partial\rho_j}
	 +\sum_\mu\frac{\partial^2}{\partial\rho_i\partial\rho_\mu}\right)\right.\nonumber \\[0.2em]
	&+\left.\frac{m_N^2}{M^2}\left(\sum_{jk}\frac{\partial^2}{\partial\rho_j\partial\rho_k}
	 +\sum_{\mu\nu}\frac{\partial^2}{\partial\rho_\mu\partial\rho_\nu}
 	 +2\sum_{j\mu}\frac{\partial^2}{\partial\rho_j\partial\rho_\mu}\right)\right]f(\rho_N,\rho_\Lambda) \;, \label{eq:dd_CM_N}\\[0.5em]
	\frac{\partial^2}{\partial r_\lambda^2}f(\rho_N,\rho_\Lambda) 
	&=\left[\frac{\partial^2}{\partial\rho_\lambda^2}-2\frac{m_\Lambda}{M}\left(\sum_\mu\frac{\partial^2}{\partial\rho_\lambda\partial\rho_\mu}
	 +\sum_j\frac{\partial^2}{\partial\rho_\lambda\partial\rho_j}\right)\right.\nonumber \\[0.2em]
	&\left.+\frac{m_\Lambda^2}{M^2}\left(\sum_{\mu\nu}\frac{\partial^2}{\partial\rho_\mu\partial\rho_\nu}
	 +\sum_{jk}\frac{\partial^2}{\partial\rho_j\partial\rho_k}
 	 +2\sum_{\mu j}\frac{\partial^2}{\partial\rho_\mu\partial\rho_j}\right)\right]f(\rho_N,\rho_\Lambda) \;.\label{eq:dd_CM_L}
\end{align}

Consider now the hypernuclear wave function of Eq.~(\ref{eq:Psi_T}) and assume the compact notation:
\begin{align}
	\psi_T&=\prod_{\lambda i}f_c^{\Lambda N}(r_{\lambda i})\,\psi_T^N(R_N,S_N)\,\psi_T^\Lambda(R_\Lambda,S_\Lambda)\;,\nonumber\\[0.4em]
	&=\prod_{\lambda i}f_c^{\Lambda N}(r_{\lambda i})\prod_{i<j}f_c^{NN}(r_{ij})\prod_{\lambda<\mu}f_c^{\Lambda\Lambda}(r_{\lambda\mu})
	\det\Bigl\{\varphi_\epsilon^N(\bm r_i,s_i)\Bigr\}\det\Bigl\{\varphi_\epsilon^\Lambda(\bm r_\lambda,s_\lambda)\Bigr\}\;,\nonumber\\[0.4em]
	&=J_{\Lambda N}\,J_{NN}\,J_{\Lambda\Lambda}\,\text{det}_N\,\text{det}_\Lambda\;.
\end{align}
The trial wave function is written in the single particle representation and thus it should be possible to factorize the calculation of the derivatives on each component. However, when we use the relative coordinates with respect to the CM, the antisymmetric part of the wave function $\text{det}_N\,\text{det}_\Lambda$ has to be treated as a function of both nucleon and hyperon coordinates, like the function $f(\rho_N,\rho_\Lambda)$ used above. The Jastrow correlation functions instead, being functions of the distances between two particles, are not affected by the CM corrections. It is then possible to obtain in a simple way the derivatives with respect to the nucleon and hyperon coordinates by calculating the local derivatives:
\begin{align}
	\frac{\partial_p\psi_T}{\psi_T}=\frac{\frac{\partial}{\partial R_p}\psi_T}{\psi_T}\quad\quad\text{with}\quad p=N,\Lambda\;,
\end{align}
which are of particular interest in the AFDMC code for the calculation of the drift velocity of Eq.~(\ref{eq:drift}) and the local energy of Eq.~(\ref{eq:E_L}).
The first local derivatives read
\begin{align}
	\frac{\partial_N\psi_T}{\psi_T}
	&=\frac{\partial_N J_{NN}}{J_{NN}}
	+\frac{\partial_N J_{\Lambda N}}{J_{\Lambda N}}
	+\frac{\partial_N\left(\text{det}_N\text{det}_\Lambda\right)}{\text{det}_N\text{det}_\Lambda}\;,\\[1.0em]
	\frac{\partial_\Lambda\psi_T}{\psi_T}
	&=\frac{\partial_\Lambda J_{\Lambda\Lambda}}{J_{\Lambda\Lambda}}
	+\frac{\partial_\Lambda J_{\Lambda N}}{J_{\Lambda N}}
	+\frac{\partial_\Lambda\left(\text{det}_N\text{det}_\Lambda\right)}{\text{det}_N\text{det}_\Lambda}\;,
\end{align}
while the second local derivatives take the form
\begin{align}
	\frac{\partial_N^2\psi_T}{\psi_T}&=\frac{\partial_N^2 J_{NN}}{J_{NN}}
	+\frac{\partial_N^2 J_{\Lambda N}}{J_{\Lambda N}}
	+\frac{\partial_N^2\left(\text{det}_N\text{det}_\Lambda\right)}{\text{det}_N\text{det}_\Lambda}
	+2\frac{\partial_N J_{NN}}{J_{NN}}\frac{\partial_N J_{\Lambda N}}{J_{\Lambda N}}\nonumber\\[0.5em]
	&\quad\,+2\frac{\partial_N J_{NN}}{J_{NN}}\frac{\partial_N\left(\text{det}_N\text{det}_\Lambda\right)}{\text{det}_N\text{det}_\Lambda}
	+2\frac{\partial_N J_{\Lambda N}}{J_{\Lambda N}}\frac{\partial_N\left(\text{det}_N\text{det}_\Lambda\right)}{\text{det}_N\text{det}_\Lambda}\;,\\[1.0em]
	\frac{\partial_\Lambda^2\psi_T}{\psi_T}&=\frac{\partial_\Lambda^2 J_{\Lambda\Lambda}}{J_{\Lambda\Lambda}}
	+\frac{\partial_\Lambda^2 J_{\Lambda N}}{J_{\Lambda N}}
	+\frac{\partial_\Lambda^2\left(\text{det}_N\text{det}_\Lambda\right)}{\text{det}_N\text{det}_\Lambda}	
	+2\frac{\partial_\Lambda J_{\Lambda\Lambda}}{J_{\Lambda\Lambda}}\frac{\partial_\Lambda J_{\Lambda N}}{J_{\Lambda N}}\nonumber\\[0.5em]
	&\quad\,+2\frac{\partial_\Lambda J_{\Lambda\Lambda}}{J_{\Lambda\Lambda}}\frac{\partial_\Lambda\left(\text{det}_N\text{det}_\Lambda\right)}{\text{det}_N\text{det}_\Lambda}
	+2\frac{\partial_\Lambda J_{\Lambda N}}{J_{\Lambda N}}\frac{\partial_\Lambda\left(\text{det}_N\text{det}_\Lambda\right)}{\text{det}_N\text{det}_\Lambda}\;.
\end{align}

The derivatives of Jastrow correlation functions require a standard calculations, while for the derivatives of the Slater determinant (SD) we need to include CM corrections as in Eqs.~(\ref{eq:d_CM_N}), (\ref{eq:d_CM_L}), (\ref{eq:dd_CM_N}) and (\ref{eq:dd_CM_L}). Moreover, the derivative of a SD is typically rather computationally expensive and in the above relations many terms, also with mixed derivatives, are involved. An efficiently way to deal with derivatives of a SD is described in the next section.

\newpage
\section{Derivatives of a Slater determinant}
\label{app:d_SD}

Consider a Slater determinant $|\mathcal A|$. Let us define $A_{ij}=f_i(j)$, so that $\partial_j A_{ij}=f'_i(j)$. Assume $^i B$ a matrix equal to $A$ but with the column $i$ replaced by the derivative of $f$: $^i B_{ki}=f'_k(i)$ and $^i B_{kj}=f_k(j)$ for $j\neq i$. Consider then the trivial identity 
\begin{align}
	|Q|=|Q|\sum_i Q_{ij} Q_{ji}^{-1}=\sum_i Q_{ij}(Q_{ji}^{-1}|Q|)\;,
\end{align}
and the following relation 
\begin{align}
	Q_{ji}^{-1}|Q|=(-1)^{i+j}|Q^{(ij)}| \;,
\end{align} 
where the minor $Q^{(ij)}$ is, by definition, $j$-independent. The first derivative of a SD takes the form
\begin{align}
	\partial_j|A|=|A|\sum_i A_{ji}^{-1}(\partial_j A_{ij})=|A|\sum_i A_{ji}^{-1}f'_i(j) \;,
	\label{eq:d_SD}
\end{align}
and the second derivative reads: 
\begin{align}
	\partial_j^2|A|=|A|\sum_i A_{ji}^{-1}(\partial^2_j A_{ij})=|A|\sum_i A_{ji}^{-1}f''_i(j) \;.
	\label{eq:d2_SD_i}
\end{align}

An efficient way to compute the second mixed derivative of a SD $\partial_j\partial_i|A|$ is to write the first derivative as $|^j B|=\partial_j|A|$, i.e. 
\begin{align}
	|^j B|=|A|\sum_i A_{ji}^{-1}f'_i(j) \;.
	\label{eq:i_B}
\end{align}
Using the relation (\ref{eq:d_SD}) for $|^i B|$, we can write 
\begin{align}
	\partial_j\partial_i|A|=\partial_j|^i B|=|^i B| \sum_k(^i B)_{jk}^{-1}(\partial_j{^i B_{kj}}) \;.
\end{align}
Choosing $j\neq i$ we have that $(\partial_j{^i B_{kj}})=(\partial_j A_{kj})=f'_k(j)$ and, using (\ref{eq:i_B}), it is possible to rewrite the previous equation as:
\begin{align}
	\partial_j\partial_i|A|=|A|\left(\sum_k(^i B)_{jk}^{-1}f'_k(j)\right)\left(\sum_k A_{ik}^{-1}f'_k(i)\right) \;.
\end{align}
Consider now the Sherman-Morrison formula
\begin{align}
	(A+\bm u\,\bm v^T)^{-1}= A^{-1}-\frac{( A^{-1}\bm u\,\bm v^T A^{-1})}{1+\bm v^T A^{-1}\bm u}\;,
\end{align}
with $\bm u,\bm v$ vectors. If we choose $(A+\bm u\,\bm v^T)={^i B}$, i.e.
\begin{align}
	u_k=f'_k(i)-f_k(i) \qquad 
	\left\{\begin{array}{ll}
	 	v_k=0	&	k\neq i \\
		v_k=1	&	k=i
	\end{array} \right.							
\end{align}
we can use the Sherman-Morrison relation to to compute $(^i B)^{-1}$:
\begin{align}
	(^i B)_{jk}^{-1}= A_{jk}^{-1}- A_{ik}^{-1}\frac{\displaystyle\left(\sum_k A_{jk}^{-1}f'_k(i)\right)
	-\left(\sum_k A_{jk}^{-1}f_k(i)\right)}{\displaystyle1+\left(\sum_k A_{ik}^{-1}f'_k(i)\right)
	-\left(\sum_k A_{ik}^{-1}f_k(i)\right)} \;.
\end{align}
Recalling that $f_k(i)=A_{ki}$ and assuming $j\neq i$ we have
\begin{align}
	(^i B)_{jk}^{-1}= A_{jk}^{-1}- A_{ik}^{-1}\frac{\displaystyle\left(\sum_k A_{jk}^{-1}f'_k(i)\right)
	-\cancelto{0}{\left(\sum_k A_{jk}^{-1} A_{ki}\right)}}{\displaystyle\cancel{1}+\left(\sum_k A_{ik}^{-1}f'_k(i)\right)
	-\cancel{\left(\sum_k A_{ik}^{-1} A_{ki}\right)}} \;.
\end{align}
Finally the second mixed derivative ($j\neq i$) of a SD results:
\begin{align}
		\partial_j\partial_i|A|&=|A|\!\left\{\left[\sum_k A_{ik}^{-1}f'_k(i)\right]\!\!\left[\sum_k A_{jk}^{-1}f'_k(j)\right]\!-\!\left[\sum_k A_{ik}^{-1}f'_k(j)\right]\!\!\left[\sum_k A_{jk}^{-1}f'_k(i)\right]\right\} \,.
	\label{eq:d2_SD_ij}
\end{align}

Eqs.~(\ref{eq:d_SD}), (\ref{eq:d2_SD_i}) and (\ref{eq:d2_SD_ij}) are used to calculate the derivatives with all the CM corrections of the Slater determinant $f(\rho_N,\rho_\Lambda)=\text{det}_N\text{det}_\Lambda$. The derivation of these equations is actually valid for any single particle operator $\mathcal O_j$. Eqs.~(\ref{eq:d_SD}), (\ref{eq:d2_SD_i}) and (\ref{eq:d2_SD_ij}) can be thus used to describe the linear or quadratic action of a single particle operator on a SD, that can be expressed as a local operator:
\begin{align}
	\frac{\mathcal O_j|A|}{|A|}&=\sum_i A_{ji}^{-1}(\mathcal O_j A_{ij}) \;,\\[0.5em]
	\frac{\mathcal O_j^2|A|}{|A|}&=\sum_i A_{ji}^{-1}(\mathcal O^2_j A_{ij})\;,\\[0.5em]
	\frac{\mathcal O_j\mathcal O_i|A|}{|A|}&=\left\{\left[\sum_k A_{ik}^{-1}(\mathcal O_i A_{ki})\right]\!\!\left[\sum_k A_{jk}^{-1}(\mathcal O_j A_{kj})\right]\right.\nonumber\\[0.2em]
	&\hspace{0.32cm}-\left.\left[\sum_k A_{ik}^{-1}(\mathcal O_j A_{kj})\right]\!\!\left[\sum_k A_{jk}^{-1}(\mathcal O_i A_{ki})\right]\right\} \;.
\end{align}
For example, considering the spin term of Eq.~(\ref{eq:V_NN_SD}) we have:
\begin{align}
	\frac{\sigma_{i\alpha}\,\sigma_{j\beta}|A|}{|A|}&=\left\{\left[\sum_k A_{jk}^{-1}\sigma_{j\beta} A_{kj} \right]\!\!\left[\sum_k A_{ik}^{-1}\sigma_{i\alpha} A_{ki}\right]\right.\nonumber\\[0.2em]
	&\hspace{0.32cm}-\left.\left[\sum_k A_{jk}^{-1}\sigma_{i\alpha} A_{ki} \right]\!\!\left[\sum_k A_{ik}^{-1}\sigma_{j\beta} A_{kj} \right]\right\}\;,
\end{align}
where $|A|$ could be again the SD $\text{det}_N\text{det}_\Lambda$ of the trial wave function.

\chapter{$\Lambda N$ space exchange potential}
\label{app:Px}

As proposed by Armani in his Ph.D. thesis~\cite{Armani:2011_thesis}, the inclusion of the $\mathcal P_x$ operator in the AFDMC propagator can be possibly realized by a mathematical extension of the isospin of nucleons
\begin{align}
	\left(\begin{array}{c} p \\ n \end{array}\right)\otimes\Bigl(\Lambda\Bigr)\quad\longrightarrow\quad\left(\begin{array}{c} p \\ n \\ \Lambda \end{array}\right)\;,
\end{align}
such that in the wave function hyperon and nucleon states can be mixed, referring now to indistinguishable particles. An antisymmetric wave function with respect to particle exchange must be an eigenstate of the pair exchange operator $\mathcal P_{pair}$ with eigenvalue $-1$:
\begin{align}
	-1=\mathcal P_{pair}=\mathcal P_x\,\mathcal P_\sigma\,\mathcal P_\tau \quad\Rightarrow\quad \mathcal P_x=-\mathcal P_\sigma\,\mathcal P_\tau \;,
\end{align}
where $\mathcal P_x$ exchanges the coordinates of the pair, $\mathcal P_\sigma$ the spins and $\mathcal P_\tau$ the extended isospins:
\begin{align}
	\mathcal P_\sigma(i\longleftrightarrow j)&=\frac{1}{2}\left(1+\sum_{\alpha=1}^3\sigma_{i\alpha}\,\sigma_{j\alpha}\right)\;,\\[0.5em]
	\mathcal P_\tau(i\longleftrightarrow j)&=\frac{1}{2}\left(\frac{2}{3}+\sum_{\alpha=1}^8\lambda_{i\alpha}\,\lambda_{j\alpha}\right)\;.
\end{align}
The particle indices $i$ and $j$ run over nucleons and hyperons and the $\lambda_{i\alpha}$ are the eight Gell-Mann matrices. $\mathcal P_x$ takes now a suitable form (square operators) for the implementation in the AFDMC propagator. The technical difficulty in such approach is that we need to deeply modify the structure of the code. The hypernuclear wave function has to be written as a single Slater determinant including nucleons and hyperons states, matched with the new 3-component isospinor and 2-component spinors, so a global 6-component vector. All the potential operators must be represented as $6\times 6$ matrices and the ones acting on nucleons and hyperons separately must be projected on the correct extended isospin states:
\begin{align}
	\mathcal P_N &=\frac{2+\sqrt{3}\,\lambda_8}{3}\left(\begin{array}{ccc}
		1 & 0 & 0 \\
		0 & 1 & 0 \\
		0 & 0 & 0
	\end{array}\right)\;,\\[0.5em]
	\mathcal P_\Lambda &=\frac{1-\sqrt{3}\,\lambda_8}{3}\left(\begin{array}{ccc}
		0 & 0 & 0 \\
		0 & 0 & 0 \\
		0 & 0 & 1
	\end{array}\right)\;.
\end{align}
In addition, due to the non negligible mass difference between nucleons and hyperons, also the kinetic operator must be splitted for states with different mass:
\begin{align}
	\e^{-d\tau\frac{\hbar^2}{2}\sum_i\mathcal O_{m_i}\nabla_i^2 }\quad\quad\text{with}\quad
	\mathcal O_{m_i}=\left(\begin{array}{ccc}
		1/m_N & 0 & 0 \\
		0 & 1/m_N & 0 \\
		0 & 0 & 1/m_\Lambda
	\end{array}\right)\;.
\end{align}
Finally, it is not even clear if all the operators of the two- and three-body hyperon-nucleon interaction will be still written in a suitable form for the application of the the Hubbard-Stratonovich transformation. For pure neutron systems this approach might simply reduce to an analog of the nucleonic case. The extended spin-isospin vector will have four components and all the operators will be represented as $4\times 4$ matrices coupled with the $\mathcal P_N$ and $\mathcal P_\Lambda$ on the reduced space. The $\mathcal O_{m_i}$ operator will have just two diagonal elements with the mass of the neutron and the hyperon. Although this purely mathematical approach could be applied, many questions arise from the physical point of view. By considering an extended isospin vector, states with different strangeness (0 for nucleons and $-1$ for the $\Lambda$~particle) will mix during the imaginary time evolution. This violates the conservation of strangeness that should be instead verified by the strong interaction. The picture becomes even less clear if we consider the $\Lambda\Lambda$ interaction of Eq.~(\ref{eq:V_LL}), because strangeness will be distributed among all the particles but the potential is explicitly developed for hyperon-hyperon pairs. Thus, for the phenomenological interactions introduced in Chapter~\ref{chap:hamiltonians}, this mathematical approach is not feasible and it has not been investigated in this work.

	\backmatter
	% Bibliography

\cleardoublepage

\fancyhead[LO]{\emph{\nouppercase{\rightmark}}}	% back to main page style preferences for left and right header fields
\fancyhead[RE]{\emph{\nouppercase{\leftmark}}}

\phantomsection
\addcontentsline{toc}{chapter}{Bibliography}
\bibliographystyle{myapsrev4-1}
\bibliography{Contents/Bibliography}
\newpage
\phantom{Empty page}

	% Acknowledgements

\chapter{Acknowledgements}

\fancyhead[LO]{\emph{Acknowledgements}}					% redefinition of left and right header fields for this chapter
\fancyhead[RE]{\emph{Acknowledgements}} 

\selectlanguage{italian}

Se scrivere una tesi di fisica in inglese è già di per sé un arduo compito, tentare di riportare i ringraziamenti in lingua anglofona è un'impresa impossibile, almeno per un veronese quadratico medio come me. Seguirà dunque uno sproloquio in quella che più si avvicina alla mia lingua madre, nel quale mi auguro di non dimenticare nessuno, anche se so che sarà inevitabile, abbiate pazienza.

Inizio col ringraziare la mia famiglia, mamma Graziella e papà Franco in primis. Nonostante abbia fatto di tutto per rendermi odioso e insopportabile, soprattutto in periodi di scadenze e consegne, mi hanno sempre sostenuto e incoraggiato, spingendomi ad andare avanti. Sempre pronti ad ascoltarmi, continuamente mi chiedevano ``Come va a Trento? I tuoi studi?''. E nonostante poi avessero le idee ancora più confuse di prima al sentire le mie fumose spiegazioni su iperoni e stelle di neutroni, ogni volta tornavano a informarsi sul mio lavoro per avere anche solo una vaga idea di quello che facevo per portare a casa quei quattro euro della borsa di dottorato. Ringrazio Simone e Sara: sarà la lontananza, sarà che superata la soglia degli ``enta'' uno inizia anche a maturare (ahahah), sarà lo snowboard o il downhill ma negli ultimi anni ci siamo ri-avvicinati parecchio, finendo addirittura in vacanza a Dublino assieme! Nonostante non sia più un ragazzetto sbarbatello (no speta, quello lo sono ancora), mi è stato utile avere il supporto, i consigli e la complicità del fratello maggiore che, anche se non lo ammetterà mai pubblicamente, so che mi vuole bene. E quindi, grazie! Meritano altrettanti ringraziamenti i nonni, gli zii e cugini di Caprino e dintorni, e quelli più geograficamente ``lontani'' di Verona, che non hanno mai smesso di credere in me. E perché per i libri, le trasferte, i soggiorni all'estero c'è MasterCard, ma sapere che nel paesello vengo pubblicizzato con espressioni del tipo ``Varda che me neòdo l'è 'n sciensiato!'' non ha prezzo!

Accademicamente parlando non posso non esser grato a Francesco, che mi ha seguito in questi tre anni di dottorato (e ancor prima durante la laurea magistrale), con spiegazioni, discussioni, consigli tecnici o anche solo chiacchierate, soprattutto in questi ultimi mesi parecchio impegnativi sotto tutti i punti di vista. Nonostante i suoi mille impegni e viaggi, è sempre stato un punto di riferimento. Aggiungiamo lo Stefano, senza l'aiuto del quale probabilmente avrei dovuto trovare lavoro come operatore ecologico in quel di Verona. A parte le mille questioni di fisica o le discussioni su quel cavolo di codice, lo ringrazio per la vagonata di consigli in generale, per l'ospitalità, le battute del piffero, le (forse troppe) birrette e le partite a biliardo super professionali... Assieme a lui è d'obbligo ringraziare la mitica Serena, che diciamolo, è la persona che porta i pantaloni in quella famiglia e detto questo ho già detto tutto! Un grazie anche alle due belvette, che mi hanno fatto un sacco ridere finché ero ospite in casa (e che casa!) Gandolfi. Tornando un po' indietro nel tempo devo sicuramente ringraziare il buon Paolo ``Ormoni'', che mi iniziò all'AFDMC e mise le basi per quello che sarebbe stato poi il mio progetto sui sistemi ``strani''. Senza di lui credo non avrei mai potuto affrontare quel codice e la Bash in generale. Per chiudere la parte accademica ringrazio poi tutti i LISCers, che hanno contribuito a creare un ambiente di lavoro intellettualmente stimolante, e tutti coloro con i quali ho avuto modo di parlare di fisica, Kevin, Steve, Bob, Ben, Abhi, i due Alessandro e i colleghi di ufficio, i quali però meritano un paragrafo a parte. Ah sì, non posso certo dimenticare l'infinita pazienza di Micaela che con la fisica centra poco, ma in merito a burocrazia e organizzazione è insuperabile.

Veniamo dunque al reparto amicizie: qui potrei dilungarmi fin troppo ma ho scelto di limitarmi un po', dividendo il campione in due sottoinsiemi geografici, quello trentino (in senso lato) e quello più storico veronese, seguendo un percorso un po' random (deformazione professionale).

Iniziamo con la completa e incontrollabile degenerazione del mio ufficio, dall'insostituibile (e dico sul serio $\heartsuit$) Roberto allo shallissimo Alessandro, dallo ``svizzerooooo'' Elia al ``miserabile'' Paolo (con nostro grande divertimento in perenne lotta per il titolo di maschio omega). E l'ormai santa donna Giorgia che ha recentemente installato una serie di filtri per escludere le nostre impertinenti voci. Non dimentichiamo coloro che in principio colonizzarano l'open space al LISC: il canterino Emmanuel, il già citato Paolo ``Ormoni'' e il mitico Enrico (che quando leggerà queste righe inizierà a riprodurre senza sosta una delle parodie dei prodotti Apple). Aggiungiamo i colleghi di FBK naturalizzati LISC, quali il Mostarda, l'Amadori, il Fossati con la fortissima Saini al seguito (ho volutamente messo i cognomi per subrazzarvi un po'), i personaggi di ``passaggio'' come Marco e gli adottati da altri atenei come l'Alessandro (Lovato). Quest'ultimo (eccellente) fisico merita un ringraziamento particolare (oltre ad una già preventivata cena in quel di Chicago) per l'estrema ospitalità e il supporto che mi ha dato (e che spero continuerà a darmi) oltreoceano, non solo per questioni di fisica. In realtà ognuna delle persone qui citate meriterebbe un grazie su misura, ma non è facile (e probabilmente nemmeno opportuno) riportare tutto su queste pagine. Chi mi è stato particolarmente vicino sa già che gli sono grato per tutto, non servono molte parole...

Uscendo dall'ufficio la cosa si complica perché il numero di persone da ringraziare cresce di molto. E quindi un caloroso grazie a Giuseppe, Paolo e Chiara, Alessia, Nicolò, Irena e Nicolò, Sergio, Mattia, Roberta, Nicola, Cinzia, Giada, Marco, Giovanni, Sebastiano, Fernando, Eleonora, Letizia, Nikolina, David, Eleonora, Federica, Beatrice, Marta, Fata e a questo punto sono costretto a mettere un politico \emph{et~al.}, non abbiatene a male. Con alcune di queste persone ho convissuto, con altre si usciva a fare festa, altre ancora erano e sono ``semplicemente'' amici, ma tutti hanno contribuito in qualche modo a farmi trascorrere momenti fantastici in questi tre anni. Essendo l'autore di questo lavoro mi riservo il diritto di ringraziare in separata sede Marianna e Gemma: nonostante ci sarebbero molte cose da dire in merito, mi limiterò ad un semplice ma profondo ``grazie!''. Per lo stesso motivo della precedente proposizione, estendo temporalmente e geograficamente un ringraziamento anche a Francesco a al Bazza, che col mio dottorato non centrano un tubo ma che sono stati elementi portanti della mia lunga avventura trentina e la coda (in termini probabilistici) della loro influenza si fa tuttora sentire.

Nelle lande veronesi è d'obbligo citare tutti gli amici storici e meno storici, che nell'ultimo periodo ho avuto modo di vedere più spesso perché, sarà la moda del momento o qualche virus contagioso, ma qui si stanno sposando tutti! E dunque grazie ad Andrea, Marco e Jessica, Alice e Francesco, Davide ed Elisa, Roberta e Alberto, Matteo, Erika, Letizia, Daniela, Mirko, Silvia e tutti gli altri con cui ho bevuto $n$ birrette (con $n$ spesso troppo grande) in quel di Caprino e dintorni. Sono particolarmente grato all'IIIIIIIIIING. Giacomo e alla gnocca Giulia: nell'ultimo periodo non c'è più stato modo di vedersi spesso ma le serate passate in vostra compagnia mi accompagneranno sempre col sorriso. Infine, non certo per ordine di importanza, devo ringraziare di cuore Alessandra (e con lei tutta la famiglia), che per molti anni è stata al mio fianco sostenendomi, sopportandomi, incoraggiandomi, facendomi arrabbiare e divertire allo stesso tempo, ma che il destino (o chi/cosa per esso) ha voluto le nostre strade prendessero due direzioni diverse, ma nulla o nessuno potrà mai cancellare tutto ciò che di bello e buono c'è stato. Per cui grazie!

Eccoci dunque alla fine del mio sproloquio. Non mi resta che ringraziare tutte quelle cose che, pur essendo inanimate, mi hanno fatto penare ma al tempo stesso esaltare non poco, fra cui meritano un posto di eccellenza Gnuplot, \LaTeX\, e gli script Bash. Chiudo (stavolta sul serio) ringraziando questo pazzo 2013 che mi ha portato immense soddisfazioni e altrettante sofferenze, ma che con il suo carico di grandi (a volte fin troppo) novità mi ha stupito e mi ha spinto a reagire con coraggio facendomi sentire veramente vivo...

\vspace{1cm}
\begin{quotation}
	\emph{Meglio aggiungere vita ai giorni che non giorni alla vita.}
	\flushright{Rita Levi Montalcini}
\end{quotation}

\newpage
\phantom{Empty page}

\end{document}